\documentclass[aps,prd,showpacs,onecolumn,nofootinbib,superscriptaddress]{revtex4-2}

\usepackage{epsfig}
\usepackage{graphicx}

\usepackage{subcaption}

\usepackage{bm,dsfont,braket,mathrsfs}
\usepackage{color}
\newcommand{\zz}[1]{ { #1}}

\newcommand{\zzz}[1]{ {{ #1}}}

\newcommand{\zzzz}[1]{ {{ #1}}}

\newcommand{\zc}[1]{ { #1}}

\newcommand{\zu}[1]{{#1}}

\newcommand{\zd}[1]{{#1}}

\newcommand{\zf}[1]{{#1}}

\newcommand{\corAOP}[1]{{#1}}
\newcommand{\corr}[1]{{#1}}
\newcommand{\corrgpt}[1]{{#1}}
\newcommand{\corrclaude}[1]{{#1}}
\newcommand{\coromega}[1]{{\bfseries #1}}
\newcommand{\corrbel}[1]{{#1}}
\newcommand{\cormag}[1]{{#1}}
\newcommand{\corrteal}[1]{{#1}}
\DeclareRobustCommand{\corGPTl}[1]{{#1}}
\DeclareRobustCommand{\GPT}[1]{{#1}}
\newcommand{\corrcltwo}[1]{{#1}}
\DeclareRobustCommand{\corCLAUDEr}[1]{{#1}}
\definecolor{ClaudeBrown}{rgb}{0.55,0.27,0.07}
\DeclareRobustCommand{\corClaude}[2]{{#1}}

\newcommand{\cl}[1]{{#1}}

\newcommand{\jer}[1]{ {{#1}}}

\usepackage{definitions}%
\usepackage[title,titletoc]{appendix}

\usepackage{array} % for better arrays (eg matrices) in maths
\usepackage{braket}

\usepackage{tikz}
\usetikzlibrary{fadings}
\usetikzlibrary{patterns}
\usetikzlibrary{shadows.blur}
\usetikzlibrary{shapes}

\usepackage{amsmath,amssymb,amsfonts}
\usepackage{bm}
\usepackage{verbatim}
\usepackage{hyperref}
\usepackage{slashed,braket}
\usepackage{ulem}
\usepackage{cancel}
\usepackage{color}
\usepackage{graphicx,graphics}
\usepackage[title,titletoc]{appendix}
\usepackage{stackengine}
\usepackage{mathrsfs}
\usepackage{tikz-feynman}
\usepackage{amsthm}

\usepackage{tikz}
\usepackage{pgfplots}
\usepackage{lineno}
\usepackage[T1]{fontenc} % if needed
\usepackage{braket}

\usepackage{bbold}
\usepackage{comment}
\excludecomment{confidential}
\usepackage{makecell}
\usepackage{bm}
\usepackage{braket}
\usepackage{slashed}
\usepackage{mathtools}

\usepackage{amsthm}

\usepackage{setspace}

\newtheorem{theorem}{Theorem}

\newcommand{\rv}[1]{{#1}}

\def\rA{{\rm A}}
\def\rR{{\rm R}}

\def\rM{{\rm M}}

\begin{document}

\title{Topological invariant responsible for the integer QHE and non-commutative geometry}

\author{G. Kovyrshin}
\email{glebkovyrshin@gmail.com}
\affiliation{Physics Department, Ariel University, Ariel 40700, Israel}

\author{A. Mekrami}
\email{ abderrahim.mekrami@gmail.com}
\affiliation{Physics Department, Ariel University, Ariel 40700, Israel}

\author{J. Miller}
\email{ jeremysharonmiller@gmail.com}
\affiliation{Shamoon College of Engineering, Beer Sheva, Israel}

\author{M.A. Zubkov}
\email{mikhailzu@ariel.ac.il}
\affiliation{Physics Department, Ariel University, Ariel 40700, Israel}

\author{A. Zuevsky}
\email{zuevsky@math.cas.cz}
\affiliation{Institute of Mathematics,
{Academy of Sciences of the Czech Republic},
Zitna, 25, Prague, Czech republic }

\date{\today}

\begin{abstract}
We consider  a wide class of  $2D$ tight - binding models of solid state physics.
These models are, in the most general case, non - homogeneous. Lattice Dirac operator $\hat{Q}$ (i.e. the inverse two point Green function) of the considered systems contains a rather general dependence on Matsubara frequency that is not reduced to the Hamiltonian case $\hat{Q} = i \omega - \hat{H}$ with a Hamiltonian $\hat{H}$. 
The topological invariant ${\cal N}_3$ responsible for the quantization of the Hall conductivity,
for the specific case of the integer quantum Hall effect in $2D$,
is expressed through the Wigner transformation of the two-point electron Matsubara Green function.
We express this invariant as a pairing of the element of the $K^{-1}$ group (generated by  the Green function)  with the specific element of the cyclic cohomology group $HC^3$.
According to a set of local index theorems  the values of ${\cal N}_3$
 can be shown to be integer for a wide but limited class of  the lattice systems. In the Hamiltonian case our construction is reduced to that of Bellissard and co - authors. 
\GPT{We concentrate on systems with a spectral gap, with the chemical potential inside the gap and an invertible Dirac operator $\hat Q$ throughout the Matsubara-frequency domain. The mobility-gap case is not considered here and will be addressed in forthcoming publications.}
\end{abstract}
%\pacs{}

\maketitle
\tableofcontents
%\section{Introduction}
%\label{SectIntro}

\section{Introduction}

\zzzz{The integer quantum Hall effect (IQHE) was discovered by von Klitzing and
collaborators in 1980. It is characterized by the quantization of the
Hall conductance in integer multiples of an inverse Klitzing constant. According to the experimental results it is not affected by sample geometry or
disorder~\cite{vonKlitzing1980}. Laughlin attempted to explain this based on gauge-invariance arguments ~\cite{Laughlin1981}. The  topological origin of the quantization was
considered for the first time by Thouless, Kohmoto, Nightingale, and den Nijs (TKNN), who showed
that the Hall conductance is proportional to an integer-valued topological
invariant---the Chern number---of the filled Bloch bands~\cite{Thouless1982}. This
invariant is given by integration of the Berry curvature over the magnetic
Brillouin zone, and it remains robust to smooth deformations of the  Hamiltonian ~\cite{Kohmoto1985}. This  \jer{consideration} was based on the geometry
of fiber bundles and the Berry phase ~\cite{Berry1984,Simon1983}. Niu, Thouless, and Wu attempted to generalize the
topological characterization to interacting and disordered systems by expressing
the invariant in terms of twisted boundary conditions~\cite{Niu1985}. Unfortunately, the expression presented in \cite{Niu1985} is so complicated, that it cannot be used in practical calculations.
Avron, Seiler, and Simon considered carefully homotopy and quantization of the TKNN construction  ~\cite{Avron1983}. The role of disorder was discussed in ~\cite{Halperin1982,Streda1982}.
Hatsugai established the bulk-boundary correspondence for the systems with boundary, and explained how boundary consideration of the QHE is related to its bulk description based on the TKNN invariant ~\cite{Hatsugai1993}. \jer{A} deep link between the TKNN invariant and  noncommutative geometry has been 
\jer{shown} by Bellissard and
collaborators~\cite{Bellissard1994}.
\jer{A standard review  of all these results can be found for example, in}  ~\cite{PrangeGirvin1990,Xiao2010},
while links to the theory of
topological insulators and superconductors  have been reviewed by Hasan
and Kane and by Qi and Zhang
~\cite{Hasan2010,Qi2011}.
}

\zzzz{It is worth mentioning that the same integer-valued Chern number
reappears in the absence of an external magnetic field, as the
quantum anomalous Hall effect in  Chern
insulators~\cite{Chang2023,ChiMoodera2022,Weber2024}. The corresponding  quantized plateaus have been measured directly in  particular systems ~\cite{Bai2024,Bosnar2023}.
 \jer{The} Chern invariant in real space instead of momentum space has been discussed in ~\cite{Bau2024}. Interestingly, this topological consideration can be extended also to the fractional quantum anomalous Hall effect
~\cite{Park2023,Cai2023,Lu2024,Ju2024}.}

As      mentioned above,  the first topological invariant proposed for {a} description of {the} quantum Hall effect in the absence of interactions between electrons \zzzz{and in the absence of inhomogeneity} is known as the  TKNN invariant \cite{Thouless1982}. It {only takes integer values}  \cite{Kaufmann:2015lga}.
The corresponding topology is discussed, for example, in  \cite{Avron1983,Fradkin1991,Tong:2016kpv,Hatsugai1997,Qi2008}.
{However, in real systems interactions between electrons \zzzz{and inhomogeneity} do indeed exist,
and
the above mentioned TKNN invariant is not defined.}
{An explicit expression for the TKNN invariant \zzzz{in terms of the Green functions was given} in \cite{IshikawaMatsuyama1986}.
It was later pointed out in \cite{Volovik1988,Volovik2003a} that \zzzz{this} formula  is valid
when there is no magnetic field.
This expression for the TKNN invariant is}
\begin{equation}
{\cal N}
=  -\frac{ \epsilon_{ijk}}{  \,3!\,4\pi^2}\, \int d^3p \Tr
\Bigl[
{G}(p ) \frac{\partial {G}^{-1}(p )}{\partial p_i}  \frac{\partial  {G}(p )}{\partial p_j}  \frac{\partial  {G}^{-1}(p )}{\partial p_k}
\Bigr].
\label{cal0}
\end{equation}
Here $G$ is {the} one particle Green function
\cite{IshikawaMatsuyama1986,Volovik1988,Volovik2003a,parity_anomaly,parity_anomaly_}.
{It has been shown in \cite{Zubkov2018a,ZZ2019} that to obtain the analogue of  Eq. (\ref{cal0}) that
includes interactions, the Green function should be replaced by the Green function that includes interactions.}
\corAOP{At the same time, the limitations of the Green-function representations of topological invariants in interacting systems have been actively discussed recently. In particular, it has been demonstrated in \cite{Zhao2023} (see also \cite{Gurarie2011,EssinGurarie2011}) that in strongly correlated matter the invariants composed of the interacting one-particle Green function may fail: the Green function may develop zeros, on which such invariants lose their topological protection and cease to reproduce the Hall conductivity. In the present paper we consider systems without interelectron interactions, although with arbitrarily strong inhomogeneity. For such systems, in the presence of a spectral gap, the Matsubara Green function is regular and has no zeros on the imaginary frequency axis, so that the pathologies discussed in \cite{Zhao2023} do not arise. The extension of the present construction to the interacting case, where the zeros of the Green function become possible, remains an important open problem\corrteal{; it is addressed here only at the formal level --- through the general class of Dirac operators introduced in Sect. \ref{SectKeldysh} and the equilibrium static-gauge derivation of Sect. \ref{SectPeriodized} --- and through the exactly solvable example of Appendix \ref{AppHK}, which illustrates both the interaction-induced frequency dependence of $Q$ and the mechanism of the failure at the Green-function zeros}.}
Topological invariants reveal an intimate relationship between solid state systems and high energy physics  \cite{zubkov2018momentum,zubkov2012momentum,volovik2017standard,zubkov2012momentum,zubkov2017topology}\corrclaude{; in particular, the relation between the momentum space topology, the Wigner transformations, and the chiral anomaly in lattice models was considered in \cite{ZubkovKhaidukov2017}}. Even more,
there exists a relationship between relativistic quantum field theory and fermionic systems in condensed matter physics,
including superfluids \cite{volovik2013nambu,katsnelson2013euler,volovik2015scalar}, and more.
It is worth mentioning that when  interactions are included, nonperturbative effects  {become significant}.
In particular, QHE systems undergo a transition to fractional Hall states \cite{selch2025non}, while on the side of high energy physics, various topological defects dominate the dynamics  \cite{bakker1999central,bakker2005standard}.

{When the system is inhomogeneous, due to disorder or an external magnetic field,
the corresponding topological invariant is then given by an extension of Eq.(\ref{cal0}) \cite{ZW2019,ZZ2022}.}
Specifically, the topological invariant responsible for the QHE is given by Eq. (\ref{cal0}) but with
an added averaging over the sample area, and the two-point Green function replaced by its Wigner transformation $G_W$.
{In \cite{ZW2019,ZZ2022} for the lattice tight - binding models the Wigner transformation  is defined {\it as} in the continuum theory approach, which is called {\it  approximate} Wigner-Weyl calculus
\cite{FZ2019,chernodub2017scale,zhang2020influence,suleymanov2019wigner,zhang2019hall}
}
% Correspondingly, the products are understood as Moyal products and by $G^{-1}_W$ we understand the inverse $Q_W$ with respect to Moyal product \cite{ZW2019}:
Accordingly, products between operators become Moyal products, and $G^{-1}_W$ denotes the inverse of $Q_W$ with respect to the Moyal product \cite{ZW2019}.
In all, the resulting expression for the topological invariant is:
\begin{eqnarray}
{\cal N}
&=&  -\frac{ \epsilon_{ijk}}{  \,3!\,4\pi^2 |{\bf V}|}\,\int d^2x\, \int d^3p \Tr
\Bigl[
{G}_W(x,p)\star \frac{\partial {Q}_W(x,p)}{\partial p_i} \nonumber\\&&\star \frac{\partial  {G}_W(x,p)}{\partial p_j} \star \frac{\partial  {Q}_W(x,p)}{\partial p_k}
\Bigr]\ ,
\label{cal01}
\end{eqnarray}
where $\star$ is Moyal product and $|{\bf V}|$ is the sample area.
{This expression is valid for  a lattice system when the sum over lattice points can be replaced by an integral,
i.e. when inhomogeneity is negligible  at the distance of the order of lattice spacing.}
A similar expression has been used  also to describe
other nondissipative transport effects \cite{zubkov2023effect,abramchuk2018anatomy}.

In the more complex case where inhomogeneities are sufficiently large at such distances,
we should use the precise Wigner-Weyl calculus \cite{FZ2019_2}.
In the particular case of a $2+1$ dimensional gapped system,
the Hall conductivity is given by $\sigma_H = \dfrac{{\cal N}_3}{2\pi}$ with the topological invariant
\be
{\mathcal N}_3
\equiv \frac{(2\pi)^3\epsilon^{lmk}  }{24 \pi^2 {\beta |{\bf V}|}}
%\int d^{D+1}p d^{D+1}x
\Tr \(
G_W \star \partial_{p_l} Q_W \star
G_W \star \partial_{p_m} Q_W \star
G_W \star \partial_{p_k} Q_W
\).
\label{N3}
\ee
Here $Q_W$ and $G_W$ are the Weyl symbols defined on the rectangular infinite lattice according to \cite{FZ2019_2}.

\medskip 
{
%%%%
One can see that this expression contains the (formally)
infinite system area $|{\bf V}|$. This necessitates an appropriate
finite-volume regularization, as proposed in \cite{Z2023}.
In the present paper, we build upon the construction of \cite{Z2023} 
and reformulate it in the language of noncommutative geometry \cite{Connes}. 
This framework enables us to apply standard results of cyclic cohomology 
theory to condensed matter systems exhibiting the integer quantum Hall effect.

\cormag{Noncommutative geometry has already been applied to the quantum Hall effect, notably by Bellissard and co-authors \cite{Bellissard1994,Connes}, whose crossed-product construction includes covariant disordered and aperiodic systems. The present paper uses a different, deterministic Wigner--Weyl description and allows explicitly prescribed impurities and spatially varying magnetic fields.}

\cormag{A further distinction concerns the frequency dependence of the Dirac operator $\hat Q$, whose inverse is the Green function. Bellissard's conductivity pairing starts from a Hamiltonian, corresponding here to $Q=i\omega-H$; Sect. \ref{SectIndexRed} shows that, with the \corrcltwo{orientation conventions} stated there, the frequency integral then reduces to the Fermi-projector pairing. We also define a formal odd invariant for admissible frequency-dependent operators built from the temporal factors \corrcltwo{(\ref{F0def}) of the class (\ref{Q_general})}. This includes compact-frequency Euclidean regularizations and may represent interacting inverse Green functions. The equality of this generalized invariant to the physical Hall conductivity is known in the restricted perturbative regimes cited below, but is not proved here for \corrcltwo{arbitrarily} strong or inhomogeneous backgrounds\corrteal{; see, however, Sect. \ref{SectPeriodized}, where the equilibrium static-gauge derivation is given for the periodized systems of the general class \corGPTl{under the additional equilibrium, coupling, and limiting assumptions stated there}}. Moreover, an admissible general $Q$ need not possess a canonical Hamiltonian Fermi projector, although the abstract Bott suspension remains available when the required frequency asymptotics hold.}

\cormag{Throughout the paper ``disorder'' and ``inhomogeneity'' refer first to a prescribed deterministic configuration. The formal symbol invariant and, under locality and a gap, the Fredholm compression may be defined configuration-wise. Their identification with a trace-per-volume Chern pairing, however, additionally requires a cyclic bulk trace, supplied for example by periodization or by a covariant ergodic setting. The mobility-gap regime requires further localization hypotheses and lies beyond the main scope of this work.}

\zzzz{Noncommutative geometry (NCG) appeared due to the duality, made precise by
	the Gelfand--Naimark theorem, between commutative C*-algebras and topological
	spaces. The noncommutative algebra may then be viewed as the algebra
	of functions on a ``quantum space''~\cite{GelfandNaimark1943}.Connes turned this analogy into a full geometric
	framework \cite{Connes1985,Connes}. The analogues of Atiyah - Singer index theorem and the notions of  characteristic classes were considered in ~\cite{Connes1995reality,ConnesMoscovici1995}. The corresponding machinery has been applied to  the Standard Model of particle physics  ~\cite{ConnesLott1991,ChamseddineConnes1997,ConnesGravityMatter1996,ChamseddineConnesMarcolli2007}.
	The proposal
	that spacetime itself acquires noncommuting coordinates at the Planck scale was discussed, for example, in ~\cite{Woronowicz1987,DoplicherFredenhagenRoberts1995,Madore1992}.
	Certain applications of noncommutative geometry exist  in string and matrix theory ~\cite{SeibergWitten1999,ConnesDouglasSchwarz1998,GraciaBondiaVarillyFigueroa2001}.}
	
	\zzzz{As mentioned above,  Bellissard and collaborators showed that the TKNN invariant responsible for the integer QHE Hall can be represented in terms of noncommutative geometry ~\cite{Bellissard1986,Bellissard1994}. This relation has been mentioned also in  ~\cite{AvronSeilerSimon1994,SchulzBaldesKellendonkRichter2000,ProdanSchulzBaldes2016,DeNittisSandoval2020}. Relation of  \jer{non-commutative} geometry to
	random-matrix, bootstrap, and Coulomb-gas
	techniques has been discussed in ~\cite{KhalkhaliPagliaroli2021,HessamKhalkhaliPagliaroli2022,KhalkhaliPagliaroli2022,HessamEtal2022review}.
	Possible links to Riemann hypothesis were discussed in  ~\cite{ConnesConsani2023,ConnesMoscovici2022,ConnesConsaniMoscovici2024}.
	Recent advances of the theory were discussed, for example, in  ~\cite{ChamseddineConnesvanSuijlekom2023,vanSuijlekom2024,Stoiber2025}.}

} % \corGPTl{Close the unmatched group opened in Introduction, original line 376.}

\section{Wigner Weyl calculus for the finite lattice}

\subsection{Definition of Weyl symbol and its properties}
\label{SectList}
For simplicity, {here}  we consider  {a} $D$-dimensional rectangular lattice, ${\cal O}$.
{Similar calculations apply to other lattices (for example the honeycomb lattice in \cite{CZ2024}).}
{Specifically, a $D$-dimensional rectangular lattice, denoted  ${\cal O}$,
is defined as the following
set of lattice sites:
}
$$
{\cal O} = \left\{(m_1\ ,\ldots\ ,m_D ) \ |\  m_i \in \{0,1,2\ ,\ldots\ ,N-1\}\right\}
$$
The lattice ${\cal O}$ contains $N^D$ sites.

{Every} function defined {on the lattice sites}, $\cal O$
is subject to periodic boundary conditions.

{Let the momentum space conjugate of $\cal O$
be defined as the following discrete set of  $N^D$ points:}
$$
{\cal M} = \left\{\left(\frac{2\pi m_1 }{N}\ ,\ldots\ , \frac{2\pi m_D }{N}\right) \ \lvert\  m_i \in \{0,1,2\ ,\ldots\ ,N-1\}\right\}
$$
\par
% {Just like in all tight-binding models, let a basis of states to each of these discrete coordinates.}
{Let the} one-particle Hilbert space, ${\cal H}$ be defined
to be
spanned by the set of ket vectors:
% $$
% | q \rangle, q \in {\cal O}
% $$
{$$ \left\{| q \rangle\ \big|  q \in {\cal O}\right\}\ .$$}
{The same one-particle Hilbert space $\cal H$ is spanned by the set of momentum eigenstates:}
% $$
% | p \rangle, p \in {\cal M}
% $$
{$$ \left\{| p \rangle\ \big|  p \in {\cal M}\right\}\ .$$}
The above mentioned periodic boundary conditions  {are}
$| q +N \rangle = | q \rangle$  {for ket vectors defined over $\cal{O}$} and $| p + 2\pi \rangle  = | p \rangle$
{for ket vectors defined over $\cal{M}$}.

\begin{figure}[h]
\centering  %
\includegraphics[width=16cm,height=16cm]{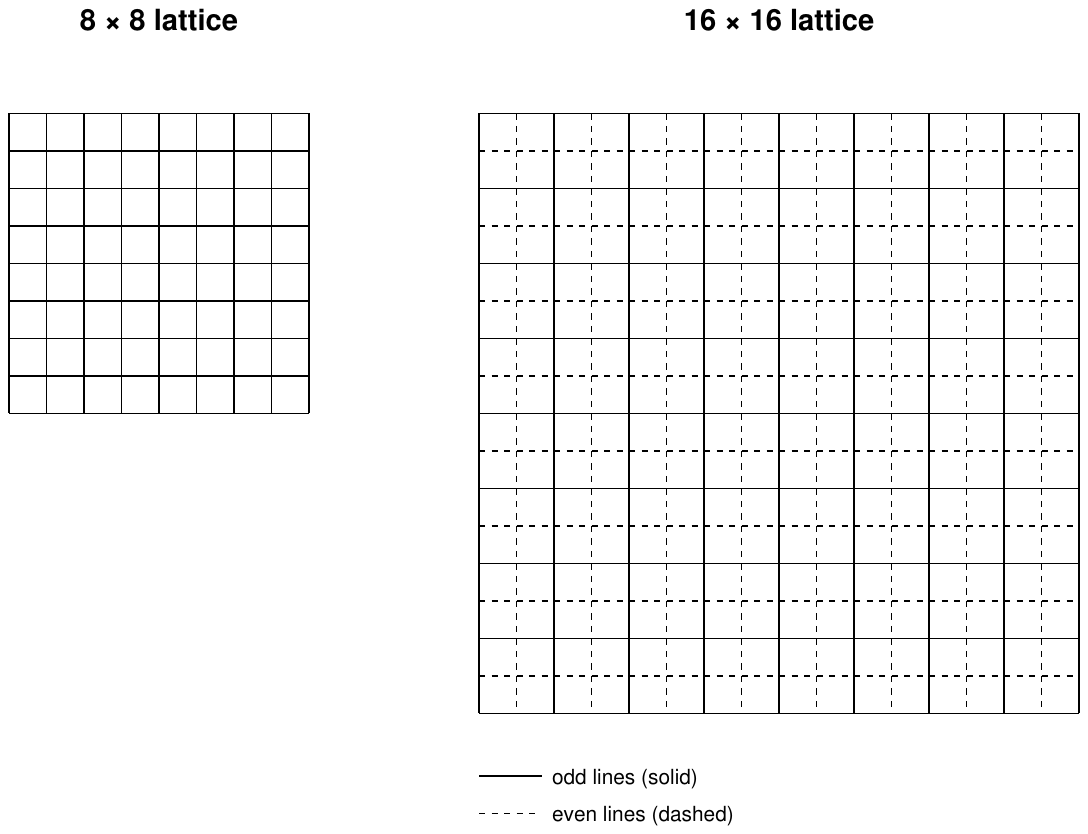} \vspace{-6cm} %
\caption{In this figure we represent schematically  ${\cal O}$ and ${\cal O}'$. In the case of ${\cal O}'$ we indentify the sublattice with solid lines with ${\cal O}$.}
\label{fig}   %
\end{figure}

Further,
we define the more dense lattice:
$$
{\cal O}^\prime = \left\{\ \left(m_1  ,\ldots,m_D \right) \ | \   m_i \in \{0,\tfrac1{2},1,\ldots,N-\tfrac1{2}\}\ \right\}
$$
where the number of {lattice} points {in $\cal O'$} is $2^D$ times larger {than those in $\cal O$}.
We also define the refined momentum space:
% $$
% {\cal M}^\prime = \left\{(2\pi m_1/N,...,2 \pi m_D/N) | m_i \in \{0,\tfrac1{2},1,\ldots,N-\tfrac1{2}\}\right\}\ ,
% $$
{
$$
{\cal M}^\prime = \left\{\left(\frac{2\pi m_1}{N},\ldots,\frac{2\pi m_D}{N}\right)\  |\  m_i \in \{0,\tfrac1{2},1,\ldots,N-\tfrac1{2}\}\right\}\ ,
$$}
\noindent
\zz{the extended lattice:
$$
{\cal O}^{2} = \{\ (m_1  ,...,m_D ) | m_i \in \{0,1,2,...,2N-1\}\ \}\ ,
$$
the extended momentum space:
% $$
% {\cal M}^{2} = \{\left(\frac{2\pi m_1}N,\ldots,\frac{2 \pi m_D}N\right) | m_i \in \{0,1,2,...,2N-1\}\}\ ,
% $$
}
{$$
{\cal M}^{2} = \left\{\ \left(\frac{2\pi m_1}{N},\ldots,\frac{2 \pi m_D}{N}\right) \ |\  m_i \in \{0,1,2,...,2N-1\}\ \right\}\ ,
$$}
\noindent
the extended refined lattice:
% $$
% {\cal O}^{2\prime} = \{(m_1  ,...,m_D ) | m_i \in \{0,\tfrac1{2},1,\ldots,2N-\tfrac1{2}\}\}\ ,
% $$
{$$
{\cal O}^{2\prime} = \{\ (m_1  ,...,m_D )\  |\  m_i \in \{0,\tfrac{1}{2},1,\ldots,2N-\tfrac{1}{2}\}\ \}\ ,
$$}
and the corresponding momentum space:
% $$
% {\cal M}^{2\prime} = \{\left(\frac{2\pi m_1}N,\ldots,\frac{2 \pi m_D}N\right) | m_i \in \{0,\tfrac1{2},1,\ldots,2N-\tfrac1{2}\}\}\ .
% $$
{
$$
{\cal M}^{2\prime} = \left\{\ \left(\frac{2\pi m_1}{N},\ldots,\frac{2 \pi m_D}{N}\right) \ |\  m_i \in \{0,\tfrac{1}{2},1,\ldots,2N-\tfrac{1}{2}\}\ \right\}\ .
$$
}
In Figs.~\ref{fig} {examples of} the lattices ${\cal O}, {\cal O}^\prime$ are shown.

{Similar structures for lattices in  \jer{momentum} space $\cal M$ can be found  by a straightforward re-scaling of the lattice
separation by a factor of $2\pi/N$, where $N$ is the lattice size.}

The
{\bf Weyl symbol} of an operator $\hat{A}$ is defined for all $q \in {\cal O}^{\prime}, p \in {\cal M}^\prime$ as:
\begin{eqnarray}
A_{{ W}}(p,q)  & = &
\sum_{\substack{n_i \in {0,1} \\   u \in {\cal M}^{\prime} }}  \,
\langle p + \pi n/N - u|\hat{A}|p + \pi n/N + u \rangle \,  e^{-2 i u q}\nonumber\\
&&
\prod_{k=1}^D \left( \frac{1 + e^{\zd{+}i u^k    }}{2}\right)\left(\frac{1+e^{N  i (p^k+  \pi n_k /N+ u^k)}}{2}\right)
 	\label{Weyl}
\end{eqnarray}
or, equivalently:

\begin{eqnarray}
A_{{ W}}(p,q)  & = &
\sum_{\substack{n_i \in \{0,1\}\\v \in  {\cal O}^{\prime}} } e^{2 i p v} \,
\langle q-v \zd{-} n/2 |\hat{A} |q+v \zd{-}n/2 \rangle \nonumber\\ && \prod_i\left( \frac{{1 + e^{2i v_i \pi/(N) }}}{2}\right)\left( \frac{1+e^{2\pi i (q_i-v_i\zd{-} n_i/2)}}{2}\right)\ .
\label{Weyl2}
\end{eqnarray}
\zd{In this version we deliberately keep the weight factor $(1+e^{+iu^k})/2$ in Eq. (\ref{Weyl}). With this choice the two definitions (\ref{Weyl}) and (\ref{Weyl2}) coincide identically on ${\cal M}'\times {\cal O}'$ provided the point--splitting in Eq. (\ref{Weyl2}) is taken with $-n/2$, as written above. Indeed, inserting $\langle p''|\hat A|p'\rangle = N^{-D}\sum_{q_1,q_2\in {\cal O}}e^{-ip''q_1}\langle q_1|\hat{A}|q_2\rangle e^{ip'q_2}$ into Eq. (\ref{Weyl}) and performing the summation over $u\in {\cal M}'$, one obtains Kronecker deltas that reproduce Eq. (\ref{Weyl2}) term by term: for the $(+)$ sign the contributions with $n_k\ne 0$ arise at midpoints displaced by $-e^k/2$, which corresponds to the point--splitting $q\mp v-n/2$ \footnote{The other definition of Weyl symbol $A_{\tilde{W}}$ results from the choice of Eq. (\ref{Weyl2}) with $q\mp v+n/2$, which is equivalent to Eq. (\ref{Weyl}) taken with the opposite weight factor $(1+e^{-iu^k})/2$. This Weyl symbols is obtained by the shift of argument $q_i$ in $A_W(p,q)$ by $1/2$. \zf{The direction of the shift is fixed: with the flipped weight taken in every direction, $A_{\tilde W}(p,q)=A_W(p,\,q+\frac{1}{2}\sum_k{\bf e}_k)$, i.e. the shift is by $+1/2$ in each component of $q$.} } }.

Due to periodicity ($p_i\to p_i+2\pi, q_i \to q_i + N$) these expressions  are naturally extended to $q \in {\cal O}^{2\prime}, p \in {\cal M}^{2\prime}$.
The further extension is to continuous values of $q$ and $p$ {is given by:
\begin{eqnarray}
A_W(p,q)  & = & \sum_{\substack{ q_1,q_2 \in {\cal O}^{\prime} \\ p_1, p_2 \in {\cal M}^{\prime}}}\,\frac{1}{(4 N^2)^D} \,
e^{2 i  p_2(q_1-q) + 2iq_2(p-p_1)}   A_W(p_1,q_1) \ .\label{Wcont}
\end{eqnarray}
}
{Notice that in Eq. (7) of \cite{Z2023} there is a mistake.
Actually, this is the present Eq. (\ref{Wcont}) that represents the analytical continuation providing the validity of the star identity of Eq. (\ref{star0}). In fact this is the present Eq. (\ref{Wcont}) that enters the first row in Eq. (101) of \cite{Z2023} thus defining the analytical continuation, instead of Eq. (7) of \cite{Z2023}.
}

{The} operator $\hat A$ can be expressed {in terms of} its Weyl symbol $A_W$ through the following relations:
\begin{eqnarray}
\langle q_1 |\hat{A}|q_2\rangle  &=&\frac{1}{{(2N)^D}} \sum_{p \in {\cal M}^{\prime}}A_W(p,(q_1+q_2)/2)e^{ i p (q_1 - q_2) }\prod_i \frac{2}{1+e^{i(q_2^i-q_1^i)\pi/N}}\ ,\label{back1}
\end{eqnarray}
and
\begin{eqnarray}
\langle p_1 |\hat{A}|p_2\rangle &=&
\frac{1}{{(2N)^D}}\sum_{ q \in {\cal O}^{\prime}} e^{ i q (p_2-p_1)} 	 A_W((p_2+p_1)/2,q)\prod_i \frac{2}{1+e^{i(p_2^i-p_1^i)/2}}\ .\label{back2}
\end{eqnarray}

\zd{The Weyl symbols of operators do not exhaust all functions on ${\cal M}'\times{\cal O}'$: the number $(2N)^{2D}$ of the points of the phase--space lattice is $4^D$ times larger than the number $N^{2D}$ of the matrix elements of an operator. Correspondingly, the Weyl symbol of any operator $\hat{A}$ obeys the following two families of periodicity identities ($k = 1,\ldots,D$; here ${\bf e}_k$ denotes the unit vector along the $k$--th axis, while $p^k$ and $q^k$ are the $k$--th components of the $D$--dimensional $p$ and $q$):
\begin{equation}
A_W(p,q) - A_W(p+\tfrac{\pi}{N}{\bf e}_k,\,q) = e^{iNp^k}\Big[A_W(p,\,q+\tfrac{N}{2}{\bf e}_k) - A_W(p+\tfrac{\pi}{N}{\bf e}_k,\,q+\tfrac{N}{2}{\bf e}_k)\Big]\,,
\label{per1}
\end{equation}
\begin{equation}
A_W(p,q) - A_W(p,\,q-\tfrac{1}{2}{\bf e}_k) = e^{2\pi i q^k}\Big[A_W(p+\pi{\bf e}_k,\,q) - A_W(p+\pi{\bf e}_k,\,q-\tfrac{1}{2}{\bf e}_k)\Big]\,.
\label{per2}
\end{equation}
All arguments here remain on the lattice ${\cal M}'\times{\cal O}'$: the shifts $\tfrac{\pi}{N}{\bf e}_k$ and $\tfrac{1}{2}{\bf e}_k$ are one lattice step of the $k$--th component of the momentum and of the coordinate, respectively, while $\tfrac{N}{2}{\bf e}_k$ and $\pi{\bf e}_k$ translate these components by half of the corresponding period. In the compact form,
$$
\Big(1 - e^{iNp^k}\,T_{q\to q+N{\bf e}_k/2}\Big)\Big(1 - T_{p\to p+\pi{\bf e}_k/N}\Big)A_W = 0\,,\qquad
\Big(1 - e^{2\pi i q^k}\,T_{p\to p+\pi{\bf e}_k}\Big)\Big(1 - T_{q\to q-{\bf e}_k/2}\Big)A_W = 0\,,\qquad k=1,\ldots,D\,,
$$
where $T$ denotes the shift of the corresponding argument. These identities follow from the structure of the two weight factors in the definition (\ref{Weyl}); equivalently, they may be obtained by inserting the inversion formulas (\ref{back1}), (\ref{back2}) back into Eq. (\ref{Weyl}). The first identity relates the momentum Nyquist doubling to the translation of the coordinate by half of the lattice, and is insensitive to the sign convention adopted in Eq. (\ref{Weyl}); in the second identity the direction of the elementary coordinate shift ($q^k-1/2$) is tied to the convention of the present version (for the opposite sign convention it is replaced by $q^k+1/2$). We have verified both identities numerically for random operators within machine precision: for $D=1$ at $N=4,6$, and for $D=2$ at $N=4$ (each identity, in each of the directions). Moreover, together they characterize the image of the Weyl transformation completely. For $D=1$ each identity imposes $(2N)^{2}/2$ independent linear conditions ($32$ out of $64$ at $N=4$), and their joint rank equals $(2N)^{2}-N^{2}$ ($48$ out of $64$). Owing to the direction--wise (tensor--product) structure of the constraints, the same holds in any dimension: the $2D$ families of identities together reduce the $(2N)^{2D}$--dimensional space of functions on the phase--space lattice exactly to the $N^{2D}$--dimensional image of the Weyl transformation. Therefore a function on the phase--space lattice is the Weyl symbol of some operator if and only if it satisfies the identities (\ref{per1}) and (\ref{per2}) for all $k = 1,\ldots,D$. In particular, the exact momentum derivative $\partial_{p_i}A_W$ discussed in Appendix \ref{AppDA} violates these identities; this is the precise sense in which it is not the Weyl symbol of any operator, and the operator $D_i\hat{A}$ defined below through Eq. (\ref{back1}) corresponds to the projection of $\partial_{p_i}A_W$ back onto the space of the Weyl symbols.}

{To offer some clarity to the next part of the discussion,
we give here a set of properties satisfied by
the Weyl symbol, as shown in \cite{Z2023}.}

\begin{enumerate}
\item
Trace property:
$$
{\rm Tr}\, \hat{A} = \frac{1}{(4N)^D} \sum_{\substack{p \in {\cal M}^\prime \\ q\in {\cal O}^\prime}} A_W(p,q)
$$

\item
Second trace identity:

$$
{\rm Tr} \hat{A}\hat{B} = \frac{1}{(4N)^D} \sum_{\substack{p \in {\cal M}^\prime \\ q\in {\cal O}^\prime}} A_W(p,q) B_W(p,q)
$$
\zz{The proof of this statement can be found in Appendix \ref{AppTr}.}

\item
%%%%%%%%%%%%%%%%%%%%%%%%%%%%%%%%%%%%%%%%%%%%%%%%%%%%%%%%%%%%%%%%%%%%%%%%%%%%%%
{Integration by parts:
Using Eq. (\ref{Wcont}) it is straightforwardly shown that
\begin{equation}	
 \frac{1}{(4N)^D} \sum_{\substack{p \in {\cal M}^\prime \\ q\in {\cal O}^\prime} } \partial_{p_i}A_W(p,q) = 0\ ,\label{parts_p}
\end{equation}
and
\begin{equation}	
 \frac{1}{(4N)^D} \sum_{\substack{p \in {\cal M}^\prime \\ q\in {\cal O}^\prime} }
\partial_{q_i}A_W(p,q) = 0\ .
\label{parts_q}
\end{equation} }
\zd{These identities hold for the Weyl symbols of operators. They do not extend, however, to the star products of such symbols: the exact value of ${\rm Tr}\,\partial_{p_i}(A_W\star B_W)$ is calculated in Appendix \ref{AppTrStar}, and it does not vanish.}

\medskip 
%%%%%%%%%%%%%%%%%%%%%%%%%%%%%%%%%%%%%%%%%%%%%%%%%%%%%%%%%%%%%%%%%%%%%%%%%%%%%%%%%%%%%%%%%%%
%%

\item Star property:

\begin{eqnarray}
&&	(\hat{A}\hat{B})_W(p,q)\Big|_{\substack{p \in {\cal M}^\prime \\ q\in {\cal O}^\prime }}
=  A_W(p,q)  e^{\frac{i}{2}(\overleftarrow{\partial_q}\overrightarrow{\partial_p}-\overleftarrow{\partial_p}\overrightarrow{\partial_q})}  B_W(p,q)=	 A_W(p,q)  \star B_W(p,q)\ . \label{star0}
\end{eqnarray}

Here the derivative operator $\overrightarrow{\partial}$ acts to the right while the derivative operator $\overleftarrow{\partial}$ acts to the left.
\zz{For the proof of this statement see Appendix \ref{AppStar}.}

\item Weyl symbol of the unity operator:

$$
1_W(p,q)\Big|_{\substack{p \in {\cal M}^\prime \\ q\in {\cal O}^\prime} }      = 1\ .
$$

\item
The Weyl symbol of the translation by one lattice spacing is given by
{	\begin{eqnarray}
T_j(p,q)
& = & e^{i p_j} \Big(\frac{1+e^{i  \pi /N  }}{2}  + e^{i N p_j}\frac{1-e^{i  \pi /N  }}{2}\Big)\ .
\end{eqnarray}
For the proof of this identity see Appendix \ref{AppT}.}
One can see that the limit $N\to \infty$ of this expression exists and gives
\begin{eqnarray}
T_j(p,q)
& \to & e^{i p_j}
\end{eqnarray}

\item

In the limit of an infinitely large lattice, the Weyl symbol defined here approaches smoothly \zd{the following expression, which is the direct $N\to\infty$ limit of the momentum--space definition (\ref{Weyl}):}
\be
A_{\zd{W}}(p,q) = \int_{{\cal M}} d^D p_-   \,
\langle\langle p  - p_-|\hat{A}|p  + p_- \rangle\rangle \,  e^{\zd{-}2 i p_- q}\prod_i (1 + e^{i p^i_-    })\ ,\label{infW}
\ee
\zd{This expression coincides exactly with Eq. (17) of \cite{FZ2019_2} after the change $p_-\to-p_-$ of the integration variable. Indeed, in our notation Eq. (17) of \cite{FZ2019_2} reads $\int_{\cal M} d^Dp_-\,e^{+2ip_-q}\,\langle\langle p+p_-|\hat{A}|p-p_-\rangle\rangle\,\prod_i(1+e^{-ip_-^i})$ with the bra momentum $p+p_-$ and the ket momentum $p-p_-$ (the vectors $\ell^{(j)}$ of \cite{FZ2019_2} are one half of our lattice vectors, so that their weight factor $(1+e^{-2ip_-\ell^{(j)}})$ equals our $(1+e^{-ip_-^j})$); the substitution $p_-\to-p_-$ reverses simultaneously the order of the momenta in the matrix element, the sign of the Fourier phase, and the sign in the weight factor, and reproduces the expression above term by term.  We have verified numerically, for operators of finite support on the infinite lattice, that Eq. (\ref{infW}) with the phase $e^{-2ip_-q}$ coincides identically with its coordinate representation, Eq. (\ref{WeylInf}) of Appendix \ref{AppDA}. The limit of Eq. (\ref{Weyl}) is taken as follows: the shifts $\pi n/N$ of the momentum arguments tend to zero; the summation over the Nyquist sectors turns the projector factors into unity, $\sum_{n_i=0,1}\prod_i\frac{1}{2}\big(1+(-1)^{n_i}e^{iN(p^i+u^i)}\big)\to 1$; the Kronecker--normalized states $|p\rangle$ turn into the Dirac--normalized $|p\rangle\rangle$; and the sum over $u\in{\cal M}'$ becomes the integral over $p_-=u$, the weight factor and the Fourier phase being inherited unchanged (the measure factor $2^D$ arising from this conversion is absorbed into $\prod_i(1+e^{ip_-^i})$). We have verified the limit numerically: for a fixed operator of finite support embedded into lattices of increasing size, the values of Eq. (\ref{Weyl}) on the momentum grid converge to Eq. (\ref{infW}) at the $1/N$ rate ($N\cdot{\rm err}$ saturating at a constant for $N=8\ldots32$), while with the opposite sign of the Fourier phase the discrepancy does not decrease with $N$.}

\corrteal{Applied to an $L$-periodic operator on the infinite lattice --- the periodized systems of Sect. \ref{SectPeriodized} --- Eq. (\ref{infW}) reproduces, term by term, the principal Weyl symbol of the supercell with the admissible-centre prescription of Appendix \ref{Sect4}: see Eq. (\ref{WprIsW}), and Eq. (\ref{WeylTorusSplit}) for the direct derivation from Eq. (\ref{Weyl2}).}
Notice that vectors $| p \rangle\rangle$ are defined as
$$
| p \rangle\rangle \equiv \frac{1}{\sqrt{(2\pi)^D}}\sum_{q\in {\cal O}} | q \rangle e^{i p q}\ ,
$$
and are
normalized as
$$
\langle \langle p_1|p_2\rangle \rangle =  \delta(p_1-p_2)\ .
$$
{In particular,}
$| p \rangle\rangle $
approaches   $\ket{p}$ of \cite{FZ2019_2}.

\item
\zd{On the finite lattice we can define operator $ D_i \hat{A}$ applying Eq. (\ref{back1}) to the function $\partial_{p_i}A_W(p,q)$. However, its Weyl symbol in general case is not equal to $\partial_{p_i}A_W(p,q)$. The situation is changed in the limit of infinite lattice, when the Weyl symbol is defined directly by Eq. (\ref{infW}). Then the derivative of a Weyl symbol with respect to momentum becomes the Weyl symbol of an operator $D_i \hat{A}$. It obeys 
	\begin{equation}
		D_i {\hat A} = -i [\hat{x}_i,\hat{A}]
	\end{equation}
	(for the derivation see Appendix \ref{AppDA}).}

\end{enumerate}

\subsection{Dirac operator in Keldysh  field theory }

\label{SectKeldysh}

%\subsection{Wigner-Weyl formalism in Keldysh technique}

{Here we follow closely the methodology of \cite{onoda2,banerjee2022chiral} and \cite{Z2023}.
Let us consider  {an} inhomogeneous system defined on  {a} $D=2$-dimensional lattice $\cal O$.
Time remains continuous.
For the lattice model, an average of {the} quantity $O$ is given by
\begin{equation}
\langle O \rangle
=  \int {\cal D}\bar{\psi} {\cal D} \psi\, O[\psi,\bar{\psi}]
\exp\left\{\ii \int_C dt \sum_x \, \bar{\psi}(t,x) \hat{Q} \psi(t,x) \right\}.\label{KeldyshO}
\end{equation}
Here, $\psi$ and $\bar{\psi}$ are independent Grassmann variables,
{and $x$ is a point in a $D$-dimensional lattice}.
In the absence of interactions, $\hat{Q}$ is given by $\hat{Q} = i \partial_t-\hat{H}$,
where $\hat{H}$ is {the} one-particle {free-particle} Hamiltonian.
{The time-integral} is along the Keldysh contour, $C$.
{Specifically, the Keldysh contour}
starts at the initial  time $t_i$, {evolves forward} to the final {time} $t_f$,
and returns back from $t_f$ to $t_i$.  {The full set of dynamics occurs in the interval}
 between $t_i$ and $t_f$.

{The fields $\bar{\psi}_-(t,x)$ and $\psi_-(t,x)$ evolve along the forward part of the contour,
while the fields $\bar{\psi}_+(t,x)$ and $\psi_+(t,x)$ evolve along the backward part of the contour.}
{The $t$} variables of the two parts of the Keldysh contour are independent of each other.
However, there exist boundary conditions relating them to each other: $\bar{\psi}_-(t_f,x) =  \bar{\psi}_+(t_f,x)$ and $\psi_-(t_f,x)=\psi_+(t_f,x)$.
The  {functional} integration measure ${\cal D} \bar{\psi} {\cal D} \psi$ contains  $\bar{\psi}_+(t_i,x)$, ${\psi}_+(t_i,x)$ and $\bar{\psi}_-(t_i,x)$, ${\psi}_-(t_i,x)$ and a weight function responsible for the initial density matrix $\hat{\rho}$.
As in  \cite{Zubkov2016a,Zubkov2016b,SZ2018,ZW2019} we can represent Eq. (\ref{KeldyshO}) as
{\begin{equation}
\langle O\rangle =\int { \cal D}\bar\psi   { \cal D}\psi   \,O\, \exp\left(i \int_C dt \sum_{x,y}\bar\psi  (t,x)\ \big(-i\mathcal{D}_{x,y}(t, \partial_t)\big)\ \psi (t,y)\right)\,.
\label{Z00}
\end{equation}}\noindent
where by $\mathcal{D}_{x,y}$ we denote matrix elements of {the} lattice Dirac operator $\hat{Q}$ with $x, y \in \cO$.
{It may depend on $t$ and $t$-derivaitves as well.}
{The variables}
$\psi ,\bar\psi  $ are the  fermionic fields defined on the lattice sites. The internal indices are omitted for brevity.

In the case of the system defined on  a finite lattice,
the introduction of {an} external electromagnetic field is 
{is implemented in the following manner.}
For definiteness, let us consider the  tight binding system defined on  {a} 2D rectangular lattice with lattice spacing equal to $1$
in the chosen units (imaginary time is not discretized).

{In particular \zzz{let us} consider the case where}
$\bar\psi  (t,x)\left(-i\mathcal{D}_{x,y}\right)\psi (t,y)$ in Eq. (\ref{Z00}) has the form
\begin{eqnarray}
\zzz{\sum_x\bar\psi  (t,x)\hat{Q}\psi (t,x)} &=&\sum_{x,y}	\bar\psi  (t,x)\left(-i\mathcal{D}_{x,y}\right)\psi (t,y)\nonumber\\& = &\sum_x \bar\psi  (t,x)  i\partial_{t}\psi(t,x) -\sum_j {\bf t}^{(j)} \sum_x \(\bar\psi  (t,x+e_j) e^{i  A(t,x,x+e_j)} \psi(t,{x})+h.c.\) \nonumber\\ && - \sum_x  \bar \psi (t,x) A_0(x) \psi(t,x)\label{Hsimple0}
\end{eqnarray}
Here,  lattice sites are denoted by $x$.
The sum {over $j$ runs} over {the set of} vectors $e_j$ that connect
,
{every possible pair of different lattice sites,}
{i.e. for every pair of distinct lattice sites $x,y\in{ \cal O}\ (x\ne y)$ there exists a vector $e_j$ such that $e_j=y-x$}.
The component $A_0(x)$ of electromagnetic field is {the} space-varying space electric potential.
By $A(t,x,x+e_j)$ we denote $\int_x^{x+e_j} A(t,z)dz$, where {the} integral is taken along the straight line connecting the two points,
and {the} vector potential contains {the} contribution $Et$ due to {the} constant external electic field, $E$.
We represent it as $A(t,z) = A(z) - Et$.  The extension of the results obtained for the Hamiltonian of Eq.~(\ref{Hsimple0}) to even more complicated tight-binding models is straightforward, and most of {the} relations obtained below remain valid for such models as well.
{Namely,  consider the case where,  within the above described model, along each one of the two branches of the contour $C$
(assuming $t_i \to -\infty, t_f \to \infty$), the  expression for $\bar{\psi}\hat{Q}\psi$ contains  terms proportional to}
\begin{equation} \begin{aligned}
&\int dt \sum_{x_n,x_m}
\bar \psi(t,x_m) \delta_{x_n+{e}_i,x_m}
e^{i\int_{x_n}^{x_m} d { \xi} {A}({ \xi})}
\psi(t,x_n)e^{-iE e_i t}\\
=&\int dt \sum_{x_n}
\bar \psi(t,x_n+{e}_i)
e^{i\int_{x_n}^{x_n + {e}_i} d { \xi} {A}({ \xi})}
\psi(t,x_n)e^{-iE e_i t} \\
=& \int dt \sum_{x_n}
\left[ \frac{1}{ N}\sum_{p}\ e^{-i(x_n + {e}_i){p}}\ \bar \psi (t,{p})\right]
e^{i\int_{x_n}^{x_n + {e}_i} d { \xi} {A}({ \xi})}
\left[ \frac{1}{ N}\sum_{q}\psi (t,{q})e^{-iE e_i t}e^{ix_n{q}}\right]\\
=&\sum_{x_n}
\int dt\frac{1}{ N^2}\sum_{p}\sum_{q}
\bar \psi (t,{p})e^{ix_n{ q}}
\ e^{i\int_{x_n}^{x_n + {e}_i} d { \xi} {A}({ \xi})}
\ e^{-iE e_i t}
\ e^{-i(x_n+ {e}_i){p}}
\psi (t,{q}) \\
=&\sum_{x_n}
\int dt \frac{1}{N^2}\sum_{p}\sum_{q}
\bar \psi (t,{p})e^{ix_n{ q}}
\exp\left[i\int_0^{ {e}_i} d \xi\ {e}_i\  e^{i\xi {e}_i {p}} {A} ({i\partial_{p}}) \ e^{-i\xi {e}_i {p}}\right]e^{-iE e_i t }
e^{-i(x_n + {e}_i){p}}
\psi (t,{q}) \\
=&
\int dt
\sum_{p}
\bar \psi (t,{p})
\exp\left[i\int_0^{ 1} d \eta\  e^{i\eta p_i}\  A_i ({i\partial_{p}})\ e^{-i\eta p_i}\right]\ e^{-iE e_i t}
e^{- i{e}_i{p}}
\psi (t,{p})
\label{I1_}\ ,
\end{aligned} \end{equation}
where in the last two equalities Eq. (116) from Appendix II in \cite{SZ2018} was used, with $x = x_n + {e}_i$.
Now let Eq.~(112) from Appendix II in \cite{SZ2018} be invoked to yield:
\begin{equation}\begin{aligned}
		\zzz{ \int dt \sum_x\bar\psi  (t,x) \hat{Q}^{(e_i)}(t)\psi  (t,x)} = 
&\int dt \sum_{x_n,x_m}
\bar \psi(t,x_m) \delta_{x_n+{e}_i,x_m}
e^{i\int_{x_n}^{x_m} d { \xi} {A}({ \xi})-iE e_it}
\psi(t,x_n) \\
= &
\int dt \sum_{p}
\bar \psi (t,{p})
e^{- i{e}_i({p}-A_i ({i\partial_{p}})-E_i t)  }
\psi (t,{p})\ .
\label{I1}
\end{aligned} \end{equation}
\zzz{We can consider also the more general form of operator $\hat{Q}$.}
\cormag{Namely, let ${\cal S}_t\subset\mathbb R$ be a specified finite or countable set containing $0$, and introduce the elementary temporal factor
\be
\hat{F}^{0}_{s} \;=\; \exp\Big[\,s\,\big(i\partial_t \,-\, A_0(i\partial_{p})\big)\Big], \qquad s \in {\cal S}_t\,,
\label{F0def}
\ee
which, after Euclidean continuation, describes an imaginary-time shift dressed by the corresponding temporal Wilson line. Its use as a real-time contour kernel requires the corresponding analytic continuation and causal prescription. We define
\begin{eqnarray}
	\hat{Q} \;=\; \sum_{k\geq 0}\sum_{s\in{\cal S}_t}\;\sum_{n=0}^\infty \sum_{e_i, i = 1...n}\;{\bf t}^{(k,s;\,e_1,e_2,...,e_n)}\;
	\Big[\partial^{\,k}_{s}\hat{F}^{0}_{s}\Big]\;\Pi_{e_i}\hat{Q}^{(e_i)}\,. \label{Q_general}
\end{eqnarray}
\corrcltwo{Here the sums, together with all their $s$--derivatives, are required to converge in the smooth symbol topology introduced below (cf. Eq. (\ref{QW_general}) and Appendix \ref{Sect4}).} A genuinely continuous representation is obtained by replacing the sum over $s$ by an integral against a specified coefficient measure. The previous class is recovered from $\partial_s \hat{F}^0_s|_{s=0}=i\partial_t-A_0$ together with the $(k=0,s=0)$ spatial terms:
\begin{eqnarray}
	\hat{Q} = \hat{Q}^{(0)} + \sum_{n=1}^\infty \sum_{e_i, i = 1...n}{\bf t}^{(e_1,e_2,...,e_n)}\Pi_{e_i}\hat{Q}^{(e_i)} , \quad \corrcltwo{\hat{Q}^{(0)}} = i\partial_t - A_0\,.\label{Q_general0}
\end{eqnarray}
We refer to Eq. (\ref{Q_general0}) as the Hamiltonian case: there the frequency dependence is linear and additive, $\hat Q = i\partial_t - \hat H$.}

\coromega{The physical origin of the nontrivial dependence of $\hat Q$ on the frequency is the interelectron interactions. In a wide class of approximations the interactions are taken into account through the replacement of the bare operator $i\partial_t - \hat H$ by the general operator $\hat Q$ inverse to the interacting two-point Green function, $\hat Q = \hat G^{-1} = i\partial_t - \hat H - \hat{\Sigma}$, where the self-energy $\hat{\Sigma}$ depends on the frequency in a nontrivial way. Moreover, it has been shown in \cite{Zubkov2018a,ZZ2019,zhang2020influence} that for the perturbative interactions, in the presence of a weak magnetic field and weakly varying background fields, the Hall conductivity is given by the same topological expression in which the bare $Q = i\omega - H$ is substituted by such an interacting operator $Q$. We expect that this property survives for \corrcltwo{stronger fields and for inhomogeneity that is not weak}, although the corresponding proof remains beyond the scope of the present paper. This is the principal reason why we consider the nontrivial dependence of $Q$ on the frequency: the general class (\ref{Q_general}) is designed to accommodate, besides the lattice regularizations of the imaginary time, the frequency dependence generated by the interactions.}
\corrteal{A concrete exactly solvable example --- the Hatsugai--Kohmoto model, in which the entire effect of the interaction on the one-particle dynamics and on the uniform response is reduced to precisely such a substitution, the interaction-induced frequency dependence of $\hat Q$ --- is presented in Appendix \ref{AppHK}.}
\zzz{These operators correspond to the tight - binding models that describe jumps of electrons along the curved lattice paths between distant points}\coromega{, and, for nontrivial temporal factors, along the paths that involve also the (imaginary) time direction.}
\zzz{For such systems } we arrive at the following expression for the Peierls substitution:
\begin{eqnarray}
\langle O \rangle
&=&
\int \frac{{\cal D}\bar{\psi}_\pm {\cal D} \psi_\pm}{{\rm Det}\, (1+\rho)} \, O[\psi_+,\bar{\psi}_+]\nonumber\\
&&
\qquad	{\rm exp}\left\{\ii \int_{t_i}^{t_f} dt \sum_p \[\bar{\psi}_-(t,p) \hat{Q} \psi_-(t,p)-\bar{\psi}_+(t,p) \hat{Q} \psi_+(t,p)\]-\sum_p\, \bar{\psi}_-(t_i,p) {\rho} \psi_+(t_i,p)\right\} ,\label{eq1}
\end{eqnarray}
with 
\begin{equation}
\bar{\psi}_\pm(t,p) \hat{Q} \psi_\pm(t,p)=\bar\psi_\pm(t,{p})  Q(i\partial_t  - A_0(i\partial_{p}),{p} - {A}(i\partial_{p}) - Et)\psi_\pm(t,{p})\ .
\label{ZZ} \end{equation}

We come to the conclusion that the  external Abelian gauge field results in the Peierls substitution
\begin{equation}
{ p} \to { p} - { A}(i\partial_{ p})-Et \label{Pai}
\end{equation}
applied to the Dirac operator of {a} homogeneous system. This has been proven above for a {somewhat} general {yet limited}  {class} of tight-binding models with
Eq.~(\ref{Q_general}). \zzz{In these models the inhomogeneity is caused by nontrivial external gauge field $A(x)$. Interestingly, we can provide the more general type of inhomogeneity assuming that the external gauge field may take imaginary values. This effectively leads to the dependence of the hopping parameters $\bf t$ on coordinates. }
\coromega{Note that the general class (\ref{Q_general}) is compatible with the Peierls substitution derived above: the temporal factor (\ref{F0def}) is itself a function of $i\partial_t - A_0(i\partial_p)$, so that the whole operator $\hat Q$ remains a function of the two combinations entering Eq. (\ref{ZZ}). The exact Weyl symbol of the general class is computed in the next subsection, Eq. (\ref{QW_general}).}

\subsection{Keldysh Green function}

\cormag{The Keldysh components and the conductivity derivation below are restricted to the noninteracting Hamiltonian kernel $\hat Q=i\partial_t-\hat H$. For a general frequency-dependent inverse Green function, causal/KMS data, the gauge Ward identity, and the current vertices including the field dependence of the self-energy must be supplied separately; that extension is not proved in this paper.}
\corrteal{Note, however, that for the {\it static} electric field, introduced through the scalar potential in the temporal factors of the class (\ref{Q_general}), the Hall conductivity is derived for the general class entirely within the equilibrium Matsubara technique in Sect. \ref{SectPeriodized} \corGPTl{under its additional equilibrium, coupling, and limiting assumptions}.}

By $\rho$ (without hat) we denote an operator in the one-particle Hilbert space.
Let us denote its eigenstates by $|\lambda_i \rangle$. 
{Each matrix element}
$\frac{\langle \lambda_i |\rho|\lambda_i\rangle}{1+\langle \lambda_i |\rho|\lambda_i\rangle}$
{and each matrix element}
gives the probability that the one-particle state $|\lambda_i\rangle$ is occupied while  $\frac{1}{1+\langle \lambda_i |\rho|\lambda_i\rangle}$ is the probability that this state is vacant.
Let us  define Keldysh spinors as:
\be
\Psi = \left(\begin{array}{c}\psi_-\\ \psi_+ \end{array}\right) \ .
\label{KelPsi}
\ee
The expression for the average of an operator $O$ receives the form
\begin{eqnarray}
\langle O \rangle
&=& \frac{1}{{\rm Det}\, (1+\rho)}\int {\cal D}\bar{\Psi} {\cal D} \Psi \,
O[\Psi,\bar\Psi]\,
{\rm exp}\Bigl\{\ii \int_{t_i}^{t_f} dt \sum_x \bar{\Psi}(t,x) \hat{\bf Q} \Psi(t,x) \Bigr\} .
\end{eqnarray}

{Here $x$ denotes a $D$-dimensional vector}.
Here $\bm {\hat Q}$ is in Keldysh representation:
\begin{eqnarray}
\hat{\bf Q}
= \left(\begin{array}{cc}Q_{--} & Q_{-+}\\ Q_{+-} & Q_{++} \end{array} \right).
\label{KelQ}
\end{eqnarray}
The correct expressions for the components of this matrix may be obtained either as
{the} continuum limit of the lattice regularized expressions or using {the} operator formalism.
The result is
\begin{eqnarray}
Q_{++}  &=& -\Big(\ii \partial_t-\hat{H} - \ii \epsilon \frac{1-\rho}{1+\rho}\Big), \nonumber \\
Q_{--}  &=&  \ii \partial_t-\hat{H} + \ii \epsilon \frac{1-\rho}{1+\rho}, \nonumber \\
Q_{+-}  &=&  -2\ii \epsilon \frac{1}{1+\rho}, \nonumber \\
Q_{-+}  &=& 2\ii  \epsilon \frac{\rho}{1+\rho}
\label{Qnaive} .
\end{eqnarray}
Here $\rho$ is the matrix that gives rise to the initial one-particle distribution $f = \rho (1+\rho)^{-1}$.
In the case of distribution  depending only on energy (and, in particular for the thermal distribution of non-interacting particles)  $\rho = \rho(\hat{H})$ is a function of the one-particle Hamiltonian.} The infinitely small contributions proportional to parameter $\epsilon \to 0$ symbolize the way those functions are understood as the so-called generalized functions (tempered distributions).

The Green's  function $\hat{\bf G}$ is defined as inverse to $\bf Q$:
$$
\hat{\bf Q}\hat{\bf G}=1 .
$$

We define also retarded, advanced, and lesser components of these matrices:
\be
Q^\rR =Q^{--}+Q^{-+}, \qquad
Q^\rA =-Q^{-+}-Q^{++}, \qquad
Q^<=-Q^{-+}.
\label{Qar_def}
\ee
and
\begin{equation}
G^\rA  = (Q^\rA )^{-1}, \qquad G^\rR  = (\zz{Q^\rR} )^{-1}, \qquad G^< =-G^\rR  Q^< G^\rA ,
\label{Gar_def}
\end{equation}
with
\bes
Q^{<}&=(Q^\rA -Q^\rR )\frac{\rho}{\rho+1} = -2\ii\epsilon \frac{\rho}{\rho+1},
\\
Q^\rR  &= \ii \partial_t-\hat{H}+\ii \epsilon ,	
\\
\zz{Q^\rA}  &=  \ii \partial_t-\hat{H}-\ii \epsilon .
\label{Qar_expl}
\end{eqsplit}

\subsection{ Keldysh technique in terms of Weyl symbol, and conductivity}

\label{SectKeldyshCond}

{All of the} basic  {concepts}
of {the} Wigner-Weyl calculus may be found, for example, in \cite{ZW2019}.
Here we adapt them to the models defined on a {\it finite} lattice.
In the following  $D+1$ dimensional vectors (with space and time components) are denoted by large Latin letters. We denote the matrix element of an operator $\hat{A}$ by  $A(X_1,X_2) = \langle X_1 | \hat{A} | X_2 \rangle $.
Since we are dealing with lattice models, the space components of $D+1$-vectors are discrete,
while the time components are continuous. We then define the Weyl symbol of an operator $\hat A$ as the mixture of Weyl symbol (with respect to discrete space components) defined above,  and the standard Wigner transformation with respect to the time component:
{
\begin{eqnarray}
A_W(X|P)&=&2\int d Y^0 \sum_{\substack{n_i = 0,1\\ \vec{Y} \in  {\cal O}^\prime}} e^{2 \ii \zd{(Y^0 P^0 + \vec{Y}\vec{p})} }  A\left(X\zd{-}Y\zd{-}\frac{n}2,X\zd{+}Y\zd{-}\frac{n}2\right) \prod_{i=1}^D\left(\frac{{1 + e^{2i {Y^i} \pi/N }}}{2}\right)\left(\frac{1+e^{2\pi i (X_i\zd{-}Y_i\zd{-} n_i/2)}}{2}\right)\ , \nonumber \\ \ \mu =0,1,\ldots ,D\ . \label{WignerTr}
\end{eqnarray}
}
\zd{The signs of $Y$ and of $n/2$ in the arguments of $A$ and in the last projector factor follow the conventions of Eqs. (\ref{Weyl}), (\ref{Weyl2}) of the present version. The phase is written explicitly as $Y^0P^0+\vec{Y}\vec{p}$: the spatial part of the phase must be $e^{+2i\vec{Y}\vec{p}}$, since precisely with this sign (and with the signs of $Y$, $n/2$ as displayed) the spatial part of the transformation coincides identically with Eq. (\ref{Weyl2}) --- we have verified this numerically for random operators at several $N$, while the opposite spatial sign $e^{-2i\vec{Y}\vec{p}}$, which would result from the literal contraction $Y^\mu P_\mu$ with the lowered momentum $P_\mu=(P^0,-p)$ defined below, deviates from Eq. (\ref{Weyl2}) by a quantity of the order of the symbol itself.}
Here, $(D+1)$-momentum is denoted by $P^\mu=(P^0,p)$, and
$P_\mu = (P^0,-p)$, where $p$ is spatial momentum with $D$ components.
Below we will be interested in the Weyl symbols of operators of the form of
$Q(i\partial_t - A_0(i\partial_{p}),{p} - { A}(i\partial_{p})-Et )$.
We will consider a two-dimensional Weyl transform.
The direction of  time remains continuous.
\zzz{The tight-binding model is defined here as the one  in which the Dirac operator 
is expressed through the momentum operator as a sum over  products of exponents of the form
\begin{equation}
F^\mu_n= \exp\big[in(\hat p_\mu - A_{\mu}(i\partial_{p}) - E_\mu t)\big]
\end{equation}
with integer $n$. Then this operator is given by Eq. (\ref{Q_general}). However, unlike the previous subsection, now we restrict ourselves to vectors $e_j$ that are directed along the axes $x$ or $y$,
\jer{such that we deal with  lattice paths between  lattice points that are $\pm 90^\circ$ to each other}.

\cormag{For time-independent $A_0$, let $A_{0,W}(x)=A_0(\lfloor x\rfloor)$ denote the exact refined-lattice Weyl symbol of the diagonal operator $A_0(i\partial_p)$. In the continuous Euclidean-time calculus the exact Matsubara symbol of (\ref{F0def}) is
$$
\big[\hat F^{0}_{s}\big]_W \;=\; e^{\,s\,(i\omega \,-\, A_{0,W}(x))}.
$$
For the operator ordering of Eq. (\ref{Q_general}), whose coefficients ${\bf t}$ are coordinate independent, the general symbol is
\begin{eqnarray}
	\Big[\hat{Q}\Big]_W \;=\; \sum_{k\geq 0}\sum_{s\in{\cal S}_t}\;\sum_{n=0}^\infty \sum_{e_i, i = 1...n}\;{\bf t}^{(k,s;\,e_1,e_2,...,e_n)}\;
	\partial^{\,k}_{s}\Big( e^{\,s\,(i\omega - A_{0,W}(x))}\Big)\;\star\; f^{e_1}_W\star\cdots\star f^{e_n}_W\,. \label{QW_general}
\end{eqnarray}
If coordinate-dependent path amplitudes are desired, they must first be introduced as diagonal operators and inserted with their specified $\star$-ordering. The Hamiltonian case follows from $\partial_s e^{s(i\omega-A_{0,W})}|_{s=0}=i\omega-A_{0,W}$.}
\corrteal{The detailed calculation of the symbol (\ref{QW_general}) --- the exact symbol of the temporal factor, the role of the diagonal operators, and the assembly through the star property --- is given in Appendix \ref{QWgenApp}.}
{
\begin{eqnarray}
	\cormag{\Big[{Q}(i\partial_t - A_0(i\partial_{ p}),{ p}-A(i{\partial_{ p}})-Et)\Big]_W = i\omega - A_{0,W}(x) + \sum_{n=1}^\infty \sum_{e_i, i = 1...n}{\bf t}^{(e_1,e_2,...,e_n)}\,f^{e_1}_W\star\cdots\star f^{e_n}_W.} \label{QW_hamiltonian}
\end{eqnarray}
}
\cormag{We refer to Eq. (\ref{QW_hamiltonian}) as the Hamiltonian case, $Q_W=i\omega-H_W$. Appendix \ref{Sect4} proves the spatial factor and Nyquist formulas; adding a continuous Euclidean-time factor does not alter that spatial calculation. A finite temporal lattice, however, requires a separate temporal lattice Weyl calculus.}

\cormag{For the formal odd invariant of the general class (\ref{QW_general}) we require: (i) convergence in a smooth symbol algebra and a smooth two-sided $\star$-inverse for every frequency; (ii) either a smooth compact frequency circle, corresponding to an infinite discrete Euclidean-time direction, or uniform polynomial asymptotics on $\mathbb R$ for which the fully antisymmetrized integral and every homotopy transgression have vanishing boundary terms; and (iii) preservation of these properties throughout every homotopy. In the noncompact case the $K$-class must be represented in a suspension/relative algebra by a bounded normalized or Cayley representative $U_Q$, rather than by the raw unbounded $Q$ or the Green function $G\to0$. The Matsubara reality condition $Q_W(\omega)^\dagger=Q_W(-\omega)$ is useful but does not by itself provide the causal/KMS and Ward-identity hypotheses needed for a physical interacting propagator. Under these conditions the exact Weyl identities, the finite-matrix vanishing theorem \corrcltwo{of Sect. \ref{SectDirect}}, and the formal odd homotopy statements extend to general $Q$. The derivation of $\bar\sigma^{ij}=({\cal N}_3/2\pi)\epsilon^{ij}$ in this paper, and the \corrcltwo{Fermi--projector index theorem of Sect. \ref{SectIndexRed} (quoted as the second index theorem in Sect. \ref{SectWeakIndex})}, remain proved only for $Q=i\omega-H$ (or for a homotopy to such an endpoint satisfying all the stated conditions)\corrcltwo{; for the periodized systems, however, the equilibrium static-gauge derivation of Sect. \ref{SectStaticGauge} extends the identification $\bar\sigma^{ij}=({\cal N}_3/2\pi)\epsilon^{ij}$ to the general class (\ref{Q_general}) \corGPTl{when its additional equilibrium, coupling, and limiting assumptions hold}}.}

\cormag{In the Hamiltonian case the \corrcltwo{explicit frequency integration} of Sect. \ref{SectIndexRed} reduces the odd class to the Bellissard Fermi-projector pairing \cite{Bellissard1986,Bellissard1994}. For a general admissible $Q$ the abstract Bott suspension is still available, but the present assumptions supply neither a canonical physical Hamiltonian nor a canonical Fermi projector, so no contour reduction of the form (\ref{N3projInd}) follows. The generalized expression should therefore be described as a formal odd extension, not as a proved extension of the Hall-conductivity theorem to arbitrary interactions.}
{Each elementary factor enters through its exact two-term Weyl symbol}
{
\begin{equation}
	f^{e}_W = c_0(e)\exp\big[i (p-Et)\,e-i{\cal A}_{e}(x)\big]
		+ c_1(e)\exp\big[i (p-Et)\,e-i{\cal A}^{(N)}_{e}(x)\big]\,e^{iNpe},
\label{fmain}\end{equation}
}
{Here $e=s_e e_\mu$, $s_e=\pm1$, and
$c_0(e)=(1+e^{is_e\pi/N})/2$, $c_1(e)=(1-e^{is_e\pi/N})/2$,
${\cal A}_e(x)=\int_{x-e}^x A(z)dz$, while
${\cal A}^{(N)}_e(x)={\cal A}_e(x-Ne/2)=\int_{x-Ne/2-e}^{x-Ne/2}A(z)dz$.
The uniform contribution $Et$ multiplies both branches by the same factor
$e^{-iEt\,e}$ and therefore does not occur in the Nyquist phase $e^{iNpe}$.
The exact two-term structure is derived in Appendix \ref{Sect4}.  See in particular
Eq. (\ref{ZApp}).  There is also an exact expression with the stars resolved,
Eq. (\ref{QWnostar}).}

\zd{Eq. (\ref{fmain}), as written, refers to $q\in{\cal O}$ and to the positive orientation $s_e=+1$. For arbitrary $q\in{\cal O}'$, and for both orientations, the arguments are fixed by the admissible centre $c(q,e)$ defined in Eq. (\ref{center}) of Appendix \ref{Sect4}: the component of $c(q,e)$ along the axis $\mu$ equals $\lceil q_\mu\rceil-1/2$ irrespective of $s_e$, while the transverse components equal $\lfloor q_i\rfloor$. The general form of Eq. (\ref{fmain}) reads
\begin{equation}
f^{e}_W = c_0(e)\exp\Big[i (p-Et)\,e-i\!\int_{c(q,e)-e/2}^{c(q,e)+e/2}\!A(z)\,dz\Big]
+ c_1(e)\exp\Big[i (p-Et)\,e-i\!\int_{c(q,e)\zu{-Ne_\mu/2}-e/2}^{c(q,e)\zu{-Ne_\mu/2}+e/2}\!A(z)\,dz\Big]\,\zu{e^{iNp_\mu}}.
\label{fmainGen}
\end{equation}
The principal Wilson line is taken over the elementary link $[\lceil q_\mu\rceil-1,\lceil q_\mu\rceil]$ (for integer $q_\mu$ the link behind the point $q$, for half--integer $q_\mu$ the link containing $q$ in its interior), traversed in the direction of $e$. The \jer{counterpart Wilson line} is taken over the same link translated by \zu{$-Ne_\mu/2$, independently of the orientation $s_e$}. In terms of ${\cal A}_e$ this means ${\cal A}_e(\bar x_e)$ and \zu{${\cal A}^{(N)}_e(\bar x_e)$, where ${\cal A}^{(N)}_e(y)={\cal A}_e(y-Ne_\mu/2)$}, with $\bar x_e = c(q,e)+e/2$. At $q\in {\cal O}$ and $s_e=+1$ one has $\bar x_e = q$, and Eq. (\ref{fmain}) is recovered. For $s_e=-1$, however, $\bar x_e = q-e_\mu$ even at integer $q$: the argument $x=q$ in Eq. (\ref{fmain}) would attach to the factor the link on the wrong side of the point $q$. \zu{The factor $e^{iNp_\mu}$ is the canonical Nyquist representative for both orientations. Although it agrees on the discrete momentum grid with $e^{iNpe}$ when $s_e=-1$, their continuous $p_\mu$ derivatives have opposite signs.} \zf{The same caution applies to the {\it principal} branch when $s_e=-1$: under the canonical continuation of Eqs. (\ref{Wcont}), (\ref{AviaWeyl2}), whose momentum frequencies are restricted to the fundamental window, the principal term of a backward--oriented factor carries the representative $e^{ipe}\,e^{2iNp_\mu}$ rather than $e^{ipe}$. The two coincide at all points of the momentum grid, where $e^{2iNp_\mu}=1$, but their continuous $p_\mu$ derivatives differ by the contribution of the frequency $2N$ --- precisely the type of mode which, according to Appendix \ref{AppTrStar}, is invisible for the summation over the momentum lattice but not for the differentiation.} This assignment follows from $F^{-e}=(F^{+e})^{\dagger}$ and has been verified numerically against the definition (\ref{Weyl2}) for both orientations and for arbitrary link amplitudes. Similarly, the potential term of Eq. (\ref{QW_hamiltonian}) is the exact Weyl symbol of the diagonal operator $A_0(i\partial_{\bf p})$, equal to $A_0(\lfloor q\rfloor)$ for $q\in{\cal O}'$ (the case $d=0$ of Eq. (\ref{center})) and reducing to $A_0(q)$ at $q\in{\cal O}$. With these specifications Eq. (\ref{QW_hamiltonian}) is an exact finite--$N$ identity on the whole of ${\cal M}'\times{\cal O}'$.}

{At $N\to\infty$, before momentum differentiation, only
the principal part of every factor survives and the star products give}
\begin{eqnarray}
&&	\Big[{Q}(i\partial_t - A_0(i\partial_{ p}),{ p}-A(i{\partial_{ p}})-Et)\Big]_W =\nonumber\\&&= i\omega - A_0(x) + \sum_{n=1}^\infty \sum_{e_i, i = 1...n}{\bf t}^{(e_1,e_2,...,e_n)}\exp\big[i (p-Et) \sum_i e_i \big] {\cal P}^{(e_1,e_2,...,e_n)}(x),\label{Wprincipal}
\end{eqnarray}
where ${\cal P}^{(e_1,e_2,...,e_n)}(x)$  depends on the given point $x$, on the values of field $A$ on the whole lattice and on the vectors $e_i$. {The terms proportional to $e^{iNpe}$, which are omitted in this $N\to\infty$ form,
 must be retained in the exact finite-$N$ Weyl symbol.}
\zd{For arbitrary $q\in{\cal O}'$ the function ${\cal P}$ is given explicitly by Eq. (\ref{PdefGen}) of Appendix \ref{Sect4}: the Wilson line of the $i$--th factor is centred at $c(q,\sum_j e_j)+\bar\Delta_i$ with $\bar\Delta_i=\frac12\sum_{j<i}e_j-\frac12\sum_{j>i}e_j$, where the admissible centre is built out of the {\it total} displacement of the chain, and the potential term is $A_0(\lfloor q\rfloor)$, as in Eq. (\ref{QW_hamiltonian}). We stress that the base point of the Wilson lines is $c(q,\sum_j e_j)$ rather than $x=q$ even on the coarse lattice ${\cal O}$; see the discussion after Eq. (\ref{PdefGen}) and, for constant field strength, after Eq. (\ref{PconstGen}).}

\cormag{For the general class (\ref{QW_general}) the temporal factors $\partial_s^k e^{s(i\omega-A_{0,W}(x))}$ precede the spatial chains with the ordering displayed there. The spatial Nyquist bookkeeping of Appendix \ref{Sect4} is unchanged for continuous Euclidean time; no claim about the additional temporal aliases of a finite time lattice is made.}
}

Below, the  Weyl symbol of the Keldysh Green function, $\hat{\bf G}$ is denoted by $\hat{G}$,
while the Weyl symbol of the Keldysh $\hat{\bf Q}$ is $\hat{Q}$. We omit the subscript $W$ for brevity in this section. The Weyl symbols $\hat{G}$ and $\hat{Q}$ obey Groenewold equation
\begin{equation}
\hat{Q} * \hat{G} = 1_W.
\end{equation}
Here the Moyal product $*$ is defined as
\begin{equation}
\left(A* B\right)(X|P) = A(X|P)\,e^{\rv{-}\ii(\overleftarrow{\partial}_{X^{\mu}}\overrightarrow{\partial}_{P_{\mu}}-\overleftarrow{\partial}_{P_{\mu}}\overrightarrow{\partial}_{X^{\mu}})/2}B(X|P).
\end{equation}
In the present paper we consider a specific  case where the electromagnetic potential
$A$ corresponds to constant components of the field strength ${\cal F}^{\mu\nu}$.

{Moreover, a particular gauge is chosen, in which the spatial part of the potential is time-dependent  but does not depend on spatial coordinates.}
An expansion in powers of
${\cal F}^{\mu \nu}$ will be  assumed up to the leading order, {which is} proportional to {the} electric  field. The introduction of such a form of the external gauge potential \zc{at $N\to \infty$} results in the
Peierls substitution $P \to \pi = P- A$. Here $\pi^\mu$ are the components of a $D+1$  dimensional vector, similar to the components $P^\mu$. When the index is lowered, its spatial components change  sign.
The Moyal product may be decomposed as
\begin{equation}
* =  \star~ e^{\rv{-}\ii  \mathcal{F}^{\mu\nu}\overleftarrow{\partial}_{\pi^{\mu}}\overrightarrow{\partial}_{\pi^{\nu}}/2}.
\end{equation}
with
\begin{equation}
\left(A\star B\right)(X|\pi) = A(X|\pi)\,e^{\rv{-\ii(\overleftarrow{\partial}_{X^{\mu}}\overrightarrow{\partial}_{\pi_{\mu}}-\overleftarrow{\partial}_{\pi_{\mu}}\overrightarrow{\partial}_{X^{\mu}})/2}}B(X|\pi).
\end{equation}
This expression remains valid specifically for the case of a given external field $A$ that does not depend on spatial coordinates, but depends on time, giving rise to an external electric field.
The case of an external field depending on spatial coordinates is more involved due to the presence of
a spatial lattice.

Next,  {we expand} $\hat{Q}$ and $\hat{G}$ in powers of  $\mathcal{F}^{\mu\nu}$  {through leading order:}
\begin{equation}
\hat{Q} = \hat{Q}^{(0)}  +\frac{1}{2}\mathcal{F}^{\mu\nu}\hat{Q}_{\mu\nu}^{(1)},\quad
\hat{G} = \hat{G}^{(0)}  +\frac{1}{2}\mathcal{F}^{\mu\nu}\hat{G}_{\mu\nu}^{(1)}.\label{QGK}
\end{equation}
Besides, below we omit for simplicity the superscript $^{(0)}$ on the zeroth order contribution to both $G$ and $Q$. For the case of Hamiltonian that does not depend on time, and the initial distribution $f(\pi_0)$
that depends on energy only, we have
\bes
G^\rR
&=(\pi_0-\hat{H}(\vec{\pi},x)+\ii \epsilon )^{-1},
\\
G^\rA  &= ( \pi_0-\hat{H}(\vec{\pi},x)-\ii \epsilon )^{-1},
\\
G^< &=(G^\rA -G^\rR ) f(\pi_0) = 2\pi i \delta(\pi_0-\hat{H}(\vec{\pi}))f(\pi_0).
\label{Gar_expl}
\end{eqsplit}
The matrix $\hat{\bf Q}^<$ is the inverse of $\hat{\bf G}^<$ (with respect to the Moyal product):
\bes
Q^{<}&=(Q^\rA -Q^\rR )f(\pi_0) = -2\ii\epsilon f(\pi_0),
\\
Q^\rR  &= \pi_0-\hat{H}(\vec{\pi},x)+\ii \epsilon ,	
\\
Q^\rA  &=  \pi_0-\hat{H}(\vec{\pi},x)-\ii \epsilon .
\end{eqsplit}
The Groenewold equation can be written as
\begin{equation}
\left(\hat{Q}  +\frac{1}{2}\mathcal{F}^{\mu\nu}\hat{Q}_{\mu\nu}^{(1)}\right)\star~ e^{\rv{-}i  \mathcal{F}^{\mu\nu}\overleftarrow{\partial}_{\pi^{\mu}}\overrightarrow{\partial}_{\pi^{\nu}}/2}\left(\hat{G}  +\frac{1}{2}\mathcal{F}^{\mu\nu}\hat{G}_{\mu\nu}^{(1)}\right) = 1_W.
\label{Groe-F}
\end{equation}
 At zeroth order in $\cal F$ the Groenewold equation is  $\hat{Q} \star \hat{G}  = 1_W$,
while the first order {contribution} reads
$\hat{Q} \star \hat{G}^{(1)}+\hat{Q}^{(1)}\star\hat{G} \rv{-} \ii \hat{Q} \star \overleftarrow{\partial}_{\pi^{\mu}}\overrightarrow{\partial}_{\pi^{\nu}} \hat{G}  = 0$. We come to
\begin{equation}
\hat{G}_{\mu\nu}^{(1)} =-\hat{G} \star  \hat{Q}_{\mu\nu}^{(1)}\star \hat{G}   \rv{-}   \ii\left(\hat{G} \star \partial_{\pi^{\mu}}\hat{Q}  \star\hat{G} \star \partial_{\pi^{\nu}}\hat{Q} \star \hat{G} -(\mu\leftrightarrow \nu)\right)/{2}
.
\label{QGK1}
\end{equation}

The derivation presented here is similar to that of \cite{onoda2}. However, it differs essentially because we consider the model defined on the finite lattice. In the case of a uniform system, {the} Weyl symbol of {the} operator $\hat{Q}$ does not depend on $x\in {\cal O}^\prime$. It may depend on $P^0$, $X^0$ and $p \in {\cal M}^\prime$. {The} electric current density on the lattice may be calculated as:
\be
J^i(t) = \rv{-}\frac{\ii }{2}\frac{1}{(2N)^{2D}} \int \frac{dP^0}{2\pi}  \sum_{\substack{p \in   {\cal M}^\prime \\ x\in {\cal O}^\prime}}
\tr\left(\hat{G} (\partial_{\pi_{i}}\hat{Q})\right)^{<}
\rv{-}\frac{\ii }{2}\frac{1}{(2N)^{2D}} \int \frac{dP^0}{2\pi}  \sum_{\substack{p \in   {\cal M}^\prime \\ x\in {\cal O}^\prime}}
\tr\left((\partial_{\pi_{i}}\hat{Q}) \hat{G}\right)^{<} .
\label{J Wigner}
\ee
Applying Eqs. (\ref{QGK})-(\ref{QGK1}), we calculate the contribution to the electric current proportional to external field strength $\mathcal{F}^{\mu\nu}$:
\begin{eqnarray}
{J}^i
&=&    -\frac{1}{4}\frac{1}{(2N)^{2D}} \int \frac{dP^0}{2\pi}  \sum_{\substack{p \in   {\cal M}^\prime \\ x\in {\cal O}^\prime}} \tr\Bigl(\hat{G} \star \partial_{\pi^{\mu}}\hat{Q}  \star\hat{G} \star \partial_{\pi^{\nu}}\hat{Q} \star \hat{G}  \partial_{\pi_{i}}\hat{Q} \Bigr)^{<}\mathcal{F}^{\mu\nu}\nonumber\\
&&
-\frac{1}{4}\frac{1}{(2N)^{2D}} \int \frac{dP^0}{2\pi}  \sum_{\substack{p \in   {\cal M}^\prime \\ x\in {\cal O}^\prime}} \tr\Bigl(\partial_{\pi_{i}}\hat{Q}  \hat{G} \star \partial_{\pi^{\mu}}\hat{Q}  \star\hat{G} \star \partial_{\pi^{\nu}}\hat{Q} \star \hat{G}  \Bigr)^{<}\mathcal{F}^{\mu\nu}.
\end{eqnarray}

Assuming that $\cal F$ includes only {the} electric field, we represent this expression  {for} two-dimensional systems as:
$$
{J}^i = \sigma^{ij}  \mathcal{F}_{0j} ,
$$
where the conductivity tensor $\sigma^{ij}$ may be given as follows:
\begin{equation}
\sigma^{ij} =  {\frac{1}{4}} \frac{1}{(2N)^{2D}} \int \frac{dP^0}{2\pi}  \sum_{\substack{p \in   {\cal M}^\prime \\ x\in {\cal O}^\prime}} \tr\left(\partial_{\pi_{i}}\hat{Q}_W  \left[\hat{G}_W \star \partial_{\rv{\pi_{[0}}}\hat{Q}_W  \star \partial_{\rv{\pi_{j]}}}\hat{G}_W  \right]\right)^< +{\rm c.c.}\label{MAIN}
\end{equation}
In this expression we restore {the} subscript  $W$ for the Weyl symbols.
Here $(...)_{[0} (...)_{ j]} =(...)_{0} (...)_{ j} -(...)_{j} (...)_{ 0}  $ means anti-symmetrization.
{The} conductivity {tensor} may be expressed as a sum of symmetric and anti-symmetric parts
$\sigma^{ij} = \sigma^{ij}_{H} + \sigma^{ij}_S$.
Here, {the} asymmetric part is  $\sigma^{ij}_H = (\sigma^{ij}-\sigma^{ji})/2$ and {is equal to} the Hall  conductivity,
while {the symmetric part is} $\sigma^{ij}_S = (\sigma^{ij}+\sigma^{ji})/2$, {and is equal to the} conventional conductivity.

\label{SectEquilibrium}

{One of the most fundamental properties of the Weyl symbol is that
if there is a  sum over all phase space,
the star may be inserted between  two Weyl symbols within the argument of a trace operator \ref{MAIN}.}
Our lattice version of Weyl symbol also obeys this property. Therefore, after averaging the conductivity over the whole volume of the system and over the overall time of the process, we obtain
\be
\bar \sigma^{ij}
=  \rv{-}{\frac{1}{4}} \frac{1}{(2N)^{2D}} \int \frac{dP^0 dX^0}{2\pi (t_f-t_i)}  \sum_{\substack{p \in   {\cal M}^\prime \\ x\in {\cal O}^\prime}}
\tr\left(\partial_{\pi_{i}}\hat{Q}_W \star \hat{G}_W \star \partial_{\pi_{[0}}\hat{Q}_W  \star \hat{G}_W\star \partial_{\pi_{j]}}\hat{Q}_W\star\hat{G}_W\right)^< +{\rm c.c.}\label{Sigma}
\ee
Recall that the external electromagnetic potential depends on time, giving rise to {an} electric field. \zz{ If we are interested in the linear term of the expansion in powers of external the electric field, then we can omit the time dependence in $Q_W$ and $G_W$ entering the above expression. We will see in the next sections, using {the} machinery of cyclic cohomology, that when the initial distribution is equilibrium thermal distribution with zero temperature, the imaginary time is discretized,  and the space-time lattice is of finite size, } \zc{ expression of Eq. (\ref{Sigma}) for  the Hall current vanishes identically if we calculate it on the finite lattice. However, this does not mean that the Hall conductivity itself vanishes on the finite lattice. The point is that representation for the conductivity of Eq. (\ref{Sigma}) is valid only in the limit $N \to \infty$ while at finite $N$ in order to calculate the conductivity we should substitute in Eq. (\ref{Sigma}) derivative with respect to $\pi_i$ by derivative with respect to $\theta_i$, where $\theta_i$ is a small constant addition to external field $A$, to be set to zero after differentiation. }

\zzzz{We will see that the situation is changed drastically if we take the limit $N\to \infty$ directly (but partially) in Eq. (\ref{Sigma}).}
Namely, in this limit the sum over momenta becomes an integral: $\frac{(2\pi)^D}{(2N)^D}\sum_{p\in {\cal M}'} \to \int_{{\cal M}'} d^D p$. {We take the principal part of Eq. (\ref{QW_general}) that  contains  all elementary factors kept on their principal branch, as the asymptotic bulk symbol, neglecting the terms proportional to $e^{iNpe}$.
These are precisely the terms proportional to $e^{iNp}$ in Eq. (\ref{ZNp})].
Their contributions to normalized traces vanish under the uniform-smoothness condition stated in Appendix \ref{Sect4}.} \corrteal{The principal part so defined is precisely the exact Weyl symbol (\ref{infW}) of the periodized operator of Sect. \ref{SectPeriodized}: see Eq. (\ref{WprIsW}) there, and Eq. (\ref{WeylTorusSplit}) for the direct derivation from Eq. (\ref{Weyl2}) with the explicit form of the discarded terms.}
At the same time, we may leave  the sum over ${\cal O}'$ with finite $N$.
Inside the resulting expression we may perform {a} change of variables $p + Et \to p$.  \zzzz{In the final expression for the observable quantities we should take the remaining limit $N\to \infty$.}

{In both cases mentioned above: 1) when we keep the lattice size, $N$ finite and consider the linear response of
the electric current to the electric field, 2) when we take  Eq. (\ref{Sigma}) in the limit $N\to \infty$, we can treat $G_W$ and $Q_W$ as
independent of time. Then, for the initial thermal distribution, we are able to represent the integral over the frequency as a sum over Matsubara frequencies. Now by $p$ we denote  Euclidean $D+1=3$-momentum, i.e. $p^{3} = \omega$ is the Matsubara frequency, while $p^i = \pi^i$ for $i=1,2$. Notice that $\partial_{\pi^0} = -\ii \partial_{p^{3}}$. Substituting $i\omega$ instead of $\pi^0$, we obtain the Matsubara Green function $G^M$ instead of the advanced or retarded Green function.}

More specifically, for the system with the one-particle Hamiltonian $\hat{H}$, one can define the real-time Green function as
$
G(x_1,x_2,\omega) \equiv \langle x_1|(\omega - \hat{H})^{-1}|x_2\rangle
$. 
The Matsubara Green's function $G^\rM $ is then defined as
\be
G^\rM ( x, x^{\prime},\omega_n) = G( x, x^{\prime},\ii\omega_n),
\label{Mats}
\ee
Here, $\omega_n = (2n+1)\pi/{\beta}$ is the Matsubara frequency while $\beta= 1/T$ is the inverse  temperature.
\cormag{For the noninteracting Hamiltonian system treated in the preceding Keldysh derivation, the zero-temperature conductivity averaged over the area is}
\cormag{\begin{equation}
\bar{\sigma}^{ij}  = \frac{{\cal N}_{\zc{3}}}{2\pi}\epsilon^{ij},\label{sigmaHall}
\end{equation}}
\corrcltwo{where ${\cal N}_3$ is given by Eq. (\ref{N3true}) below. In it, $D_0=\partial_{p^3}$, while $D_j=\partial_{\theta^j}$ ($j=1,2$) denotes the derivative with respect to the uniform constant addition $\theta^j$ to the external vector potential ($A_j\to A_j+\theta^j$, i.e. every elementary hopping with the displacement $d$ is multiplied by $e^{i\theta\cdot d}$), taken at $\theta=0$: this is the exact Peierls (current) vertex of the finite lattice. \corrteal{It must be distinguished from the inner derivation $i[\hat x_j,\,\cdot\,]$ generated by the coordinate operator $\hat x_j$ of the finite torus. The two coincide on the hoppings inside the torus, for which the displacement $d_j$ equals the difference $x_j(X)-x_j(Y)$ of the coordinates of the final and the initial sites, but they differ on the hoppings crossing the boundary of the torus in the $\pm j$ direction, for which $x_j(X)-x_j(Y)=d_j\mp N$. Explicitly, $D_j\hat Q = i[\hat x_j,\hat Q] + iN\big(\hat T^{+}_j-\hat T^{-}_j\big)$, where $\hat T^{\pm}_j$ is the part of $\hat Q$ formed by the hoppings crossing the boundary in the $\pm j$ direction (the overall sign convention of $\theta$ drops out of Eq. (\ref{N3true}), since every term there contains exactly two spatial derivatives). The boundary term is essential: were $D_j$ the inner derivation $i[\hat x_j,\,\cdot\,]$ of the finite matrix algebra, Eq. (\ref{N3true}) would vanish identically at any finite $N$ by the theorem of Sect. \ref{SectDirect}; with the Peierls vertex it does not vanish, but it is not quantized either. (This is the boundary correction to the relation $D_j\hat A=-i[\hat x_j,\hat A]$ on the finite lattice discussed in Appendix \ref{AppDA}: on the finite lattice the coordinate labels are reduced modulo $N$, while the displacement brought down by the momentum derivative is not.) In the language of Sect. \ref{SectPeriodized}, $D_j$ is the outer derivation $-i[\hat X_j,\,\cdot\,]$ of the periodized crystal --- $\hat X_j$ being the position operator of the infinite superlattice, which is not reduced modulo $N$ --- restricted to one cell: in the uniform gauge its fiber is $\pm\partial_{\theta^j}$, in the boundary gauge it is given by Eq. (\ref{commfiber}).} The relation of $\theta$ to the twisted boundary conditions --- with the twist angles equal to $L\theta$ up to orientation --- is explained in Sect. \ref{SectPeriodized}:}
\cormag{\begin{eqnarray}
		{\cal N}_3 &=& \frac{1}{3!\,}\epsilon^{\mu\nu\rho} \frac{1}{(2N)^{2D}} \int {dp^3}  \sum_{\substack{p \in   {\cal M}^\prime \\ x\in {\cal O}^\prime}}
		\tr\left(D_{\mu}\hat{Q}_W ^\rM  \star\hat{G}_W ^\rM \star D_{\nu}\hat{Q}_W ^\rM  \star\hat{G}_W ^\rM \star D_{\rho}\hat{Q}_W ^\rM \star \hat{G}_W ^\rM \right)\Big|_{\theta = 0}\label{N3true}
\end{eqnarray}}
(finite lattice, linear response of electric current to electric field, the dependence on $Et$ inside $Q_W$ and $G_W$ is disregarded). This expression ceases to be topological invariant for finite $N$. We can also consider expression 
\begin{eqnarray}
{\cal N}_3 &=& \frac{1}{3!\,}\epsilon^{\mu\nu\rho} \frac{1}{(2N)^{2D}} \int {dp^3}  \sum_{\substack{p \in   {\cal M}^\prime \\ x\in {\cal O}^\prime}}
\tr\left(\partial_{\rv{p^{\mu}}}\hat{Q}_W ^\rM  \star\hat{G}_W ^\rM \star \partial_{\rv{p^{\nu}}}\hat{Q}_W ^\rM  \star\hat{G}_W ^\rM \star \partial_{\rv{p^{\rho}}}\hat{Q}_W ^\rM \star \hat{G}_W ^\rM \right),
\label{NEQD}
\end{eqnarray}
\zd{in which precise Weyl symbol $Q_W$ is replaced by the principal symbol of Eq. (\ref{Wprincipal}), while  precise $G_W$ is replaced by the inverse to Eq. (\ref{Wprincipal}) with respect to the star product.  This expression gives a reasonable approximaton to the ${\cal N}_3$ of Eq. (\ref{N3true}) entering Eq. (\ref{sigmaHall}).  } 

  Another option is to take
\zz{\begin{eqnarray}
{\cal N}_3 &=& \frac{1}{3!\,}\epsilon^{\mu\nu\rho} \frac{1}{(2N)^{D}} \int \frac{d^3p}{(2\pi)^2}  \sum_{ x\in {\cal O}^\prime}
\tr\left(\partial_{\rv{p^{\mu}}}\hat{Q}_W ^\rM  \star\hat{G}_W ^\rM \star \partial_{\rv{p^{\nu}}}\hat{Q}_W ^\rM  \star\hat{G}_W ^\rM \star \partial_{\rv{p^{\rho}}}\hat{Q}_W ^\rM \star \hat{G}_W ^\rM \right)
\label{NEQ}
\end{eqnarray}
(we replace the sum over momenta by the integral in the limit of infinite $N$, and then in Eq. (\ref{Sigma}), $Et$ is absorbed by momenta). \zc{Moreover, we neglect in expression for the Weyl symbol of operators the terms proportional $(1-e^{i\pi/N})e^{iNp}$ entering, for example, Eq. (\ref{ZNp}), and use Eq. (\ref{Wprincipal}) as the definition of Weyl symbol of an operator, which means that the limit $N\to \infty $ in $Q_W$ is taken before we perform differentiation with respect to momentum. With this definition Eq. (\ref{NEQ})  represents another approximation to Hall conductivity of a system with large $N$ according to $\bar{\sigma}^{ij}  = \frac{{\cal N}_{\zc{3}}}{2\pi}\epsilon^{ij}$.}
\cormag{The last physical identification is proved here for the Hamiltonian kernel; for a general frequency-dependent $Q$, Eq. (\ref{NEQ}) is the formal odd expression unless the additional real-time and Ward-identity hypotheses stated above are supplied.} \corrteal{For the periodized systems the identification is extended to the general class (\ref{Q_general}) in Sect. \ref{SectStaticGauge}, \corGPTl{subject to its additional equilibrium, coupling, and limiting assumptions}. Let us also stress that Eq. (\ref{NEQ}) is well defined at every finite $N$, not only for large $N$, and is there a topological invariant equal to the twist-averaged invariant (\ref{NEQnum}) of the periodized crystal --- the Proposition of Sect. \ref{SectPeriodized}\,B; the fixed-twist quantity (\ref{N3true}), which is not quantized, is approximated by it with the accuracy improving with $N$ (Sect. \ref{SectNumerics}).}
Here  $\epsilon^{ij}$ and $\epsilon^{\mu\nu\rho}$  are standard antisymmetric tensors. $Q_W^M$ is {the} inverse  {of the} Matsubara Green function $G_W^M$:
$$
Q_W^M \star G_W^M = 1_W  = 1
$$
In the following we will omit for brevity the superscript $\rM$ for the Matsubara Green function.}

%%%%%%%%%%%%%%%%%%%%%%%%%%%%%%%%%%%%%%%%%%%%%%%%%%%%%%%%%%%%%%%%%%%%%%%%%%%%%%%%%%%%%%%%%%%%%%%
%%%%%%%%%%%%%%%%%%%%%%%%%%%%%%%%%%%%%%%%%%%%%%%%%%%%%%%%%%%%%%%%%%%%%%%%%%%%%%%%%%%%%%%%%%%%%%%
\section{\corrteal{Periodized systems: supercells, twisted boundary conditions, and the Hall conductivity in the static gauge}}
\label{SectPeriodized}

\corrteal{In this section we introduce the operator counterpart of the partial thermodynamic limit: the periodization of a finite inhomogeneous configuration to an infinite crystal with a supercell, its Bloch decomposition into the fibers labelled by the twisted boundary conditions, and the corresponding exact operator form of the invariant (\ref{NEQ}). Throughout the section the Dirac operator is {\it not} assumed to be of the Hamiltonian form: the constructions apply to the general class (\ref{Q_general}), the Hamiltonian case (\ref{Q_general0}) being a particular example. We then use this machinery to derive the topological expression for the Hall conductivity \corGPTl{under the additional equilibrium, coupling, and limiting assumptions stated in Sect. \ref{SectStaticGauge},} entirely within the equilibrium Matsubara technique, in the static gauge in which the constant electric field enters through the scalar potential of the superlattice, $A_0 \to A_0 - E X_1$. The derivation follows the strategy of Sects. III and IV of Ref. \cite{FSWZZ2020}, with one essential improvement: there the approximate lattice Wigner--Weyl calculus restricted the result to slowly varying external fields, whereas here every step is an exact identity of the precise calculus of the periodized lattice, valid for arbitrarily strong inhomogeneity --- and, unlike the Keldysh derivation of Sect. \ref{SectKeldyshCond}, not restricted to $Q = i\omega - H$.}

\subsection{\corrteal{Periodization and the supercell Bloch decomposition}}

\corrteal{The construction is formulated directly for the general Dirac operator of the class (\ref{Q_general}). At a fixed value of the Matsubara frequency $\omega$ the operator $\hat Q$ acts on the functions of the lattice coordinates, and \corGPTl{we additionally restrict the chains with nonzero coefficients in Eq. (\ref{Q_general}) to finitely many total displacements, each of magnitude less than $L/2$ in every coordinate. This finite-range assumption does not follow from the infinite sum over chain lengths. The temporal factors (\ref{F0def}) are diagonal in the coordinates.} \corrteal{(The restriction serves the reconstruction of the periodized operator from the torus one; how it is lifted by the chain-wise periodization is explained in the remark on the range condition at the end of Sect. \ref{SectPeriodized}\,A.)} The construction below uses nothing but this finite range; the frequency is a spectator, so that all the formulas hold at every $\omega$. The Hamiltonian case $Q = i\omega - H + \mu$ is contained as a particular example, and the reader may keep it in mind for the comparison with Sect. \ref{SectNumerics}.}

\corrteal{For the convenience of the reader we repeat the definition (\ref{Q_general}):
\be
\hat{Q} \;=\; \sum_{k\geq 0}\;\sum_{s\in{\cal S}_t}\;\sum_{n=0}^{\infty}\;\sum_{e_1\dots e_n}\;{\bf t}^{(k,s;\,e_1,\dots,e_n)}\;\Big[\partial^{\,k}_{s}\hat{F}^{0}_{s}\Big]\;\hat{Q}^{(e_1)}\cdots\,\hat{Q}^{(e_n)}\,,
\qquad
\hat{F}^{0}_{s}\;=\;\exp\Big[\,s\,\big(i\partial_t-A_0(i\partial_{p})\big)\Big]\,,
\label{QgenRecap}
\ee
with the constant coefficients ${\bf t}$, the elementary temporal factors $\hat F^{0}_{s}$, $s\in{\cal S}_t$, and the elementary magnetic translations $\hat Q^{(e)}$ dressed by the Wilson lines of the external gauge field. The Hamiltonian case is the contribution of the first $s$-derivative at $s=0$, $\partial_s\hat F^{0}_{s}\big|_{s=0}=i\partial_t-A_0$, together with the purely spatial chains.}

\corrteal{Let us characterize the resulting dependence on the frequency. At a fixed $x$ the temporal factors span the exponential polynomials $z^{k}e^{sz}$ of $z=i\omega-A_{0}$; no specific functional form of the $\omega$-dependence is thereby prescribed. \corGPTl{On the compact Matsubara circle of Sect. \ref{SectKeldyshCond}, choosing $s\in\mathbb Z$ and $k=0$ supplies the Fourier modes, whose span is dense in the smooth periodic functions of $\omega$ (with the temporal lattice spacing set to one). A general prescribed set ${\cal S}_t$ need not have this completeness property; terms with $k>0$ require the complete sum to be smooth and periodic.} On the noncompact frequency line an arbitrary dependence, analytic in a neighbourhood of the contour, is admissible whenever it satisfies the three structural requirements of Sect. \ref{SectKeldyshCond}: the invertibility along the contour, the polynomial-type asymptotics $Q\sim(i\omega)^{m}\times(\hbox{invertible})$ with the vanishing boundary terms, and \corGPTl{preservation of these properties along the homotopies. The additional Matsubara reality condition is $Q(i\omega)^{\dagger}=Q(-i\omega)$} --- the rational dependence generated by the interactions (Appendix \ref{AppHK}) being the paradigm. We stress that analyticity itself is not what the topological constructions use: smoothness and invertibility on the contour, together with the asymptotics, suffice; the analyticity in a strip is what the physical Matsubara propagators supply for the causal (KMS) reasons.}

\corAOP{The invariant is evaluated through the following periodization construction. Let $Q$ be the \corrteal{Dirac operator of the class (\ref{Q_general}) for} the given $L\times L$ configuration (a concrete realization of the random potential, the given smooth potential, the given pattern of magnetic fluxes)\corrteal{, taken at a fixed value of the Matsubara frequency, which enters below as a spectator parameter and is suppressed in the notation}. Define the operator $\tilde{Q}$ on the infinite lattice $\mathbb{Z}^2$ by repeating this configuration with the period $L$ in both directions:
\be
\corr{\tilde{Q}_{X,\,Y} \;=\;
\begin{cases}
Q_{X\,{\rm mod}\,L,\;\,Y\,{\rm mod}\,L}\,, & X-Y\,\in\,\left(-\frac{L}{2},\,\frac{L}{2}\right)^{2},\\[4pt]
\;0\,, & {\rm otherwise}\,,
\end{cases}
\qquad X,\,Y\,\in\,\mathbb{Z}^{2}\,,}
\label{Hper}
\ee
\corr{Here $Q_{x,y}$ ($x,y \in {\rm cell}$) are the matrix elements of the \corrteal{Dirac operator} of the $L\times L$ torus, and $X\,{\rm mod}\,L \in {\rm cell}$ denotes the componentwise residue. Thus every matrix element of $Q$ is repeated with the period $L$ and is placed at the minimal separation $X-Y$ of the two sites compatible with their positions inside the cell; in particular, the matrix elements of $Q$ that connect the opposite boundaries of the torus connect in $\tilde{Q}$ the neighboring copies of the cell. The construction is unambiguous provided the range of $Q$ is smaller than $L/2$\corrteal{ (for the general class the range is the maximal total displacement of the chains entering Eq. (\ref{Q_general}))}; for the nearest--neighbor models considered \corrteal{in Sect. \ref{SectNumerics}} the nonzero matrix elements of $\tilde{Q}$ have $X-Y \in \{0,\,\pm e_1,\,\pm e_2\}$. The operator $\tilde{Q}$ is \corrteal{periodic (the Hermiticity of the Hamiltonian case is replaced by the reality condition $\tilde{Q}(\omega)^\dagger = \tilde{Q}(-\omega)$, \corGPTl{when this additional reality condition is imposed as discussed in Sect. \ref{SectKeldyshCond}})}, $\tilde{Q}_{X+L{\bf m},\;Y+L{\bf m}} = \tilde{Q}_{X,Y}$, with the {\it same} shift ${\bf m}\in\mathbb{Z}^2$ of both arguments. Let us stress that the matrix elements of $\tilde{Q}$ between {\it different} copies of the cell do not all vanish --- only those between non--adjacent copies do. By Eq. (\ref{Hper}) the element $\tilde{Q}_{X,Y}$ is nonzero whenever the separation $X-Y$ lies within the range of $Q$, irrespective of how the two sites are situated with respect to the partition of $\mathbb{Z}^2$ into the cells. In particular, for $x$ and $y$ near the opposite boundaries of the cell the nonvanishing element is $\tilde{Q}_{x+L{\bf n},\,y}$ with ${\bf n}\neq 0$, connecting the neighboring copies, whereas the same--copy element $\tilde{Q}_{x,y}$ vanishes. E.g., in one dimension, for $x=0$, $y=L-1$: the sites $x$ and $y$ are $L-1$ steps apart, so that $\tilde{Q}_{x,y}=0$, while the sites $x+L$ and $y$ are nearest neighbors, and $\tilde{Q}_{x+L,\,y} = Q_{0,\,L-1}$ is the matrix element of the torus \corrteal{operator} connecting its opposite boundaries. These inter--copy elements are essential: if all the matrix elements between different copies vanished, one would obtain $\langle \phi,x|\tilde{M}|\theta,y\rangle = (2\pi)^2\delta^{(2)}(\phi-\theta)\,\langle x|\tilde{M}|y\rangle$ in the Bloch basis introduced after Eq. (\ref{trfiber}), the fiber matrices would not depend on the quasimomentum at all, the boundary conditions of the fiber problem would never be twisted, and the average over $\theta$ would be trivial. (It would be incorrect to write $\tilde{Q}_{x+L{\bf m},\;y+L{\bf m}'} = Q_{x,y}$ with independent ${\bf m},{\bf m}'$: such a relation would couple every copy of the cell to every other copy with equal amplitude, thus describing a nonlocal operator rather than a crystal; we return to this point after Eq. (\ref{trfiber}).) The inverse relation, expressing $Q$ through $\tilde{Q}$, will also be given there.}
This is an ordinary crystal whose unit cell contains $L^2$ sites carrying the entire inhomogeneity pattern. To such a system the calculus of \cite{FZ2019_2} applies verbatim: the position operator $\hat{x}$ is well defined, the Weyl symbols contain no Nyquist companion terms, and the momentum derivative of a symbol is exactly the symbol of the commutator with the position operator, $D_j a = -i[\hat{x}_j, a]$ (cf. the beginning of Sect. \ref{SectDirect}). The invariant contains the trace per unit volume rather than the bare trace (cf. Eq. (\ref{cal01})); for a periodic operator $\tilde{M}$ the trace per unit volume is finite and reduces to one period:
\be
\corrgpt{\tau(\tilde M) \;:=\; \lim_{\Lambda\nearrow\mathbb Z^2}\,
\frac{1}{|\Lambda|}\sum_{x\in\Lambda}\langle x|\tilde M|x\rangle
\;=\;\frac{1}{L^2}\sum_{x\in{\rm cell}}\langle x|\tilde M|x\rangle .}
\label{trper}
\ee
\corGPTl{The limit is taken along expanding boxes (or a F\o lner sequence).}
Thus $|{\cal O}| = L^2$ below refers to the supercell, and no infinite quantity appears. Next, $\tilde{Q}$ commutes with the translations by $(L,0)$ and $(0,L)$, so that the Bloch theory with respect to the superlattice applies. The Hilbert space decomposes into \corGPTl{$rL^2$-dimensional fibers when there are $r$ internal states per site; all traces below include those internal states, and the subscript $L^2$ counts sites. The fibers are} labelled by the quasimomentum of the supercell, which we parametrize by $\theta \in [0,2\pi)^2$; the fiber \corrteal{operator} is precisely the \corrteal{Dirac operator} of the $L\times L$ torus with the twisted boundary conditions, $Q(\theta)$. For any operator $\tilde{M}$ built out of $\tilde{Q}$ the normalized trace becomes the average over the twists of the fiber traces:
\be
\frac{1}{L^2}\sum_{x\,\in\, {\rm cell}} \langle x |\tilde{M}|x\rangle \;=\; \int_{[0,2\pi)^2} \frac{d^2\theta}{(2\pi)^2}\; \frac{1}{L^2}\, {\rm Tr}_{L^2} \big[ M(\theta)\big] .
\label{trfiber}
\ee
\corr{Let us spell out the right--hand side of Eq. (\ref{trfiber}). The Bloch states of the superlattice are $|\theta,x\rangle = \sum_{{\bf m}\in\mathbb{Z}^2}e^{i\theta{\bf m}}|x+L{\bf m}\rangle$, where $x$ runs over the $L^2$ sites of one supercell, while $\theta \in [0,2\pi)^2$ is the quasimomentum of the supercell (the superlattice period is taken as the unit of $\theta$). Any operator $\tilde M$ commuting with the translations by $L$ acts within the fibers spanned by these states, $\tilde M |\theta, y\rangle = \sum_{x} M(\theta)_{x,y}\,|\theta,x\rangle$, with the $L^2\times L^2$ fiber matrix
$$
M(\theta)_{x,y} \;=\; \sum_{{\bf n}\in \mathbb{Z}^2} e^{-i\theta{\bf n}}\; \langle x + L{\bf n}\,|\tilde M |\, y \rangle\,, \qquad x,\,y \,\in\, {\rm cell}\,.
$$
Equivalently, in terms of the matrix elements between the Bloch states,
$$
\langle \phi, x\,|\,\tilde M\,|\,\theta, y\rangle \;=\; (2\pi)^2\,\delta^{(2)}(\phi-\theta)\; M(\theta)_{x,y}\,.
$$
Indeed, using the periodicity $\langle x+L{\bf m}|\tilde M|y+L{\bf m}'\rangle = \langle x+L({\bf m}-{\bf m}')|\tilde M|y\rangle$ and summing at fixed ${\bf n}={\bf m}-{\bf m}'$, one finds $\sum_{{\bf m}'}e^{i(\theta-\phi){\bf m}'} = (2\pi)^2\delta^{(2)}(\theta-\phi)$ multiplied by the sum over ${\bf n}$ written above. We stress that the diagonal (in the quasimomentum) block is the full fiber matrix $M(\theta)_{x,y}$, and not the single matrix element $\langle x|\tilde M|y\rangle$: the sum over ${\bf n}$ does not collapse, because $\tilde M$ possesses matrix elements between different copies of the cell --- already $\tilde Q$ connects the neighboring copies (Eq. (\ref{Hper})), while, e.g., its resolvent connects all of them, with exponentially decaying amplitudes. It is precisely the surviving sum over ${\bf n}$ that generates the $\theta$--dependence of the fibers: an operator with no matrix elements between different copies would have $\theta$--independent fibers, and the average over the twists would be trivial. (We have verified the displayed relation numerically as well, within machine precision, both for the periodized \corrteal{Hamiltonian of Sect. \ref{SectNumerics}} and for its resolvent.) By ${\rm Tr}_{L^2}$ we denote the ordinary matrix trace over the $L^2$--dimensional fiber space (the space of functions on one supercell)\corrteal{, of dimension $rL^2$ in the presence of $r$ internal states per site, the trace over the internal indices being included}, ${\rm Tr}_{L^2} M(\theta) = \sum_{x\in{\rm cell}} M(\theta)_{x,x}$. At every fixed $\theta$ the map $\tilde M \mapsto M(\theta)$ is a homomorphism of operator algebras: it takes sums to sums, products to products, and inverses to inverses. \corrgpt{Therefore products and resolvents can be computed in the fibers. For expressions containing position commutators, the commutators must also be fibered according to Eq. (\ref{commfiber}); replacing the infinite position operator by the diagonal position matrix of one cell would not give the correct current vertices.} For $\tilde{Q}$ itself Eq. (\ref{Hper}) implies that for every pair $(x,y)$ at most one term of the sum over ${\bf n}$ is nonzero --- the one with $x+L{\bf n}-y \in (-L/2,\,L/2)^2$, to be denoted ${\bf n}(x,y)$ --- so that $Q(\theta)_{x,y} = e^{-i\theta\,{\bf n}(x,y)}\,Q_{x,y}$: the matrix elements with ${\bf n}(x,y)=0$ are unchanged, while those connecting the opposite boundaries of the cell (${\bf n}(x,y) = \pm e_j$) acquire the factors $e^{\mp i\theta_j}$. This $Q(\theta)$ is precisely the \corrteal{Dirac operator} of the $L\times L$ torus with the twisted boundary conditions $\psi(x+L e_j) = e^{i\theta_j}\psi(x)$, and $Q(0)=Q$; equivalently, Eq. (\ref{Hper}) is inverted as $Q_{x,y} = \sum_{{\bf n}} \tilde{Q}_{x+L{\bf n},\;y}$. (This also makes precise the warning given after Eq. (\ref{Hper}): for the incorrect periodization with independent shifts, $\tilde{Q}_{x+L{\bf m},\;y+L{\bf m}'}=Q_{x,y}$, all the terms of the sum over ${\bf n}$ would be equal, and the fiber matrix would degenerate into $\sum_{{\bf n}} e^{-i\theta{\bf n}}\, Q_{x,y} = (2\pi)^2 \delta^{(2)}(\theta)\, Q_{x,y}$ --- a delta function of the quasimomentum instead of a family of twisted Hamiltonians.) \corGPTl{For nearest-neighbor hopping along the coordinate axes, the sum over ${\bf n}$ for the fiber operator contains only ${\bf n}=0,\,\pm e_1,\,\pm e_2$, and $Q(\theta)$ is the finite combination}
$$
Q(\theta) \;=\; Q_0 \;+\; \sum_{j=1,2}\Big( e^{-i\theta_j}\, T_j \;+\; e^{+i\theta_j}\, \bar{T}_j\Big)\,,
$$
where $Q_0$ collects the matrix elements of $\tilde{Q}$ inside one copy, while $T_j$ collects those connecting the copy to its neighbor in the $+j$ direction, $(T_j)_{x,y} = \tilde{Q}_{x+Le_j,\,y}$\corrteal{, and $\bar{T}_j$ those connecting it to the neighbor in the $-j$ direction, $(\bar{T}_j)_{x,y} = \tilde{Q}_{x-Le_j,\,y}$; for a Hermitian operator $\bar{T}_j = T_j^{\dagger}$} (for a hopping range exceeding one lattice spacing, but still smaller than $L/2$, the corner terms $e^{-i(\theta_1\pm\theta_2)}(\ldots)$ would be added). The same finite form, \corGPTl{with corner terms when present,} applies to every operator whose range is smaller than $L/2$ \corrteal{(for a larger range, or for the chain-wise periodization of the remark on the range condition below, the sum over the windings ${\bf n}$ simply contains more terms)} --- in particular to the current vertices $-i[\hat{x}_j,\tilde{Q}]$ of Eq. (\ref{commfiber}) below. It does {\it not} apply to general functions of $\tilde{Q}$: e.g. the Green function connects arbitrarily distant copies, and its fiber is an infinite Fourier series in $e^{\pm i\theta_j}$. The latter, however, need not be computed elementwise: by the homomorphism property, \corrteal{$G_b(\theta) = Q(\theta)^{-1}$}, so that the entire $\theta$--dependence enters through the finite combination displayed above, and the fiber Green function is a rational function of $e^{\pm i\theta_j}$ --- the fact used in the discussion of the twist average \corrteal{in Sect. \ref{SectNumerics} below}. The equality (\ref{trfiber}) itself is the exact Fourier inversion: in the diagonal fiber element integrated over $\theta$ only the term with ${\bf n}=0$ survives, since $\int_{[0,2\pi)^2}\frac{d^2\theta}{(2\pi)^2}\,e^{-i\theta{\bf n}} = \delta_{{\bf n},0}$, so that
$$
\int_{[0,2\pi)^2} \frac{d^2\theta}{(2\pi)^2}\; M(\theta)_{x,x} \;=\; \langle x |\tilde M| x\rangle\,,
$$
and the sum over the sites of the cell yields Eq. (\ref{trfiber}). Thus Eq. (\ref{trfiber}) involves no approximation: it is an identity valid at any finite $L$. (We have also confirmed it numerically \corrteal{for the Hamiltonian example} --- together with the homomorphism property, the coincidence of the union of the fiber spectra with the spectrum of $\tilde H$, and the identification of $H(\theta)$ with the twisted--boundary--condition Hamiltonian --- within machine precision, $\sim 10^{-15}$, for random $L$--periodic tight--binding models.)}
}

\corrteal{{\bf The range condition, and how to lift it.} The condition ``range smaller than $L/2$'' is not a condition on the periodized construction itself, but on the reconstruction of the periodized crystal from the torus operator, and it can be lifted by changing what is taken as the primary object. Given an $L$-periodic operator $\widetilde Q$ on $\mathbb{Z}^2$ of arbitrary --- even infinite --- range, with absolutely convergent sums, the Bloch decomposition, the fibers
$$
Q(\theta)_{x,y}\;=\;\sum_{{\bf n}\in\mathbb{Z}^2}e^{-i\theta\cdot{\bf n}}\,\widetilde Q_{x+L{\bf n},\;y}
$$
with all the windings ${\bf n}$, the trace identity (\ref{trfiber}), the current vertices (\ref{commfiber}), the map $W_{\rm pr}$ and the Proposition of Sect. \ref{SectPeriodized}\,B hold as they stand; the Green function, whose range is infinite, has already been treated in this way above. A matrix element connecting $y$ to $x+L{\bf n}$ is the hop from $y$ to $x$ ``through the boundary of the supercell'' ${\bf n}$ times, and it carries the phase $e^{-i\theta\cdot{\bf n}}$. The range enters only when one starts from the torus operator. The latter is the $\theta=0$ fiber, $Q_{x,y}=\sum_{\bf n}\widetilde Q_{x+L{\bf n},\,y}$, which sums all the windings, and the windings can be disentangled from it only if at most one of them contributes to each matrix element --- which is the case for the range smaller than $L/2$, when the minimal image is forced. Without this condition the torus operator does not determine the crystal: assigning one and the same element to the winding ${\bf n}$ or to the winding ${\bf n}'$ produces fibers differing by the factor $e^{-i\theta\cdot({\bf n}-{\bf n}')}$ at $\theta\neq0$, i.e. different periodic crystals sharing the $\theta=0$ fiber and having, in general, different twist averages. (The simplest illustration is the ring of two sites: assigning the hop from the site $1$ back to the site $0$ the displacement $+1$ produces the uniform infinite chain, assigning it $-1$ produces decoupled dimers; the torus operator is the same, the twist dependence is not.) For the Dirac operators of the class (\ref{Q_general}) the winding is never ambiguous, since every chain has its own path on $\mathbb{Z}^2$, with the total displacement $d=\sum_i e_i$ and the Wilson line along that path. The natural definition of the periodized operator is therefore {\it chain-wise}: every chain is placed at its actual displacement on the infinite lattice, in the $L$-periodic pattern of the external fields, and the winding number of the chain is part of the data. With this definition no range condition is needed --- the coefficients ${\bf t}$ may decay with the length of the chain rather than terminate, the sums being absolutely convergent --- and $W_{\rm pr}(\widetilde Q)$ coincides with the principal symbol (\ref{QW_general}) by construction, Eq. (\ref{PdefGen}) being itself a sum over the chains labelled by their total displacement, with the admissible centre. The external gauge field is treated in the same way. The Wilson line of a chain winding around the torus is taken along its actual path, so that the long and the short way between the same two sites of the torus differ by the Wilson loop around the noncontractible cycle --- in a magnetic field a position-dependent phase, $e^{2\pi i k\,x_\perp/L}$ in the Landau gauge for $k$ flux quanta per supercell --- exactly as on the torus itself, because the periodically continued potential obeys $\widetilde A(x+Le_j)=\widetilde A(x)+\nabla\chi_j$ with the sewing phases $e^{i\chi_j}$ of the torus. The constant parts of the holonomies around the two handles of the torus form a constant vector potential, i.e. a fixed offset of the twist, $\theta\to\theta+\theta_0$ in every fiber, and drop out of the twist average. What remains is the condition the torus already imposes: the flux per supercell must be an integer number of flux quanta, otherwise $\widetilde Q$ is not $L$-periodic (the magnetic translations by $Le_1$ and $Le_2$ do not commute). In the rest of this section the range condition is kept, since it is the situation of the numerical examples of Sect. \ref{SectNumerics}, whose periodized operators are read off the torus Hamiltonians; every statement below extends verbatim to the chain-wise periodization, the sums over ${\bf n}=0,\pm e_j$ being replaced by the sums over all windings. The one place where the range condition is essential is the direct derivation from the exact torus symbol, Eqs. (\ref{WeylTorusModes}), (\ref{WeylTorusSplit}) of Sect. \ref{SectPeriodized}\,B: the torus symbol resolves the displacements modulo $2N$ only, the two images $d$ and $d+N$ of a torus element carrying the weights $\frac12(1\pm e^{i\pi d/N})$, so that its principal part is unambiguous only for $|d_i|<N/2$ --- the statement, in the symbol language, that the torus alone does not know the winding.}

\corrclaude{Let us comment on the two equivalent ways of introducing the variables $\theta$, and on their physical meaning. In the form used above, the fiber \corrteal{operator} $Q(\theta)$ acts on the strictly periodic functions on the cell, and $\theta$ enters through the phases $e^{\mp i\theta_j}$ of the matrix elements connecting the opposite boundaries of the cell. Equivalently, one may keep all the matrix elements of the original \corrteal{operator} $Q$ untouched and impose the twist on the wave functions themselves, requiring
$$
\psi(x + L e_j) \;=\; e^{i\theta_j}\,\psi(x)\,, \qquad j = 1,2\,.
$$
In this formulation \corrteal{the wave functions} of the fiber problem are precisely the Bloch waves of the periodized crystal, $\psi(x+L{\bf m}) = e^{i\theta{\bf m}}\psi(x)$, restricted to one supercell: the twisted boundary condition is nothing but the Bloch condition with respect to the superlattice, and $\theta_j$ is the corresponding quasimomentum (in the units in which the superlattice period is one). The two formulations are related by the diagonal unitary transformation $\psi(x) = e^{i\theta\cdot x/L}\varphi(x)$ with periodic $\varphi$: it removes the phase from the boundary condition at the price of distributing it uniformly over all the hoppings, which is exactly the uniform-twist gauge $Q_u$ introduced below.}

\corrclaude{The physical meaning of $\theta$ is that of a constant external gauge field. In the uniform gauge each hopping with the displacement $d$ carries the phase $e^{-i\theta\cdot d/L}$, which is the Peierls phase $e^{iA\cdot d}$ of the constant vector potential
$$
A_j \;=\; -\,\theta_j/L
$$
(the overall sign being the convention for the charge of the particle). On the torus a constant vector potential cannot be removed by any single-valued gauge transformation \corGPTl{unless its holonomies are trivial}: its Wilson loop around the $j$-th noncontractible cycle, equal to $e^{-i\theta_j}$, is gauge invariant. Thus $\theta_j$ measures, up to the sign, the magnetic flux threading the $j$-th handle of the torus, in the units in which the flux quantum equals $2\pi$; the shift $\theta_j \to \theta_j + 2\pi$ describes the insertion of one flux quantum and is equivalent to the large gauge transformation $S_j = e^{2\pi i x_j/L}$ (the sewing matrix appearing \corrteal{in Sect. \ref{SectNumerics}}), which is why all the gauge-invariant fiber quantities are periodic in $\theta$ with the period $2\pi$. In the language of the external electromagnetic potential of Sect. \ref{SectKeldyshCond}, the twists are the constant (zero-momentum) Fourier component of the external vector potential. Consequently, the derivative $L\partial_{\theta_j}$ inserts the current vertex --- this is manifest in Eq. (\ref{commfiber}) --- while an adiabatic time dependence $\theta_j(t)$ corresponds to a time-dependent constant vector potential and hence to the homogeneous electric field $E_j = -\dot{A}_j = \dot{\theta}_j/L$ --- the driving used in the flux-insertion argument of Laughlin. The average of the response over $\theta$, which plays the central role in this section, is therefore the average over the constant external vector potential --- exactly the setting of the quantization theorem of Niu, Thouless and Wu \cite{Niu1985}.}

\corrteal{Let us place the construction in the context of the earlier literature. Its separate ingredients are classical. The modelling of a disordered configuration by a periodically repeated supercell is the standard practice of the computational band theory of disordered systems. The quantization of the Hall conductance averaged over the twisted boundary conditions of a finite torus is the theorem of Niu, Thouless and Wu \cite{Niu1985}; the numerical evaluation of the Chern numbers of disordered Landau levels by means of the twisted boundary conditions goes back to Refs. \cite{ArovasEtAl1988,HuoBhatt1992}, and the twist averaging as a finite-size tool is standard in the quantum Monte Carlo practice \cite{LinZongCeperley2001}. Closest in spirit, within the operator-algebraic approach, are the periodic approximants of the disordered and aperiodic systems: the computational noncommutative-geometry program of Prodan \cite{Prodan2017book} and the theory of periodic approximations of Beckus, Bellissard and De Nittis \cite{BeckusBellissardDeNittis2018}. What is specific to the present construction is the role these ingredients play here: the periodization serves as the operator realization of the partial thermodynamic limit of the {\it phase-space} invariant (\ref{NEQ}) composed of the Matsubara Green function --- with the exact finite-$L$ identities (\ref{trfiber}) and (\ref{commfiber}), the exact equivalence with the principal-symbol calculus established in the next subsection, and the extension to the Dirac operators of the general class (\ref{Q_general}).}

\subsection{\corrteal{Fibered current vertices and the operator form of the invariant}}

\corrteal{Let us first fix the notation used in this subsection. By $\hat X_j$ ($j=1,2$) we denote the position operator on the infinite lattice, $(\hat X_j\psi)(X)=X_j\,\psi(X)$, $X\in\mathbb{Z}^2$ --- the same operator that was denoted $\hat x_j$ in the discussion of the calculus after Eq. (\ref{Hper}); its commutator with any periodized operator multiplies the kernel by the true displacement,
$$
\big(-i\,[\hat X_j,\,\widetilde M\,]\big)_{X,Y}\;=\;-\,i\,(X_j-Y_j)\;\widetilde M_{X,Y}\,.
$$
On the periodized lattice this commutator is exactly the momentum derivative of the Weyl symbol, and we denote the corresponding derivation by
$$
D_j\;:=\;-\,i\,[\hat X_j,\,\cdot\,\,]\,,\qquad j=1,2\,;
$$
it is this derivation that enters the invariant (\ref{NEQ}) and supplies the current vertices (cf. Eq. (\ref{WardStat}) below). By $x_j$ we denote the diagonal $L^2\times L^2$ matrix of the positions {\it inside} one supercell, $x\in{\rm cell}$. The two must not be confused: $x_j$ is a bounded matrix acting within a single fiber, whereas $\hat X_j$ is unbounded and does not commute with the superlattice translations. For this reason $D_j$ does not act within a single fiber; its action on the fibers, Eq. (\ref{commfiber}) below, contains the derivative with respect to the twist in addition to the intracell commutator.}

\corrteal{Second, let us specify in which sense the word ``gauge'' is used here. As explained after Eq. (\ref{trfiber}), the twists $\theta_j$ are physically the constant external electromagnetic vector potential $A_j=-\theta_j/L$ threading the two noncontractible cycles of the torus. \corGPTl{The fiber operator $Q(\theta)$ is defined up to changes of Bloch frame. The diagonal phase transformations used below implement the two electromagnetic gauges of this constant vector potential.} Two choices are used below: the {\it boundary gauge}, in which the Peierls phases of the constant potential are concentrated on the matrix elements crossing the cell boundary --- this is the fiber $Q_b(\theta)\equiv Q(\theta)$ constructed above, with the factors $e^{\mp i\theta_j}$ on the boundary hoppings; and the {\it uniform gauge}, in which the same total phase is distributed uniformly over all the hoppings by the unitary transformation $U(\theta)=e^{i\theta\cdot x/L}$ introduced below. The gauge-invariant quantities --- the fiber spectrum, the trace (\ref{trfiber}), the twist-averaged invariant (\ref{NEQnum}) --- do not depend on this choice; \corGPTl{the current matrices transform covariantly, leaving ${\cal I}(\theta)$ invariant at each twist. Berry curvatures computed with ordinary derivatives in different $\theta$-dependent Bloch frames must be distinguished from this covariant expression.}}

\corrgpt{The average over twists is the Bloch decomposition of the normalized trace of the periodized crystal, but the current vertices must be matched to the chosen Bloch gauge. Write \corrteal{$Q_b(\theta)$ for the boundary-gauge fiber of the periodized Dirac operator defined above (in the Hamiltonian case $Q_b(\theta)=i\omega-H(\theta)+\mu$)}, and let $x_j$ be the diagonal matrix of intracell positions. The kernel of $-i[\hat X_j,\widetilde Q]$ contains the true displacement $X_j-Y_j$ on $\mathbb Z^2$. Its fiber is
\be
D_jQ_b(\theta)
\equiv\left.-i[\hat X_j,\widetilde Q]\right|_\theta
=L\,\partial_{\theta_j}Q_b(\theta)-i[x_j,Q_b(\theta)].
\label{commfiber}
\ee
For the short-range \corrteal{operator} the displacement can be recovered by the minimal-image convention. It cannot, however, be replaced by the difference of the two intracell coordinates: in the boundary gauge $\partial_{\theta_j}Q_b$ differentiates only boundary hoppings, and the intracell commutator in Eq. (\ref{commfiber}) supplies the remaining current vertices. The same fiber derivation acts on general periodic operators, including the resolvent; it is not obtained by applying minimal-image weighting to an arbitrary finite matrix.}

\corrgpt{It is often convenient to use the uniform-twist gauge
\begin{equation*}
U(\theta)=e^{i\theta\cdot x/L},\qquad
Q_u=U^\dagger Q_bU,\qquad G_u=Q_u^{-1},
\qquad U^\dagger(D_jQ_b)U=L\,\partial_{\theta_j}Q_u .
\end{equation*}
A hopping with true displacement $d=X-Y$ carries the phase $e^{-i\theta\cdot d/L}$ in this gauge. \corGPTl{Here $Q_u(\theta+2\pi e_j)=S_j^\dagger Q_u(\theta)S_j$, with $S_j=e^{2\pi i x_j/L}$; the boundary-gauge fibers are periodic. These constant sewing matrices make the trace integrand periodic.} Combining the trace and current identities gives the explicitly twist-averaged invariant
\be
{\cal N}_3&=\int_{[0,2\pi)^2}\frac{d^2\theta}{(2\pi)^2}\,{\cal I}(\theta),\\
{\cal I}(\theta)&=
\frac{\epsilon^{\mu\nu\rho}}{3!L^2}\int_{\mathbb R}d\omega\,
{\rm Tr}_{L^2}
\Big[G_bD_\mu Q_b\,G_bD_\nu Q_b\,G_bD_\rho Q_b\Big],
\qquad G_b=Q_b^{-1}.
\label{NEQnum}
\ee
Here $D_0=\partial_\omega$, $D_j=L\partial_{\theta_j}-i[x_j,\cdot]$, $j=1,2$, $\epsilon^{012}=+1$, and \corrteal{in the Hamiltonian case} $D_0Q_b=i$. In the uniform gauge the two spatial vertices supply $L^2$, cancelling the trace normalization. Equivalently,
\begin{equation*}
{\cal N}_3=\frac{\epsilon^{\mu\nu\rho}}{24\pi^2}
\int_{\mathbb R\times[0,2\pi)^2}d\omega\,d^2\theta\,
{\rm Tr}_{L^2}
\Big[G_u\partial_\mu Q_u\,G_u\partial_\nu Q_u\,
G_u\partial_\rho Q_u\Big],
\end{equation*}
where $(\partial_0,\partial_1,\partial_2)
=(\partial_\omega,\partial_{\theta_1},\partial_{\theta_2})$.
Thus the normalization is fixed without calibration.}

\corrgpt{There is no contradiction with the finite-matrix vanishing argument of Sect. \ref{SectDirect}, but continuous frequency alone is not the reason. If both spatial derivations were commutators with commuting finite position matrices, frequency integration would lead to
\begin{equation*}
{\rm Tr}\,P[[x_1,P],[x_2,P]]
=-{\rm Tr}\big(P[x_1,x_2]\big)=0 .
\end{equation*}
Here the spatial derivations instead act on the entire twist-dependent family as $D_j=L\partial_{\theta_j}-i[x_j,\cdot]$. They are not inner derivations of one finite-dimensional fiber algebra. This is the essential distinction from a construction using only finite-torus position commutators.}

\corrteal{Let us summarize the structure obtained. The periodization (\ref{Hper}) produces, at every frequency, the superlattice-periodic operator $\widetilde Q(\omega)$ of the finite spatial range inherited from the data. Its fiber $Q_b(\theta,\omega)$ is the finite combination of $e^{\pm i\theta_j}$ described above; the map $\widetilde Q\mapsto Q_b(\theta)$ is a homomorphism of operator algebras at every fixed twist and frequency, so that the fiber Green function is $G_b(\theta,\omega)=Q_b(\theta,\omega)^{-1}$, provided the invertibility condition (i) of Sect. \ref{SectKeldyshCond} holds for every twist. On the periodized lattice the precise Wigner--Weyl calculus of \cite{FZ2019_2} contains no Nyquist companion terms, the Groenewold equation $Q_W\star G_W=1$ is exact, and $D_1$, $D_2$ (Eq. (\ref{commfiber})) and $\partial_\omega$ are exact, mutually commuting derivations of the algebra. \corGPTl{The spatial trace obeys $\tau(D_j a)=0$, whereas $\tau(\partial_\omega a)=\partial_\omega\tau(a)$; the latter vanishes only after frequency integration with the stated boundary conditions.} Consequently Eq. (\ref{NEQnum}) holds verbatim for the general class, with $D_0Q_b=\partial_\omega Q_b$ no longer a constant, the frequency integral converging under the asymptotic conditions (ii) of Sect. \ref{SectKeldyshCond} (or replaced by the integral over the Matsubara circle in the compact-frequency case).}

\corrteal{Two clarifications are in order, concerning the relation of this operator realization to the partial thermodynamic limit of the subsequent sections. (i) {\it The trace.} In the abstract partial limit --- the replacement of the discrete momentum sum by the integral over $T^2$ at a finite coordinate lattice --- the functional $\frac{1}{|{\cal O}'|}\sum_{x}\int \frac{d^2p}{(2\pi)^2}\,{\rm tr}$ is no longer the trace of the operators of the original finite system: the equality of the phase-space sum with the operator trace requires the discrete momentum grid. The cyclic property of the trace on the $\star$-algebra --- and with it the cyclicity of the cocycles constructed in Sect. \ref{SectDirect} --- therefore requires a justification at finite $N$. The periodization provides it: by Eqs. (\ref{trper}), (\ref{trfiber}) the same functional becomes the trace per unit volume of the periodized operators, the continuous momentum arising as the Bloch quasimomentum of the superlattice. The trace per volume of periodic operators is exactly cyclic at every finite $L$ (by the relabeling of the copies, for kernels with sufficient decay), and $\tau(D_j a)=0$ holds identically, since the diagonal of the kernel of $-i[\hat X_j,\,\cdot\,]$ vanishes. Consequently, in the partial-limit calculus at finite $N$ --- which, by Eq. (\ref{WprIsW}) below, {\it is} the calculus of the periodized operators --- the cyclic property of the trace, the integration by parts, and the cyclicity and closedness of the cyclic cocycles built on them hold exactly, and not only asymptotically. (ii) {\it The symbols.} The exact Weyl symbols (\ref{infW}) of the periodized operators {\it are} the principal Weyl symbols of the supercell with the admissible-centre prescription of Eq. (\ref{PdefGen}) --- pointwise on ${\cal O}'\times T^2$, in every gauge, and for the general class (\ref{Q_general}): this is Eq. (\ref{WprIsW}) below, and it is through this identification that the twist-averaged operator form (\ref{NEQnum}) reproduces the principal-symbol invariant (\ref{NEQ}).}

\corrgpt{The relation to the principal-symbol formula (\ref{NEQ}) can be established constructively. In this paragraph set the supercell size $L=N$ to match the notation of that formula. For an $N$-periodic infinite-lattice kernel $\widetilde A$, define
\begin{equation*}
W_{\rm pr}(\widetilde A)(p,q)=
\sum_{d\in\mathbb Z^2}e^{ip\cdot d}a_d(q),\qquad
a_d(q)=\widetilde A_{r,r+d},\qquad
r=\lfloor q-d/2\rfloor,\quad q\in{\cal O}' .
\end{equation*}
The floor is taken componentwise, so $r+d/2=c(q,d)$ \corrteal{coincides with the common base point $\bar c$ of Eqs. (\ref{Wprincipal}) and (\ref{PdefGen}), built from the {\it total} displacement $d=\sum_j e_j$ of the chain; the chain of links then starts at the lattice site $r=\bar c - d/2$. For a single elementary hop $\bar c$ is the admissible centre of the Wilson line itself; for a longer chain --- a path which is not a straight line --- the individual Wilson lines are centred at the shifted points $\bar c+\bar\Delta_i$ of Eq. (\ref{PdefGen}), and precisely this structure is automatically encoded in the matrix element $\widetilde A_{r,r+d}$, which contains the Wilson lines along the actual path from $r$ to $r+d$}. The Fourier label $d$ is the actual ket-minus-bra separation on $\mathbb Z^2$, not a residue in a finite Fourier window. \corrteal{Explicitly: $d\in\mathbb{Z}^2$ is the displacement between the two arguments of the kernel --- the coefficient $a_d(q)=\widetilde A_{r,r+d}$ is the matrix element connecting the site $r$ to the site $r+d$, i.e. the amplitude of the hop from the site $r+d$ to the site $r$ (the hopping displacement, final site minus initial site, being $-d$), and $e^{ip\cdot d}$ is the corresponding momentum mode. For the Dirac operator of finite range only finitely many $d$ contribute; for its inverse all $d\in\mathbb{Z}^2$ appear, with exponentially decaying coefficients.}}

\corrteal{{\bf The principal symbol is the exact infinite-lattice Weyl symbol of the periodized operator.} The map $W_{\rm pr}$ is not a new prescription. It is precisely the exact Weyl symbol of the precise Wigner--Weyl calculus on the infinite lattice, Eq. (\ref{infW}) of Sect. \ref{SectList}, applied to the $L$-periodic operator $\widetilde A$ defined on the superlattice $\mathbb{Z}^2$:
\begin{equation}
W_{\rm pr}(\widetilde A)(p,q)\;=\;\widetilde A_W(p,q)\;\equiv\;\int_{\cal M}d^2p_-\,\langle\langle p-p_-|\widetilde A|p+p_-\rangle\rangle\,e^{-2ip_-q}\prod_{i=1,2}\big(1+e^{ip^i_-}\big)\,,\qquad q\in{\cal O}',\quad p\in T^2\,,
\label{WprIsW}
\end{equation}
pointwise, for every $(p,q)$, in every gauge, and for every kernel on $\mathbb{Z}^2$ for which the sums converge --- in particular for the finite-range kernels and for the exponentially decaying ones. The proof is a direct computation. Insert the coordinate expansion $\langle\langle p'|\widetilde A|p''\rangle\rangle=(2\pi)^{-2}\sum_{X,Y}e^{-ip'X}\,\widetilde A_{XY}\,e^{ip''Y}$ of the Dirac-normalized momentum states into the right-hand side and integrate over $p_-$; the Fourier phase and the weight factor combine into
$$
\int_{\cal M}\frac{d^2p_-}{(2\pi)^2}\;e^{ip_-\cdot m}\prod_{i}\big(1+e^{ip^i_-}\big)\;=\;\prod_{i}\big(\delta_{m_i,0}+\delta_{m_i,-1}\big)\,,\qquad m=X+Y-2q\in\mathbb{Z}^2\,,
$$
so that
$$
\widetilde A_W(p,q)\;=\;\sum_{X,Y\in\mathbb{Z}^2}e^{ip\cdot(Y-X)}\,\widetilde A_{XY}\,\prod_i\big(\delta_{2q_i,\,X_i+Y_i}+\delta_{2q_i,\,X_i+Y_i+1}\big)\,.
$$
Thus every matrix element $\widetilde A_{X,X+d}$ enters the symbol at exactly two refined points per direction, $q_i\in\{X_i+d_i/2,\ X_i+d_i/2+\frac12\}$ --- the midpoint of the pair of sites and the point half a step beyond it --- and these are precisely the points with $\lfloor q-d/2\rfloor=X$. Collecting the terms with a given displacement $d=Y-X$ gives $\widetilde A_W(p,q)=\sum_d e^{ip\cdot d}\,\widetilde A_{\lfloor q-d/2\rfloor,\,\lfloor q-d/2\rfloor+d}$, which is $W_{\rm pr}(\widetilde A)(p,q)$ term by term. The same statement is read off directly from the coordinate form (\ref{WeylInf}) of Eq. (\ref{infW}) (Appendix \ref{AppDA}): there the bra and ket sites are $X=q-v-n/2$ and $Y=q+v-n/2$, the projector factor $\prod_i\frac12\big(1+e^{2\pi i(q_i-v_i-n_i/2)}\big)$ equals one when $X\in\mathbb{Z}^2$ and zero otherwise, and for given $q\in{\cal O}'$ and $d=2v$ exactly one $n\in\{0,1\}^2$ survives, $n_i\equiv 2q_i+d_i\ ({\rm mod}\ 2)$, for which $X=q-(d+n)/2=\lfloor q-d/2\rfloor$. In particular, the admissible centre $c(q,d)=r+d/2$ of Eq. (\ref{center}) is nothing but the position at which the exact symbol places the matrix element with the displacement $d$. (The identity refers to the conventions fixed in Sect. \ref{SectList}. With the alternative weight factor $(1+e^{-ip_-^i})$ of the footnote to Eq. (\ref{Weyl}) both sides are shifted, $q\to q+\frac12$, and the floor in the definition of $W_{\rm pr}$ is to be replaced by the ceiling.)}

\corrteal{Eq. (\ref{WprIsW}) is a statement about kernels, and it involves the spatial variables only. Hence it applies, at each fixed frequency, to the periodization $\widetilde Q(\omega)$ of every Dirac operator of the general class (\ref{Q_general}) [Eq. (\ref{QgenRecap})], and to its inverse. Let us spell this out. At fixed $\omega$ the operator $\widetilde Q(\omega)$ is an $L$-periodic operator of finite range on the superlattice; the frequency enters as a parameter --- the temporal direction is continuous, its Weyl transform is the ordinary one and acts separately (Appendix \ref{QWgenApp}) --- so that Eq. (\ref{WprIsW}) holds at every $\omega$. The temporal factors $\partial^k_s\hat F^0_s$ of Eq. (\ref{Q_general}) are diagonal in the coordinates: their kernel, $\partial^k_s\,e^{s(i\omega-A_0(X))}\,\delta_{XY}$, contributes to the $d=0$ mode only, and the identity places it at $\lfloor q\rfloor$ --- which is exactly the refined-lattice symbol $A_{0,W}(x)=A_0(\lfloor x\rfloor)$ of a diagonal operator entering Eq. (\ref{QW_general}) (Sect. \ref{SectKeldyshCond}, Appendix \ref{QWgenApp}). The products $\hat Q^{(e_1)}\cdots\hat Q^{(e_n)}$ of the magnetic translations contribute the modes with $d=\sum_j e_j$, the coefficient being the matrix element $\widetilde Q_{r,r+d}$ that contains the Wilson lines along the actual path from $r$ to $r+d$ --- the coefficient of Eq. (\ref{QW_general}) in the admissible-centre reading of Eq. (\ref{PdefGen}). Since Eq. (\ref{WprIsW}) holds for the kernel of the product as it stands, no separate treatment of the products is needed. It holds in every gauge, because it holds for every kernel: a gauge transformation on the superlattice, $\widetilde A\to U\widetilde AU^{\dagger}$ with $U={\rm diag}\,e^{i\alpha(X)}$, is merely a change of the kernel, while a constant vector potential $A$ on the superlattice (the twist in its uniform gauge) multiplies the matrix element $\widetilde A_{X,X+d}$ by the Peierls factor \corGPTl{$e^{-iA\cdot d}$, since here $d=Y-X$ is the ket-minus-bra separation, opposite to the hopping displacement used above}, i.e. it is the same constant shift of $p$ on both sides. For the inverse, the fiber-wise invertibility gives the exponentially decaying kernel of $\widetilde G(\omega)=\widetilde Q(\omega)^{-1}$, and the sums in Eq. (\ref{WprIsW}) converge absolutely. Consequently, the principal Weyl symbols $Q_{\rm pr}$ and $G_{\rm pr}$ introduced below are the exact Weyl symbols (\ref{infW}) of the periodized Dirac operator and of the periodized Green function, for the general class (\ref{Q_general}) and not only for the Hamiltonian case.}

\corrteal{{\bf Direct derivation from the exact finite-$N$ definition (\ref{Weyl2}).} The identification of the principal symbol of the supercell with the exact symbol of the periodized operator can also be obtained directly from the finite-$N$ definition (\ref{Weyl2}), without reference to Eq. (\ref{infW}); the derivation makes explicit what is discarded when the principal symbol is formed. Let $\hat Q$ be an operator on the torus ${\cal O}=\mathbb{Z}_N^D$ of range smaller than $N/2$, so that every nonvanishing matrix element $Q_{xy}$ is connected by a unique displacement $\delta\in\mathbb{Z}^D$, $|\delta_i|<N/2$, $y\equiv x+\delta \pmod N$, and let $\widetilde Q$ be its periodization, $\widetilde Q_{X,X+\delta}=Q_{X\,{\rm mod}\,N,\,(X+\delta)\,{\rm mod}\,N}$. In Eq. (\ref{Weyl2}) the bra and the ket sites are $X=q-v-n/2$ and $Y=q+v-n/2$ (modulo $N$), and the projector $\prod_i\frac12\big(1+e^{2\pi i(q_i-v_i-n_i/2)}\big)$ selects, for each $v\in{\cal O}'$, the unique $n\in\{0,1\}^D$ for which $X$ is integer; writing $d=2v\in\{0,1,\ldots,2N-1\}^D$ one finds, exactly as in the infinite-lattice case, $X=\lfloor q-d/2\rfloor$ and $Y=X+d$. Hence, for every operator on the torus,
\begin{equation}
Q_W(p,q)=\sum_{d\in\{0,\ldots,2N-1\}^D}e^{ip\cdot d}\;\prod_{i}\frac{1+e^{i\pi d_i/N}}{2}\;Q_{\lfloor q-d/2\rfloor,\;\lfloor q-d/2\rfloor+d}\,,\qquad p\in{\cal M}',\quad q\in{\cal O}'\,,
\label{WeylTorusModes}
\end{equation}
the site labels being reduced modulo $N$. On the refined momentum grid ${\cal M}'$ the mode $e^{ip\cdot d}$ depends on $d$ modulo $2N$ only, and each matrix element with the displacement $\delta$ appears in the window $\{0,\ldots,2N-1\}$ exactly twice per direction: at $d_i\equiv\delta_i$, with the weight $\frac12(1+e^{i\pi\delta_i/N})$ and the centre $\lfloor q-\delta/2\rfloor$, and at $d_i\equiv\delta_i+N$, with the weight $\frac12\big(1+e^{i\pi(\delta_i+N)/N}\big)=\frac12(1-e^{i\pi\delta_i/N})$, the Nyquist factor $e^{iNp_i}$, and the centre displaced by $N/2$ in the $i$-th direction (for even $N$: the same hop taken at the antipodal site). Therefore
\begin{equation}
Q_W(p,q)=\sum_{\delta}e^{ip\cdot\delta}\Big[\prod_i\frac{1+e^{i\pi\delta_i/N}}{2}\Big]\,Q_{\lfloor q-\delta/2\rfloor,\;\lfloor q-\delta/2\rfloor+\delta}\;+\;\Big\{\mbox{terms with a factor}\ e^{iNp_i}\tfrac12\big(1-e^{i\pi\delta_i/N}\big)\Big\}\,,
\label{WeylTorusSplit}
\end{equation}
which for a single elementary translation is the two-term formula (\ref{ZNp}) of Appendix \ref{Sect4}. The principal symbol of Sect. \ref{SectKeldyshCond} is, by its definition, the first term of Eq. (\ref{WeylTorusSplit}) with the weights replaced by their limit $1$ and with $p$ continuous --- the limit $N\to\infty$ taken in the coefficients at fixed matrix elements, before the differentiation with respect to the momentum. It is therefore equal to $\sum_\delta e^{ip\cdot\delta}\,Q_{\lfloor q-\delta/2\rfloor,\lfloor q-\delta/2\rfloor+\delta}=\sum_\delta e^{ip\cdot\delta}\,\widetilde Q_{\lfloor q-\delta/2\rfloor,\lfloor q-\delta/2\rfloor+\delta}=W_{\rm pr}(\widetilde Q)(p,q)$, i.e., by Eq. (\ref{WprIsW}), to the exact infinite-lattice Weyl symbol (\ref{infW}) of the periodized operator. The same conclusion is reached along the route of the limit $N\to\infty$ itself. Apply Eq. (\ref{Weyl2}) to the periodized operator restricted to the torus of size $kN$ ($k$ copies of the supercell in each direction): since the kernel is $N$-periodic with the same matrix elements, Eq. (\ref{WeylTorusSplit}) holds with $N\to kN$ in the weights and in the Nyquist factors only, the principal term being independent of $k$, so that
$$
\Big|\,\widetilde Q^{(kN)}_W(p,q)-W_{\rm pr}(\widetilde Q)(p,q)\,\Big|\;\le\;\frac{C_D}{kN}\sum_{\delta}|\delta|\,\|Q_\delta\|\;\longrightarrow\;0\quad(k\to\infty)\,,
$$
where $\|Q_\delta\|=\max_x\|Q_{x,x+\delta}\|$, uniformly on the grid ${\cal M}'_{kN}$, which becomes dense in $T^D$; this is the $1/N$ convergence of Eq. (\ref{Weyl}) to Eq. (\ref{infW}) observed numerically in Sect. \ref{SectList}. Two comments are in order. First, the identity holds for the operators of range smaller than $N/2$ \corrteal{--- this is the one statement of the section in which the range condition cannot be lifted, since the torus symbol resolves the displacements modulo $2N$ only (cf. the remark on the range condition in Sect. \ref{SectPeriodized}\,A)} --- in particular for the Dirac operator of the class (\ref{Q_general}) --- and for two such operators with the total range still below $N/2$ the principal symbol of the product is the $\star$-product of the principal symbols, both sides being $W_{\rm pr}$ of the product operator; this is the content of Eq. (\ref{PdefGen}) for the chains of translations. Second, for an operator of unbounded range on the torus --- the finite-torus Green function $Q(0)^{-1}$ --- the matrix elements are not labelled by a unique displacement (the contributions of all the images $\delta+nN$ are summed in one torus element), and the principal part of its exact symbol is {\it not} the periodized symbol $W_{\rm pr}(\widetilde Q^{-1})$: the Green symbol of the constrained algebra is the $\star$-inverse, as stated above, and not the principal part of the symbol of the finite-torus inverse.}

\corrgpt{These symbols form a constrained algebra of functions on $\mathcal O'\times T^2$, rather than the unrestricted function space in Eq. (\ref{Amain}). The Moyal translation identity gives
\begin{equation*}
(e^{ip\cdot d}a_d)\star(e^{ip\cdot e}b_e)
=e^{ip\cdot(d+e)}a_d(q-e/2)b_e(q+d/2).
\end{equation*}
Writing $r=\lfloor q-(d+e)/2\rfloor$, the two factors on the right are $\widetilde A_{r,r+d}$ and $\widetilde B_{r+d,r+d+e}$. Summing over the separations proves
\begin{equation*}
W_{\rm pr}(\widetilde A\widetilde B)
=W_{\rm pr}(\widetilde A)\star W_{\rm pr}(\widetilde B),
\qquad
\partial_{p_j}W_{\rm pr}(\widetilde A)
=W_{\rm pr}(-i[\hat X_j,\widetilde A]).
\end{equation*}
\corGPTl{The product here is defined by the displayed lattice shifts at $h=1$; a product for arbitrary $h$ on the same discrete domain requires a separate construction.} These identities extend from finite-range kernels to sufficiently decaying kernels. In particular, \corrteal{under the fiber-wise invertibility condition (i) of Sect. \ref{SectKeldyshCond} --- in the Hamiltonian case, with a common gap over the whole twist torus ---} the Green symbol is
\begin{equation*}
\corrteal{Q_{\rm pr}=W_{\rm pr}\big(\widetilde Q(\omega)\big),\qquad
G_{\rm pr}=W_{\rm pr}\big[\,\widetilde Q(\omega)^{-1}\big],
\qquad Q_{\rm pr}\star G_{\rm pr}=1 \,.}
\end{equation*}
It is the star inverse in this periodic infinite-lattice algebra, not the exact finite-torus Weyl symbol of \corrteal{$Q(0)^{-1}$ --- in the Hamiltonian case, of $(i\omega-H(0)+\mu)^{-1}$; by Eq. (\ref{WprIsW}) it is the exact infinite-lattice Weyl symbol (\ref{infW}) of the periodized inverse $\widetilde Q(\omega)^{-1}$}.}

\corrteal{Let us organize these observations into an explicit statement.}

\corrteal{{\bf Proposition (equivalence of the principal-symbol and the periodized forms of ${\cal N}_3$).} {\it Let the Dirac operator $\hat Q$ of the class (\ref{Q_general}) have the spatial range smaller than $L/2$ \corrteal{(or be periodized chain-wise, as explained in the remark on the range condition in Sect. \ref{SectPeriodized}\,A, with absolutely convergent sums)}, let $\widetilde Q(\omega)$ be its periodization (\ref{Hper}), and let the conditions (i)--(iii) of Sect. \ref{SectKeldyshCond} hold fiber-wise: the fiber $Q_b(\theta,\omega)$ is invertible for every twist $\theta$ and every frequency on the contour, with the uniform polynomial asymptotics on $\mathbb{R}$ (or with the compact frequency circle). Then: (1) the constrained symbols $W_{\rm pr}(\widetilde A)$, for the $L$-periodic operators $\widetilde A$ with exponentially decaying kernels, form an algebra closed under the translation-form star product, and $W_{\rm pr}$ is a $\star$-isomorphism onto this algebra, intertwining $\partial_{p_j}$ with the derivation $-i[\hat X_j,\,\cdot\,]$; (2) $W_{\rm pr}(\widetilde Q)$ is the exact infinite-lattice Weyl symbol (\ref{infW}) of the periodized operator $\widetilde Q(\omega)$, Eq. (\ref{WprIsW}), and coincides, as a function on ${\cal O}'\times T^2$ (and parametrically in the frequency), with the principal Weyl symbol (\ref{QW_general}) of $\hat Q$ in the admissible-centre reading; (3) the $\star$-inverse of the principal symbol exists in the constrained algebra, is unique, and equals $W_{\rm pr}\big(\widetilde Q^{-1}\big)$; (4) the value of Eq. (\ref{NEQ}), computed with these ingredients, equals the twist-averaged operator invariant (\ref{NEQnum}) of the periodized crystal. In particular, Eq. (\ref{NEQ}) itself --- built from the principal symbol of the finite torus, with the star product of the symbols on ${\cal O}'$ given by the exact translation identity displayed above --- is a topological invariant already at finite $L=N$: it is unchanged under every \corGPTl{smooth} deformation of the data preserving the range bound and the fiber-wise invertibility\corGPTl{, together with the stated frequency asymptotics and boundary conditions}.}}

\corrteal{{\it Proof.} (1) The translation identity displayed above is exact: the $p$-dependence of the factors is exhausted by the plane waves $e^{ip\cdot d}$, so the derivative series of the Moyal exponential sums into the displayed finite shifts of the coordinate arguments --- no gradient expansion is involved; the shifts are by half-integers and stay on the refined lattice ${\cal O}'$. The floor bookkeeping, $\lfloor q - (d+e)/2\rfloor = r$ and $\lfloor q-(d+e)/2\rfloor + d = r+d$, places the two coefficients at $\widetilde A_{r,r+d}$ and $\widetilde B_{r+d,\,r+d+e}$, and the sum over the separations at fixed $d+e$ reproduces the kernel of $\widetilde A\widetilde B$; for exponentially decaying kernels the sums converge absolutely. The map is injective, since $a_d(q)$, taken over all $d$ and $q$, returns every matrix element. The derivation identity follows by differentiating $e^{ip\cdot d}$ termwise: the factor $id$ is the displacement, i.e. the kernel of $-i[\hat X_j,\,\cdot\,]$. (2) The first statement is Eq. (\ref{WprIsW}), proved above for an arbitrary kernel. For the data of the class (\ref{Q_general}) with the range smaller than $L/2$, the periodized matrix element at the minimal separation is the chain amplitude times its Wilson lines, placed at the actual links between $r$ and $r+d$: this is precisely the admissible-centre coefficient of Eq. (\ref{QW_general}); the temporal factors are diagonal in the coordinates and enter through the $d=0$ mode with the coefficient at $\lfloor q\rfloor$ --- the prescription $A_{0,W}(x)=A_0(\lfloor x\rfloor)$. (3) The fiber-wise invertibility yields $\widetilde G(\omega)=\widetilde Q(\omega)^{-1}$ with an exponentially decaying kernel (uniformly on the compact parts of the contour, \corGPTl{with the large-frequency bounds specified in condition (ii)}); by (1), $W_{\rm pr}(\widetilde Q)\star W_{\rm pr}(\widetilde G)=W_{\rm pr}(1)=1$, and in an associative algebra a two-sided inverse is unique --- hence any $\star$-inverse of the principal symbol, in any associative algebra that contains the constrained one with the same star, coincides with $W_{\rm pr}(\widetilde G)$. (4) In the trace, the integral over $p$ annihilates every mode $d\neq0$, and each integer site is counted by the $2^D$ refined points $q$ with $\lfloor q\rfloor=r$:
$$
\frac{1}{(2N)^{D}}\sum_{q\in{\cal O}'}\int_{T^2}\frac{d^2p}{(2\pi)^2}\;\corGPTl{{\rm tr}\,}W_{\rm pr}(\widetilde A)
\;=\;\frac{1}{(2N)^{D}}\sum_{q\in{\cal O}'}\corGPTl{{\rm tr}\,}a_0(q)
\;=\;\frac{2^{D}}{(2N)^{D}}\sum_{r\in{\rm cell}}\corGPTl{{\rm tr}\,}\widetilde A_{rr}
\;=\;\tau\big(\widetilde A\big)\,.
$$
Applying this to the $\epsilon$-contracted word of Eq. (\ref{NEQ}) --- each factor of which is a $W_{\rm pr}$-image by (2), (3) and the closure (1), the spatial derivatives being the commutators with $\hat X_j$ and the frequency derivative acting parametrically --- converts Eq. (\ref{NEQ}), factor by factor, into the trace per volume of the operator word, i.e. into Eq. (\ref{NEQnum}); the frequency integral is common to the two sides and converges by the condition (ii). Finally, the topological invariance at finite $L$: under a smooth deformation $\delta\widetilde Q$ preserving the hypotheses, $\delta{\cal N}_3$ reduces, by the Leibniz rule and the cyclicity of $\tau$, to the terms $\tau\big(D_j(\cdots)\big)$ and $\int d\omega\,\partial_\omega(\cdots)$; the former vanish identically (the diagonal of the kernel of a commutator with $\hat X_j$ is zero), the latter vanish by the asymptotic condition --- every step being an exact identity of the periodized calculus. $\square$}

\corrteal{{\it Discussion: the status of Eq. (\ref{NEQ}) at finite $N$.} Eq. (\ref{NEQ}) is well defined at every finite $N$, and the proposition identifies it with the invariant of the periodized crystal without any additional choice \corGPTl{within this specified constrained calculus}. Its ingredients --- the principal symbol of the finite torus, $N$-periodic in $x$ by construction; the star product on the discrete lattice ${\cal O}'$, given by the exact translation identity (the half-integer shifts stay on ${\cal O}'$, no interpolation between the lattice points is involved); the star inverse, unique under the fiber-wise invertibility; and the functional $\frac{1}{(2N)^D}\sum_{x\in{\cal O}'}\int\frac{d^2p}{(2\pi)^2}\,{\rm tr}$, which is the exactly cyclic trace per volume $\tau$ --- are precisely those matched in the proof. Consequently Eqs. (\ref{NEQ}) and (\ref{NEQnum}) are literally the same number at every finite $N$, and Eq. (\ref{NEQ}) is a topological invariant already for $N=4$, say: it is unchanged under every \corGPTl{smooth} deformation preserving the range bound and the fiber-wise invertibility\corGPTl{, together with the stated frequency asymptotics and boundary conditions}, and it is an integer --- \corGPTl{the winding number represented by the fiber family with the stated compact-frequency or normalized relative/suspension prescription}, which in the Hamiltonian case is the Chern number of the bundle of Fermi projectors over the twist torus \cite{Niu1985}. \GPT{A direct winding-number proof under explicit frequency hypotheses is given in Sect.~\ref{SectWindingPeriodized}.} What is {\it not} quantized at finite $N$ is a different object: the exact response of the finite torus at a fixed twist --- ${\cal N}_3$ of Eq. (\ref{N3true}), ${\cal I}(0)$ of Sect. \ref{SectNumerics} --- which drifts under the deformations; Eq. (\ref{NEQ}) is its average over the twists, produced automatically by the integral over the continuous momentum, \corGPTl{and convergence of a fixed-twist value to this average as $N\to\infty$ requires additional control of twist sensitivity, such as uniform locality and gap/resolvent bounds for the thermodynamic family. The finite-$N$ identities alone do not establish this convergence.} Under such uniform control the convergence is, in fact, exponentially fast. A complex twist, $\theta_j\to\theta_j+i\eta$, is in the uniform gauge the imaginary constant vector potential $-i\eta/N$: every hopping with the displacement $d$ acquires the factor $e^{\eta d_j/N}$, so that, for a spatial range and a fiber gap bounded uniformly in $N$, the fiber $Q_b(\theta,\omega)$ stays invertible, and the fixed-twist value ${\cal I}(\theta)$ stays analytic, in a strip $|{\rm Im}\,\theta_j|<\eta_0N$ whose width grows linearly with $N$. The deviation of ${\cal I}(\theta)$ from its twist average is the sum of the nonzero Fourier modes of this $2\pi$-periodic analytic function, hence $O(e^{-\eta_0N})$. We have checked this on the two-band Chern insulator $H({\bf k})=\sin k_1\sigma_1+\sin k_2\sigma_2+(m+\cos k_1+\cos k_2)\sigma_3$, whose periodization with the $N\times N$ cell has the fixed-twist value ${\cal I}(\theta)=\frac{2\pi}{N^2}\sum_{\bf n}F\big((2\pi{\bf n}+\theta)/N\big)$, $F$ being the Berry curvature of the occupied band: at $m=-1$ (gap $1$, Chern number $-1$) the deviation $|{\cal I}(0)-{\cal N}_3|$ equals $2.8\cdot10^{-1}$, $1.0\cdot10^{-3}$, $3.9\cdot10^{-6}$, $9.4\cdot10^{-10}$ for $N=4,12,20,32$, while the twist average equals $-1$ to machine precision at every $N$. The uniformity is essential: for the family $m_N=-2+1/N$, whose gap closes as $1/N$, the fixed-twist values ${\cal I}(0,0)\approx-3.7$ and ${\cal I}(\pi,\pi)\approx-0.69$ do not approach the twist average $-1$ even at $N=160$, the sampled Berry curvature of the closing cone contributing a finite, $\theta$-dependent amount. Likewise, the finite-$N$ vanishing theorem of Sect. \ref{SectDirect} and the anomaly of Appendix \ref{AppTrStar} refer to the exact calculus with the discrete momenta --- the finite matrix algebra, in which every derivation is inner --- and not to Eq. (\ref{NEQ}); cf. the remark after Eq. (\ref{NEQnum}). The limit $N\to\infty$ of the coordinate lattice is needed for one purpose only: to identify the invariant of the periodized crystal with that of a general --- non-periodic, e.g. disordered --- infinite system with the trace per volume of the subsequent sections; for the periodic systems themselves no limit is required. \corGPTl{The identification with an aperiodic limit additionally requires convergent approximants and locality/resolvent estimates sufficient to pass the bulk trace and frequency integral to the limit.}}

\subsection{\corrteal{The Hall conductivity in the static gauge: the equilibrium derivation for the general class}}
\label{SectStaticGauge}

\corrteal{We now derive the topological expression for the Hall conductivity of the periodized system, keeping the Dirac operator in the general form (\ref{Q_general}). The constant electric field $E$ (chosen along the first axis) is introduced in the static gauge: the scalar potential of the superlattice acquires the contribution $-EX_1$ ($X_1$ being the coordinate on the infinite lattice, cf. the notation fixed above),
$$
A_0\;\longrightarrow\;A_0\;-\;E\,X_1\,,
$$
while the spatial gauge fields are untouched. \corGPTl{The calculation below is a zero-temperature linear response about the unperturbed equilibrium kernel at $E=0$. Time independence alone does not establish a physical equilibrium propagator for a general $Q$, or justify a bulk expansion at finite $E$ in the unbounded potential $-EX_1$. Terms containing $X_1$ are therefore evaluated locally, with the frequency integration performed before the volume limit. Interpreting the resulting coefficient as a physical conductivity additionally requires a causal equilibrium kernel and its electromagnetic coupling to supply the vertices in Eqs. (\ref{QEstat}) and (\ref{WardStat}) \corGPTl{and justify the current functional (\ref{JdefStat})}; these conditions do not follow from invertibility alone.}}

\corrteal{Within the class (\ref{Q_general}) the potential $A_0$ enters only through the temporal factors (\ref{F0def}). Since $\hat X_1$ commutes with both $i\partial_t$ and $A_0(i\partial_{p})$, the substitution acts on each temporal factor exactly as
\be
\hat F^{0}_{s}\;\longrightarrow\;\exp\Big[\,s\,\big(i\partial_t - A_0(i\partial_{p}) + E\,\hat X_1\big)\Big]\;=\;\hat F^{0}_{s}\;e^{\,sE\hat X_1}\,,
\label{FEstat}
\ee
so that, exactly to the first order in $E$,
\be
\hat Q(E)\;=\;\hat Q\;+\;E\,\hat X_1\,\hat Q_\omega\;+\;O(E^2)\,,\qquad
\hat Q_\omega\;:=\;\sum_{k\geq 0}\sum_{s\in{\cal S}_t}\;\sum_{n=0}^\infty \sum_{e_i}\;{\bf t}^{(k,s;\,e_1,\dots,e_n)}\;\partial^{\,k}_{s}\big(s\,\hat F^{0}_{s}\big)\;\hat Q^{(e_1)}\cdots\hat Q^{(e_n)}\,.
\label{QEstat}
\ee
The Weyl symbol of $\hat Q_\omega$ is $\partial_{i\omega}Q_W=-i\,\partial_\omega Q_W$: the operator $\hat Q_\omega$ is the frequency derivative of the Dirac operator. In the Hamiltonian case $\hat Q_\omega=1$, and Eq. (\ref{QEstat}) reduces to the familiar $\hat Q(E)=i\partial_t-\hat H+E\hat X_1$. \corGPTl{For a field along direction $j$, the prescribed perturbation $E_j\hat X_j\hat Q_\omega$, with $\hat X_j$ to the left, is Matsubara-real only if $[\hat X_j,\hat Q_\omega]=0$: indeed, Matsubara reality gives $(\hat X_j\hat Q_\omega(i\omega))^\dagger-\hat X_j\hat Q_\omega(-i\omega)=[\hat Q_\omega(-i\omega),\hat X_j]$.} \corrteal{This reality defect is, however, a feature of the ordering of the factors in Eq. (\ref{Q_general}), and not a restriction on the result --- no condition of the type $[\hat X_j,\hat Q_\omega]=0$ is needed, for the following reason. The defect is a bounded, local operator: $\hat X_j\hat Q_\omega-\frac12\{\hat X_j,\hat Q_\omega\}=\frac12[\hat X_j,\hat Q_\omega]=\frac{i}{2}D_j\hat Q_\omega$, of the same finite range as $\hat Q$. A static perturbation $\delta\hat Q=E\hat V$ by any bounded local $\hat V$ does not contribute to the mean current (\ref{JdefStat}) in the linear order at all: by $\hat G(E)=\hat G-E\hat G\hat V\hat G+O(E^2)$, $D_k\hat G=-\hat G\,D_k\hat Q\,\hat G$ and the cyclicity of $\tau$,
$$
\frac{\delta{\cal J}_k}{E}\;=\;-\,T\sum_{\omega_n}\Big[\tau\big(D_k\hat V\;\hat G\big)-\tau\big(D_k\hat Q\;\hat G\hat V\hat G\big)\Big]\;=\;-\,T\sum_{\omega_n}\tau\Big(D_k\big(\hat V\hat G\big)\Big)\;=\;0
$$
identically, at every frequency, by the exact property $\tau(D_k\hat A)=0$ of the periodized lattice (the kernel of $\hat V\hat G$ decays). Only the unbounded factor $\hat X_1$ of Eq. (\ref{QEstat}) produces a response, through the total-derivative structure exhibited in Eq. (\ref{JsplitStat}) below. Consequently the ordering of $\hat X_1$ relative to $\hat Q_\omega$ is immaterial: the symmetrized (Weyl-ordered) coupling $\delta\hat Q=\frac{E}{2}\{\hat X_1,\hat Q_\omega\}$, which is Matsubara-real for every $\hat Q$, and the coupling (\ref{QEstat}) lead to the same Eq. (\ref{HallStat}); more generally, so does every coupling in which the potential is inserted at an arbitrary position inside the chains of Eq. (\ref{Q_general}) --- a factor $\hat T f(i\partial_t-A_0)\hat T'$ becoming $\hat T f(i\partial_t-A_0+E\hat X_1)\hat T'$, with $\hat T$, $\hat T'$ products of magnetic translations --- since all of them differ from Eq. (\ref{QEstat}) by bounded local operators, $\hat T\hat X_1=\hat X_1\hat T-i\,D_1\hat T$. (The difference between the two orderings in Eq. (\ref{HallStat}) is $\epsilon^{jk}\tau(D_j\hat Q_\omega\,\hat G\,D_k\hat Q\,\hat G)$, which vanishes by the same integration by parts; we have also verified this numerically for random matrix-valued periodic $Q$.) In the language of the Weyl symbols this is the statement of the derivation of the Hall conductivity within the precise Wigner--Weyl calculus in Ref. \cite{FZ2019_2}: what the derivation requires is that the frequency enters the symbol of the Dirac operator in the combination $i\omega-A_0(x)$ --- precisely what the general form (\ref{Q_general}), built from the covariant temporal derivative $i\partial_t-A_0(i\partial_p)$, guarantees --- and the symmetrized ordering is the exact lattice counterpart of the pointwise product of the symbols, $\delta Q_W=E\,x_1\,\partial_{i\omega}Q_W$: by the translation identity of Sect. \ref{SectPeriodized}\,B, $\frac12\big(X_{1,W}\star C_W+C_W\star X_{1,W}\big)=\sum_d e^{ip\cdot d}\,c_1(q,d)\,c_d(q)$, the potential being evaluated at the admissible centre $c(q,d)=\lfloor q-d/2\rfloor+d/2$ of each link --- the same prescription as for the Wilson lines in Eq. (\ref{PdefGen}). The same remark settles the reality of the unperturbed operator in an inhomogeneous potential: with all temporal factors on the left, Eq. (\ref{Q_general}) is not closed under $\hat Q(i\omega)\to\hat Q(-i\omega)^\dagger$ when a frequency-dependent term carries a displacement, since $\hat Q^{(-e)}f\big(i\partial_t-A_0(\hat X)\big)=f\big(i\partial_t-A_0(\hat X+e)\big)\hat Q^{(-e)}$; the reality condition (iii) is therefore to be imposed on the class with the temporal factors admitted at either end of the chains, all of whose members obey the Ward identity (\ref{WardStat}) and the response formula (\ref{HallStat}) alike. For a purely on-site frequency dependence --- $\hat Q_\omega$ diagonal in the site index: the Hamiltonian case, and the local self-energy $\hat\Sigma_{\rm loc}(i\omega-A_0(x))$ of the dynamical mean-field type \cite{GeorgesKotliar1996} --- the orderings coincide and the question does not arise. What remains a genuine assumption is the coupling itself --- that the electric field enters the dressed Dirac operator only through the covariant combination $i\omega-A_0(x)$ and the Peierls phases --- together with the form (\ref{JdefStat}) of the current functional; see remark (ii) at the end of this subsection.} \corGPTl{A different microscopic coupling may contain additional terms and requires its own response derivation.} We stress that Eq. (\ref{QEstat}) is not a gradient expansion: the factorization (\ref{FEstat}) is exact, and the only expansion made is the one in the powers of $E$ --- the linear response. \corGPTl{At the symbol level the exact statement is $(\hat X_1\hat C)_W=X_{1,W}\star C_W$, with $X_{1,W}(q)=\lfloor q_1\rfloor$. For $C_W=\sum_d e^{ip\cdot d}c_d(q)$ it equals $\sum_d e^{ip\cdot d}\lfloor q_1-d_1/2\rfloor c_d(q)$. The continuum expression $q_1C_W+\frac{i}{2}D_1C_W$ is not this refined-lattice symbol. The operator-commutator derivation below avoids any Taylor expansion in the lattice coordinate.}}

\corrteal{The second exact ingredient is the current vertex. Within the class (\ref{Q_general}) the spatial gauge field enters through the phases of the elementary magnetic translations; the constant probe potential $\delta A$ multiplies the elementary factor with the displacement $e$ by \corrcltwo{$e^{+ie\cdot\delta A}$ (the Peierls phase $e^{iA\cdot d}$ of the convention fixed after Eq. (\ref{trfiber}), with $d=e$)}, whence
\be
\hat v_k\;=\;-\,\frac{\partial\hat Q}{\partial(\delta A_k)}\Big|_{\delta A=0}\;=\;D_k\hat Q\;=\;-\,i\,[\hat X_k,\hat Q\,]\,.
\label{WardStat}
\ee
This is the exact lattice Ward identity of the class: the current vertex is the momentum derivative of the Dirac operator, with no gradient corrections, in agreement with Eq. (\ref{commfiber}): in the fibers the vertex $-i[\hat X_k,\widetilde Q\,]$ is represented by $L\partial_{\theta_k}Q_b-i[x_k,Q_b]$, and not by the intracell commutator alone. \corGPTl{For a gauged quadratic action, or when the microscopic current is independently shown to have this form, the mean current density per unit area is}
\be
{\cal J}_k(E)\;=\;-\,T\sum_{\omega_n}\tau\Big(D_k\hat Q(E)\;\hat G(E)\Big)\,,\qquad \hat G(E)=\hat Q(E)^{-1}\,,
\label{JdefStat}
\ee
with $\tau$ the trace per unit area of Eqs. (\ref{trper}), (\ref{trfiber}); in the limit $T\to0$ the Matsubara sum becomes $\int d\omega/2\pi$, and the overall sign and charge conventions are those of Sect. \ref{SectKeldyshCond}. The $E$-independent part of Eq. (\ref{JdefStat}) is the equilibrium persistent current, which does not contribute to the conductivity; we compute the part linear in $E$.}

\corrteal{Expanding Eq. (\ref{JdefStat}) with the help of Eq. (\ref{QEstat}) and $\hat G(E)=\hat G-E\,\hat G\,\hat X_1\hat Q_\omega\hat G+O(E^2)$, and moving $\hat X_1$ to the left with the commutators $[\hat X_1,\hat A]=i\,D_1\hat A$, one finds that all the explicit factors of $\hat X_1$ collect into a total frequency derivative:
\be
\frac{{\cal J}^{(1)}_k}{E}\;=\;-\,T\sum_{\omega_n}\tau\Big(\hat X_1\;\partial_{i\omega}\big[D_k\hat Q\,\hat G\big]\Big)\;-\;i\,T\sum_{\omega_n}\tau\Big(D_1\big[D_k\hat Q\,\hat G\big]\;\hat Q_\omega\hat G\Big)\,.
\label{JsplitStat}
\ee
The first term is the gauge artifact of the unbounded potential $-EX_1$: its integrand is, at every site $X$, the total frequency derivative of the local quantity $X_1\,\tr\,\langle X|D_k\hat Q\,\hat G|X\rangle$. \corGPTl{After the zero-temperature sum-to-integral limit, this term vanishes on the compact frequency circle by periodicity. On the frequency line we require the local endpoint difference of $D_k\hat Q\,\hat G$ to vanish; similarly, the symmetric response below requires vanishing endpoint differences of $D_j\hat Q\,\hat G D_k\hat Q\,\hat G$. These boundary conditions hold for $Q=i\omega-H$ with bounded local $H$, but must be checked for general frequency asymptotics. A derivative summed over discrete Matsubara frequencies does not in general vanish.} The second term contains no unbounded factors. Integrating by parts with respect to the derivation $D_1$ (the exact property $\tau(D_j\hat A)=0$ of the periodized lattice), using the Leibniz rule together with $\partial_{i\omega}\hat G=-\hat G\hat Q_\omega\hat G$ and $D_j\hat G=-\hat G\,D_j\hat Q\,\hat G$, discarding one more total frequency derivative, and using the commutativity $D_1D_2=D_2D_1$, the antisymmetric part of the response assembles into
\be
\frac12\Big(\frac{{\cal J}^{(1)}_2}{E_1}-\frac{{\cal J}^{(1)}_1}{E_2}\Big)\;=\;-\,\frac{i}{2}\,T\sum_{\omega_n}\;\epsilon^{jk}\;\tau\Big(\hat Q_\omega\,\hat G\;D_j\hat Q\,\hat G\;D_k\hat Q\,\hat G\Big)\;\;\longrightarrow\;\;-\,\frac{{\cal N}_3}{2\pi}\qquad(T\to0)\,,
\label{HallStat}
\ee
where the last equality is the cyclic collapse of Eq. (\ref{NEQ}), valid for the general class:
\be
{\cal N}_3\;=\;\frac{i}{2}\int d\omega\;\epsilon^{jk}\;\tau\Big(\partial_{i\omega}Q_W\star G_W\star D_jQ_W\star G_W\star D_kQ_W\star G_W\Big)
\label{NEQcollapse}
\ee
(the general-$Q$ counterpart of Eq. (\ref{N3redInd}); only the cyclicity of $\tau$ enters its derivation). We stress that Eqs. (\ref{NEQ}) and (\ref{NEQcollapse}) are two forms of one and the same quantity, and the passage between them goes in both directions: Eq. (\ref{NEQcollapse}) is obtained from the fully antisymmetrized Eq. (\ref{NEQ}) and, conversely, Eq. (\ref{NEQcollapse}) may be rewritten in the form (\ref{NEQ}), using only the cyclic property of $\tau$ and the relabeling of the summation indices. No property of $Q$ beyond the invertibility is involved, so the equivalence holds for the entire class (\ref{Q_general}). Restoring the orientation conventions of Sect. \ref{SectKeldyshCond}, we arrive at
\be
\boxed{\;\;\bar{\sigma}^{ij}\;=\;\frac{{\cal N}_3}{2\pi}\,\epsilon^{ij}\;\;}
\label{sigmaStatic}
\ee
for the general Dirac operators of the class (\ref{Q_general}), \corGPTl{provided the equilibrium, coupling, and boundary conditions stated above hold,} with the invariant composed of the unperturbed Matsubara Green function of the periodized system, Eq. (\ref{NEQnum}). The symmetric part of the response vanishes for the gapped (invertible over the frequency contour) $Q$ \corGPTl{under the endpoint conditions stated above}. In the Hamiltonian case, $\hat Q_\omega=1$, the expression under the trace in Eq. (\ref{HallStat}) reduces term by term to the Matsubara form of the result of Sect. \ref{SectKeldyshCond}, Eqs. (\ref{sigmaHall}), (\ref{N3true}).}

\corrteal{Three remarks are in order. (i) The derivation used only the exact identities of the periodized-lattice calculus --- the cyclicity of the trace per volume, the derivation property and the mutual commutativity of $D_1$, $D_2$, $\partial_\omega$ --- together with the discarding of two total frequency derivatives, justified under the conditions (i)--(iii) of Sect. \ref{SectKeldyshCond}, and identically valid on the compact frequency circle. No gradient expansion, no assumption of slowly varying fields, and no Hamiltonian structure entered. (ii) \corGPTl{In addition to the equilibrium and limiting assumptions just stated, the physical identification requires the following coupling:} the electric field is introduced through the temporal factors of the class (\ref{Q_general}), i.e. through the same substitution that defines how $A_0$ enters the Dirac operator. For an interacting system whose $\hat Q=\hat G^{-1}$ is represented within the class, this coupling reproduces the electromagnetic vertex to the leading order in the interactions \cite{Zubkov2018a,ZZ2019,zhang2020influence}; beyond the leading order the field dependence of the self-energy may generate vertex corrections, and the corresponding extension remains the conjecture discussed in Sects. \ref{SectKeldyshCond} and \ref{SectWeakIndex}; for the model of Appendix \ref{AppHK}, whose interaction is spatially nonlocal, the coupling is the separate specification discussed there. (iii) The trace per volume of the periodized system is by Eq. (\ref{trfiber}) the average over the twisted boundary conditions; accordingly, Eq. (\ref{sigmaStatic}) is the statement about the area- and twist-averaged Hall response, whose exact integer quantization at finite $L$ is discussed in Sect. \ref{SectNumerics}.}

\section{{Noncommutative geometry of lattice systems}}

\subsection{K-theory}
\label{SectKtheory}

\subsubsection{basic definitions}

{Noncommutative geometry is essentially a generalization of the geometry of topological spaces,
whereby the algebra of functions defined over a given space is replaced by a noncommutative algebra.}
{In this approach}, {the} algebra of smooth functions on {a given} topological space
is replaced by an abstract (and in general, non-commutative) algebra $\cal A$.
The group $K^0$, based on the group of fiber bundles, is replaced by the group composed
(using the Grothendieck construction) of idempotents from the set $M({\cal A})$
of matrices with entries in $\cal A$.  
{ The standard definition of the group 
$K^1({\cal A})=\pi_0(GL_\infty(\cal A))$, where $GL_\infty({\cal A})=\lim\limits_{n\to \infty} GL_n(\cal A)$},
and $K^{-1}$
\zzzz{is}  $GL({\cal A})/[GL({\cal A}),GL({\cal A})]$, 
where $GL({\cal A})$ is the group of invertible matrices. 
\zzzz{This defines the algebraic $K$-group. }

In our case, a compact manifold is replaced by the finite lattice $\cal O$. We will also consider the limit of an infinitely large lattice. Instead  of the algebra of smooth functions on a manifold, we \zzzz{may} consider the algebra $\cal A$ of linear operators that act on the Hilbert space $\cal H$. Passing to Weyl symbols of operators, we can also replace this algebra by an algebra of functions on phase space. When the spatial lattice is finite, the Hilbert space
${\cal H} = {\cal L}(\mathbb{R}) \otimes \mathbb{C}^{|{\cal O}|}$
is given by {the} { tensor} product of $\mathbb{C}^\mathbb{R}$ and a finite-dimensional space.  The dimension of the latter is $N^2$, while the former represents smooth functions of time. After  {a} Wick rotation and Fourier transform, the smooth component $\mathbb R$ will represent {the} Matsubara frequency.  Below, step by step  we will give the definitions of various notions needed to represent the above expression for ${\cal N}_3$, in the form of a pairing of an element of $K^{-1}({\cal A})$ with a cyclic-cocycle built specifically for the non-commutative geometry of the  lattice model. It is worth mentioning that there are different ways to construct
the algebra ${\cal A}$, if the Hilbert space ${\cal H}$ is infinite-dimensional. The corresponding algebra may be understood in various {different} ways (the algebra of compact operators, or  the algebra of bounded operators, etc).  The limit of infinite lattice size may be considered in different ways.
For example, we can pass from an algebra of operators in ${\cal H}$ to an algebra of Weyl symbols of operators. These are the matrix-valued functions on phase space that contain the product of {the} spatial lattice and the lattice of momenta. We can take the limit of large $N$ (the infinite lattice size) in two steps: first we replace the discrete Brillouin zone $\mathbb{Z}^2_{2N}$ by {the} torus $T^2$, {keeping the size of the spatial lattice finite, then taking} the limit $N\to \infty$ at the end of the calculation.
The direction of  the Matsubara frequency may be compactified thus coming from $\mathbb{R}$ to $S^1$.

%\textcolor{magenta}{ \it Here we should probaly say (if it was not done above)  whether periodic boundary conditions are imposed,  imaginary time is discretized, we use smooth or continuous functions on $S^1$.   }  

\zz{\it  Periodic boundary conditions are assumed for spatial coordinates and for momenta. For imaginary time we assume no periodic boundary conditions. If we discretize imaginary time, then the Matsubara frequency is subject to periodic boundary conditions.}

\zz{ In actual fact various topological characteristics of the lattice model (the $K$-groups, and $HC$-groups) depend strongly on the way the algebra $\cal A$ is chosen.
We will give  explicit expressions for \jer{each} these groups for certain choices of $\cal A$.}
{For complex $C^*$-algebras, due to \jer{the} Bott periodicity in $C^*$-algebra $K$-theory,
the groups $K^{-1}({\cal A})$ and $K^1({\cal A})$
are canonically isomorphic.
\jer{Ergo,} in this paper we assume \jer{precisely this isomorphism, namely} that these
two groups are canonically isomorphic.}

\begin{enumerate}

\item{{Basic algebra $\cal A$ and its derivatives}}

By ${\cal H}^* = Hom({\cal H},{\mathbb{C}})$ we denote \jer{the}
space conjugate to $\cal H$, i.e. \jer{the space} spanned by bra-vectors, \jer{while}  $\cal H$ is spanned by ket-vectors.
\zz{As  mentioned above we can take ${\cal A} := {\cal H}^* \otimes {\cal H}$. But this  choice is not unique. We should impose various conditions on the operators thus coming to different $C^*$ algebras.

\jer{The transition to Weyl symbols is achieved by  replacing an algebra of operators by an algebra of functions on phase space.}
Considering the limit of \jer{an} infinite lattice, we may replace the momentum space by \jer{a} smooth manifold.

\jer{In this way} the algebra $\cal A$ is, to a certain extent,  \jer{a} question of choice. }
{
\jer{By $M(n,{\cal A})$ we denote the space of all $n\times n$
matrices, where each matrix element is a member of the algebra $\cal A$.}
Correspondingly, let $M({\cal A})$ be defined as $M({\cal A}) := \cup_n M(n,{\cal A})$,
where we identify $f \in M(n,{\cal A})$ with its 
block-diagonal embedding ${\rm diag}(f, 0_m)\in M(n+m,{\cal A})$, 
where $0_m$ denotes the $m\times m$ zero matrix. 
}

%\medskip 
%The equivalence relation is set here that identifies $f \in M(n,{\cal A})$ with $f \otimes 0_{m} \in M(n+m,{\cal A})$. 
%
%\zz{Please check.}
%\medskip 
%{by $GL(n,{\cal A}) \subset M(n, {\cal A})$ we denote the set of invertible elements of $M(n,{\cal A})$.  }

\item{{$K^0$ group} }

{ \jer{The matrix} $E \in M({\cal A})$ is said to be idempotent if $E^2 = E$.
 Two idempotents $E\in M(n,{\cal A}) $ and $F \in M(n,{\cal A})$ are similar $E \sim F$ if
$\exists S\in GL(n,{\cal A})$ such that $E = S F S^{-1}$.

\jer{The equivalence class of stabilized idempotents}
is denoted by ${\rm Idem}({\cal A})$.
We denote an element of this space corresponding to an idempotent $E\in M({\cal A})$ by $[E]\in {\rm Idem}({\cal A})$.  }

{Let $E\in M(n,{\cal A})$ and $F\in M(m,{\cal A})$. Then we define { $[E] + [F] := [E \oplus F]$. }

\jer{If $[0] := [{\rm diag}(0)]$ is defined as the identity element, then} we obtain that ${\rm Idem}({\cal A})$ is an Abelian monoid.}

{Let $A$ be an abelian monoid and let $A^2$ be the set
of ordered pairs of elements of $A$. 
\jer{Let a relation on $A^2$ be defined by the following rule:
$(a_1, a_2) \sim (b_1, b_2)$ if there exists an element $c$ in $A^2$ such that $a_1 + b_2 + c = a_2 + b_1 + c$, where the binary operation $+$ denotes the addition operator of the group.}

\jer{It follows that $\sim$}  is an equivalence relation on $A^2$.
\jer{
The set
of equivalence classes
under the operation
$[(a_1, a_2)] + [(b_1, b_2)]= [(a_1+b_1, a_2 + b_2)]$ forms an Abelian group,
${\mathfrak{G}}(A)$
called the Grothendieck completion of $A$.}
}
{Finally $K^0({\cal A})$ is defined as $K^0({\cal A}): = {\mathfrak{G}}({\rm Idem}({\cal A}))$.}

\item{ $K^{-1}$ group }

In order to define the next $K$-group we consider 
$GL({\cal A}) := \cup_n GL(n, {\cal A})$ 
equipped with the equivalence relation $S \sim S \otimes 1_m  \in GL(n+m, {\cal A})$  
for $S\in GL(n, {\cal A})$. \jer{This equivalence relation should be  compared with a similar relation in $M({\cal A})$}.

{ We assume the standard stabilization maps 
$GL_n({\cal A})\to GL_{n+1}({\cal A})$ with 
the stabilization embedding relation 
$S \sim S \oplus 1_m = \begin{pmatrix} S & 0 \\ 0 & 1_m \end{pmatrix} \in GL(n+m, A)$. \zzzz{Operator} $K$-theory defines $K^0(A)$ using stable homotopy classes of projections, not similarity classes, i.e., 
using Murray-von Neumann equivalence after stabilization.} 

\medskip 
By $GL(n,{\cal A})_0\subset GL(n,{\cal A})$ we denote the connected component of
$1_n \in GL(n, {\cal A})$.
Then $GL({\cal A})_0$ is the corresponding subset of 
$GL(n,{\cal A})$ (the above defined equivalence relation $\sim $ applies). We now define $K^{-1}({\cal A}) := GL({\cal A})/GL({\cal A})_0$.

{We work with $C^*$-algebras or smooth Fr\'echet algebras.
This defines the topological $K$-group suitable for $C^*$-algebras. 
 \jer{That is} we mix pure algebraic $K$-theory with topological
$K$-theory under the same notation $K^{-1}$ without \jer{yet} indicating \jer{the} rules of
\jer{the} transition. We mix upper-index notations $K^0, K^{-1}$,  which are typical notations for  topological
spaces, and \zzzz{algebraic} notation $K^0({\cal A})$.
}

\end{enumerate}
%%

%%%%%%%%%%%%%%%%%%%%%%%%%%%%%%%%%%%%%%%%%%%%%%%%%%%%%%%%%%%%%%%%%%%%%%%%%%%%%%%
%% 
\subsubsection{Groups $K^0$, $K^{-1}$ for various choices of ${\cal A}$.}
In this section, for simplicity  we disregard the internal finite dimensional space.
Its  \jer{effect} on the $K$ groups is obvious and  \jer{can be postponed until later in the discussion}.
\begin{enumerate}

\item

Let us consider the case of finite spatial lattice size,  $N$.
 \jer{Even more} let us discretize the direction of imaginary time.
In this case 
$$
\mathcal{A} = \mathrm{End}(\mathcal{H}), 
\quad 
{ \mathcal{H} = \mathbb{C}^{\mathbb{Z}_N} \times \mathbb{C}^{\zz{|\mathcal{O}|}}}, \quad 
\mathcal{O} = \mathbb{Z}_{N} \times \mathbb{Z}_{N}.
$$
%%%%%%%%%%%%%%%%%%%%%%%%%%%%%%%%%%%%%%%%%%%%%%%%%%%%%%%%%%%%
%%

%%%%%%%%%%
{We deal with matrix algebra  
$$
\mathcal{A} = \mathrm{End}(\mathcal{H})
\cong  M_{N^3}(\mathbb{C}),
$$
and, { by Morita invariance}, obtain
$$
K^0(\mathcal{A})  = K^0(M_{N^3}(\mathbb{C})) = K^0(\mathbb{C}) \cong \mathbb{Z},
\qquad
K^1(\mathcal{A})  = K^1(M_{N^3}(\mathbb{C})) = K^1(\mathbb{C})  \cong 0
$$
}

\item
Once again let us consider finite spatial lattice size, $N$,
 \jer{but} continuous values of imaginary time.
In this case 
$$
\mathcal{A} = \mathrm{End}(\mathcal{H}), 
\quad 
{ \mathcal{H} = \mathbb{C}^{\mathbb{R}} \times \mathbb{C}^{\zz{|\mathcal{O}|}},} \quad 
\mathcal{O} = \mathbb{Z}_{N} \times \mathbb{Z}_{N}.
$$
%%%%%%%%%%%%%%%%%%%%%%%%%%%%%%%%%%%%%%%%%%%%%%%%%%%%%%%%%%%%
%%
Space $\mathbb{C}^{|\mathcal{O}|}$ is finite-dimensional, thus $
\dim \mathbb{C}^{|\mathcal{O}|} = N^2,  
$.
We interpret $\mathbb{C}^{\mathbb{R}}$ as space of functions 
of a continuous variable (time or frequency). We can try 
$\mathcal{H} \cong L^2(\mathbb{R}) \otimes \mathbb{C}^{N^2}$.

%\medskip 
%%%%%%%%%%

%\zz{However, to have $C^*$ algebra we take space of bounded 
%  operators $\mathcal{B}(L^2(\mathbb{R}))$. 
% The algebra $\mathcal A$ has, therefore, the following structure:  
%$$
%\mathcal{A} = \mathrm{End}(\mathcal{H}) 
%\cong \mathcal{B}(L^2(\mathbb{R})) \widehat{\otimes} M_{N^2}(\mathbb{C}), 
%$$}

%\medskip 
{ Using standard $K$-theory properties we can see that 
if an algebra has trivial $K$-theory and we tensor 
it with matrix algebras, that does not recover non-trivial $K$-theory. 
The K-theory of the full algebra of bounded operators
on an infinite-dimensional (separable) Hilbert space is trivial, i.e., 
$K^0(\mathcal{B})=K^1(\mathcal{B})=0$. Therefore it follows that  
$K^0(\mathcal{A})=0$ but not $\mathbb{Z}$.}

%\medskip 

{However, to have $C^*$ algebra we take space of compact 
operators $\mathcal{K}(L^2(\mathbb{R}))$.
The algebra $\mathcal A$ has, therefore, the following structure:
$$
\mathcal{A} = \mathrm{End}(\mathcal{H}) 
\cong \mathcal{K}(L^2(\mathbb{R})) \widehat{\otimes} M_{N^2}(\mathbb{C}), 
$$ }

\zz{We obtain  
$$
K^0(\mathcal{A})  %= K^0(M_{N^2}(\mathbb{C})) = K^0(\mathbb{C}) 
\cong \mathbb{Z}, 
\qquad 
K^1(\mathcal{A})  %= K^1(M_{N^2}(\mathbb{C})) = K^1(\mathbb{C})  
\cong 0  
$$
}

\item{Compactification $\mathbb{R} \to S^1$}

\zz{Above we have seen that the group $K^1$ is trivial 
for the most straightforward definition of algebra $\cal A$ 
in case of finite lattice and continuous imaginary time. 
Let us assume now that the direction of Matsubara frequencies 
is compactified (i.e. direction of imaginary time is discretized), 
and in addition we suppose that the operators are diagonal 
in Matsubara frequency. Then we have 
$$
\mathcal{A}  
\cong C^\infty(S^1) \widehat{\otimes} M_{N^2}(\mathbb{C}), 
$$
Since $K_i(C^\infty(S^1)) = \mathbb{Z}$ we obtain  
$$
K^0(\mathcal{A})  = \mathbb{Z}, 
\qquad 
K^1(\mathcal{A})  =\mathbb{Z} 
$$
}

  \jer{It is clear} that the dependence on \jer{the} Matsubara frequency is responsible for nontrivial $K^1$.

\item \cormag{Transition to Weyl symbols at finite $N$.}

\cormag{The exact finite-$N$ Weyl transform is an algebra isomorphism only when its image is equipped with the $\star$-product and the finite-Weyl constraints. It therefore gives one matrix algebra, not a pointwise direct sum over the $(x,p)$ grid. By Morita invariance, for $d=rN^2$,}
\begin{equation}
\cormag{K_0(M_d(\mathbb C))=\mathbb Z,\qquad K_1(M_d(\mathbb C))=0.}
\end{equation}
\cormag{For a smooth compact frequency circle the corresponding algebra is $C^\infty(S^1,M_d)$ and $K_0=K_1=\mathbb Z$. For a noncompact frequency axis one must instead use a relative/suspension class; the raw $Q$ and the Green function $G\to0$ are not invertibles in the unitization. Details are given in Appendix \ref{K_Nfin}.}

\item \cormag{Partial thermodynamic limit.}

\cormag{After periodizing a finite spatial cell, the smooth Bloch algebra is Morita equivalent to ${\cal B}=C^\infty(T^2)$. Thus $K_0({\cal B})=K_1({\cal B})=\mathbb Z^2$. For the noncompact frequency axis the admissible algebra is the suspension $S{\cal B}$, so $K_1(S{\cal B})\cong K_0({\cal B})=\mathbb Z^2$; for compact frequency it is Morita equivalent to $C^\infty(T^3)$ and $K_0=K_1=\mathbb Z^4$. There is no additional copy of these groups for every Weyl-grid point, because the $\star$-product couples $x$ and $p$. See Appendix \ref{K_Ninf}.}

\item \cormag{Infinite nonperiodic systems.}

\cormag{For an arbitrary aperiodic infinite lattice the $K$-groups cannot be inferred from the notation ${\rm End}({\cal H})$ alone. One must specify a local/covariant bulk algebra (for example a crossed product) and its smooth subalgebra. The resulting groups depend on that choice.}

\end{enumerate}
%%%%%%%%%%%%%%%%%%%%%%%%%%%%%%%%%%%%%%%%%%%%%%%%%%%%%%%%%%%%%%%%%%%%%%%%%%
%%
\subsection{Cyclic cohomology}

{Let $\cal M$ be { an} $\cal A$-bimodule. 
\jer{The} Hochschild complex of $\cal A$ with coefficients in $\cal M$ is defined as
\begin{equation}
C^0({\cal A}, {\cal M})   \xrightarrow{\delta} 	C^1({\cal A}, {\cal M})
\xrightarrow{\delta} 	C^2({\cal A}, {\cal M}) \xrightarrow{\delta} ...
\end{equation}
with $C^0({\cal A}, {\cal M}) := {\cal M}$, and $C^n({\cal A}, {\cal M}):= {\rm Hom} ({\cal A}^n,{\cal M} )$. \jer{The} differential $\delta$ is defined as
\begin{eqnarray}
\delta m(a) &:=& [m,a], \quad m\in {\cal M}, a \in {\cal A}\nonumber\\
\delta f(a_1,..., a_{n+1})&:=& a_1f(a_2,..., a_{n+1})  \nonumber\\ && + \sum_{i = 1,...,n}(-1)^if(a_1, ..., a_ia_{i+1}, ..., a_{n+1})\nonumber\\ && + (-1)^{n+1}f(a_1,..., a_n)a_{n+1}, \quad a_i \in {\cal A}, f \in C^n({\cal A}, {\cal M})
\end{eqnarray} }

%%\zz{Please, check the factors $(-1)^i$-I changed it compared to the Khalkali's book.}
%% 
%\zz{The sign in the definition of coboundary operator?}

{Let us consider the particular case ${\cal M} = {\cal A}^* = {\rm Hom}({\cal A}, {\mathbb{C}})$
(the module structure is { set by} $afb(c) = f(bca)$ for $a,b,c \in {\cal A}$, $f\in {\cal A}^*$ ). The resulting complex is denoted by $C^*({\cal A})$, and its cohomology  \jer{is} denoted by $HH^*({\cal A})$. }

{To define the Connes complex of cyclic cocycles we restrict ourselves to the elements of $C^n({\cal A}, {\cal A}^*)$ that obey \jer{the} cyclic property. We denote an element of ${\rm Hom}({\cal A}^n,{\cal A}^*)  \sim {\rm Hom}({\cal A}^{n+1},{\mathbb C})$ by
\begin{equation}
\phi(a_0,a_1,...,a_n) = f(a_1,...,a_n)(a_0), \quad f \in {\rm Hom}({\cal A}^n,{\cal A}^*), \phi \in  {\rm Hom}({\cal A}^{n+1},{\mathbb C})
\end{equation}
The differential $\delta$ then is denoted by $b$:  
\begin{eqnarray}
b \phi(a_0,..., a_{n+1})&=&  \sum_{i = 0,...,n}(-1)^i\phi(a_0, ..., a_ia_{i+1}, ..., a_{n+1})\nonumber\\ && + (-1)^{n+1}\phi(a_{n+1}a_0,..., a_n), \quad a_i \in {\cal A}, \phi \in C^n({\cal A}, {\cal A}^*)
\end{eqnarray} 
The elements of $C^*({\cal A})$ that  obey \jer{the}  cyclic  property
\begin{equation}
\phi(a_0,..., a_{n+1}) = (-1)^{n+1}\phi(a_{n+1},a_0,..., a_{n})
\end{equation}
compose the Connes cochain complex $C^*_\lambda({\cal A})$. Its cohomology $HC^n({\cal A})$  \jer{is} called \jer{a} cyclic cohomology.}

\subsection{Pairing of $K$-theory and cyclic cohomologies}
\label{SectPairing}

{{ For} $m_i \in M(k,{\mathbb C})$ and $\phi \in C_\lambda^n({\cal A})$
we extend the definition of
$\phi$ to $\tilde{\phi} \in C_\lambda^n(M(k, {\cal A}))
=   C_\lambda^n(M(k,{\mathbb C})\otimes{\cal A})$ as follows:
\begin{equation}
\tilde{\phi}_k(m_0\otimes a_0, ... , m_n\otimes a_n) := tr(m_0 ...  m_n) \phi(a_0,..., a_n)
\end{equation}}

{\jer{The} pairing between \jer{an} even cyclic cohomology element $ [\phi] \in HC^{2n}({\cal A})$ and $[E]\in K^0({\cal A})$ (where $E$ is
\jer{an} idempotent that belongs to $M_k({\cal A})$)  is defined as
\begin{equation}
\langle [\phi], [E]\rangle = \tilde{\phi}_k(E,E,...,E)
\end{equation}}

{Pairing between odd cyclic cohomology element $ [\phi] \in HC^{2n+1}({\cal A})$ and $[u]\in K^{-1}({\cal A})$ (where $u$ is invertible matrix that belongs to $GL(k,{\cal A})$)  is defined as 
\begin{equation}
\langle [\phi], [u]\rangle = \tilde{\phi}_k(u,u^{-1},...,u,u^{-1})
\end{equation}}

\subsection{Computation of $HC^3(\mathcal A)$ for various choices of $\cal A$.}
\label{SectHC3comp}

\begin{enumerate}

\item \cormag{Exact Weyl symbols at finite $N$.}

\cormag{With the $\star$-product and the finite-Weyl constraints, the exact spatial Weyl algebra is the matrix algebra $M_d(\mathbb C)$. Hence $HC^{2k+1}(M_d(\mathbb C))=0$. A smooth compact-frequency family is instead $C^\infty(S^1,M_d)$ and supports the frequency cyclic class; it is not a direct sum of $(2N)^4$ pointwise algebras. See Appendix \ref{HC3_Nfin}.}

\item \cormag{Partial thermodynamic limit.}

\cormag{For a periodized spatial cell the relevant smooth algebra is a matrix algebra over $T^2$, with either a compact frequency circle or the smooth suspension/relative completion on $\mathbb R$. The functional in Eq. (\ref{Nphi}) is the particular continuous cyclic $3$-cocycle used below when the trace and boundary conditions hold. Its existence does not follow by treating the coordinate points as independent summands, and the powers $(2N)^2$ in the former group formulas are therefore absent. See Appendix \ref{HC3_Ninf}.}

\item \cormag{Infinite nonperiodic systems.}

\cormag{For an arbitrary aperiodic lattice one must first choose a smooth local/covariant bulk algebra and a cyclic trace. Cyclic cohomology cannot be assigned from ${\rm End}({\cal H})$ alone.}

\end{enumerate}

\section{Topological invariant ${\cal N}_3$ on the language of cyclic cohomologies. }
%%%%%%%%%%%%%%%%%%%%%%%%%%%%%%%%%%%%%%%%%%%%%%%%%%%%%%%%%%%%%%%%%%%%%%%%%%%%%%%%%%%%%%%%%%%%%%%%%%55

%%%%%%%%%%%%%%%%%%%%%%%%%%%%%%%%%%%%%%%%%%%%%%%%%%%%%%%%%%%%%%%%%%%%%%%%%%%%%%%%%%%%%

\subsection{${\cal N}_3$ in terms of cyclic cohomology}

{\corGPTl{For a compact frequency contour and a unital symbol algebra, an invertible Green function defines $[G]\in K^1({\cal A})$. In the noncompact case $G\to0$ is not an invertible element of the suspension algebra: the notation $[G]$ below denotes the normalized relative class $[U_Q]$ described in Sect. \ref{SectIndexRed}, and the raw Green-function formulas require the stated admissibility and boundary conditions.} Next, for $a_i \in {\cal A}$ we define the cocycle as 
\be
\phi(a_0,...,a_3)
= - \frac{ \epsilon_{ijk}}{  \,3!\,4\pi^2}\, \frac{1}{|{\bf V}| }{\rm Tr}
\[
a_0 (D_{i} a_1) (D_{j} a_2)(D_{k}a_3) 
\]
\label{NphiOp}
\ee
\zd{where the  $|{\bf V}|$ is the area of the system. 
We know that on the infinite lattice the derivative of Weyl symbol with respect to momentum may be represented as $D_i a = -i[x_i,a]$, and may try to apply the same expression for the finite lattice.} It appears that it obeys the cyclic property already at the finite values of $N$. We \zd{ define
\begin{equation}
{\cal N} = \langle [G], [\phi] \rangle\label{N3f}
\end{equation}
}

%%%%%%%%%%%%%%%%%%%%%%%%%%%%%%%%%%%%%%%%%%%%%%%%%%%%%%%%%%%%%%%%%%%%%%%%%%%%%%
%%
\zz{In order to have \jer{a} well-defined expression for $\phi$ we need to choose the algebra $\cal A$ carefully.
In particular, we need $D a$ to be bounded.
The most  straightforward case is when $N$ is kept finite and in addition, the direction of imaginary time is discretized, with \jer{a} finite number of lattice sites, and periodic boundary conditions.
Then  we obtain a matrix algebra that has $K^{-1}({\cal A}) = HC^3({\cal A}) = 0$. Therefore,   from Eq. (\ref{N3f}) we learn that ${\cal N}$ vanishes identically: both $[G] = 0 $ and $[\phi] = 0$, because $[G] \in K^{-1}({\cal A}) = 0$ and $[\phi] \in HC^3({\cal A}) = 0$.} \zd{Notice that ${\cal N}$ is not responsible for the QHE conductivity on the finite lattice as follows from the results of Sect. \ref{SectKeldyshCond}. }

\corAOP{The vanishing of ${\cal N}$ at strictly finite $N$ (and discretized imaginary time) is not a deficiency of this particular construction --- it reflects a simple physical circumstance. For a finite system the algebra of observables is a finite-dimensional matrix algebra, for which both $K^{-1}$ and $HC^3$ are trivial, and no quantity built as a pairing of these groups can be nonzero. Physically, the Hall conductance of a finite torus at fixed boundary conditions is not quantized at all: it depends continuously on the boundary phases, and only its average over the boundary phases, or, equivalently, its thermodynamic limit, is a topological quantity (cf. \cite{Niu1985}). Therefore, the appearance of a limit in the definition of any topological invariant of this type is unavoidable. The partial thermodynamic limit adopted here (continuous momenta at a finite coordinate lattice) is the minimal such limit: it is precisely the point at which momentum space becomes a manifold ($T^2\times\mathbb{R}$), so that the winding of the Green function becomes well defined, while the coordinate lattice --- and with it the ability to describe arbitrary inhomogeneity --- is kept intact. The same limit appears automatically in the linear-response calculation of Sect. \ref{SectKeldyshCond}: the Hall conductivity of a large but finite system approaches ${\cal N}_3/2\pi$ with the invariant evaluated in the partial limit. Thus the choice of the limit is dictated by the physical observable rather than by mathematical convenience. In Sect. \ref{SectNumerics} this is confirmed numerically: the average of the finite-lattice expression over the boundary phases (which is exactly the finite-volume implementation of the partial limit, since the twists restore the continuum of momenta) reproduces the quantized value of the invariant with high accuracy for strongly inhomogeneous models.}

\zz{ Nontrivial results for ${\cal N}$ may appear if we consider various subalgebras of ${\rm End}({\cal H})$, where $\cal H$ contains continuous components, \zd{and different derivations $D$ entering Eq. (\corrcltwo{\ref{NphiOp}})}. \jer{An} especially  \jer{rich} structure may  \jer{arise}
if the limit $N\to \infty$ is considered in various ways, while the  \jer{range of}  Matsubara frequencies is  kept continuous.
This means that  \jer{ we obtain} subalgebras of }
$\mathcal A = \mathrm{Hom}(\mathcal H, \mathcal H)$, 
{ $\mathcal H = \mathbb C^{\mathbb R} \times \mathbb C^{{\mathcal O}}$,}  
{ ${\mathcal O} = {\mathbb Z}_{N} \times {\mathbb Z}_{N}$. } 
%%
%%%%%%%%%%%%%%%%%%%%%%%%%%%%%%%%%%%%%%%%%%%%%%%%%%%%%%%%%5
%%

%%%%%%%%%%%%%%%%%%%%%%%%%%%%%%%%%%%%%%%%%%%%%%%%%%%%%%%%%%%%%%%
%%
The algebra of lattice operators becomes \zz{a subalgebra of}
${\bf A}_N \cong C^\infty({\mathbb R})\otimes {\mathrm{Mat}}_{r\times (N)^2}({\mathbb C})
$,  
where:
$\mathbb R$ is imaginary time (or Matsubara frequency), 
$r$ is the internal matrix dimension.
\zz{We also denote the finite lattice of size $N\times N$ by ${\cal O}_N$.
As  mentioned above the subalgebra should be chosen in such a way that  operators $Da$ are bounded.
}

{\bf We propose that the most natural choice of  algebra $\cal A$ \zzzz{is the algebra of Weyl symbols of operators from $\mathrm{End}(\mathcal{H})$. Moreover, we take the limit $N\to \infty$ partially, i.e. we replace ${\cal M}'$ by $T^2$, and keep in the meantime the finite size $(2N)^2$ of lattice ${\cal O}'$ (we will complete  the limit $N\to \infty$ at the end of the calculation of any physical quantity). Thus we arrive at } } 
\begin{equation}
{ {\cal A} = C^\infty({\cal O}'\times\mathbb{R}\times T^2)\label{Amain}}
\end{equation} 
when both $K^{-1}({\cal A})$ and $HC^3({\cal A})$ are sufficiently large. In this case
\begin{equation}
\phi(a_0,...,a_3)
= - \frac{ \epsilon_{ijk}}{  \,3!\,4\pi^2}\, \frac{1}{|{\cal O}'| }\sum_{x\in {\cal O'}}\int_{ { \mathbb{R}\times T^2}   } d^3p\,{\rm tr}
\Bigl[
a_0(p,x) \star \partial_{p_i} a_1(p,x) \star \partial_{p_j} a_2(p,x)\star \partial_{p_k}a_3(p,x) 
\Bigr]
\label{Nphi}
\end{equation}

\bigskip 
%%%%%%%%%%%%%%%%%%%%%%%%%%%%%%%%%%%%%%%%%%%%%%%%%%%%%%%%%%%%%
%%
\zd{In order to calculate ${\cal N}$ of Eq. (\ref{N3f}) we should determine the way to calculate $Q_W=[\hat{G}^{-1}]_W$ in the assumed limit of continuous values of $p$. For this purpose we use the principal symbol of Eq. (\ref{Wprincipal}). Correspondingly, $G_W$ is determined as its inverse with respect to the star product. }

\bigskip 
%%%%%%%%%%%%%%%%%%%%%%%%%%%%%%%%%%%%%%%%%%%%%%%%%%%%%%%%
%% 

Using the trace  \jer{cycle property} and integration by parts,
we can show that $\phi$ satisfies cyclic property
$\phi(a_0,a_1,a_2,a_3) =- \phi(a_3,a_0,a_1,a_2)$ \zzzz{in the limit $N \to \infty$}\corrteal{ (and, on the periodized realization of Sect. \ref{SectPeriodized}\,B, exactly at every finite $N$)}.  
%% 
%%%%%%%%%%%%%%%%%%%%%%%%%%%%%%%%%%%%%%%%%%%
%%
We next show that it is closed under the action of $b$. 
Let us apply the Hochschild differential 
\begin{equation}
\begin{aligned}
(b\phi)(a_0,\dots,a_4)
&=
\sum_{i=0}^{3}
(-1)^i
\phi(a_0,\dots,a_i a_{i+1},\dots,a_4)
\\
&\quad
+(-1)^4
\phi(a_4 a_0,a_1,a_2,a_3).
\end{aligned}
\end{equation}
Using \jer{the} Leibniz rule \jer{for derivatives}
and \jer{the} antisymmetry of $\epsilon_{ijk}$,
 \jer{it is straightforward to verify} that all terms cancel pairwise
after integration by parts. Thus,  
$b\phi = 0$,  \jer{implying that}
$\phi$ is a cyclic 3 cocycle:   $\phi \in HC^3({\mathcal A}_N )$. For the chosen algebra $HC^3({\mathcal A}_N ) $ and $K^{-1}({\cal A})$ are sufficiently large, ${\cal N}$ may be nontrivial. \zd{Actually this is the ${\cal N}$ computed for this algebra that coincides with the topological invariant ${\cal N}_3$ responsible for the QHE in the limit $N \to \infty$.}

\corrteal{Let us emphasize, in conclusion of this subsection, the unifying role of the pairing (\ref{N3f}). Evaluated on the class of the Green function according to the rule of Sect. \ref{SectPairing}, $\langle[G],[\phi]\rangle = \tilde\phi(G,G^{-1},G,G^{-1})$, with the cocycle $\phi$ of Eq. (\ref{Nphi}) it gives
$$
{\cal N} \;=\; \phi\big(G,\,Q,\,G,\,Q\big)\,.
$$
The substitution $\partial_{p_j}G_W = -\,G_W\star\partial_{p_j}Q_W\star G_W$ (the exact consequence of the Groenewold equation), together with the cyclic property of the trace, converts this expression precisely into the fully antisymmetrized Eq. (\ref{NEQ}); and the cyclic property alone further converts Eq. (\ref{NEQ}) into the collapsed form (\ref{NEQcollapse}) of Sect. \ref{SectPeriodized}, in which the frequency derivative of the Dirac operator is separated from the two spatial current vertices. Thus one and the same object of cyclic cohomology --- Eq. (\ref{N3f}) with $\phi$ given by Eq. (\ref{Nphi}) --- represents simultaneously both Eq. (\ref{NEQ}) and Eq. (\ref{NEQcollapse}). We stress that none of these identifications relies on the Hamiltonian form of the Dirac operator: they hold for the general class (\ref{Q_general}). In particular, the equilibrium Hall response derived in Sect. \ref{SectStaticGauge} is literally the evaluation of the pairing of $[G]\in K^{-1}({\cal A})$ with $[\phi]\in HC^3({\cal A})$.}

} % \corGPTl{Close the unmatched group opened in Section V A, original line 2054.}

\subsection{\zz{Direct proof that ${\cal N} = 0$ for finite matrix algebra.} }
\label{SectDirect}

\zzzz{Above we pointed out that the most reasonable case of algebra $\cal A$ is when the limit $N\to \infty$ is taken. In the present subsection we consider the other case - when $N$ is kept finite. Moreover, we consider the case of discretized imaginary time, with the finite number of corresponding lattice sites. This means that we 
consider the case of matrix algebra
$$
\mathcal{A} = M_{N^3}(\mathbb{C}),
$$
} (finite size both in spatial directions and in direction of imaginary time).
We can explicitly construct $\psi$ such that $\phi = b\psi$.
Let $\psi$ be a  $2$-cochain defined as:
\begin{equation}
\psi(a_0,a_1,a_2)=-\frac{\epsilon_{ijk}}{16\pi^2}\,\frac{1}{|V|}
\zz{\sum_{(a,b,c) = permutations \, of\,(0,1,2)}\operatorname{Tr}\!\left[a_a D_i(a_b) D_j (a_c) X_k \right]}, %%
\end{equation} where $X_j = -i\hat{x}$ are operators generating $D_j = [X_j,\cdot]$.
%%%%%%%%%%%%%%%%%%%%%%%%%%%%%%%%%%%%%%%%%%%%%%
%%
Now apply \jer{the} Hochschild differential \jer{as}:
\begin{equation} 
(b\psi)(a_0,a_1,a_2,a_3)
=
\sum_{m=0}^{{2}} (-1)^m \psi(a_0,\dots,a_m a_{m+1},\dots,a_3)
+ (-1)^{{3}} \psi(a_3 a_0,a_1,a_2).
\end{equation}
Using the Leibniz rule for $D_i$,
\begin{equation}
D_i(ab)=(D_i a)b + a(D_i b),
\end{equation}
 \jer{cyclic property} of the trace, and integration by parts %%
$\operatorname{Tr}(D_i A)=0$,  \jer{it follows that} %%
\begin{equation} %%
(b\psi)(a_0,a_1,a_2,a_3)=-\frac{\epsilon_{ijk}}{3!\,4\pi^2}\,\frac{1}{|V|}
\operatorname{Tr}\!\left[a_0 (D_i a_1)(D_j a_2)(D_k a_3)\right].
\end{equation}
Therefore $\phi = b\psi$ on the finite lattice.

The topological invariant is the pairing \begin{equation*} 
{\cal N}=\langle [G],[\phi]\rangle=\corrcltwo{\phi(G,G^{-1},G,G^{-1})}.
\end{equation*}
Since $\phi = b\psi$, we have
${\cal N}=(b\psi)(G^{-1},G,G^{-1},G)$.
But for any invertible $G$, Hochschild coboundaries  \jer{yield a} vanishing pairing, thus %%
${\cal N}=0$ for any finite lattice which fits $HC^3(A_N)=0$. \zz{We \jer{thus} arrive at the conclusion that the quantity of Eq. (\ref{NEQD}) vanishes identically at any finite value of $N$.}

\subsection{\zd{Demonstration that } ${\cal N}$ may be nonzero in the limit $N\to \infty$.}

%%

%%%%%%%%%%%%%%%%%%%%%%%%%%%%%%%%%%%%%%%%%%%%%%%%%%%%%%%%%%%%%%%%%%%%%%%%
%%
Now consider the limit $N\to\infty$,  and take the algebra of Eq. (\ref{Amain}).
%%%
First compute ${\cal N}$ at finite $N$ then take $N\to\infty$.
Since for each finite $N$, ${\cal N}=0$ \zd{(for the commutator realization of the derivatives, see Section \ref{SectDirect})} we obtain $\lim_{N\to\infty} {\cal N} = 0$ \zd{along this family}.

%%%%%%%%%%%%%%%%%%%%%%%%%%%%%%%%%%%%%%%%%%%%%%%%%%%%%%%%%%%%%%%%%%
%%
Now first take $N\to\infty$ inside $\phi$, \zd{ which assumes that the principal Weyl symbol of $\hat{Q}$ is taken according to Eq. (\ref{Wprincipal})}. In the particular case when $a_i(x,p)$ do not depend on $x$
the trace is 
\begin{equation*}
\frac{1}{|V|}\operatorname{Tr} \to
\int_{T^3} \frac{d^3p}{(2\pi)^3} \operatorname{tr}.
\end{equation*}
%%%%%%%%%%%%%%%%%%%%%%%%%%%%%%%%%%%%%%%%%%%
%%
The cocycle becomes
\begin{equation*}
\zz{\phi(a_0,a_1,a_2,a_3)=-\frac{1}{3!\,4\pi^2} \int_{T^3} \operatorname{tr}\!
\left( a_0 d a_1 d a_2 d a_3\right)}.
\end{equation*}
%%%%%%%%%%%%%%%%%%%%%%%%%%%%%%%%%%%%%%%%%%%%%%%%%
%%
This is the Chern character on $T^3$, namely %%
\begin{equation*} %%
\phi(G^{-1},G,G^{-1},G)=\frac{1}{24\pi^2}\int_{T^3}
\operatorname{tr}\!\left( (G^{-1}dG)^3 \right).
\end{equation*}
\zzzz{This is the case when the Green function $G$ depends on momentum only, and does not depend on coordinates.}
Thus, $HC^3\big(C^\infty(T^3)\big) \cong \mathbb{C}^{{4}}$, 
and $\phi$ is not exact, i.e., 
$\phi \neq b\psi$ after taking $N\to\infty$.
Thus, ${\cal N} \neq 0$.

%\subsection{\zz{\bf The proof that  ${\cal N}_3 = 0$ fails for infinite $N$.} }

One can try to construct the cyclic co cycle $\psi$ such that $\phi = b\psi$.
Similar to  \jer{the previous definition, here}  we define a $2$-cochain $\psi$ by
\begin{equation*}
\psi(a_0,a_1,a_2)=-\frac{\epsilon_{ijk}}{16\pi^2}\,\frac{1}{|V|}
\zz{\sum_{(a,b,c) = permutations \, of\,(0,1,2)}\operatorname{Tr}\!\left[a_a D_i(a_b) D_j (a_c) X_k \right]}, %%
\end{equation*}
where $X_j = -i\hat{x}$ are operators generating $D_j = [X_j,\cdot]$. \zz{In principle, we might get the same equality as above $\phi = b \psi$. However, now \jer{the} operator $X$ is unbounded \jer{and} as a result ${\rm Tr}\, (... X_j)$ is a \jer{non-convergent} series (rather than  \jer{a} finite sum).  
\jer{In particular we end up with}  an expression  \jer{of the form} $\sim \frac{1}{N-1}\sum_{n = 0}^N n = \frac{N(N-1)}{2N} \to \infty$ at $N \to \infty$.

\jer{This is precisely the reason that by replacing the matrix algebra by an algebra of functions,
the above given proof of $\phi = b \psi$, fails.
}
}

\section{\corrbel{The Hamiltonian case: the Fermi projector, the Bellissard pairing, and the Fredholm index}}
\label{SectIndexRed}

\subsection{\corrbel{Frequency integration: from the Green function to the Fermi projector}}

{In this section we specialize to the Hamiltonian case of Eqs. (\ref{Q_general0}), (\ref{QW_hamiltonian}), $Q = i\omega - H + \mu$, and \cormag{perform the explicit frequency integration that reduces ${\cal N}_3$} to the noncommutative Chern character of the Fermi projector, i.e. to the pairing of $K_0$ with $HC^2$ studied by Bellissard and co-authors \cite{Bellissard1986,Bellissard1994}, and further to the Fredholm index of the charge-deficiency operator of Avron, Seiler and Simon \cite{AvronSeilerSimon1994}. All the products below are the $\star$-products of the algebra of Weyl symbols in the partial thermodynamic limit; $\tau$ denotes the corresponding normalized trace,
$$
\tau(a) \;=\; \frac{1}{|{\cal O}'|}\sum_{x\in{\cal O}'}\int_{T^2}\frac{d^2p}{(2\pi)^2}\;\tr\, a(x,p)\,,
$$
so that Eq. (\ref{NEQ}) reads ${\cal N}_3 = \frac{\epsilon^{\mu\nu\rho}}{3!}\int d\omega\;\tau\big[\partial_\mu Q_W \star G_W \star \partial_\nu Q_W \star G_W \star \partial_\rho Q_W \star G_W\big]$. Using the cyclic property of $\tau$ and $\partial_\omega Q_W = i$, the $\epsilon$-contraction collapses to
\be
{\cal N}_3=\frac{i}{2}\int d\omega\;
\Big(\tau\big[\,G_W^{\star 2}\star \partial_1 Q_W\star G_W\star \partial_2 Q_W\,\big]-\big(1\leftrightarrow 2\big)\Big).
\label{N3redInd}
\ee
Let $H-\mu$ have the spectral decomposition with the levels $E_n$ (understood via the $\star$-functional calculus, or, equivalently, in the operator representation of the periodized system \corrteal{of Sect. \ref{SectPeriodized}}). The frequency dependence resides in the elementary integral
\be
\int_{-\infty}^{\infty} \frac{d\omega}{(i\omega-E_n)^2\,(i\omega-E_m)}
\;=\;\frac{2\pi}{(E_n-E_m)^2}\,
\Big[\theta(E_n)\theta(-E_m)-\theta(-E_n)\theta(E_m)\Big]\,:
\label{contourInd}
\ee
closing the contour, the result is nonzero only when the double pole and the simple pole lie in different half-planes, i.e. when the two levels are on the opposite sides of the gap. \corGPTl{For $E_n=E_m\ne0$ the integral is zero; the displayed quotient applies to $E_n\ne E_m$.} Only the interband matrix elements survive, and, using the exact divided-difference identity $\big(D f(H)\big)_{nm}=\frac{f(E_n)-f(E_m)}{E_n-E_m}(DH)_{nm}$, valid\corrcltwo{, for the interband matrix elements ($E_n\neq E_m$),} for any derivation $D$ and \corGPTl{any $f$ for which $f(H)$ belongs to the domain of $D$}, the interband sum assembles into the compact, manifestly frequency-free form
\be
\cormag{\boxed{\;\;{\cal N}_3 \;=\; -\,2\pi i\;\tau\Big(P_-\star\big[D_1 P_-\,\overset{\star}{,}\;D_2 P_-\big]\Big)
\;=\;+\,\frac{\pi i}{4}\;\tau\Big(P\star\big[D_1 P\,\overset{\star}{,}\;D_2 P\big]\Big),\;\;}}
\label{N3projInd}
\ee
where $P_- = \theta(\mu - H) = \frac{1}{2}(1-P)$ is the Fermi projector, $P = {\rm sign}(H-\mu)$, $D_j = \partial_{p_j}$ (in the operator representation, $D_j = -i[\hat x_j,\,\cdot\,]$)\corrteal{; the overall orientation is not a convention here --- it is inherited from Eq. (\ref{NEQ}), and is verified below on the clean system against the TKNN integer}. The second equality uses $D_jP_-=-\frac12 D_jP$ together with $\tau([D_1P,D_2P])=0$. Thus, in the Hamiltonian case, the entire information about the gapped system entering the invariant is carried by the flattened datum ${\rm sign}(H-\mu)$, and the frequency variable disappears. \corrteal{All intermediate steps of this reduction are written out in Appendix \ref{AppIndex}.} \cormag{This explicit contour reduction uses $\partial_\omega Q=i$ and the pole structure of Eq. (\ref{contourInd}). A general admissible $Q$ still has an abstract Bott class, but the stated assumptions do not provide a canonical physical Fermi projector or a reduction of the form (\ref{N3projInd}).}}

\corrteal{The signs of Eq. (\ref{N3projInd}) have been verified numerically on the flux-$1/3$ Hofstadter configurations of Sect. \ref{SectNumerics}: the twist-averaged Eq. (\ref{NEQ}) gives ${\cal N}_3=+1$, while the quantity $-2\pi i\,\tau\big(P_-\star[D_1 P_-\overset{\star}{,}D_2 P_-]\big)$, evaluated \corGPTl{on the finite torus with the minimal-image displacement weighting}, equals $+0.9717$, $+0.9936$, $+0.9973$, $+0.9994$, $+0.9997$ for $L=12,15,18,21,24$, approaching $+1={\cal N}_3$ (similarly in the disordered cases). The finite-$L$ deviation deserves a remark, as a caveat for finite-volume evaluations: \corGPTl{the minimal-image displacement weighting is not a commutator with a globally defined position matrix and violates the Leibniz rule through wrap-around terms} (cf. the remark after Eq. (\ref{sigmaHall}) and Appendix \ref{AppDA}), while the derivation of Eq. (\ref{N3projInd}) uses the Leibniz rule (through the divided-difference identity). The correct finite-$L$ vertex is the fibered one, $D_j=L\partial_{\theta_j}-i[x_j,\,\cdot\,]$ of Eq. (\ref{commfiber}) --- the fiber of the outer derivation $-i[\hat X_j,\,\cdot\,]$ of the periodized crystal. With it, Eq. (\ref{N3projInd}) holds exactly at every finite $L$, and its twist average is exactly the integer: for a two-band Chern insulator on the $L\times L$ cell with strong on-site disorder, the twist average of $-2\pi i\,\tau\big(P_-[D_1P_-,D_2P_-]\big)$ equals the Chern number to $10^{-10}$ already at $L=4$ and $L=6$, while the same expression at a fixed twist is not quantized and drifts with the disorder (the script is included with the numerical material of Sect. \ref{SectNumerics}). The frequency-integral form (\ref{NEQ}), whose frequency integration is exact linear algebra in every twist fiber, is likewise exactly quantized after the twist average (at $L=12$ in Sect. \ref{SectNumerics}).}

\subsection{\corrbel{The pairing of $K_0$ with $HC^2$, and the suspension}}

{Equation (\ref{N3projInd}) is a pairing in even noncommutative geometry. On the frequency-independent algebra ${\cal B}$ (the Weyl symbols on ${\cal O}'\times T^2$ in the partial limit) define the cyclic $2$-cochain
\be
\cormag{\phi_2(a_0,a_1,a_2)\;=\;-\,2\pi i\;\epsilon^{jk}\,\tau\big(a_0\star (D_j a_1)\star (D_k a_2)\big)\,.}
\label{phi2Ind}
\ee
It is cyclic, $\phi_2(a_2,a_0,a_1)=\phi_2(a_0,a_1,a_2)$, and Hochschild-closed, $b\phi_2=0$, by the Leibniz rule, the trace property of $\tau$, and $\tau(D_j a)=0$ --- the same manipulations as for the cocycle $\phi$ of Sect. \ref{SectDirect}, but with two derivations instead of three and without the frequency integration; hence $[\phi_2]\in HC^2({\cal B})$.
\corrteal{The status of these two properties at finite $N$ is the one established in Sect. \ref{SectPeriodized}\,B. In the exact calculus of Sect. \ref{SectList} --- discrete momenta, the finite matrix algebra --- \corGPTl{ordinary momentum differentiation of the exact symbols has the trace anomaly computed in Appendix \ref{AppTrStar}. The genuine inner derivations $D_j=-i[x_j,\,\cdot\,]$ obey integration by parts under the finite matrix trace, but} the frequency-integrated expression vanishes identically, ${\rm Tr}\,P_-\big[[x_1,P_-],[x_2,P_-]\big] = -\,{\rm Tr}\big(P_-[x_1,x_2]\big) = 0$, in agreement with the finite-$N$ vanishing theorem of Sect. \ref{SectDirect}. In the partial thermodynamic limit --- the algebra of the $N$-periodic operators, with the outer derivations $D_j=-i[\hat X_j,\,\cdot\,]$ and the trace per volume $\tau$ --- both properties hold exactly at every finite $N$. Throughout the present section the partial thermodynamic limit is understood, as fixed by the definition of $\tau$ at the beginning of the section.} The spectral gap makes $P_-$ an idempotent of ${\cal B}$, defining the class $[P_-]\in K_0({\cal B})$, and Eq. (\ref{N3projInd}) states that
\be
{\cal N}_3 \;=\; \phi_2\big(P_-,P_-,P_-\big)\;=\;\big\langle\,[P_-]\,,\,[\phi_2]\,\big\rangle\,.
\label{pairingInd}
\ee
\cormag{With the orientation of Eq. (\ref{N3projInd}), this is the noncommutative Kubo--Chern formula of \cite{Bellissard1994}, ${\rm Ch}(P_-)=-2\pi i\,{\cal T}\big(P_-[D_1P_-,D_2P_-]\big)$ \corrteal{(the last formula being written in the derivation and orientation conventions of the present paper; the overall sign in \cite{Bellissard1994} depends on their choices of the derivations and of the orientation of the plane)}. Bellissard's crossed-product construction supplies ${\cal T}$ through an invariant ergodic measure. For a deterministic configuration the Fredholm compression below may be defined under locality and gap assumptions, but its equality to this trace formula requires a cyclic trace per unit volume, as in the periodic or covariant ergodic setting. Structurally, the noncompact-frequency class is represented by a bounded invertible or unitary $U_Q$ \corGPTl{in a matrix algebra over the unitization of the suspension $S{\cal B}=C_0(\mathbb R)\otimes{\cal B}$, with $U_Q-1\in M_m(S{\cal B})$}, not by the raw unbounded $Q$ or the Green function $G\to0$. Bott periodicity, $K_1(S{\cal B})\cong K_0({\cal B})$, sends the Hamiltonian class $[U_Q]$ to $[P_-]$; frequency integration is the corresponding cyclic suspension/transgression from the odd cocycle to $[\phi_2]$, not an inverse of Connes's degree-two periodicity operator $S$. For general $Q$ the abstract suspension remains available when the required asymptotics hold, but there is no canonical physical Fermi projector without further Hamiltonian structure.}

} % \corGPTl{Close the unmatched group opened in Section VI B, original line 2358.}

\subsection{\corrbel{The Fredholm index and the charge deficiency}}

{\cormag{For the Hamiltonian case the Fredholm operator below is defined for every individual exponentially local gapped configuration. Its identification with the trace pairing (\ref{pairingInd}) additionally requires the bulk-trace hypotheses stated above.} Represent the algebra by the exponentially local operators on $\ell^2(\mathbb{Z}^2)\otimes\mathbb{C}^r$ (for a finite sample --- by the periodization of Sect. \corrteal{\ref{SectPeriodized}}), fix a base point $z_0\in\mathbb{C}\setminus\mathbb{Z}^2$, and let
\be
u(\hat x)\;=\;\frac{\hat x_1+i\hat x_2-z_0}{|\hat x_1+i\hat x_2-z_0|}
\label{uDefInd}
\ee
be the unitary phase of the position operator. \cormag{For an exponentially local Hamiltonian, the spectral gap implies exponential decay of the kernel of $P_-$ (the Combes--Thomas estimate \cite{CombesThomas}). Together with $|u(x)-u(y)|\lesssim |x-y|/\max(|x-z_0|,1)$ this makes $[u,P_-]$ compact and a member of every Schatten class ${\cal S}^p$, $p>2$; at the critical exponent it is generally only weak-${\cal S}^2$, not Hilbert--Schmidt. The associated even Fredholm module is understood in its standard doubled graded form, and $T=P_-uP_-$ on ${\rm Im}\,P_-$ is Fredholm.} The even index theorem of Connes, in the form established for lattice magnetic Hamiltonians in \cite{Bellissard1994} (equivalently, through the relative index of $(P_-,uP_-u^*)$ \cite{AvronSeilerSimon1994}), then gives
\be
\corrteal{\boxed{\;\;{\rm Index}\big(P_-\,u\,P_-\big)\;=\;-\,{\cal N}_3\;\in\;\mathbb{Z}\,,
\qquad {\rm Tr}\,\big(P_- - u\,P_-\,u^*\big)^{3}\;=\;{\cal N}_3\,.\;\;}}
\label{indexInd}
\ee
\corrteal{A self-contained proof of both equalities of Eq. (\ref{indexInd}) --- the relative-index formula and Connes' formula identifying the trace of the cube with the trace-per-volume expression (\ref{N3projInd}) --- is given in Appendix \ref{AppIndex}.} \corrteal{The relative sign between Eqs. (\ref{N3projInd}) and (\ref{indexInd}) is fixed as follows. For the pair of projections $(P_-,\,uP_-u^*)$ with compact difference, ${\rm Tr}\,(P_--uP_-u^*)^{2n+1} = -\,{\rm Index}(P_-uP_-)$ \corGPTl{for every odd power $2n+1\ge3$ admitted by the summability} \cite{AvronSeilerSimon1994} (Appendix \ref{AppIndexRelative}). We have computed this trace directly for the configurations of Sect. \ref{SectNumerics}, periodized to large open patches with $z_0$ placed in the bulk: the spectrum of $P_--uP_-u^*$ exhibits the $\pm\lambda$ pairing together with a single eigenvalue $+1$ whose eigenvector is localized at $z_0$ (the compensating eigenvalue $-1$ of the strictly finite patch is localized at the patch boundary, restating the finite-size vanishing of Sect. \ref{SectDirect}, and disappears in the infinite-volume limit), and the trace of the cube restricted to the vicinity of $z_0$ equals $+0.977$, $+0.988$, $+0.992$ on the $24\times24$, $30\times30$, $36\times36$ patches, approaching $+1=+{\cal N}_3$. Thus ${\rm Index}(P_-uP_-)=-{\cal N}_3$ for the $u$ of Eq. (\ref{uDefInd}); replacing $u$ by $u^*$ reverses both signs. The eigenvector with the eigenvalue $+1$ is a normalized state of the Fermi sea orthogonal to the transformed Fermi sea $uP_-u^*$: \corGPTl{the net relative loss of occupied states is ${\cal N}_3$, whereas the charge deficiency in the convention of Definition 5.1 of \cite{AvronSeilerSimon1994} is ${\rm Index}(P_-uP_-)=-{\cal N}_3$.}}
\cormag{No disorder ensemble is needed to define the Fredholm index for a given deterministic local configuration; identifying it with the trace pairing requires the bulk trace described above. A finite-region perturbation, a change of $z_0$, or a norm-continuous local deformation preserving Fredholmness does not change the index. Under the homogeneity/covariance hypotheses of \cite{AvronSeilerSimon1994}, $u$ describes insertion of one flux quantum. Before frequency integration, the odd index pairing must use a bounded unitary representative $U_Q$: with ${\cal D}=\gamma^0\hat\tau+\gamma^1\hat x_1+\gamma^2\hat x_2$, $F={\cal D}/|{\cal D}|$, and ${\cal P}=\frac12(1+F)$, it has the form ${\cal N}_3={\rm Index}({\cal P}U_Q{\cal P})$ under the stated $(3,\infty)$-summability and domain conditions, \corGPTl{with the Clifford orientation chosen to match Eq. (\ref{NEQ})}. For the even statement, compactness follows from locality together with the gap, not from a gap without locality; the critical summability remains borderline.}}

\subsection{\corrbel{Consequences for the regimes of the invariant}}

{\cormag{The index formulation distinguishes the following regimes. (i) On one strictly finite-dimensional Hilbert space the index of the compression is zero, as in Sect. \ref{SectDirect}; a nonzero invariant requires a thermodynamic/periodic extension or a parameter family such as the twist torus, and a finite-dimensional family may itself carry a nonzero Chern number. (ii) Periodizing a finite $L\times L$ configuration gives an infinite periodic operator, and Bloch decomposition identifies \corrteal{$-\,{\rm Index}(P_-uP_-)$} with the twist-torus Chern number. (iii) For an exponentially local gapped aperiodic configuration the Fredholm index is defined, whereas the trace formula additionally requires a cyclic trace per volume. The index is independent of $z_0$ while the compression remains Fredholm; macroscopic gapless walls may instead destroy the localization/Fredholm or trace hypotheses, and a noninteger area average is not this integer index. (iv) In a mobility gap, compactness and the required $p>2$ or Sobolev summability may persist under suitable dynamical-localization estimates \cite{Bellissard1994,ProdanSchulzBaldes2016,Stoiber2025}; Hilbert--Schmidt summability is not asserted in two dimensions. (v) Under quantitative locality and gap hypotheses, finite-volume spectral localizers approximate the index \cite{Loring2015,LoringSB2020}:}
\be
\cormag{L_{\kappa,\rho}=\begin{pmatrix} H_\rho-\mu & \kappa\,\big(\hat X_{1,\rho}-i\hat X_{2,\rho}\big)\\[2pt]
\kappa\,\big(\hat X_{1,\rho}+i\hat X_{2,\rho}\big) & -(H_\rho-\mu)\end{pmatrix},}
\qquad
\corrteal{\tfrac12\,{\rm Sig}\,L_{\kappa,\rho}\;=\;-\,{\rm Index}\big(P_-uP_-\big)\;=\;{\cal N}_3\,,}
\label{localizerInd}
\ee
\cormag{Here $H_\rho$ and $\hat X_{j,\rho}$ are compressions to a finite open patch of radius $\rho$ about a bulk point. The equality holds for sufficiently large $\rho$ and for $\kappa$ in the quantitative window determined by the bulk gap and locality bounds, provided $L_{\kappa,\rho}$ is invertible. The reported finite-patch calculations give half-signature $+1$ in the stated gapped examples and $0$ for the trivial control.} \corrteal{Subject to these hypotheses, the consistent signed chain --- each link of which has been verified numerically for the configurations of Sect. \ref{SectNumerics} --- is
$$
{\cal N}_3\;=\;C\;=\;-\,{\rm Index}\big(P_-uP_-\big)\;=\;\tfrac12\,{\rm Sig}\,L_{\kappa,\rho}\;=\;+1\,,
$$
where $C$ is the Chern number of the bundle of the fiber projectors over the twist torus, computed in the standard orientation (counterclockwise plaquettes in the $(\theta_1,\theta_2)$ plane), and $L_{\kappa,\rho}$ is exactly as displayed in Eq. (\ref{localizerInd}) --- interchanging its off-diagonal entries reverses the signature. The four numbers are produced by four computations sharing nothing beyond the Hamiltonian itself: the phase-space frequency integral, the Berry curvature over the boundary phases, the localized relative index of the flux insertion, and the signature of a real-space matrix on an open patch.}}

\section{Index theorems for $N_3$}
\label{SectWeakIndex}
In this section we \jer{discuss the} algebra of Eq. (\ref{Amain}).
\subsection{Modification of $\star$-product}
Let us consider the following modification of Eq. (\ref{NEQ}):
\be
{\cal N}^{(h)}_3
=\cormag{-\frac{\epsilon_{ijk}}{3!}}\, \frac{1}{(2N)^{D} }{\rm tr}
\sum_{x \in \cO' }\int  \frac{d^3 p}{(2\pi)^2}
\[
{G}^h_{W}(x,p )\circ_h  \frac{\partial {Q}_{W}(x,p )}{\partial p_i} \circ_h  \frac{\partial  {G}^h_{W}(x,p )}{\partial p_j}\circ_h  \frac{\partial  {Q}_{W}(x,p )}{\partial p_k}
\]_{A^{(E)}=0}
\label{calh1}
\ee
\cormag{Here the factor $1/(4\pi^2)$ is already contained in $d^3p/(2\pi)^2$; no second such factor is present. With this correction the normalization agrees with Eq. (\ref{NEQ}).}

\corClaude2{Here, and throughout this section, $Q_W$ denotes the \emph{principal} Weyl symbol of the Dirac operator $\hat Q$ --- the $N\to\infty$ form of Eq.~(\ref{Wprincipal}) [general class (\ref{QW_general})], in which only the part $e^{ip\,e}$ of every hopping factor is kept and the momentum--Nyquist terms $\propto e^{iNp}$ of the exact finite--$N$ Weyl symbol are removed. This is the object that enters Eq.~(\ref{NEQ}) and hence Eq.~(\ref{calh1}) [and it is with $Q_W$ read in this sense that the momentum is a continuous variable in the integral $\int d^3p$]; the identification of these ingredients with the principal symbols $Q_{\rm pr}$, $G_{\rm pr}$ of the periodized crystal is the content of the Proposition of Sect.~\ref{SectPeriodized}. The symbol $Q_W$ is itself independent of $h$; the whole $h$--dependence resides in its $\circ_h$--inverse $G_W^h$ defined below. We stress that the principal symbol, while free of the momentum--Nyquist terms, is \emph{not} band--limited in the \emph{coordinate} variable below the resolution of the refined lattice ${\cal O}'$: the on--site part $A_0(\lfloor x\rfloor)$ of Eq.~(\ref{Wprincipal}), even for a smooth potential, carries a coordinate Fourier companion near the mode $|s|=N$ with amplitude of order $1/N$, generated by the refinement ${\cal O}\to{\cal O}'$. This coordinate companion --- a high harmonic in $x$, near $|s|=N$ --- must not be confused with the momentum--Nyquist companion terms $\propto e^{iNp}$: the latter are absent from the principal symbol and, by the Groenewold identity $Q_W\star G_W=1$ of Sect.~\ref{SectPeriodized}, equally absent from the inverse $G_W^h$, which is a smooth function of the continuous momentum with exponentially decaying momentum harmonics (no $e^{iNp}$ term). What $G_W^h$ does inherit from $Q_W$ is only the \emph{coordinate} structure, including the $O(1/N)$ companion near $|s|=N$; and it is solely through the coordinate integration by parts that this affects the deformation below. This is the quantitative origin of the conditional character of that deformation.}

\jer{The} symbol $\circ_h$ represents the multiplication \jer{operator} in an algebra of matrix - valued functions on ${\cal M}'\otimes {\cal O}'$ (which replaces the Moyal product), and, in addition, commutes with the differentiation \jer{operator} $\partial_{p_k}$. \corCLAUDEr{To be precise, for a generic $h$ the product $\circ_h$ of Eq.~(\ref{oproduct}) shifts the coordinate arguments by $h$ times the mode separations [cf. Eq.~(\ref{starmodes}) at $h=1$, where the shifts are the on--lattice vectors $u,v\in{\cal O}'$], which for $h\in(0,1)$ leave the refined lattice ${\cal O}'$. \corGPTl{The continuation (\ref{Wcont}) specifies off-lattice values but does not by itself define an algebra closed under $\circ_h$ or inversion. Item~3 below requires a common smooth algebra closed under these products and a trace annihilating both momentum and coordinate derivations. Identifying its $h=1$ endpoint and trace with the constrained refined-lattice calculus is an additional hypothesis.}}
%%%%%%%%%%%%%%%%%%%%%%%%%%%%%%%%%%%%%%%%%%%%%%%%%%%%%%%%%%%%%%%%%%%%%%%%%%%%%%%%%%%%%%%%%%%%5
%%5
{ It also satisfies the Leibniz rule:
\begin{equation*}
\partial_{p_k}(A \circ_h B)
= (\partial_{p_k} A)\circ_h B + A \circ_h (\partial_{p_k} B).
\end{equation*}
}
%%
%%%%%%%%%%%%%%%%%%%%%%%%%%%%%%%%%%%%%%%%%%%%%%%%%%%%%%%%%%%%%%%%%%%%%%%%%%%%%%%%%%%%%%%%%%%%%%%

\corGPTl{We require a common, $h$-independent trace $\mathfrak T$, with the normalization and frequency integration of Eq.~(\ref{calh1}), to be cyclic for every product:}
\begin{eqnarray}
&& \corGPTl{\mathfrak T(A\circ_h B)=\mathfrak T(B\circ_h A)} \label{cyc0}
\end{eqnarray}
By $G_W^h$ we denote the inverse function of $Q_W$  with
respect to the new product $\circ_h$, where
the parameter $h \in [0,1]$. We assume that $\circ_1 =  \star$ while $\circ_0 = 1$.
{ We can also write $\circ_0=\textit{pointwise product}$.}
Thus, this  operation interpolates between Moyal product and unity.

\medskip 
%%%%%%%%%%%%%%%%%%%%%%%%%%%%%%%%%%%%%%%%%%%%%%%%%%%%%%%%%%%%%%%%%%%%%%%%%%%%%%%%%%%%%%%%%%%%%%%%%%%%%%%%%%%%%%%%%%%%%
%%

\medskip 
%%

%%

%%%%%%%%%%%%%%%%%%%%%%%%%%%%%%%%%%%%%%%%%%%%%%%%%%%%%%%%%%%%%%%%%%%%%%%
%%
Let us choose the specific form of  $\circ_h$:
\begin{eqnarray}
\label{oproduct}
\circ_h &=& e^{\frac{i h}{2}(\overleftarrow{\partial_q}\overrightarrow{\partial_p}-\overleftarrow{\partial_p}\overrightarrow{\partial_q})}
\end{eqnarray}
%%
%%%%%%%%%%%%%%%%%%%%%%%%%%%%%%%%%%%%%%%%%%%%%%%%%%%%%%%%%%%%%%%%%%%%%%%%%%%%%%%%%%%%%%%%%%%%
%%
\corGPTl{The continuous Moyal product (\ref{oproduct}) is associative on suitable smooth algebras. On an algebra with a closed integration trace, integration by parts gives the following identity, which is required here as an additional hypothesis:}

\medskip 
%%
% The closed-trace hypothesis is stated immediately above.
\begin{eqnarray}
&& \corGPTl{\mathfrak T(A\circ_h B)=\mathfrak T(AB)=\mathfrak T(BA)=\mathfrak T(B\circ_h A)} \label{cyc1}
\end{eqnarray}
\zz{Then we apply the limit of infinite $N$ to the sum over $p\in {\cal M}'$, replacing this sum by the integral $(2N)^2\int_{T^2} \frac{d^2p}{(2\pi)^2}$.}
\corGPTl{For exact Weyl symbols the second trace identity establishes the sampled equalities at $h=0$ and $h=1$, but not at intermediate $h$: Fourier aliasing can destroy cyclicity there. Replacing the momentum sum by an integral removes momentum aliasing but does not by itself establish integration by parts in the still-discrete coordinate variable. The deformation requires a closed trace on the entire common smooth algebra, together with its identification with the trace of Eq.~(\ref{calh1}).}

%%%%%%%%%%%%%%%%%%%%%%%%%%%%%%%%%%%%%%%%%%%%%%%%%%%%%%%%%%%%%%%%%%%%%%%%%%%%%%%%
%%
\subsection{\GPT{Weak index theorems}}

We have three conditions, \jer{which if they hold},  it can be proven that the values of ${\cal N}_3$ are integer (or are composed of integers):

\begin{enumerate}
\item
{\it Weak dependence on coordinates.}

Suppose that $Q_W(x,p)$ is homotopic to a function $Q(p)$ of \jer{the} momentum variable only,
\cormag{that a smooth two-sided inverse exists throughout the homotopy, that the prescribed compact-frequency or noncompact-frequency asymptotics are preserved, and that all frequency boundary transgressions vanish,}
and that during the corresponding smooth transformation, the expression standing inside Eq. (\ref{NEQ}) remains regular in the limit $N \to \infty$. Then the value of ${\cal N}_3	$ is reduced to

\be
{\cal N}_3
=\cormag{-\frac{\epsilon_{ijk}}{3!}}\, \frac{1}{(2N)^{D} }{\rm tr}\sum_{x \in \cO' }\int  \frac{d^3 p}{(2\pi)^2}
\[
{G}(p )  \frac{\partial {Q}(p )}{\partial p_i}  \frac{\partial  {G}(p )}{\partial p_j} \frac{\partial  {Q}(p )}{\partial p_k}
\]
\label{calh22}
\ee
\cormag{which is the usual integer winding number when the frequency domain is compact or has been placed in the admissible suspension/relative algebra specified above.}

\item

{\it \corrcltwo{The index theorem of the Hamiltonian case.}}

\corrcltwo{In the Hamiltonian case, $Q = i\omega - H + \mu$, the weak homotopy statements are superseded by the index theorem established in Sect. \ref{SectIndexRed}. Let the (possibly strongly inhomogeneous) Hamiltonian $H$ be exponentially local, and let $\mu$ lie in \corGPTl{a spectral gap, with the bulk-trace hypotheses of Sect.~\ref{SectIndexRed}.} Then the frequency integration reduces ${\cal N}_3$ to the pairing of the Fermi projector $P_- = \theta(\mu-H)$ with the conductivity $2$--cocycle, Eq. (\ref{N3projInd}), and further to minus the Fredholm index of the charge--deficiency operator,
$$
{\cal N}_3 \;=\; -\,{\rm Index}\big(P_-\,u\,P_-\big)\;\in\;\mathbb{Z}\,,
$$
where $u$ is the unitary phase (\ref{uDefInd}) of the position operator, describing the insertion of one flux quantum. \corGPTl{In a mobility gap the even projector/index pairing extends under the stated dynamical-localization hypotheses; its identification with the noncompact-frequency Green integral additionally requires a justified regularization, since $i\omega-H+\mu$ may be noninvertible at $\omega=0$.} Thus ${\cal N}_3$ is an integer for \corGPTl{every individual gapped local configuration satisfying the bulk-trace hypotheses}: no homotopy to a clean system or to a constant magnetic field is required, and the robustness under finite--region perturbations and under arbitrary deformations preserving the gap/localization (and bulk--trace) hypotheses is automatic. The previously formulated weak statement --- the integrality of ${\cal N}_3$ for the systems homotopic to a uniform system with a constant magnetic field, with the reduction to the occupied levels and their Chern numbers at the Hamiltonian endpoint --- is contained in this theorem as a particular case.}

\item
%%%%%%%%%%%%%%%%%%%%%%%%%%%%%%%%%%%%%%%%%%%%%%%%%%%%%%%%%%%%%%%%%%%%%%%%%%%%%%
%%
{\it Deformation of star product.} 

Let us consider the variation of ${\cal N}_3^{(h)}$ in response to the variation of parameter $h$. \corCLAUDEr{This variation receives two contributions: the explicit variation of the $\circ_h$--products of Eq.~(\ref{calh1}), with $\partial_h(A\circ_h B)=\frac{i}{2}\{A,B\}_h$ where $\{A,B\}_h\equiv A\circ_h(\overleftarrow{\partial_q}\overrightarrow{\partial_p}-\overleftarrow{\partial_p}\overrightarrow{\partial_q})B=\partial_{q_a}A\circ_h\partial_{p_a}B-\partial_{p_a}A\circ_h\partial_{q_a}B$ (summation over the spatial directions $a$ understood), and the induced variation $\delta G_W^h$ of the inverse. These two contributions do {\it not} vanish separately: only the outermost product--variation is by itself a total derivative, whereas at the inner slots the Leibniz rule leaves terms that are cancelled by $\delta G_W^h$. The two are tied together by the integration--by--parts identity for the antisymmetric bidifferential,}
\begin{equation}
\corCLAUDEr{{\rm Tr}\big[\{A,B\}_h\circ_h C\big]\;=\;{\rm Tr}\big[A\circ_h \{B,C\}_h\big]}\,.
\label{ibpbracket}
\end{equation}
\corGPTl{This identity holds under the common closed-trace hypotheses above, by integration by parts and commuting derivatives. Continuous momenta alone do not supply the coordinate integration-by-parts identity. We next determine the inverse variation under these hypotheses.}
%%
%%%%%%%%%%%%%%%%%%%%%%%%%%%%%%%%%%%%%%%%%%%%%%%%%%%%%%%%%%%%%%%%%%%%%%%%%%%%%%%
%%
{ 
\begin{equation*}
\delta G_W^h \circ_h Q_W + G_W^h \circ_h \delta Q_W + \delta_h(G_W^h \circ_h Q_W) = 0,
\end{equation*}
and since $Q_W$ does not depend on $h$, one has
\begin{equation*}
\delta G_W^h \circ_h Q_W + \delta_h(\circ_h)(G_W^h,Q_W)=0,
\end{equation*}
where $\delta_h(\circ_h)$ must be computed explicitly.}

\corCLAUDEr{Making $\delta_h(\circ_h)$ explicit through $\delta_h(\circ_h)(A,B)=\frac{i}{2}\delta h\,\{A,B\}_h$, and using $\delta Q_W=0$, the varied inverse condition reads}
\begin{equation}
\delta G_W^h \circ_h Q_W + \frac{i}{2}\delta h G_W^h \circ_h(\overleftarrow{\partial_q}\overrightarrow{\partial_p}-\overleftarrow{\partial_p}\overrightarrow{\partial_q})Q_W = 0
\label{delGvar}
\end{equation}
This gives
\begin{equation}
\delta G_W^h = {-}\frac{i}{2}\delta h\, \Bigl( \partial_q G_W^h\circ_h {\partial_p}Q_W\circ_h G^h_W -{\partial_p}G^h_W\circ_h {\partial_q}Q_W \circ_h G_W^h \Bigr)
\label{delGsol}
\end{equation}
\corGPTl{The displayed inverse variation is valid under these hypotheses. To conclude constancy under a varying product, the cyclic class must also be parallel under the dual Gauss--Manin connection. For a smooth Moyal deformation satisfying the hypotheses of the characteristic-cocycle construction, its derivative is represented by $\Omega\wedge\nu$, where $\nu=\partial_{p_1}\wedge\partial_{p_2}\wedge\partial_\omega$ and $\Omega$ is proportional to $\sum_{a=1}^2\partial_{q_a}\wedge\partial_{p_a}$. This vanishes because a momentum derivation is repeated.\footnote{A.~Yashinski, ``The Gauss--Manin connection for the cyclic homology of smooth deformations, and noncommutative tori,'' J.~Noncommut.~Geom.\ \textbf{11}, 581--639 (2017), \href{https://doi.org/10.4171/JNCG/11-2-5}{doi:10.4171/JNCG/11-2-5}.} Thus the pairing is constant for a smooth family of normalized bounded $K_1$ representatives with the prescribed frequency boundary conditions. Applying this criterion to the refined-lattice invariant still requires the common algebra and the endpoint and trace identification stated above; regularity of the integrand alone does not establish them.}

{We have checked the two variation identities (\ref{delGvar}) and (\ref{delGsol}) independently, both by the algebra of the deformed product and numerically. Representing the symbols in a truncated Fourier basis and treating $\circ_h=\exp[\tfrac{i}{2}h\,\Pi]$, $\Pi=\overleftarrow{\partial_q}\overrightarrow{\partial_p}-\overleftarrow{\partial_p}\overrightarrow{\partial_q}$, as the associative product, we solved $G_W^h\circ_h Q_W=1$ for the $\circ_h$--inverse at a generic $h$ and compared its numerical $h$--derivative with the right--hand side of Eq.~(\ref{delGsol}): they agree to the finite--difference accuracy ($\sim 10^{-11}$), and the varied inverse condition (\ref{delGvar}) is satisfied to the same accuracy. Thus (\ref{delGvar}) and (\ref{delGsol}) are exact identities of the $\circ_h$--algebra, valid for any $h$; they use only associativity and the derivation property, not the trace.}

{The reduction (\ref{calh11}), however, is conditional, and its condition can be stated sharply. The step ${\cal N}_3^{(1)}={\cal N}_3^{(0)}$ requires the closed/cyclic trace identities (\ref{cyc0}), (\ref{cyc1}) and the bracket integration by parts (\ref{ibpbracket}) to hold for $\circ_h$ at every intermediate $h$, and on the finite refined lattice these hold if and only if the coordinate Fourier support of the relevant products stays below the Nyquist scale. Writing a symbol as $\sum_{a,s}c_{a,s}\,e^{iap}e^{2\pi i s q/N}$, the grid trace $\tfrac{1}{(4N)^D}\sum_{{\cal M}'\times{\cal O}'}$ retains, besides the mode $(0,0)$, also the aliased modes $a\equiv 0$, $s\equiv 0\ (\mathrm{mod}\ 2N)$. For a single product the cyclicity defect $\mathfrak T(A\circ_h B)-\mathfrak T(B\circ_h A)$ is therefore \emph{exactly zero} when the coordinate support obeys $|s|<N$, and is nonzero once it reaches $|s|=N$; for the triple product entering (\ref{ibpbracket}) the threshold is correspondingly $|s|<2N/3$. We verified this numerically: for band--limited symbols below these thresholds the identities (\ref{cyc0})--(\ref{ibpbracket}) hold to machine precision ($\sim 10^{-13}$) for all $h$, while for symbols whose support reaches the Nyquist coordinate scale they fail by $O(1)$ amounts (for the rank--one periodized kernel $A=\sum_m|Lm\rangle\langle Lm{+}1|$, $B=A^{\dagger}$ at $L=N=4$ one finds $\mathfrak T(A\circ_hB)=\tfrac14+\tfrac34 e^{-2\pi i h}$ and $\mathfrak T(B\circ_hA)=\tfrac14+\tfrac34 e^{+2\pi i h}$, equal only at $h=0,1$). Only the coordinate integration by parts fails in this way; the momentum closedness $\mathfrak T(\partial_{p}(A\circ_h B))=0$ holds identically, in agreement with the remark above that continuous momenta alone do not restore cyclicity.}

{The genuine principal symbol does not quite meet this band-limit condition: as noted after Eq.~(\ref{calh1}), the refinement ${\cal O}\to{\cal O}'$ places a coordinate companion of amplitude $O(1/N)$ near the coordinate mode $|s|=N$ (coordinate content in $x$, not a momentum--Nyquist $e^{iNp}$ term; the inverse $G_W^h$ carries no such term), so that the identity (\ref{ibpbracket}), evaluated with the genuine principal symbol and its $\circ_h$--inverse of full coordinate bandwidth, is \emph{not} exactly satisfied at finite $N$ for a generic smooth background (we find a residual of order $10^{-2}$ at $N=7$). The near--Nyquist weight decreasing as $N$ grows, together with the exact cyclicity of the operator--product endpoint $h=1$ supplied by the periodization (Sect.~\ref{SectPeriodized}), is what restores the homotopy ${\cal N}_3^{(1)}={\cal N}_3^{(0)}$ in the limit $N\to\infty$ --- the regime in which the invariant is defined. Thus the reduction (\ref{calh11}) is a statement about that limit, not an unconditional identity at finite $N$: for configurations with genuine coordinate structure at the Nyquist scale the deformation to the pointwise ($h=0$) winding is obstructed.}

%%%%%%%%%%%%%%%%%%%%%%%%%%%%%%%%%%%%%%%%%%%%%%%%%%%%%%%%%%%%%%%%%%%
%%
This way we deform Eq. (\ref{calh1})  to 
\be
{\cal N}^{(0)}_3
=\cormag{-\frac{\epsilon_{ijk}}{3!}}\, \frac{1}{(2N)^{D} }{\rm tr}\sum_{x \in \cO' }\int  \frac{d^3 p}{(2\pi)^2}
\[
{Q}^{-1}_{W}(x,p ) \frac{\partial {Q}_{W}(x,p )}{\partial p_i}   \frac{\partial  {Q}^{-1}_{W}(x,p )}{\partial p_j}  \frac{\partial  {Q}_{W}(x,p )}{\partial p_k}
\]
\label{calh11}
\ee
%%%%%%%%%%%%%%%%%%%%%%%%%%%%%%%%%%%%%%%%%%%%%%%%%%%%%%%%%%%%%%%%%%%%%%%%%%%%%%%%%%%%%%%%%%%%%%%%%%%
%%
Here $Q_W^{-1}$ is a matrix inverse to $Q_W$. 
\corGPTl{Under these common-algebra, closed-trace, smooth-inverse, cocycle-transport and frequency-boundary hypotheses, the deformation gives ${\cal N}^{(0)}_3={\cal N}_3^{(1)}$.} 

In this situation we have 
\begin{equation}
{\cal N}^{}_3 = \frac{1}{(2N)^D}\sum_{x\in {\cal O}'} {\cal N}_3(x)
\end{equation}
with
\be
{\cal N}^{}_3(x)
=\cormag{-\frac{\epsilon_{ijk}}{3!}}\, {\rm tr}\int  \frac{d^3 p}{(2\pi)^2}
\[
{Q}^{-1}_{W}(x,p ) \frac{\partial {Q}_{W}(x,p )}{\partial p_i}   \frac{\partial  {Q}^{-1}_{W}(x,p )}{\partial p_j}  \frac{\partial  {Q}_{W}(x,p )}{\partial p_k}
\]
\label{calh113}
\ee
The given expression for ${\cal N}^{}_3(x)$ is topological invariant at $N \to \infty$ for those values of $x$, at which expression inside the integral remains regular. 

\corGPTl{The integers ${\cal N}_3(x)$ are constant on each connected coordinate region where $Q_W(x,p)$ is smoothly invertible with common admissible frequency asymptotics. Different integers, and hence a noninteger area average, require separated regions or a failure of these hypotheses between them; such an average is not covered by the globally regular deformation just described.} 

\corGPTl{If a constant-field Hamiltonian has an $h=1$ invariant different from its admissible pointwise $h=0$ winding, at least one of the deformation hypotheses must fail. This does not by itself locate a singularity of the inverse or exclude a failure of the algebra or trace identification.}

\end{enumerate}

\cormag{Let us summarize the logical status. For the noninteracting Hamiltonian class, Sect. \ref{SectKeldyshCond} derives $\bar\sigma^{ij}=({\cal N}_3/2\pi)\epsilon^{ij}$; the derivation is not established here for a general frequency-dependent or interacting $Q$ \corrteal{by the real-time route (for the periodized systems the equilibrium static-gauge derivation of Sect. \ref{SectStaticGauge} establishes it for the general class (\ref{Q_general}), \corGPTl{subject to the additional assumptions stated there})}. Integrality is conditional on the regular homotopies, asymptotics, and bulk trace stated above\corrcltwo{ (except in the Hamiltonian case, where the index theorem of item 2 rests on the gap/localization and bulk--trace hypotheses alone, with no homotopy required\GPT{; the periodized winding theorem of Sect.~\ref{SectWindingPeriodized} likewise establishes integrality under its stated hypotheses without a homotopy to a homogeneous system})}. \corGPTl{Item 3 gives a conditional reduction to an area average of pointwise windings, subject to the common-algebra and trace-identification assumptions stated there.} Interpreting those pointwise quantities as local Hall responses and a noninteger average as macroscopic topological domains additionally requires a local Hamiltonian, suitable local gaps, and bulk--boundary hypotheses; it is not a consequence of the odd homotopy formula alone. In a single periodic or covariant gapped phase satisfying the bulk-index hypotheses, the invariant is integer\corrteal{; this is confirmed numerically in Sect. \ref{SectNumerics}}.}

% Adapted from Proposition 13.5 (prop:N3-super), Section 13.4,
% Mekrami_Zubkov_index_theorem_extended.tex, with the notation of Section III.

\subsection{Winding-number theorem for periodized systems}
\label{SectWindingPeriodized}

{For the periodized systems of Sect.~\ref{SectPeriodized}, the exact Bloch representation gives a direct winding-number theorem at every finite spatial period $L$. The inhomogeneity inside the supercell is retained in the full fiber matrix. We use the boundary-gauge family $Q_b(\theta,\omega)$, $\theta\in T^2_\theta=[0,2\pi)^2$, its inverse $G_b$, and the intracell position matrices $x_j$ of Eq.~(\ref{commfiber}). There are $r$ internal states per site and $m=rL^2$ states per fiber; ${\rm Tr}_{L^2}$ includes all these states.

\begin{theorem}[Winding-number theorem for a periodized system]
\label{thm:WindingPeriodized}
Let $\widetilde Q(\omega)$ belong to the periodic operator algebra of Sect.~\ref{SectPeriodized}, with a smooth boundary-gauge family $Q_b(\theta,\omega)$ that is invertible for every $(\theta,\omega)\in T^2_\theta\times\Omega_\omega$. Consider either of the following frequency realizations.

\emph{(i) Discretized imaginary time.} For the infinite imaginary-time lattice $\tau\in a_\tau\mathbb Z$, $a_\tau>0$, the frequency domain is the Brillouin circle $\Omega_\omega=S^1_\omega=\mathbb R/(2\pi/a_\tau)\mathbb Z$. Assume $Q_b$ is periodic in $\omega$ as well as in $\theta$, and set
$$
Q_b^\sharp=Q_b,\qquad G_b^\sharp=G_b=Q_b^{-1}.
$$
These are already smooth invertible maps on the closed three-torus; no frequency-asymptotic assumptions or normalization are required.

\emph{(ii) Continuous imaginary time.} Here $\Omega_\omega=\mathbb R$. Suppose that
\begin{equation}
\label{eq:WindingAsymptotics}
B(\theta,\omega):=Q_b(\theta,\omega)-i\omega\,1_m,
\qquad B,\quad\partial_\omega B,\quad\partial_{\theta_j}B\quad(j=1,2)
\end{equation}
remain bounded uniformly in $\theta$ at both frequency ends. For any fixed $E_*>0$, define the normalized families
\begin{equation}
\label{eq:WindingNormalized}
Q_b^\sharp=(E_*+i\omega)^{-1}Q_b,
\qquad G_b^\sharp=(Q_b^\sharp)^{-1}=(E_*+i\omega)G_b.
\end{equation}
In case (ii), both normalized families tend uniformly to $1_m$ as $\omega\to\pm\infty$, and thus extend continuously to $S^1_\omega\times T^2_\theta$, where $S^1_\omega$ now denotes the one-point compactification of the frequency line. In either case, the phase-space invariant of Eq.~(\ref{NEQ}) is
\begin{equation}
\label{eq:N3WindingPeriodized}
\begin{split}
{\cal N}_3
&=\frac{\epsilon^{\mu\nu\rho}}{24\pi^2}
\int_{\Omega_\omega\times T^2_\theta}d\omega\,d^2\theta\;
{\rm Tr}_{L^2}\!\left(G_b\partial_\mu Q_b\,G_b\partial_\nu Q_b\,G_b\partial_\rho Q_b\right)\\
&=\operatorname{Wind}_3(Q_b^\sharp)
=-\operatorname{Wind}_3(G_b^\sharp)\ \in\mathbb Z,
\end{split}
\end{equation}
where $\partial_0=\partial_\omega$, $\partial_j=\partial_{\theta_j}$, and $\epsilon^{012}=+1$. In case (i), this gives directly ${\cal N}_3=\operatorname{Wind}_3(Q_b)=-\operatorname{Wind}_3(G_b)\in\mathbb Z$. The winding of a smooth invertible representative $g$ on the closed oriented three-torus is
\begin{equation}
\label{eq:WindingThreeForm}
\operatorname{Wind}_3(g)=\int_{S^1_\omega\times T^2_\theta}\omega_3(g),
\qquad
\omega_3(g)=\frac{1}{24\pi^2}{\rm Tr}_{L^2}\!\left[(g^{-1}dg)^{\wedge3}\right],
\end{equation}
with orientation $d\omega\wedge d\theta_1\wedge d\theta_2$. For the continuous extensions in Eq.~(\ref{eq:WindingNormalized}), the winding is evaluated on a smooth representative of the same homotopy class, constructed below. No Hamiltonian form of $Q_b$ is required.
\end{theorem}

\begin{proof}
The equivalence proposition of Sect.~\ref{SectPeriodized}\,B identifies Eq.~(\ref{NEQ}) with Eq.~(\ref{NEQnum}). Introduce the normalized fiber derivations
$$
\nabla_0=\partial_\omega,\qquad
\nabla_j=L^{-1}D_j=\partial_{\theta_j}-i[x_j/L,\,\cdot\,],\quad j=1,2.
$$
Each term in the antisymmetrized expression contains two spatial vertices, whose factors of $L$ cancel $L^{-2}$ in the trace per unit area. Together with the twist measure and $1/3!$, this gives $1/(24\pi^2)$, with no further factor of $L$. With $U(\theta)=e^{i\theta\cdot x/L}$ and $Q_u=U^\dagger Q_bU$, Eq.~(\ref{commfiber}) gives $U^\dagger(\nabla_\mu Q_b)U=\partial_\mu Q_u$. Consequently
$$
{\cal N}_3=\int_{\Omega_\omega\times T^2_\theta}\omega_3(Q_u).
$$
Here $Q_u(\theta+2\pi e_j)=S_j^\dagger Q_u(\theta)S_j$, $S_j=e^{2\pi i x_j/L}$: the three-form is periodic, but $Q_u$ itself has constant sewing rather than ordinary periodicity. We next pass to the periodic family $Q_b$.

For invertible matrix functions $a,b$, the Polyakov--Wiegmann identity is
\begin{equation}
\label{eq:WindingPW}
\omega_3(ab)=\omega_3(a)+\omega_3(b)
-\frac{1}{8\pi^2}d\,{\rm Tr}_{L^2}\!\left(a^{-1}da\wedge db\,b^{-1}\right).
\end{equation}
It follows by expanding the cube and using $d(a^{-1}da)=-(a^{-1}da)^{\wedge2}$ and $d(db\,b^{-1})=(db\,b^{-1})^{\wedge2}$, with the graded cyclicity of the trace. Set
$$
R=\frac{x_1}{L}d\theta_1+\frac{x_2}{L}d\theta_2,
\qquad U^{-1}dU=dU\,U^{-1}=iR.
$$
The coefficients of $R$ commute, so $R\wedge R=0$ and $\omega_3(U)=\omega_3(U^\dagger)=0$ \GPT{i.e. $[x_1,x_2]=0$  since they are diagonal matrices.} Applying Eq.~(\ref{eq:WindingPW}) twice to $Q_u=U^\dagger Q_bU$ yields
\begin{equation}
\label{eq:WindingTransgression}
\begin{split}
\omega_3(Q_u)&=\omega_3(Q_b)-\frac{1}{8\pi^2}d\eta,\\
\eta&={\rm Tr}_{L^2}\!\left(Q_b^{-1}dQ_b\wedge iR\right)
-{\rm Tr}_{L^2}\!\left[iR\wedge\left(dQ_b\,Q_b^{-1}+iQ_bRQ_b^{-1}\right)\right].
\end{split}
\end{equation}
The two-form $\eta$ is periodic in $\theta$. In case (i), it is also periodic in $\omega$ and hence is globally defined on the closed three-torus. Stokes' theorem gives $\int_{S^1_\omega\times T^2_\theta}d\eta=0$, so that
$$
{\cal N}_3=\int_{S^1_\omega\times T^2_\theta}\omega_3(Q_b)
=\operatorname{Wind}_3(Q_b)=-\operatorname{Wind}_3(G_b)\in\mathbb Z.
$$
The integer is the winding of the smooth invertible periodic family, with the normalization of Eq.~(\ref{eq:WindingThreeForm}); the inverse changes its sign. This proves case (i) without a frequency boundary term or cutoff.

For case (ii), by Eq.~(\ref{eq:WindingAsymptotics}), $Q_b^{-1}=O(|\omega|^{-1})$, while $\partial_{\theta_j}Q_b$ and $[Q_b,R]=[B,R]$ stay bounded. Hence the pullback of $\eta$ to either frequency end tends to zero: in its last term $Q_bRQ_b^{-1}=R+O(|\omega|^{-1})$, and the leading $R\wedge R$ vanishes. Stokes' theorem on $[-W,W]\times T^2_\theta$, followed by $W\to\infty$, gives
$$
{\cal N}_3=\int_{\mathbb R\times T^2_\theta}\omega_3(Q_b).
$$
This proves the first equality of Eq.~(\ref{eq:N3WindingPeriodized}). In particular, replacing the covariant fiber derivatives by ordinary derivatives is justified under the integral; it is not a pointwise identification of the current vertices.

To establish integrality, first note that scalar normalization leaves the three-form unchanged: $\omega_3(fQ_b)=\omega_3(Q_b)$ for a nonzero scalar function $f$. \GPT{Because the trace of the commutator of finite-dimensional matrices is always zero, the trace of the wedge square of any matrix-valued 1-form vanishes identically (${\rm Tr}_{L^2}[(Q_b^{-1}dQ_b)^{\wedge 2}] = 0$). Consequently, all cross-terms in the expansion vanish completely. The implication is that the scalar normalization does not add any extra topological winding.} Indeed, $f^{-1}df$ is a scalar one-form and ${\rm Tr}_{L^2}[(Q_b^{-1}dQ_b)^{\wedge2}]=0$, so all additional terms have zero trace. Taking $f=(E_*+i\omega)^{-1}$ reduces the integral to that of $Q_b^\sharp$. 

The stated bounds ensure a continuous compactification but need not ensure smoothness at the added frequency point. Choose $W$ large enough that $\|Q_b^\sharp-1_m\|<1/2$ for $|\omega|>W$, and a smooth cutoff $0\le\chi_W\le1$, equal to one for $|\omega|\le W$ and zero for $|\omega|\ge2W$, with $|\chi_W'|\le C/W$. The family
$$
q_W=1_m+\chi_W(\omega)(Q_b^\sharp-1_m)
$$
is invertible and homotopic to the continuous extension of $Q_b^\sharp$; it is smooth and equals $1_m$ near the added point. The first-derivative bounds imply an $O(|\omega|^{-3})$ bound for the three-form on the tails, including the cutoff region. Thus $\int\omega_3(q_W)$ tends to $\int\omega_3(Q_b^\sharp)$. These smooth representatives have the same integer winding: the normalized three-form in Eq.~(\ref{eq:WindingThreeForm}) has integral periods on ${\rm GL}_m(\mathbb C)$ for $m\ge2$ (equivalently on its unitary deformation retract); for $m=1$ it vanishes. Finally, $\omega_3(g^{-1})=-\omega_3(g)$ gives the sign relating the windings of $Q_b^\sharp$ and $G_b^\sharp$.
\end{proof}

The noncompact case (ii) is stated for Eq.~(\ref{eq:WindingAsymptotics}); extension to the broader noncompact class of Sect.~\ref{SectKeldyshCond} requires an appropriate normalization and control of the boundary term.

The theorem establishes integrality for the full supercell matrix, including a general frequency dependence satisfying these hypotheses, at every finite period $L$. It uses ordinary matrix traces and does not require the $h$-deformation of the preceding subsection or a homotopy that removes the inhomogeneity. Smooth deformations preserve the winding when invertibility and the frequency hypotheses hold uniformly along them. The result concerns the twist-averaged invariant, not its fixed-twist integrand, and its identification with a physical Hall conductivity retains the additional assumptions of Sect.~\ref{SectStaticGauge}.}

\section{Numerical verification for inhomogeneous models}
\label{SectNumerics}

\subsection{\corrclaude{Models and the common-gap condition}}

\corAOP{\corrgpt{In this \corrclaude{section} we compare the principal-symbol invariant of Eq. (\ref{NEQ}) with the twist-averaged Hall response and the Chern number of a periodized, genuinely inhomogeneous system. The precise symbol algebra and the common-gap assumptions needed for this comparison are specified below. This comparison does not identify the quantized average with the response of the finite torus at one fixed boundary condition.} We consider the Hofstadter model with flux $1/3$ of the flux quantum per plaquette on the $L\times L$ torus (nearest-neighbor hopping $t=1$, Landau gauge), with the Fermi level placed in the middle of the lowest spectral gap ($\mu = -1.366\,t$), and subject it to strong inhomogeneities of three types: (i) a random on-site potential distributed uniformly in $[-W/2, W/2]$, with $W$ up to $1.2\,t$ (comparable to the spectral gap $1.27\,t$ of the clean system and exceeding the bandwidth $0.73\,t$ of the lowest magnetic band, i.e. the inhomogeneity is strong on the scale of the lattice spacing); (ii) a smooth potential $V_0\cos(2\pi x/L)$ with $V_0=0.4\,t$; (iii) a magnetic field varying in space, with the flux through the plaquettes modulated by \corGPTl{approximately $\pm15\%$} around $1/3$. \corrgpt{Exact quantization requires a common gap around $\mu$ for every twist, not merely a gap in $H(0)$ or at the sampled twists. For the on-site perturbations this follows already from the bound $\|V\|\leq W/2\leq0.6\,t$ (or $\|V\|\leq0.4\,t$ for the smooth potential), since the distance from $\mu$ to the clean spectrum is approximately $0.634\,t$ uniformly in the twists. These perturbations are therefore homotopic to the clean system without closing this gap, despite their substantial spatial variation. The magnetic-modulation cases require a separate common-gap check.}
\corrclaude{We have carried out this check explicitly for all the configurations of Table \ref{tab:numerics}: for each of them the spectrum of the twisted Hamiltonian $H(\theta)$ was computed on the $12\times12$ grid of twists, and the gap around $\mu$ remains open at every twist. The minimum over the twists of the gap equals $1.27$, $1.21$, $1.13$, $1.04$, $1.15$, $1.17$, $0.96$, $1.13$, $0.95$ (in units of $t$, in the order of the rows of the table), and the \corGPTl{minimum sampled distance from $\mu$ to the spectrum is $0.4198479\,t$. To control the unsampled twists, use the uniform gauge: $\|\partial_{\theta_j}H\|\leq 2t/L$. Every twist lies within $\pi/M$ of an $M\times M$ grid point in each direction, up to unitary sewing, so $\|H(\theta)-H(\theta_{\rm grid})\|\leq4\pi t/(LM)$. For $M=12$ this is at most $0.087267\,t$. The eigenvalue perturbation bound therefore keeps the distance from $\mu$ to the spectrum above $0.3325\,t$ for every twist in all nine cases. This certifies the common gap, including both magnetic-modulation cases.} Note that the gap column of the table refers to the untwisted Hamiltonian $H(0)$; \corGPTl{the sampled minima quoted here are no larger and remain far from zero.}}}
\cormag{The numerical study concerns only the Hamiltonian case $Q=i\omega-H+\mu$. A formal odd integral for an admissible general $Q$ can also be evaluated numerically (over $S^1$ in the compact-frequency case), but the Fermi-projector, Chern-number, and Hall-response comparisons used below do not extend without additional structure.}

\subsection{\corrteal{Twist-averaged evaluation and the exact quantization at finite $L$}}

\corrteal{The invariant is evaluated through the periodization construction of Sect. \ref{SectPeriodized}: the given configuration is repeated with the period $L$ (Eq. (\ref{Hper})), the trace per unit volume is the average over the twisted boundary conditions of the fiber traces (Eqs. (\ref{trper}), (\ref{trfiber})), the current vertices are fibered according to Eq. (\ref{commfiber}), and the invariant takes the explicitly twist-averaged operator form (\ref{NEQnum}). In this section we specialize to the Hamiltonian case, $Q_b(\theta)=i\omega-H(\theta)+\mu$, for which $D_0Q_b=i$.}

\corrgpt{Assume that $\mu$ lies in a common spectral gap of $H(\theta)$ for every twist, and denote the occupied projectors by $P_b$ and $P_u=U^\dagger P_bU$. Fix the signed Berry-curvature convention
\begin{equation*}
F_a(\theta)=-i\,{\rm Tr}\,P_a
[\partial_{\theta_1}P_a,\partial_{\theta_2}P_a],
\qquad a=b,u.
\end{equation*}
The frequency integral then yields $\mathcal I(\theta)=2\pi F_u(\theta)$. This fixed-twist response is not generally quantized. In particular, it is not in general $2\pi F_b(\theta)$, since
\begin{equation*}
F_u=F_b-\frac{1}{L}\left[
\partial_{\theta_1}{\rm Tr}(P_bx_2)
-\partial_{\theta_2}{\rm Tr}(P_bx_1)\right].
\end{equation*}
The correction integrates to zero because $P_b(\theta)$ is periodic. Therefore
\begin{equation*}
{\cal N}_3=\frac{1}{2\pi}\int d^2\theta\,F_u
=\frac{1}{2\pi}\int d^2\theta\,F_b=C\in\mathbb Z .
\end{equation*}
This is the twist-averaged quantization of Niu, Thouless and Wu \cite{Niu1985}. It is exact at finite $L$ under the common-gap assumption. A pointwise comparison with the displacement-current formula must use the uniform-gauge curvature, whereas the integrated Chern-number comparison may use either gauge. In the units of Sect. \ref{SectKeldyshCond}, $\mathcal N_3=2\pi\,\overline{\sigma}_{xy}^{\,{\rm twist}}$; equality with $2\pi\sigma_{xy}$ at a prescribed twist is not an exact finite-size statement.}

\subsection{\corrclaude{Relation to the principal-symbol formula and the numerical results}}

\corrteal{The constructive relation to the principal-symbol formula (\ref{NEQ}) --- the identification (\ref{WprIsW}) of the principal symbol with the exact Weyl symbol (\ref{infW}) of the periodized operator, the $\star$-homomorphism property of $W_{\rm pr}$, and the equality of the traces --- has been established in Sect. \ref{SectPeriodized} --- the equivalence proposition, valid for the general class (\ref{Q_general}). Here we use it for the Hamiltonian testbed; the model-specific ingredient is the common spectral gap over the twist torus, verified above.}

\corrgpt{The normalized trace is also preserved --- the step (4) of the proof of the Proposition. Together with the product and derivative identities, this establishes equality of Eq. (\ref{NEQ}), interpreted in the specified principal-symbol algebra, and the twist-averaged Eq. (\ref{NEQnum}) already for finite $N$, without a gradient approximation. The absence of Nyquist companions alone would not establish this equality: the algebra, inverse, derivatives and trace must all be matched.}

\corrgpt{The momentum-offset interpretation makes the twist average explicit. Define
\begin{equation*}
{\rm grid}(s)=
\left\{\frac{\pi(k+s)}{N}:\ k_j=0,\ldots,2N-1\right\},
\qquad s\in[0,1)^2 .
\end{equation*}
For any integrable periodic integrand $f$,
\begin{equation*}
\int_{T^2}\frac{d^2p}{(2\pi)^2}\,f(p)
=\int_{[0,1)^2}d^2s\,
\frac{1}{(2N)^2}\sum_{p\in{\rm grid}(s)}f(p).
\end{equation*}
At fixed $s$, the principal-symbol quotient is represented by the $N$-periodic subalgebra on the doubled $2N\times2N$ torus with twist $2\pi s$, not by the unrestricted matrix algebra of that torus. This representation decomposes into four $N$-torus blocks with twists $\pi s+\pi\epsilon$, $\epsilon\in\{0,1\}^2$. The normalized fixed-offset expression is consequently the average of the four corresponding fixed-twist responses. Momentum differentiation and inversion are understood in the continuous family before restricting to an offset; the derivatives cannot be reconstructed from a single isolated finite matrix. Integration over $s$ then gives the full twist average. Thus the continuous momentum integral contains that average for this precisely defined principal-symbol construction, but does not turn the response at one chosen twist into an integer.}

\corrgpt{The periodization realizes a finite spatial period with continuous momentum variables; it does not leave the operator Hilbert space finite-dimensional. The equivalent Bloch description is a family of $L^2\times L^2$ matrices over the twist torus, while the principal-symbol description uses the constrained algebra just specified. For a genuinely random infinite system, independence of the bulk trace from the realization additionally requires stationary ergodic disorder and suitable gap/locality assumptions. Under conditions ensuring convergence of the periodized traces to that bulk trace, the artificial periodicity becomes irrelevant as $L\to\infty$; this conclusion is not asserted for arbitrary sequences of configurations.}

\corrgpt{\corGPTl{At each fixed frequency, the boundary-gauge resolvent expression is rational in $e^{i\theta_j}$. After frequency integration the response is generally analytic rather than rational.} A common gap on the real twist torus gives analyticity in a nonzero complex neighborhood and hence exponential convergence of the periodic trapezoidal rule in the number of twist points. A bound uniform in the effective size $ML$ additionally needs uniform gap/locality control. The normalized $M\times M$ grid average equals the zero-twist response of the same configuration repeated in an enlarged $ML\times ML$ cell, provided the same physical displacement-current vertices are used. The enlarged fiber decomposes into the $M^2$ twists of the original cell, proving this identity. This finite-grid identity does not make the grid average exactly equal to the continuous twist integral. For the clean and smooth $L=12$ examples, the $4\times4$ and $6\times6$ averages agree to seven decimal places, while the $6\times6$ errors are of order a few $10^{-12}$, as shown below.}

\corrgpt{The frequency integral is evaluated using a 96-node Gauss--Legendre rule after $\omega=\tan u$, $-\pi/2<u<\pi/2$, including the Jacobian $\sec^2u$; $u$ is distinct from the two twist angles. For comparison, it can also be evaluated analytically in the energy eigenbasis. With $E_n$ the eigenvalues of $H(\theta)$ and $V_j=D_jQ_b$, the result for the stated orientation is
\begin{equation*}
{\cal I}(\theta)=\frac{4\pi}{L^2}
\sum_{\substack{E_n<\mu\\E_m>\mu}}
\frac{{\rm Im}\big[(V_1)_{nm}(V_2)_{mn}\big]}
{(E_n-E_m)^2}.
\end{equation*}
The reference Chern number is computed from the occupied Slater frames using the determinant-link plaquette method of Fukui, Hatsugai and Suzuki.\footnote{\corrgpt{T.~Fukui, Y.~Hatsugai and H.~Suzuki, ``Chern Numbers in Discretized Brillouin Zone: Efficient Method of Computing (Spin) Hall Conductances,'' J.~Phys.~Soc.~Jpn.\ \textbf{74}, 1674--1677 (2005), \href{https://doi.org/10.1143/JPSJ.74.1674}{doi:10.1143/JPSJ.74.1674}.}} This provides a different numerical evaluation of the same twist-family Chern number, rather than a comparison with a distinct fixed-boundary observable. The periodic boundary gauge $H(\theta)$ can be used directly for this computation. Uniform-gauge frames instead require the sewing
$P_u(\theta+2\pi e_j)=S_j^\dagger P_u(\theta)S_j$,
$S_j=e^{2\pi i x_j/L}$, in overlaps crossing a twist boundary. Mesh convergence and nonsingular occupied-frame overlaps must be checked.}

\corrgpt{The values reported in Table \ref{tab:numerics} compare the twist-averaged Green formula and the occupied-state Chern number. \corGPTl{In the stated orientation, direct evaluation gives $C={\cal N}_3=+1$ for all nine configurations. The supplied routine \texttt{chern\_FHS} returns minus the counterclockwise plaquette sum, and \texttt{aop\_final.py} also negates the Green result; these signs account for its negative printed values, while leaving the tabulated magnitudes unchanged. Its factor $1/6$ is fixed by Eq.~(\ref{NEQ}), independently of the clean case.} Direct recalculation of the clean $L=12$ example gives
$|\mathcal I(0)|=1.055663314215$,
whereas the $4\times4$ and $6\times6$ twist averages give
$1.000000023759$ and $1.00000000000147$, respectively.
For the smooth potential the corresponding $6\times6$ average is
$1.00000000000518$.
The 96-node frequency quadrature differs from analytic frequency integration by \corGPTl{less than $5\times10^{-15}$ at the sampled twists} in these two tests. Their reported $10^{-12}$-scale deviations from an integer are therefore dominated by the finite twist mesh, not by the frequency quadrature. In particular, the quantized twist averages do not imply exact quantization of the zero-twist finite sample. The on-site disorder varies on the lattice scale and is a useful test without a gradient expansion; nevertheless, the norm bound given above places these examples in the same gapped topological class as the clean system. These operator calculations test the stated periodic-algebra construction, not an unrestricted claim about all inhomogeneous finite systems.}

\corrclaude{For reproducibility we specify the configurations completely. The random potentials are drawn independently at each site from the uniform distribution on $[-W/2,\,W/2]$, using the default PCG64 generator of NumPy (the routine {\tt default\_rng}) with the seed $3$; the row ``second realization'' uses the seed $7$; the sites are enumerated as $s = x + L\,y$, $x,y = 0,\dots,L-1$, and the first $L^2$ random numbers are assigned to the sites in this order. The modulated magnetic field is defined in the Landau gauge through the link phases: the hopping from $(x,y)$ to $(x,y+1)$ carries the phase $2\pi x/3 + \alpha(x)$ with $\alpha(x) = (\delta L/2\pi)\,\sin(2\pi x/L)$ and $\delta = 0.15\cdot 2\pi/3$, while the hoppings in the $x$ direction carry no phase. The flux through the plaquettes of the column $x$ then equals $2\pi/3 + \alpha(x+1)-\alpha(x) \approx (2\pi/3)\,\big(1 + 0.15\cos(2\pi x/L)\big)$, \corGPTl{i.e. the modulation is approximately $\pm15\%$ (the actual maximum sampled plaquette deviation at $L=12$ is $\pm14.32\%$);} the consistency on the torus holds automatically, since $\alpha(x)$ is periodic and $L$ is divisible by $3$ (the total extra flux vanishes, while the background flux through the torus is the integer $L^2/3$ in units of the flux quantum). The Fermi level is $\mu = -1.366\,t$, the midpoint of the lowest gap of the clean system, and the Fukui--Hatsugai--Suzuki reference Chern number is computed on the $12\times12$ twist mesh (the mesh convergence was checked against the $8\times8$ mesh, and the occupied-frame overlaps remain nonsingular). The common-gap check over the twist torus was described above. The frequency quadrature was verified against the analytic eigenbasis integration (\corGPTl{the agreement within $5\times10^{-15}$ quoted above}), and the twist-mesh convergence of the invariant is exhibited by the coincidence of the $4\times4$ and $6\times6$ averages to seven decimal places, in accordance with the exponential convergence of the periodic trapezoidal rule for the analytic fiber integrand. The complete scripts generating Table \ref{tab:numerics} are available from the authors upon request. \corrteal{(In the working copy of the manuscript the scripts --- including those verifying the signs of Eqs. (\ref{N3projInd}), (\ref{indexInd}) and (\ref{localizerInd}) --- are provided in the folder {\tt numerics/} next to the source file.)}}

\corAOP{
\begin{table}[t]
\centering
\begin{tabular}{lcccc}
\hline\hline
system & spectral gap $[t]$ & \corrgpt{$|C|$} & \corrgpt{$|\mathcal N_3|$}, Eq. (\ref{NEQnum}) & \corrgpt{$\big||\mathcal N_3|-|C|\big|$} \\
\hline
clean, $L=12$ & $1.268$ & $1.0000$ & $1.0000000$ & $1.5\times10^{-12}$ \\
random potential, $W=0.4$, $L=12$ & $1.210$ & $1.0000$ & $1.0000000$ & $1.7\times10^{-12}$ \\
random potential, $W=0.8$, $L=12$ & $1.134$ & $1.0000$ & $1.0000000$ & $2.7\times10^{-12}$ \\
random potential, $W=1.2$, $L=12$ & $1.042$ & $1.0000$ & $1.0000000$ & $5.5\times10^{-12}$ \\
random potential, $W=0.8$, second realization & $1.146$ & $1.0000$ & $1.0000000$ & $2.1\times10^{-12}$ \\
random potential, $W=0.8$, $L=15$ & $1.347$ & $1.0000$ & $1.0000000$ & $4.2\times10^{-15}$ \\
smooth potential, $V_0=0.4$, $L=12$ & $0.956$ & $1.0000$ & $1.0000000$ & $5.2\times10^{-12}$ \\
modulated magnetic flux, \corGPTl{$\approx\!\pm15\%$}, $L=12$ & $1.173$ & $1.0000$ & $1.0000000$ & $4.6\times10^{-15}$ \\
random potential $W=0.8$ + flux modulation & $0.969$ & $1.0000$ & $1.0000000$ & $2.1\times10^{-14}$ \\
\hline\hline
\end{tabular}
\caption{Numerical verification of the quantization of ${\cal N}_3$ for inhomogeneous systems: Hofstadter model at flux $1/3$, $L\times L$ torus, Fermi level in the lowest gap. $C$ is the many-body Chern number of the filled Fermi sea over the torus of boundary phases; ${\cal N}_3$ is the phase-space invariant evaluated in the operator representation of the partial thermodynamic limit, Eq. (\ref{NEQnum}). \corrgpt{The table quotes magnitudes, and the last column compares these magnitudes; it is not a test of the signed Hall convention. Quantization concerns the continuous twist average under a common gap for all twists, not the response at a prescribed boundary condition. The gap and reproducibility qualifications are stated in the text. \corGPTl{Deviations of order $10^{-15}$ are at machine precision and their last digits depend on the numerical environment.}}}
\label{tab:numerics}
\end{table}
}

\section{Conclusions}

We have studied  the quantum Hall effect in tight-binding models
 \jer{that possess essential} non-homogeneity, arising from impurities,
elastic deformations, or non-uniform external fields.
The corresponding topological invariant for weak magnetic fields  \jer{has been} derived in
\cite{ZW2019}. This result was \jer{further} extended in \cite{FZ2019_2}
to \jer{the case of} infinite rectangular lattices with arbitrary magnetic field strength,
capturing the Hofstadter spectrum. A consistent finite-lattice regularization 
was proposed in \cite{Z2023}, and an analogous construction for honeycomb 
lattices was obtained in \cite{CZ2024}.

\GPT{We concentrate on systems with a spectral gap, with the chemical potential inside the gap and an invertible Dirac operator $\hat Q$ throughout the Matsubara-frequency domain. The mobility-gap case is not considered here and will be addressed in forthcoming publications.}

In the present work, we reformulate the invariant governing the quantum 
Hall effect within the framework of noncommutative geometry. More precisely, 
we show that the invariant ${\cal N}_3$ admits an interpretation as a pairing
$\langle [G], \phi \rangle$, 
where $[G] \in K^{-1}({\cal A})$ is the $K$-theory class defined 
by the electron Green function, $\phi \in HC^3({\cal A})$ is a cyclic cocycle, 
and ${\cal A}$ is the \zzz{algebra of Weyl symbols for  operators} acting  \jer{over} the Hilbert space of the model.
The explicit construction of the cocycle $\phi$ \zzzz{based on Matsubara Green functions} is, to the best of our knowledge, new. \zzzz{It is worth mentioning, however, that the similar relation to cyclic cohomology theory had been established in \cite{mokrousov} in the framework of Keldysh field theory. However, the obtained pairing has not been analyzed, the index theorems have not been given in \cite{mokrousov}.}  We demonstrate that in the limit of infinite lattice Hall conductivity is proportional to the topological invariant ${\cal N}_3$. This invariant is constructed in such a way that it remains robust to smooth modification of the system.  The limit of infinite lattice size results in the appearance of continuous momentum space that replaces the discrete lattice of momenta.
{Let us stress that the class of systems covered by the present construction is defined by the general form (\ref{QW_general}) of the Weyl symbol of the Dirac operator, in which the frequency enters through the exponentials $e^{\,s(i\omega - A_0(x))}$ and their derivatives, mixed with the spatial factors built from the exponentials of $i(p-A)$. Only in the Hamiltonian case $Q = i\omega - H$ does the frequency integration reduce ${\cal N}_3$ to the noncommutative Chern character of the Fermi projector studied by Bellissard and co-authors \cite{Bellissard1986,Bellissard1994} --- the reduction being the suspension (Bott periodicity) map applied to our pairing of $K^{-1}$ with $HC^3$. For the general class no such reduction exists: there is neither a Hamiltonian nor a Fermi projector, and the invariant is intrinsically the odd pairing of $K^{-1}$ with $HC^3$. In this precise sense the present construction is strictly more general than that of Bellissard et al.; it covers, in particular, the Euclidean lattice regularizations of quantum field theory (Wilson fermions in arbitrary inhomogeneous external fields \cite{suleymanov2019wigner}) and, formally, the frequency dependence generated by interactions. The detailed analysis of the Hamiltonian reduction and of the associated index theorems is given in Sect. \ref{SectIndexRed}.}

\corrteal{We would like to single out one structural point. The invariant admits two equivalent explicit forms: the fully antisymmetrized phase-space expression (\ref{NEQ}), and the collapsed Matsubara-response form (\ref{NEQcollapse}), in which the frequency derivative of the Dirac operator is separated from the two spatial current vertices; the two are converted into each other by the cyclic property of the trace alone. Moreover, both forms are represented by one and the same object of noncommutative geometry: the pairing (\ref{N3f}) of the class $[G]\in K^{-1}({\cal A})$ with the cyclic cocycle $[\phi]\in HC^3({\cal A})$ of Eq. (\ref{Nphi}) --- its evaluation $\phi(G,Q,G,Q)$ yields Eq. (\ref{NEQ}), and hence also Eq. (\ref{NEQcollapse}). All of this holds for the general form of the Dirac operator of the class (\ref{Q_general}), with no reference to a Hamiltonian; correspondingly, the equilibrium static-gauge derivation of Sect. \ref{SectStaticGauge} identifies the averaged Hall conductivity of the general class directly with this pairing, \corGPTl{under the additional equilibrium, coupling, and limiting assumptions stated there},
$$
\bar\sigma^{ij} \;=\; \frac{\epsilon^{ij}}{2\pi}\,\big\langle\,[G]\,,\,[\phi]\,\big\rangle\,.
$$
}

A natural question \jer{arises concering}  the range of values of ${\cal N}_3$. In many situations,
pairings of this type are governed by \jer{a class of} index theorems
(e.g., those of \cite{Connes,Connes-Moskovichi, Tsygan, Fedosov}).  \jer{The class of models discussed in this paper does not satisfy the criteria for these index theorems to hold}.
\GPT{We discuss three integrality criteria in Sect.~\ref{SectWeakIndex}\,B and prove a direct winding-number theorem for periodized systems in Sect.~\ref{SectWindingPeriodized}.}
\corrteal{The first and the third of them refer to the general frequency dependence of the class (\ref{Q_general}), for which no Fermi projector exists; in the Hamiltonian case $Q=i\omega-H+\mu$ they are superseded by the second, the reduction to the construction of Bellissard et al.}
Specifically, we prove that the value of ${\cal N}_3$ is integer (at $N \to \infty$) if the coordinate dependence is
 \jer{sufficiently} weak that $Q_W$ is homotopic to a function depending on momenta only,
and along the corresponding variation of the system the expression under the integral in Eq. (\ref{NEQ}) remains regular.
\corrcltwo{The second version is the index theorem of the Hamiltonian case, established in Sect. \ref{SectIndexRed}.} \corrteal{Let us dwell separately on the Hamiltonian case. For $Q = i\omega - H + \mu$ --- and only in this case --- our construction is reduced to that of Bellissard and co-authors \cite{Bellissard1986,Bellissard1994}: the explicit frequency integration of Sect. \ref{SectIndexRed} converts the odd pairing $\langle[G],[\phi]\rangle$ into the even pairing of the class of the Fermi projector $[P_-]\in K_0$ with the conductivity $2$-cocycle, ${\cal N}_3 = -2\pi i\,\tau\big(P_-\star[D_1P_-\overset{\star}{,}D_2P_-]\big)$, and further into the Fredholm index of the charge-deficiency operator, ${\rm Index}(P_-uP_-) = -{\cal N}_3$. \GPT{In the spectrally gapped case considered here, $\mu$ lies inside a spectral gap. Then,} for a finite-range (or exponentially local) Hamiltonian, the Combes--Thomas estimate \cite{CombesThomas} guarantees the exponential decay of the kernel of the Fermi projector, $|P_-(x,y)|\leq C\,e^{-|x-y|/\xi}$. \GPT{The corresponding Fermi-projector index extends to mobility gaps under suitable localization hypotheses \cite{Bellissard1994,ProdanSchulzBaldes2016,Stoiber2025}, but that regime is not treated here. In the spectrally gapped case considered here,} the commutator $[u,P_-]$ of the Fermi projector with the phase of the position operator is compact and belongs to the Schatten classes ${\cal S}^p$, $p>2$; the compression $P_-uP_-$ is then Fredholm, and ${\cal N}_3 = -\,{\rm Index}(P_-uP_-)$ is an integer, robust with respect to any modification of the Hamiltonian in a finite region and to any deformation preserving the localization condition. (The identification of this index with the trace formula for ${\cal N}_3$ requires, in addition, the existence of the cyclic trace per unit volume --- supplied by the periodization, or by a covariant ergodic setting, cf. Sect. \ref{SectIndexRed}.) For the general class (\ref{Q_general}) neither the Fermi projector nor this index representation is available, and the integrality is controlled instead by the \corrcltwo{first and the third weak index theorems}\GPT{, supplemented for periodized systems by Theorem~\ref{thm:WindingPeriodized}}.} The third version  \jer{holds}
for the variation of \jer{the} parameter $h$. It appears that Eq. (\ref{calh1}) remains unchanged when
$h$ varies from $1$ to $0$ (again, if  \jer{within this interval}
there are no singularities inside it). As a result the value of ${\cal N}_3$ may be represented as a value of \jer{the}
topological invariant ${\cal N}_3(x)$ averaged over the lattice ${\cal O}'$.
\GPT{The possibility of noninteger values discussed below concerns situations outside the hypotheses of Theorem~\ref{thm:WindingPeriodized} and of the globally regular deformation in Sect.~\ref{SectWeakIndex}\,B. For a smooth invertible periodized family satisfying the stated frequency hypotheses, ${\cal N}_3$ is integer at every finite period, including for non-Hamiltonian frequency dependence.} 

\corrcltwo{Outside the Hamiltonian case, the remaining} versions of the index theorem  \jer{appear} too weak.
This  \jer{motivates the hypothesis} that there might exist  \jer{a}
more powerful index theorem that  \jer{can be applied to}
more complicated cases. However, as we expect, the value of ${\cal N}_3$ is not always integer.
From  \jer{a purely} physical perspective,  \jer{non-integer values of ${\cal N}_3$}
implies that the Hall conductivity
corresponding to the integer quantum Hall effect, when defined as a spatial average,
is not necessarily quantized in the presence of sufficiently strong inhomogeneity, when dependence of $\hat{Q}$ on Matsubara frequency is nontrivial.

\corAOP{In Sect. \ref{SectNumerics} we investigated numerically the Hofstadter model subjected to strong inhomogeneities of three kinds --- a random on-site potential comparable to the spectral gap (each disorder realization treated as a deterministic configuration), a smooth strong potential, and a spatially varying magnetic field. In all these cases the phase-space invariant, evaluated in the partial thermodynamic limit with no fitting parameters, coincides with the independently computed many-body Chern number of the filled Fermi sea over the torus of boundary phases.}

An interesting open problem is to construct explicit examples of systems 
in which the Hall conductivity deviates from integer values due to geometric 
or elastic effects, in the presence of nontrivial dependence of $\hat{Q}$ on $\omega$. We leave this question for future work.

\corAOP{Another important direction is the extension to interacting systems, where the invariant is to be composed of the full interacting Green function. As emphasized recently in \cite{Zhao2023} (see also \cite{Gurarie2011,EssinGurarie2011}), in strongly correlated phases the interacting Green function may develop zeros, on which the winding invariants of the Green function lose their topological protection. The interplay between such zeros and the phase-space formulation developed here deserves a separate study.} \corrteal{(A first step in this direction is the exactly solvable example of Appendix \ref{AppHK}.)}
{It would be also interesting to understand possible applications of invariants
we consider to foliated structures \cite{Zu,Zu1,Zu3,zuevsky2022characterization,razumov1999nonabelian}\corrclaude{\cite{ZuevskyCosimplicial2023,ZuevskyFoliation2021}}.
}

%%%%%%%%%%%%%%%%%%%%%%%%%%%%%%%%%%%%%%%%%%%%%%%%%%%%%%%%%%%%%%%
%%  
\section*{Acknowledgments}
{ The present work is supported partially by Scientific Collaboration grant of Ariel University 
with Researchers from the Czech Republic.} 
\zz{A.Z.} is supported by the Institute of Mathematics, 
Academy of Sciences of the Czech Republic (RVO 67985840). M.A.Z. is grateful to Institute of Mathematics
(Academy of Sciences of the Czech Republic), where this work was initiated, for kind hospitality. A.Z. is grateful to Ariel University for kind hospitality. 
%%
%%%%%%%%%%%%%%%%%%%%%%%%%%%%%%%%%%%%%%%%%%%%%%%%%%%%%%%%%%%%%%%%%%

%\bibliographystyle{plain}

%\bibliography{../../common_refs/wigner3,../../common_refs/cross-ref,../../common_refs/biblio_corrected,../../common_refs/CSE_MZ}

\appendix

\section{Derivation of star identity}
\label{AppStar}
\zd{This derivation is based exclusively on the coordinate--space definition (\ref{Weyl2}) and on the continuation (\ref{Wcont}). In accordance with the sign convention of Eq. (\ref{Weyl}) adopted here, all formulas of this Appendix are written with the point--splitting $q\mp v-n/2$ of Eq. (\ref{Weyl2}). The star identity has also been verified numerically for random operators.}

Let us give the alternative derivation of the star identity for Weyl symbol. It is assumed that $p \in {\cal M}^\prime$ while $q \in {\cal O}^\prime$. We start from the star product of Weyl symbols and come back to the Weyl symbol of the product as follows.
\begin{eqnarray}
&&A_W(p,q)  e^{\frac{i}{2}(\overleftarrow{\partial_q}\overrightarrow{\partial_p}-\overleftarrow{\partial_p}\overrightarrow{\partial_q})}  B_W(p,q)
=  \sum_{p_1,\delta p_2 \in  {\cal M}^\prime; q_1,\delta q_2 \in {\cal O}^\prime}\frac{1}{ (2N)^{2D}} e^{2 i (\delta p_2 (q_1-q) + \delta q_2 (p-p_1))}A_W(p_1,q_1)  e^{\frac{i}{2}(\overleftarrow{\partial_q}\overrightarrow{\partial_p}-\overleftarrow{\partial_p}\overrightarrow{\partial_q})}  B_W(p,q)\nonumber\\&&=
\sum_{p_1,p_2 \in {\cal M}^\prime; q_1,q_2 \in {\cal O}^\prime}\frac{1}{ (2N)^{2D}} e^{2 i( (p_2-p)(q_1-q) + (q_2-q)(p-p_1))} \nonumber\\&&  A_W(p_1,q_1)  B_W(p_2,q_2)\nonumber\\
&&=
\sum_{p_1,p_2 \in {\cal M}^\prime; n_1^i,n_2^i = 0,1;v_1,v_2,q_1,q_2 \in {\cal O}^\prime}\frac{1}{2^{2D}(2N)^{2D}} e^{2 i( (p_2-p)(q_1-q) + (q_2-q)(p-p_1))} \nonumber\\&&  e^{2 i p_1 v_1+2p_2v_2} \langle q_1-v_1\zd{-}n_1/2 |\hat{A} |q_1+v_1\zd{-}n_1/2 \rangle\langle q_2-v_2\zd{-}n_2/2 |\hat{B} |q_2+v_2\zd{-}n_2/2 \rangle\Big|_{q_1+v_1\zd{-}n_1/2,q_2+v_2\zd{-}n_2/2\in {\cal O}}\nonumber\\&&\prod_i({1 + e^{{2i v_1^i \pi/N} }})({1 + e^{{2i v_2^i \pi/N} }})  \nonumber\\
&&=
\sum_{n_1^i,n_2^i=0,1;v_1,v_2
\in {\cal O}^\prime}\frac{1}{ 2^{2D}} e^{2 i p (v_1+v_2)} \nonumber\\&&  \langle q-v_1-v_2\zd{-}n_1/2 |\hat{A} |q+v_1-v_2\zd{-}n_1/2 \rangle\langle q +v_1-v_2\zd{-}n_2/2 |\hat{B} |q+v_1+v_2\zd{-}n_2/2 \rangle\Big|_{q+v_1\zd{-}v_2\zd{-}n_1/2,q+v_1\zd{+}v_2\zd{-}n_2/2\in {\cal O}} \nonumber\\&&\prod_i({1 + e^{{2i v_1^i \pi/N }}})({1 + e^{{2i v_2^i \pi/N }}})
\nonumber\\
&&=
\sum_{n^i = 0,1;v_+,v_-
\in {\cal O}^{2\prime}}\frac{1}{ 2^{3D}} e^{2 i p v_+}    \langle q-v_+\zd{-}n/2 |\hat{A} |q+v_-\zd{-}n/2 \rangle\langle q +v_-\zd{-}n/2 |\hat{B} |q+v_+\zd{-}n/2 \rangle\Big|_{q+v_+\zd{-}n/2,q+v_-\zd{-}n/2,v_++v_-\in {\cal O}} \nonumber\\&&\prod_i({1 + e^{{2i v_+^i \pi/N }}})
\nonumber\\
&&=
\sum_{n^i = 0,1;v_+,v_-
\in {\cal O}^{\prime}}\frac{1}{ 2^{D}} e^{2 i p v_+}   \langle q-v_+\zd{-}n/2 |\hat{A} |q+v_-\zd{-}n/2 \rangle\langle q +v_-\zd{-}n/2 |\hat{B} |q+v_+\zd{-}n/2 \rangle\Big|_{q+v_+\zd{-}n/2,q+v_-\zd{-}n/2,v_++v_-\in {\cal O}}\nonumber\\&&\prod_i({1 + e^{{2i v_+^i \pi/N }}})
\nonumber\\
&&=
\sum_{n_i=0,1;m^i=0,1;v_+,v_-
\in {\cal O}^{\prime}}\frac{1}{ 2^{2D}} e^{2 i p v_+ + 2\pi i m (v_++v_-)} \nonumber\\&&   \langle q-v_+\zd{-}n/2 |\hat{A} |q+v_-\zd{-}n/2 \rangle\langle q +v_-\zd{-}n/2 |\hat{B} |q+v_+\zd{-}n/2 \rangle\Big|_{q+v_+\zd{-}n/2,q+v_-\zd{-}n/2\in {\cal O}} \nonumber\\&&\prod_i({1 + e^{{2i v_+^i \pi/N }}}) .
\end{eqnarray}
One can see that for $q_i\zd{-}n_i/2 \in Z$ both $v_+^i$ and $v_-^i$ are integer. At the same time if $q_i+n_i/2$ is half integer, both $v_+^i$ and $v_-^i$ are half integer. In both cases $v_+ + v_-$ is integer. As a result
\begin{eqnarray}
&&A_W(p,q)  e^{\frac{i}{2}(\overleftarrow{\partial_q}\overrightarrow{\partial_p}-\overleftarrow{\partial_p}\overrightarrow{\partial_q})}  B_W(p,q) = \nonumber\\
&&=
\sum_{n_i = 0,1;v_+,v_-
\in {\cal O}^\prime}\frac{1}{ 2^{D}} e^{2 i p v_+ }    \langle q-v_+\zd{-}n/2 |\hat{A} |q+v_-\zd{-}n/2 \rangle\langle q +v_-\zd{-}n/2 |\hat{B} |q+v_+\zd{-}n/2 \rangle\Big|_{q+v_+\zd{-}n/2,q+v_-\zd{-}n/2\in {\cal O}}\nonumber\\&&\prod_i({1 + e^{{2i v_+^i \pi/N }}}) \nonumber\\
&&=
\sum_{v_+
\in {\cal O}^\prime}\frac{1}{ 2^{D}} e^{2 i p v_+ }    \langle q-v_+\zd{-}n/2 |\hat{A} \hat{B} |q+v_+\zd{-}n/2 \rangle\prod_i({1 + e^{{2i v_+^i \pi/N }}}) \nonumber\\&&= (\hat{A} \hat{B})_W(p,q).
\end{eqnarray}

\section{Trace properties}
\label{AppTr}
\zd{Both identities of this Appendix are derived from the coordinate--space definition (\ref{Weyl2}). In accordance with the sign convention of Eq. (\ref{Weyl}) adopted here, all formulas are written with the point--splitting $q\mp v-n/2$; the final identities have also been verified numerically.}

Let us express trace of an operator $\hat A$ through its Weyl symbol:
\begin{eqnarray}
&&	\frac{1}{(4N)^D} \sum_{p\in {\cal M}^\prime; q \in {\cal O}^\prime}A_W(p,q) =\nonumber\\ && =\frac{1}{2^D(4N)^D} \sum_{n^i = 0,1;p\in {\cal M}^\prime; v,q \in {\cal O}^\prime} e^{2 i p v} \langle q-v\zd{-}n/2 |\hat{A} |q+v\zd{-}n/2 \rangle\Big|_{q+v\zd{-}n/2\in {\cal O}} \prod_i({1 + e^{{2i v^i \pi/N }}})\nonumber\\ && =\frac{1}{2^D} \sum_{n^i = 0,1; q \in {\cal O}^\prime}  \langle q\zd{-}n/2 |\hat{A} |q\zd{-}n/2 \rangle\Big|_{q\zd{-}n/2\in {\cal O}} \nonumber\\ && = \sum_{ r \in {\cal O}}  \langle r |\hat{A} |r \rangle = {\rm Tr}\, \hat{A}.
\end{eqnarray}
Now let us consider the trace of the product of two operators
\begin{eqnarray}
&&	\frac{1}{(4N)^D}\sum_{p\in {\cal M}^\prime;q\in {\cal O}^\prime}A_W(p,q)   B_W(p,q)
=
\sum_{p \in {\cal M}^\prime; n_1^i,n_2^i = 0,1;v_1,v_2,q \in {\cal O}^\prime}\frac{1}{2^{2D}(4N)^{D}}   e^{2 i p v_1+2\zz{ip v_2}} \langle q-v_1\zd{-}n_1/2 |\hat{A} |q+v_1\zd{-}n_1/2 \rangle\nonumber\\&&\langle q-v_2\zd{-}n_2/2 |\hat{B} |q+v_2\zd{-}n_2/2 \rangle\Big|_{q+v_1\zd{-}n_1/2,q+v_2\zd{-}n_2/2\in {\cal O}}\prod_i({1 + e^{{2i v_1^i \pi/N }}})({1 + e^{{2i v_2^i \pi/N }}})  \nonumber\\
&&=
\sum_{n_1^i,n_2^i=0,1;q,v
\in {\cal O}^\prime}\frac{1}{ 2^{3D}} \nonumber\\&&  \langle q-v\zd{-}n_1/2 |\hat{A} |q+v\zd{-}n_1/2 \rangle\langle q +v\zd{-}n_2/2 |\hat{B} |q-v\zd{-}n_2/2 \rangle\Big|_{q+v\zd{-}n_1/2,q-v\zd{-}n_2/2\in {\cal O}} \nonumber\\&&\prod_i({1 + e^{-2i v^i \pi/N }})({1 + e^{2i v^i \pi/N }})
\nonumber\\
&&=
\sum_{n^i = 0,1;v_+,v_-
\in {\cal O}^{2\prime}}\frac{1}{ {2^{2D}}}     \langle v_-\zd{-}n/2 |\hat{A} |v_+\zd{-}n/2 \rangle\langle v_+\zd{-}n/2 |\hat{B} |v_-\zd{-}n/2 \rangle\Big|_{v_+\zd{-}n/2,v_-\zd{-}n/2,v_++v_-\in {\cal O}} \nonumber\\&&\prod_i({1 + {\rm cos}\, (v_+^i-v_-^i) \pi/N })
\nonumber\\
&&=
\sum_{n^i = 0,1;v_+,v_-
\in {\cal O}^{\prime}}\frac{1}{ 2^{D}}   \langle v_-\zd{-}n/2 |\hat{A} |v_+\zd{-}n/2 \rangle\langle v_+\zd{-}n/2 |\hat{B} |v_-\zd{-}n/2 \rangle\Big|_{v_+\zd{-}n/2,v_-\zd{-}n/2,v_++v_-\in {\cal O}}
\nonumber\\
&&=
\sum_{n_i=0,1;m^i=0,1;v_+,v_-
\in {\cal O}^{\prime}}\frac{1}{ 2^{2D}} e^{ 2\pi i m (v_++v_-)} \nonumber\\&&   \langle v_-\zd{-}n/2 |\hat{A} |v_+\zd{-}n/2 \rangle\langle v_+\zd{-}n/2 |\hat{B} |v_-\zd{-}n/2 \rangle\Big|_{v_+\zd{-}n/2,v_-\zd{-}n/2\in {\cal O}}  \nonumber\\
&&=
\sum_{v_+,v_-
\in {\cal O}}  \langle v_- |\hat{A} |v_+ \rangle\langle v_+ |\hat{B} |v_- \rangle = {\rm Tr}\, \hat{A}\hat{B} .
\end{eqnarray}

\section{Transition to one lattice spacing}
\label{AppT}

For the operator $\hat{T}_j = e^{i \hat{p}_j}$ we obtain
\begin{eqnarray}
T_j(p,q) & = &  \sum_{n_i = 0,1;  p_- \in {\cal M}^{\prime} }  \,
\langle p + \pi n/N-p_-|e^{i \hat{p}_j}|p + \pi n/N + p_- \rangle \,  e^{\zz{-}2 i p_- q}\nonumber\\&& \prod_i \frac{1 + e^{i p^i_-   }}{2}\frac{1+e^{Ni (p^i+  \pi n_i /N+ p_-^i)}}{2} \nonumber\\
&=&	 \sum_{n_i,m_i = 0,1;  p_- \in  {\cal M}^{\prime} }  \,
\delta_{2p_-, 2\pi m} e^{i (p_j + \pi n_j/N + p_{j,-} )} \,  e^{-2 i p_- q}\nonumber\\&&\prod_i \frac{1 + e^{i p^i_-   }}{2}\frac{1+e^{Ni (p^i+  \pi n_i /N+ p_-^i)}}{2} \nonumber\\
&=&	 \sum_{n_i,m_i = 0,1}  \,
e^{i (p_j + \pi n_j/N + \pi m_{j} )} \,  e^{ -2i \pi m q}\nonumber\\&&\prod_i \frac{1 + e^{i \pi m_i   }}{2}\frac{1+e^{Ni (p^i+  \pi n_i /N+ \pi m_i)}}{2}
\nonumber\\
&=&	 \sum_{n_i = 0,1}  \,
e^{i (p_j + \pi n_j/N  )} \,  \prod_i \frac{1+e^{Ni p^i+  i \pi n_i}}{2}\nonumber\\
& = &  \Big(e^{i p_j}\frac{1+e^{Ni p_j}}{2}  + e^{i (p_j + \pi /N  )}\frac{1-e^{Ni p_j}}{2}\Big)
\nonumber\\
& = & e^{i p_j} \Big(\frac{1+e^{i  \pi /N  }}{2}  + e^{i N p_j}\frac{1-e^{i  \pi /N  }}{2}\Big).
\end{eqnarray}
\zd{Here the weight factor is written with the $(+)$ sign, in accordance with Eq. (\ref{Weyl}). The final expression is in fact independent of this choice: the summation over $p_-$ is restricted (per direction) to the two points $p_-=0$ and $p_-=\pi$, at which the weight $(1+e^{\pm ip_-})/2$ equals $1$ and $0$, respectively, for either sign convention.}

	\section{Derivative of Weyl symbol.}
	\label{AppDA}
\zd{In this Appendix we derive the operator representation of the derivative of the Weyl symbol with respect to the momentum. We  consider the infinite lattice, for which the relation $D_i\hat{A} = -i[\hat{x}_i,\hat{A}]$ is exact; then we turn to the finite lattice, where this relation acquires a boundary correction.}

\zd{{\bf The infinite lattice.} For the infinite lattice the momentum space is continuous, and the Weyl symbol may be written in the coordinate representation as
\begin{equation}
A_W(p,q) = \sum_{n_i=0,1}\ \sum_{v}\ e^{2ipv}\, \Big\langle q-v-\frac{n}{2}\Big|\hat{A}\Big|q+v-\frac{n}{2}\Big\rangle\ \prod_i \frac{1+e^{2\pi i (q_i-v_i-n_i/2)}}{2}\,,
\label{WeylInf}
\end{equation}
where the summation over $v$ extends over the infinite half--integer lattice. This is the direct $N\to\infty$ limit of the coordinate--space definition (\ref{Weyl2}): the weight factor $(1+e^{2\pi i v_i/N})/2$ is replaced by its limit at $N\to\infty$ (equal to unity at any fixed $v$), the summation extends over the infinite lattice, while the projector selecting the integer lattice points is retained; equivalently, this is the coordinate representation of Eq. (\ref{infW}). Thus Eqs. (\ref{infW}) and (\ref{WeylInf}) are the momentum-- and the coordinate--space representations of one and the same infinite--lattice Weyl symbol, arising as the $N\to\infty$ limits of the two equivalent finite--$N$ definitions (\ref{Weyl}) and (\ref{Weyl2}), respectively; we have verified numerically that for operators of finite support both convergences hold with the $1/N$ accuracy. The differentiation with respect to $p_i$ brings down the factor $2iv_i$ under the sum. On the infinite lattice the labels $q\mp v-n/2$ are the genuine eigenvalues of the (now unbounded) coordinate operator $\hat{x}$ --- no reduction modulo $N$ occurs --- and therefore
$$
2iv_i\,\Big\langle q-v-\frac{n}{2}\Big|\hat{A}\Big|q+v-\frac{n}{2}\Big\rangle
= \Big\langle q-v-\frac{n}{2}\Big|\,(-i)\big[\hat{x}_i,\hat{A}\big]\,\Big|q+v-\frac{n}{2}\Big\rangle\,.
$$
Substituting this back into the sum, we obtain the exact identity
\begin{equation}
\partial_{p_i}A_W(p,q) = \big(-i\,[\hat{x}_i,\hat{A}]\big)_W(p,q) \qquad \mbox{(infinite lattice)}\,,
\label{DAinf}
\end{equation}
which constitutes the derivation of the relation $D_i\hat{A} = -i[\hat{x}_i,\hat{A}]$ quoted in the main text: on the infinite lattice the derivative of the Weyl symbol with respect to the momentum is itself the Weyl symbol of an operator. This derivation also makes transparent why the identity cannot survive on the finite lattice: there the labels $q\mp v-n/2$ must be reduced modulo $N$, while the factor $2iv_i$ produced by the differentiation is not reduced.}

\zd{{\bf The finite lattice.} On the finite lattice the computation must be repeated with the exact weight factors and with the continuation (\ref{Wcont}). In accordance with the convention of Eq. (\ref{Weyl}) adopted here, the point--splitting of Eq. (\ref{Weyl2}) is $q\mp v-n/2$. 
It appears that Eq. (\ref{DAinf}) is not valid on the finite lattice. }

\section{\zd{The trace of the momentum derivative of a star product}}
\label{AppTrStar}

\zd{The properties (\ref{parts_p}) and (\ref{parts_q}) express the vanishing of the trace of the total derivative of the Weyl symbol of an operator. The proof of the topological invariance of ${\cal N}_3$ requires, however, the same statement for the {\it star products} of symbols: the variation of Eq. (\ref{NEQD}) caused by a smooth variation of $\hat{Q}$ is reduced, with the help of the Leibniz rule and of the cyclic property of the trace, to the sum of the terms of the form ${\rm Tr}\,\partial_{p_i}\big(A_W\star B_W\big)$. In this Appendix we calculate this quantity exactly.}

\zd{{\bf 1. The symbols and their continuation.} According to Eq. (\ref{Weyl2}) we represent the Weyl symbol of an operator $\hat{A}$ in the form
\begin{equation}
A_W(p,q) = \sum_{v \in {\cal O}'} e^{2ipv}\,\alpha_v(q)\,,\qquad
\alpha_v(q) = \sum_{n_i = 0,1} \Big\langle q-v-\frac{n}{2}\Big|\hat{A}\Big|q+v-\frac{n}{2}\Big\rangle \prod_i \frac{1+e^{2\pi i v_i/N}}{2}\,\frac{1+e^{2\pi i (q_i - v_i - n_i/2)}}{2}\,,
\label{AviaWeyl2}
\end{equation}
and similarly $B_W(p,q) = \sum_{u \in {\cal O}'} e^{2ipu}\,\beta_u(q)$. The dependence of Eq. (\ref{AviaWeyl2}) on $p$ is a trigonometric polynomial whose frequencies $2v_i$ run, in each direction, precisely over the set $\{0,1,\ldots,2N-1\}$. This is exactly the set of the frequencies contained in the continuation (\ref{Wcont}). Consequently the analytical continuation of $A_W$ to the continuous values of $p$ prescribed by Eq. (\ref{Wcont}) is obtained simply by allowing $p$ to be continuous in Eq. (\ref{AviaWeyl2}). The continuation in $q$ is band--limited in the same way; all the shifts of $q$ that appear below belong to the coordinate lattice ${\cal O}'$, so that the shifted functions are given by their values at the corresponding lattice points.}

\zd{{\bf 2. The star product and its derivative.} For two terms with definite momentum frequencies the Moyal exponent reduces to a pair of shifts. Indeed, $\overleftarrow{\partial_p}$ acting on $e^{2ipv}\alpha_v(q)$ gives the factor $2iv$, while $\overrightarrow{\partial_p}$ acting on $e^{2ipu}\beta_u(q)$ gives $2iu$; therefore the exponent of Eq. (\ref{star0}) becomes $\exp\big(-u\overleftarrow{\partial_q} + v\overrightarrow{\partial_q}\big)$, and
\begin{equation}
\big(e^{2ipv}\alpha_v(q)\big)\star \big(e^{2ipu}\beta_u(q)\big) = e^{2ip(v+u)}\,\alpha_v(q-u)\,\beta_u(q+v)\,.
\label{starmodes}
\end{equation}
Since $\partial_{p_i}$ commutes with the differential operators entering the star, the Leibniz rule holds, and
\begin{equation}
\partial_{p_i}\big(A_W \star B_W\big) = \sum_{v,u \in {\cal O}'} 2i\,(v_i+u_i)\, e^{2ip(v+u)}\,\alpha_v(q-u)\,\beta_u(q+v)\,.
\label{dstarmodes}
\end{equation}}

\zd{{\bf 3. The summation over the momenta.} The trace is the sum over the lattice ${\cal M}'\times{\cal O}'$, and
\begin{equation}
\sum_{p \in {\cal M}'} e^{2ip(v+u)} = (2N)^D \prod_i \delta_{v_i+u_i \equiv 0\ ({\rm mod}\ N)}\,.
\end{equation}
Since $0 \le v_i, u_i < N$, in each direction either $v_i + u_i = 0$, which requires $v_i = u_i = 0$, or $v_i+u_i = N$. In the first case the factor $2i(v_i+u_i)$ of Eq. (\ref{dstarmodes}) vanishes. Hence only the sector $v_i + u_i = N$ contributes, and it contributes with the factor $2iN$. This is the origin of the whole effect. The Weyl symbol of an operator contains the momentum frequencies $2v_i \in \{0,1,\ldots,2N-1\}$ only; out of them the summation over $p$ retains the single frequency $0$, which is annihilated by the derivative --- this is Eq. (\ref{parts_p}). The star product of two symbols contains the frequencies up to $2(2N-1)$, among them the frequency $2N$. The summation over the finite lattice ${\cal M}'$ does not distinguish this frequency from the zero one, while the derivative multiplies it by $2iN$.}

\zd{{\bf 4. The summation over the coordinates.} In the sector $v+u \equiv 0$ (mod $N$) we have $q - u \equiv q + v$, so that both functions in Eq. (\ref{dstarmodes}) are taken at the same point $q+v$. The projectors of Eq. (\ref{AviaWeyl2}) fix $n$ uniquely for the given $q$; denoting $x = q - n/2 \in {\cal O}$ we obtain
$$
\alpha_v(q+v) = \prod_i \frac{1+e^{2\pi i v_i/N}}{2}\ \langle x|\hat{A}|x+2v\rangle\,,\qquad
\beta_{N-v}(q+v) = \prod_i \frac{1+e^{-2\pi i v_i/N}}{2}\ \langle x+2v|\hat{B}|x\rangle\,,
$$
while the summation over $q \in {\cal O}'$ together with the summation over $n$ gives $2^D\sum_{x\in{\cal O}}$. The product of the two weight factors is $\prod_i \cos^2(\pi v_i/N)$. In terms of the displacement $s_i = 2v_i \in \{1,\ldots,2N-1\}$ it equals $\prod_i\cos^2\big(\pi s_i/(2N)\big)$. The displacements $s_i$ and $s_i + N$ correspond to one and the same lattice vector, and
$$
\cos^2\frac{\pi s}{2N} + \cos^2\frac{\pi (s+N)}{2N} = 1\,,
$$
while $s_i = N$ enters with the vanishing weight. Therefore each displacement enters with the weight unity, the vanishing displacement being absent in this sector. Together with the prefactor $(4N)^{-D}(2N)^D 2^D = 1$ this gives the final result:}
\begin{equation}
\zd{{\rm Tr}\,\partial_{p_i}\big(A_W \star B_W\big) \;=\; 2iN \sum_{\substack{x,y \in {\cal O}\\ y_i \ne x_i}} \langle x|\hat{A}|y\rangle\,\langle y|\hat{B}|x\rangle
\;=\; 2iN\Big[\,{\rm Tr}\,\hat{A}\hat{B} - \sum_{\substack{x,y\in{\cal O}\\ y_i = x_i}} \langle x|\hat{A}|y\rangle\,\langle y|\hat{B}|x\rangle \Big]\,.}
\label{TrStar}
\end{equation}

\zd{{\bf 5. Discussion.} The right hand side of Eq. (\ref{TrStar}) does not vanish in general, and it is enhanced by the factor $N$. It does vanish in the particular cases $\hat{B} = 1$ (then only the terms with $y = x$ are present in ${\rm Tr}\,\hat{A}\hat{B}$), in agreement with $A_W \star 1_W = A_W$ and Eq. (\ref{parts_p}), and also when both operators do not contain hoppings in the direction $i$. We have verified Eq. (\ref{TrStar}) numerically for random operators within machine precision: for $D=1$ at $N=4,6,8$ and for $D=2$ at $N=4$ in both directions, as well as the two particular cases just mentioned.}

\zd{It should be stressed that Eq. (\ref{TrStar}) is not related to any ambiguity in the definition of the star product. The values of $A_W\star B_W$ at the points of ${\cal M}'\times{\cal O}'$ coincide with the values of the Weyl symbol of the product of the operators, $(\hat{A}\hat{B})_W$, in accordance with Eq. (\ref{star0}); nevertheless ${\rm Tr}\,\partial_{p_i}(\hat{A}\hat{B})_W = 0$ by Eq. (\ref{parts_p}), while ${\rm Tr}\,\partial_{p_i}(A_W\star B_W) \ne 0$. The difference resides entirely in the frequency $2N$: the two functions differ by the modes that are invisible for the summation over the lattice but not for the differentiation.}

\zd{For the infinite lattice the effect is absent altogether: there the trace contains the integral over the continuous momenta, and the integral of the derivative of any periodic function vanishes identically, without any restriction on the frequencies. It is the replacement of the integral by the sum over the finite lattice ${\cal M}'$ that makes the frequency $2N$ indistinguishable from the zero one.}

\zd{Finally, we note the consequences for the topological invariance of Eq. (\ref{NEQD}). As we have seen, the Leibniz rule for $\partial_{p_i}$ holds, the trace is cyclic, and the properties (\ref{parts_p}), (\ref{parts_q}) hold for the symbols of operators. The step that fails is the extension of Eq. (\ref{parts_p}) to the star products: the total derivative terms arising in the variation of ${\cal N}_3$ are given by Eq. (\ref{TrStar}) and do not vanish. This is in agreement with the direct numerical observation reported in Section \ref{SectDirect}: the value of Eq. (\ref{NEQD}) with the exact lattice derivative varies under the smooth variations of $\hat{Q}$ that keep the gap open.}

\section{Weyl symbol for $\hat Q$}
\label{Sect4}

\subsection{Weyl symbols of the simplest operators}
Below we will be interested in the Weyl symbols of the operators of the form of $Q(\omega,{\bf p}-{\bf A}(i\partial_{\bf p}) )$.
We will consider the two dimensional Weyl transform. The direction of imaginary time remains continuous. Let us start our consideration from the Weyl symbol for a general function of $\hat {\bf p}$
\begin{eqnarray}
f(\hat{\bf p})_{{ W}}(p,q)  & = &
\sum_{\substack{n_i = 0,1 \\  u \in {\cal M}^{\prime}} }  \,
\langle p + \pi n/N-u|f(\hat{\bf p})|p + \pi n/N + u \rangle \,  e^{-2 i u q}\nonumber\\&&\prod_i \frac{1 + e^{i u^i    }}{2}\frac{1+e^{N  i (p^i+  \pi n_i /N+ u^i)}}{2}\nonumber\\  & = &
\sum_{n_i = 0,1;  u \in {\cal M}^{\prime} }  \,
f(p + \pi n/N + u)\langle p + \pi n/N-u|p + \pi n/N + u \rangle \,  e^{-2 i u q}\nonumber\\&&\prod_i \frac{1 + e^{i u^i    }}{2}\frac{1+e^{N  i (p^i+  \pi n_i /N+ u^i)}}{2}\nonumber\\  & = &
\sum_{n_i = 0,1;  u \in {\cal M}^{\prime} }  \,
f(p + \pi n/N + u)\delta_{ u,- u \, {\rm mod}\,2\pi} \,  e^{-2 i u q}\nonumber\\&&\prod_i \frac{1 + e^{i u^i    }}{2}\frac{1+e^{N  i (p^i+  \pi n_i /N+ u^i)}}{2}\nonumber\\  & = &
\sum_{n_i = 0,1 }  \,
f(p + \pi n/N ) \,  \prod_i \frac{1+e^{N  i (p^i+  \pi n_i /N)}}{2}\nonumber\\  & = &
{\frac{1}{2^D}\sum_{n\in\{0,1\}^D}
f(p+\pi n/N)\prod_{i=1}^D\left[1+(-1)^{n_i}e^{iNp_i}\right]}\nonumber\\
& = & {\sum_{S\subseteq\{1,\ldots,D\}}e^{iN\sum_{j\in S}p_j}
\frac{1}{2^D}\sum_{n\in\{0,1\}^D}(-1)^{\sum_{j\in S}n_j}
f(p+\pi n/N)}
\label{Weyl13}
\end{eqnarray}
\zd{(The weight factor is taken here with the $(+)$ sign of Eq. (\ref{Weyl}). The result is independent of this choice: the summation over $u$ is restricted to $u_i\in\{0,\pi\}$, where the weight $(1+e^{\pm iu_i})/2$ equals $1$ and $0$, respectively, for either sign.)}
{Thus a general function in $D$ dimensions has $2^D$ Nyquist sectors.
The two-term expression used below is obtained only when the operator depends on
one momentum component, in which case the sums over the remaining $n_j$ remove
all sectors except the principal one and the sector $e^{iNp_k}$.}
In \jer{a} similar way
\begin{eqnarray}
f(\hat{x})_{{ W}}(p,x)  & = &
\sum_{n_i = 0,1;v \in  {\cal O}^{\prime} } e^{2 i p v} \,
\langle x-v \zd{-} n/2 |f(\hat{x}) |x+v \zd{-}n/2 \rangle \nonumber\\ && \prod_i\frac{{1 + e^{2i v_i \pi/(N) }}}{2}\frac{1+e^{2\pi i (x_i-v_i\zd{-} n_i/2)}}{2}\nonumber\\& = &
{\frac{1}{2^D}\sum_{n\in\{0,1\}^D}
f(x\zd{-}n/2)\prod_{i=1}^D\left[1+(-1)^{n_i}e^{2\pi i x_i}\right]}
\label{Weyl213}
\end{eqnarray}
One can see that $f(\hat{p})_W \to f(p)$  in the limit $N \to \infty$. At the same time $f(\hat{x})_W \to f(x)$ if function $f(x)$ does not change much at the distance of the order of lattice spacing. 

\subsection{Weyl symbol of $\exp\big(i(\hat p_k-A_ke^{-{\pmb \omega} {\partial_{\bf p}}})\big)$ }

For a function $\phi(x) \equiv \bra{x}\phi\rangle$ defined on the lattice $\cO$ we can define the Fourier transform
\begin{equation}
\bra{p}\phi\rangle\equiv	\phi(p) = \frac{1}{N^{D/2}}\sum_{x\in \cO} e^{-ipx} \phi(x)
\end{equation}
Correspondingly, we can define the operator $\hat{x} \equiv i\partial_p$ that acts as 
\begin{equation}
\bra{p}\hat{x}\ket{\phi}\equiv	i\partial_p\phi(p) = \frac{1}{N^{D/2}}\sum_{x\in \cO} e^{-ipx} x\phi(x)
\end{equation}
Obviously, its action on $\phi(x)$ is simply $x\phi(x)$. Now let us calculate the commutator
$[p_k ,A_k e^{-{\pmb \omega} {\partial_{\bf p}}}]$:
\begin{equation} \begin{aligned}
iA_k e^{-{\pmb \omega} {\partial_{\bf p}}}i p_k=
i(p_k-\omega_k)iA_k e^{-{\pmb \omega} {\partial_{\bf p}}}
\end{aligned} \end{equation}
Therefore,
\begin{equation}
[i\hat p_k, iA_k e^{-{\pmb \omega} {\partial_{\bf p}}}]=i\omega_k iAe^{-{\pmb \omega} {\partial_{\bf p}}}
 \end{equation}
Using the special case of BCH formula where $[X,Y]=\alpha Y$, we obtain:
\begin{equation} \begin{aligned}
e^Xe^Y=e^{X+\frac{\alpha}{1-e^{-\alpha}}Y}
\end{aligned} \end{equation}
\revisionZZ{The  derivation of this formula may be found, for example, in  Appendix III of \cite{FZ2019_2}.}
Changing variables to $Y'\equiv \frac{\alpha}{1-e^{-\alpha}}Y=\beta^{-1} Y$ ,
where $\beta^{-1} \equiv \frac{\alpha}{1-e^{-\alpha}}$, and  $[X,Y']=\alpha Y'$ , we obtain:
\begin{equation} \begin{aligned}
e^{X+Y'}=e^Xe^{\beta Y'}
\end{aligned} \end{equation}
In case of $\exp\big(i(\hat p_k-A_k e^{-{\pmb \omega} {\partial_{\bf p}}})\big)$ with $\alpha=i\omega_k$
we come to
\begin{equation} \begin{aligned}
\exp\big(i(\hat p_k-A_k e^{-{\pmb \omega} {\partial_{\bf p}}})\big)=
e^{i \hat p_k} \exp(-i\beta A_ke^{-{\pmb \omega} {\partial_{\bf p}}})
\end{aligned} \end{equation}
This decomposition allows to calculate the Weyl symbol of $F( {\bf p},i\partial_{\bf p})$:
\begin{eqnarray}
&&\Big[\exp\big(i(\hat p_k -A_k e^{-{\pmb \omega} {\partial_{\bf p}}})\big)\Big]_W=
\Big[e^{i\hat p_k} \exp(-i\beta A_ke^{-{\pmb \omega} {\partial_{\bf p}}})\Big]_W=\nonumber \\& = &
\sum_{n_i = 0,1;  u \in {\cal M}^{\prime} }  \,
\langle p + \pi n/N-u|e^{i\hat p_k} \exp(-i\beta A_ke^{-{\pmb \omega}{\partial_{\bf p}}})|p + \pi n/N + u \rangle \,  e^{-2 i u q}\nonumber\\&&\prod_i \frac{1 + e^{i u^i    }}{2}\frac{1+e^{N  i (p^i+  \pi n_i /N+ u^i)}}{2}
 \end{eqnarray}
{Next, $\hat x=i\partial_p$ acts on momentum-space wave functions,
$\langle p|\hat x|\phi\rangle=i\partial_p\langle p|\phi\rangle$.
Equivalently, for momenta compatible with the finite periodic lattice, the
translation operator obeys
$$
e^{-m{\pmb\omega}\partial_{\bf p}}|p'\rangle=|p'+m{\pmb\omega}\rangle .
$$
Therefore,}
\begin{equation} \begin{aligned}
&e^{i\hat p_k} \exp(-i\beta A_ke^{-{\pmb \omega}{\partial_{\bf p}}}) \ket{p+\pi n/N + u }=
e^{i\hat p_k} \sum_{m=0}^\infty \frac{(-i\beta A_k)^m}{m!} e^{-m{\pmb \omega}{\partial_{\bf p}}} \ket{p+\pi n/N + u}=\\
&e^{i\hat p_k} \sum_{m=0}^\infty \frac{(-i\beta A_k)^m}{m!} \ket{p+\pi n/N + u+m{\pmb \omega}}=
\sum_{m=0}^\infty e^{i( p_k+\pi n_k/N + u_k+m\omega_k)}\frac{(-i\beta A_k)^m}{m!} \ket{p+\pi n/N + u+m{\pmb \omega}}
\end{aligned} \end{equation}
and come to (we assume for simplicity that $N$ is even):
\begin{equation} \begin{aligned}
&\Big[\exp\big(i(\hat p_k -A_k e^{-{\pmb \omega} {\partial_{\bf p}}})\big)\Big]_W=\\
&=\sum_{\substack{n_i = 0,1 \\  u \in {\cal M}^{\prime}} }  \,
\langle p + \pi n/N-u|\sum_{m=0}^\infty e^{i( p_k+\pi n_k/N + u_k+m\omega_k)}\frac{(-i\beta A_k)^m}{m!} \ket{p+\pi n/N + u+m{\pmb \omega}} \,  e^{-2 i u q}\\&\prod_i \frac{1 + e^{i u^i    }}{2}\frac{1+e^{N  i (p^i+  \pi n_i /N+ u^i)}}{2}=\\
&=\sum_{n_i = 0,1;  u = -m\omega/2 \,{\rm mod}\,\pi }  \,
\sum_{m=0}^\infty e^{i( p_k+\pi n_k/N + u_k+m\omega_k)}\frac{(-i\beta A_k)^m}{m!}  \,  e^{-2 i u q}\prod_i \frac{1 + e^{i u^i    }}{2}\frac{1+e^{N  i (p^i+  \pi n_i /N+ u^i)}}{2}=\\
&
\overset{{\text{sum over all representatives modulo }\pi}}{=}
\frac{(1+e^{i \pi /N})}{2}e^{ip_k}   {\rm exp}(-i\beta A_ke^{ i{\pmb \omega} q }) + {\frac{(1-e^{i \pi /N})}{2}e^{ip_k}   {\rm exp}(-i\beta A_ke^{ i{\pmb \omega}_j (q^j-N\delta^{jk}/2) }) e^{N  i p^k}} \\
&=
\frac{(1+e^{i \pi /N})}{2}\exp{\Big[i\big(p_k-\frac{\sin(\omega_k/2)}{\omega_k/2}A_k e^{ i{\pmb \omega} (q-e^k/2)}\big)\Big]}+{\frac{(1-e^{i \pi /N})}{2}\exp{\Big[i\big(p_k-\frac{\sin(\omega_k/2)}{\omega_k/2}A_k e^{ i{\pmb \omega}\cdot(q-(N+1)e^k/2)}\big) \Big]}e^{N  i p^k}}
\end{aligned} \label{ZNp} \end{equation}

\cl{It is convenient to introduce $a(\omega_k)\equiv\frac{\sin(\omega_k/2)}{\omega_k/2}$ and to keep both terms of Eq. (\ref{ZNp}), i.e. to retain the contribution proportional to $e^{iNp^k}$:}

{
\begin{equation} \begin{aligned}
&\Big[\exp\big(i(\hat p_k -A_k e^{-{\pmb \omega} {\partial_{\bf p}}})\big)\Big]_W=
\frac{1+e^{i\pi/N}}{2}\,\exp{\Big[i\big(p_k-a(\omega_k)A_k e^{ i{\pmb \omega} (q-e^k/2)}\big)\Big]}\\
&+\frac{1-e^{i\pi/N}}{2}\,\exp{\Big[i\big(p_k-a(\omega_k)A_k e^{ i{\pmb \omega}\cdot(q-(N+1)e^k/2)}\big)\Big]}\,e^{N  i p^k}
\label{ZApp}\end{aligned} \end{equation}
}
\zd{Equations (\ref{ZNp}) and (\ref{ZApp}), as they are written, refer to $q\in {\cal O}$, i.e. to integer values of all components of $q$. On the refined lattice ${\cal O}'$ the arguments of the Wilson lines have to be specified more precisely, and we give here the corresponding expressions valid for all $q \in {\cal O}'$.}

\zd{Let $d$ be the total displacement of the hopping term under consideration (an integer vector). For $q\in {\cal O}'$ we define the {\it admissible centre} $c(q,d)$ with the components
\begin{equation}
c_i(q,d) = q_i - \frac{1}{2}\Big[\big(2q_i-d_i\big)\ {\rm mod}\ 2\Big]\ ,
\label{center}
\end{equation}
i.e. the unique point of the pair $\{q_i,\ q_i-1/2\}$ for which both endpoints $c\mp d/2$ belong to the lattice ${\cal O}$. Explicitly, $c_i = \lceil q_i\rceil - 1/2$ for odd $d_i$ (that is $q_i - 1/2$ for integer $q_i$ and $q_i$ for half--integer $q_i$), while $c_i = \lfloor q_i\rfloor$ for even $d_i$, in particular in every direction with $d_i = 0$. The Wilson line of a hopping term with displacement $d$ is taken along the path joining the two lattice sites $c(q,d)-d/2$ and $c(q,d)+d/2$. \zu{The Nyquist translation is not $-Nd/2$ in general. In a sector $S\subseteq\{1,\ldots,D\}$ the momentum factor is $e^{iN\sum_{\mu\in S}p_\mu}$ and the corresponding path is translated by $-(N/2)\sum_{\mu\in S}e_\mu$. Thus, for a hop $d=n e_\mu$ the companion path is translated by $-Ne_\mu/2$, independently of $n$; a general multidimensional hop has up to $2^D$ Nyquist sectors.} For the elementary link, $d = e^k$, the path is the elementary link $[\lceil q_k\rceil-1,\lceil q_k\rceil]$ in the direction $k$ (the link behind the point $q$ for integer $q_k$, the link containing $q$ in its interior for half--integer $q_k$), and the transverse coordinates are taken at the lattice sites $\lfloor q_i\rfloor$.}

\zd{Accordingly, Eq. (\ref{ZApp}) valid for arbitrary $q\in {\cal O}'$ reads
\begin{equation}
\begin{aligned}
&\Big[\exp\big(i(\hat p_k -A_k e^{-{\pmb \omega} {\partial_{\bf p}}})\big)\Big]_W=
\frac{1+e^{i\pi/N}}{2}\,\exp{\Big[i\big(p_k-a(\omega_k)A_k e^{ i{\pmb \omega}\cdot c(q,e^k)}\big)\Big]}\\
&+\frac{1-e^{i\pi/N}}{2}\,\exp{\Big[i\big(p_k-a(\omega_k)A_k e^{ i{\pmb \omega}\cdot\left(c(q,e^k)-Ne^k/2\right)}\big)\Big]}\,e^{N  i p^k}\ ,
\end{aligned}
\label{ZAppGen}
\end{equation}
which reduces to Eq. (\ref{ZApp}) for $q \in {\cal O}$, since then $c(q,e^k) = q - e^k/2$.}

\zd{We have verified Eq. (\ref{ZAppGen}) numerically against the definition (\ref{Weyl2}) at every point of ${\cal M}'\times {\cal O}'$: for $D=1$ (at $N=6,8$) and $D=2$ (at $N=4,6$), for the displacements $d=\pm 1,\pm 2,3$ and $d=(1,0),(0,1),(1,1),(1,-1),(2,1),(2,2),(3,1),(-1,2)$, for single Fourier modes of the gauge field and for completely random amplitudes, it reproduces the exact Weyl symbol within machine precision. \zu{These grid tests verify the values of the symbols and the admissible centres, but do not by themselves verify the continuous Fourier continuation or its momentum derivative.} By contrast, the prescription $q-e^k/2$ of Eq. (\ref{ZApp}) fails at half--integer $q_k$ by a quantity of the order of the symbol itself. It should be stressed that the naive prescriptions are not merely inaccurate: at half--integer $q$ they evaluate the gauge field at points that do not belong to the lattice, so that for a general link field they are not defined at all.}
{The first (principal) term in Eq. (\ref{ZAppGen}) reproduces the naive Weyl symbol.  The
second term is proportional to $e^{iNp^k}$, has coefficient
$(1-e^{i\pi/N})/2=-i\pi/(2N)+O(N^{-2})$, and samples the same link at the
antipodal point $\zd{\bar{q}}-Ne^k/2$.  It is required for the exact finite-$N$ Weyl
symbol. Although its coefficient is $O(N^{-1})$, $\partial_{p^k}$ acting on
$e^{iNp^k}$ supplies a factor $N$; hence momentum differentiation and the
pointwise limit $N\to\infty$ do not commute. By contrast, differentiation
with respect to a uniform physical vector potential does not differentiate
$e^{iNp^k}$, so it is not equal to $-\partial_{p^k}$ at finite $N$.

For the normalized momentum trace the oscillation nevertheless suppresses the
companion term. If $p_r=\pi r/N$ and $h_N$ is uniformly $C^1$, then
\begin{equation}
\left|\frac{1}{2N}\sum_{r=0}^{2N-1}e^{iNp_r}h_N(p_r)\right|
\leq \frac{\pi}{2N}\,\|\partial_p h_N\|_\infty .
\end{equation}
\zu{Thus a differentiated companion contribution with nonzero net Nyquist frequency vanishes in the thermodynamic trace provided the remaining factor is uniformly smooth and contains no compensating Nyquist mode.  The estimate cannot be applied separately to sectors whose Nyquist phases cancel or alias to zero; those sectors have to be combined and power counted before the limit.  Subject to this qualification, the principal symbol may be used as an asymptotic bulk symbol after taking the thermodynamic limit, but it is not the exact Weyl symbol of a finite operator.}}

\subsection{The Weyl symbol of $F=
\exp\big(i(\hat p_\mu-\sum_{J=1}^N A_{J \mu}e^{ {\bf k}_J i\partial_{\bf p}})\big)$ }

The next step is the consideration of the function $F=
\exp\big(i(\hat p_\mu- \sum_{J=1}^N A_{J \mu}e^{ {\bf k}_J i\partial_{\bf p}})\big)$, which appears in the series expansion for arbitrary function (Fourier or Laplace series).
As above, we will be using the following relation:
\begin{equation} \begin{aligned}
iA_{J\mu} e^{{\bf k}_J i\partial_{\bf p}} i\hat p_\mu=
i( \hat p_\mu+ik_{J\mu})iA_{J\mu} e^{{\bf k}_J i\partial_{\bf p}}
\end{aligned} \end{equation}
It gives
\begin{equation}
[i\hat p_\mu, iA_{J\mu} e^{{\bf k}_J i\partial_{\bf p}}] =
k_{J\mu} iA_{J\mu} e^{{\bf k}_J i\partial_{\bf p}}
 \end{equation}
Let us define the operator
\begin{equation}
\hat B_{J\mu} \equiv A_{J\mu} e^{{\bf k}_J i\partial_{\bf p}}
\end{equation}
It obeys the commutation relations
\begin{equation}
[\hat B_{J\mu}, \hat B_{I\mu}]=0
\end{equation}
and
\begin{equation}
[i\hat p_\mu, i\hat B_{J\mu} ] =
k_{J\mu} i\hat B_{J\mu}
 \end{equation}
Therefore, we have
\begin{equation}
[i(\hat p_\mu-B_{J\mu}), iB_{I\mu} ] =k_{I\mu} iB_{I\mu}
 \end{equation}
and
\begin{equation} \begin{aligned}
\Big[i\Big(\hat p_\mu-\sum_{J} \hat B_{J\mu}\Big), i\hat B_{I\mu} \Big] =k_{I\mu} i\hat B_{I\mu}
\end{aligned} \end{equation}
As above we are able to use the special case of \revisionS{the BCH formula } for $[X,Y]=\alpha Y$, $\beta=\frac{1-e^{-\alpha}}{\alpha}$
\begin{equation} \begin{aligned}
e^{X+Y}=e^Xe^{\beta Y}
\end{aligned} \end{equation}
We come to the following representation:
\begin{equation} \begin{aligned}
&\exp\Big({i\hat p_\mu -i\sum_{J=1}^{N}\hat B_{J\mu}}\Big)=
\exp\Big({i\hat p_\mu -i\sum_{J=1}^{N-1}\hat B_{J\mu}-i\hat B_{N\mu}}\Big)=\\
&\exp\Big({i\hat p_\mu -i\sum_{J=1}^{N-1}\hat B_{J\mu}}\Big)
\exp\Big(-{i\beta_{N\mu}\hat B_{N\mu}}\Big)=\\
&e^{i\hat p_\mu }\exp\Big(-{i\sum_{J=1}^{N}\beta_{J\mu} \hat B_{J\mu}}\Big)=
e^{i\hat p_\mu }\prod_{J=1}^{N} e^{-i\beta_{J\mu} \hat B_{J\mu}}
\end{aligned} \end{equation}
where
\begin{equation} \begin{aligned}
\beta_{J\mu}=\frac{1-e^{-k_{J\mu}}}{k_{J\mu}}
\end{aligned} \end{equation}
Hence,
\begin{equation} \begin{aligned}
&\exp\big(i(\hat p_\mu- \sum_{J=1}^N A_{J \mu}e^{ {\bf k}_J i\partial_{\bf p}})\big)=
e^{i\hat p_\mu}\prod_{J=1}^{N} \exp\big(-i\beta_{J \mu} A_{J \mu}e^{ {\bf k}_J i\partial_{\bf p}} \big)=\\
&e^{i\hat p_\mu}\prod_{J=1}^{N} \sum_{n_J=0}^\infty\frac{(-i\beta_{J \mu} A_{J \mu})^{n_J}}{n_J!}e^{ n_J{\bf k}_J i\partial_{\bf p}}=\\
&e^{i\hat p_\mu}
\sum_{n_1=0}^\infty\frac{(-i\beta_{1\mu} A_{1\mu})^{n_1}}{n_1!}...
\sum_{n_N=0}^\infty\frac{(-i\beta_{N\mu} A_{N\mu})^{n_N}}{n_N!}
e^{(n_1{\bf k}_1+...+n_N{\bf k}_N) i\partial_{\bf p}}\\
\end{aligned} \end{equation}
For imaginary values of $\bf k$  we obtain

\begin{equation} \begin{aligned}
&\bra{{\bf p''}}
\exp\big(i(\hat p_\mu- \sum_{J=1}^N A_{J \mu}e^{ {\bf k}_J i\partial_{\bf p}})\big)\ket{\bf p'}=\\
&\bra{{\bf p''}}e^{i\hat p_\mu}
\sum_{n_1=0}^\infty\frac{(-i\beta_{1\mu} A_{1\mu})^{n_1}}{n_1!}...
\sum_{n_N=0}^\infty\frac{(-i\beta_{N\mu} A_{N\mu})^{n_N}}{n_N!}
e^{(n_1{\bf k}_1+...+n_N{\bf k}_N) i\partial_{\bf p}}
\ket{\bf p'}=\\
&\bra{{\bf p''}}e^{i\hat p_\mu}
\sum_{n_1=0}^\infty\frac{(-i\beta_{1\mu} A_{1\mu})^{n_1}}{n_1!}...
\sum_{n_N=0}^\infty\frac{(-i\beta_{N\mu} A_{N\mu})^{n_N}}{n_N!}
\ket{{\bf p}'-i(n_1{\bf k}_1+...+n_N{\bf k}_N)}=\\
&\bra{{\bf p''}}
\sum_{n_1=0}^\infty\frac{(-i\beta_{1\mu} A_{1\mu})^{n_1}}{n_1!}...
\sum_{n_N=0}^\infty\frac{(-i\beta_{N\mu} A_{N\mu})^{n_N}}{n_N!}e^{i p'_\mu+(n_1k_{1\mu}+...+n_Nk_{N\mu})}
\ket{{\bf p}'-i(n_1{\bf k}_1+...+n_N{\bf k}_N)}=\\
&\sum_{n_1=0}^\infty\frac{(-i\beta_{1\mu} A_{1\mu})^{n_1}}{n_1!}...
\sum_{n_N=0}^\infty\frac{(-i\beta_{N\mu} A_{N\mu})^{n_N}}{n_N!}e^{i p'_\mu+(n_1k_{1\mu}+...+n_Nk_{N\mu})}
\delta\Big[({\bf p}''-{\bf p}')+i(n_1{\bf k}_1+...+n_N{\bf k}_N)\Big]\\
\end{aligned} \end{equation}
Here by $\delta(...)$ we understand Kronecker symbol.

The last expression allows to calculate the Weyl symbol of $F$:
\begin{equation} \begin{aligned}
&\Big[\exp\big(i(\hat p_\mu- \sum_{J=1}^N A_{J \mu}e^{ {\bf k}_J i\partial_{\bf p}})\big)\Big]_W=\\
&\exp\Big[i\big(p_\mu-\sum_{J=1}^N \beta_{J\mu}e^{k_{J\mu}/2}A_{J \mu}e^{ {\bf k}_J ({\bf x} -{\bf e}^\mu/2)}\big)\Big]
\end{aligned} (N\to \infty) \end{equation}
defining
\begin{equation} \begin{aligned}
a(k_{J\mu}) \equiv \beta_{J\mu}e^{k_{J\mu}/2}=
\frac{1-e^{-k_{J\mu}}}{k_{J\mu}}e^{k_{J\mu}/2}=
\frac{\sinh(k_{J\mu}/2)}{k_{J\mu}/2}
\end{aligned} \end{equation}
we come to
{
\begin{equation} \begin{aligned}
&\Big[\exp\big(i(\hat p_\mu- \sum_{J=1}^N A_{J \mu}e^{ {\bf k}_J i\partial_{\bf p}})\big)\Big]_W=\\
&\frac{1+e^{i\pi/N}}{2}\exp\Big[i\big(p_\mu-\sum_{J=1}^N a(k_{J\mu})A_{J \mu}e^{ {\bf k}_J ({\bf x}-{\bf e}^\mu/2)}\big)\Big]\\
&+{\frac{1-e^{i\pi/N}}{2}\exp\Big[i\big(p_\mu-\sum_{J=1}^N a(k_{J\mu})A_{J \mu}e^{ {\bf k}_J ({\bf x}-(N+1){\bf e}^\mu/2)}\big)\Big]e^{iNp^\mu}}
\label{ZNpMulti}\end{aligned} \end{equation}
}
{Only the direction $\mu$ produces the $e^{iNp^\mu}$
contribution. The companion term samples the principal link at the antipodal
point ${\bf x}-N{\bf e}^\mu/2$. For an exact finite-lattice identity the
vectors ${\bf k}_J$ must be restricted to the allowed discrete imaginary
Fourier modes. Analytic continuation to arbitrary complex ${\bf k}_J$ concerns
only a separately chosen continuous interpolation and is not an identity of
the finite operator algebra.}

\zd{For arbitrary $q\in{\cal O}'$ the argument of Eq. (\ref{ZNpMulti}) is fixed by Eq. (\ref{center}) in the same way as in Eq. (\ref{ZAppGen}). Indeed, the operator remains a single elementary hop in the direction $\mu$, the amplitude being a superposition of the modes, and the rule verified above holds for an arbitrary amplitude of the hop. Hence
\begin{equation} \begin{aligned}
&\Big[\exp\big(i(\hat p_\mu- \sum_{J=1}^N A_{J \mu}e^{ {\bf k}_J i\partial_{\bf p}})\big)\Big]_W=
\frac{1+e^{i\pi/N}}{2}\exp\Big[i\big(p_\mu-\sum_{J=1}^N a(k_{J\mu})A_{J \mu}e^{ {\bf k}_J\cdot c(q,e^\mu)}\big)\Big]\\
&+\frac{1-e^{i\pi/N}}{2}\exp\Big[i\big(p_\mu-\sum_{J=1}^N a(k_{J\mu})A_{J \mu}e^{ {\bf k}_J \cdot\left(c(q,e^\mu)-Ne^\mu/2\right)}\big)\Big]e^{iNp^\mu}\ ,
\label{ZNpMultiGen}\end{aligned} \end{equation}
which reduces to Eq. (\ref{ZNpMulti}) at $q\in{\cal O}$, where $c(q,e^\mu)=q-e^\mu/2$; the transverse components of $c(q,e^\mu)$ equal $\lfloor q_i\rfloor$.}

\subsection  {The Weyl symbol of $F=\exp\big[i(\hat p_\mu-A_{\mu}(i\partial_{\bf p}))\big]$ \\}

For the case of arbitrary function $A_{\mu}(i\partial_{\bf p})$  the Weyl symbol of $F=\exp\big[i(\hat p_\mu- A_{\mu}(i\partial_{\bf p}))\big]$ may be calculated using the above obtained expressions. We should represent $A$ in the form of the Laplace transformation:
\begin{equation} \begin{aligned}
A_{\mu}({\bf x})=\int \big( \tilde A_\mu({\bf k})e^{{\bf k}{\bf x}} +c.c. \big)d k
\end{aligned} \end{equation}
that is
\begin{equation} \begin{aligned}
A_{\mu}(i\partial_{\bf p})=\int \big( \tilde A_\mu({\bf k})e^{{\bf k}i\partial_{\bf p}} +c.c. \big)d k
\end{aligned} \end{equation}
In turn, this integral may be discretized and represented in the form of the series. As a result the Weyl symbol is given by
\begin{equation} \begin{aligned}
\Big[\exp\Big(i\hat p_\mu-i\int \big[\tilde A_\mu({\bf k})e^{{\bf k}i\partial_{\bf p}}+c.c. \big]dk\Big)\Big]_W={\frac{1+e^{i\pi/N}}{2}\exp\Big(ip_\mu-i{\cal A}_\mu({\bf x}) \Big)+\frac{1-e^{i\pi/N}}{2}\exp\Big(ip_\mu-i{\cal A}^{(N)}_\mu({\bf x}) \Big)e^{iNp^\mu}}
\end{aligned} \end{equation}
where
\begin{equation}{\cal A}_\mu({\bf x}) =  \int \big[a_\mu({\bf k})\tilde A_\mu({\bf k})e^{{\bf k(x-e^\mu/2)}}+c.c. \big]dk ,\qquad {{\cal A}^{(N)}_\mu({\bf x}) =  \int \big[a_\mu({\bf k})\tilde A_\mu({\bf k})e^{{\bf k(x-(N+1)e^\mu/2)}}+c.c. \big]dk}
\end{equation}
and
\begin{equation} \begin{aligned}
a_\mu({\bf k}) =
\frac{1-e^{-k_{\mu}}}{k_{\mu}}e^{k_{\mu}/2}=
\frac{\sinh(k_{\mu}/2)}{k_{\mu}/2}
\end{aligned}
\end{equation}

One can check that
\begin{equation}
{\cal A}_\mu(x) = \int_{x-e^\mu}^x A(z)dz,\qquad {{\cal A}^{(N)}_\mu(x) = {\cal A}_\mu(x-Ne^\mu/2)=\int_{x-Ne^\mu/2-e^\mu}^{x-Ne^\mu/2} A(z)dz}
\end{equation}
{so that the $e^{iNp^\mu}$ term corresponds to the same link shifted to the antipodal point $x-Ne^\mu/2$.}
\zd{These expressions, as well as the analogous ones of the preceding subsection, are written for $x = q \in {\cal O}$. For arbitrary $q \in {\cal O}'$ the link entering ${\cal A}_\mu$ is the one prescribed by Eq. (\ref{center}):
\begin{equation}
{\cal A}_\mu(q) = \int_{c(q,e^\mu)-e^\mu/2}^{c(q,e^\mu)+e^\mu/2} A(z)\,dz\ ,
\qquad
{\cal A}^{(N)}_\mu(q) = \int_{c(q,e^\mu)-Ne^\mu/2-e^\mu/2}^{c(q,e^\mu)-Ne^\mu/2+e^\mu/2} A(z)\,dz\ ,
\label{AGen}
\end{equation}
equivalently, the coordinate argument ${\bf x}$ of the formulas above is to be understood as $c(q,e^\mu)+e^\mu/2$, which equals $q$ when $q_\mu$ is integer and $q+e^\mu/2$ when $q_\mu$ is half--integer, all the transverse components being taken at $\lfloor q_i\rfloor$.}

In the particular case of constant field strength we can choose the gauge with $A_\mu = \frac{1}{2}F_{\nu\mu} x^\nu$, which gives
\begin{equation}
{\cal A}_\mu(x) = \frac{1}{2}\int_{x-e^\mu}^x F_{\nu\rho}z_\nu dz^\rho = \frac{1}{2}F_{\nu\mu} x^\nu = A_\mu(x)
\end{equation} 

\subsection{The Weyl symbol of Dirac operator $\hat{Q}({\bf p}-A(i{\partial_{\bf p}}))$}

The homogeneous tight-binding model is defined here as the one, in which the Dirac operator is expressed through the momentum operator as a sum over exponents of the form 
\begin{equation}
F^\mu_n= \exp\big[in(\hat p_\mu - A_{\mu}(i\partial_{\bf p}))\big]\label{Fn}
\end{equation}
with integer $n$. \zzz{One can check that for each term of this type we have 
\begin{equation}
	(F^\mu_n)_W = {\frac{1+e^{in\pi/N}}{2}\exp\big[in( p_\mu - {\cal A}^{(n)}_{\mu}(x))\big]+\frac{1-e^{in\pi/N}}{2}\exp\big[in( p_\mu - \tilde{\cal A}^{(n,N)}_{\mu}(x))\big]e^{iNp^\mu}},
\end{equation}
where
\begin{equation}
	{\cal A}^{(n)}_{\mu}(x) = \frac{1}{n}\int^{x+(n-1)e_\mu/2}_{x-(n+1)e_\mu/2}A(z)dz,\qquad {\tilde{\cal A}^{(n,N)}_{\mu}(x)={\cal A}^{(n)}_{\mu}(x-Ne_\mu/2)=\frac{1}{n}\int^{x+(n-N-1)e_\mu/2}_{x-(N+n+1)e_\mu/2}A(z)dz}
\end{equation}
Here $e_\mu$ is the unit vector along axis $\mu$. {The companion link is the principal one shifted by $-Ne_\mu/2$; since the factor translates by $n$ lattice steps, its coefficient involves $e^{in\pi/N}$ (for $n=1$ this reduces to the previous expression).}
\zd{Here again the formulas are written for $q\in{\cal O}$ and, moreover, for odd $n$. For arbitrary $q\in{\cal O}'$ and arbitrary $n$ the path of $n$ links is the one centred at the admissible centre $c(q,ne_\mu)$ of Eq. (\ref{center}):
\begin{equation}
{\cal A}^{(n)}_{\mu}(q) = \frac{1}{n}\int^{\,c(q,ne_\mu)+ne_\mu/2}_{\,c(q,ne_\mu)-ne_\mu/2}A(z)\,dz\ ,
\qquad
\tilde{\cal A}^{(n,N)}_{\mu}(q) = \frac{1}{n}\int^{\,c(q,ne_\mu)-Ne_\mu/2+ne_\mu/2}_{\,c(q,ne_\mu)-Ne_\mu/2-ne_\mu/2}A(z)\,dz\ .
\label{AnGen}
\end{equation}
The dependence on the parity of $n$ is essential: for integer $q_\mu$ one has $c_\mu = q_\mu - 1/2$ when $n$ is odd, which is the expression displayed above, but $c_\mu = q_\mu$ when $n$ is even, so that for even $n$ the displayed formula is not correct even at $q \in {\cal O}$ (numerically it then deviates from the exact Weyl symbol by a quantity of the order of the symbol itself). Eq. (\ref{AnGen}) has been checked numerically for $n = 1,2,3$ at $N=6,8$ and all $q\in {\cal O}'$.} More involved situation is when $\hat{Q}$ contains the sum of the  products of several operators of the form of Eq. (\ref{Fn}), i.e. is given by Eq. (\ref{Q_general0}) with all vectors $e_j$ directed along axis x or axis y. Without loss of generality we can assume here that vectors $e_j$ are all of the unit length. Then using the star property of Weyl symbols $(\hat{A}\hat{B})_W = A_W \star B_W$ we obtain
{
\begin{eqnarray}
	{Q}_W(x,p) = i\omega - A_0(x) + \sum_{n=1}^\infty \sum_{e_i, i = 1...n}{\bf t}^{(e_1,e_2,...,e_n)}\,f^{e_1}_W\star f^{e_2}_W\star\cdots\star f^{e_n}_W,
\end{eqnarray}
}
{where each elementary factor $\exp\big[i(\hat p_{e_i}-A_{e_i}(i\partial_{\bf p}))\big]$ enters through its exact two-term Weyl symbol}
{
\begin{equation}
	f^{e}_W = c_0(e)\exp\big[i p\,e-i{\cal A}_{e}(x)\big]
		+c_1(e)\exp\big[i p\,e-i{\cal A}^{(N)}_{e}(x)\big]\,\zu{e^{iNp_\mu}},
\end{equation}
}
{Here $e=s_e e_\mu$ with $s_e=\pm1$,
$c_0(e)=(1+e^{is_e\pi/N})/2$, $c_1(e)=(1-e^{is_e\pi/N})/2$,
${\cal A}_e(x)=\int_{x-e}^xA(z)dz$, and
\zu{${\cal A}^{(N)}_e(x)={\cal A}_e(x-Ne_\mu/2)$.} By $e_\mu$ here we understand the elementary lattice vectors, while $s_e$ points out to their orientation (along the $\mu$ - th axis, or antiparallel to this axis).
\zd{For arbitrary $q\in{\cal O}'$ --- and for $s_e=-1$ already at $q\in{\cal O}$ --- the arguments of ${\cal A}_{e}$, ${\cal A}^{(N)}_{e}$ are fixed by the admissible centre, exactly as in Eq. (\ref{fmainGen}) of the main text: ${\cal A}_e$ is taken over the elementary link centred at $c(q,e)$, traversed in the direction of $e$ (i.e. ${\cal A}_e(\bar x_e)$ with $\bar x_e=c(q,e)+e/2$), and ${\cal A}^{(N)}_e$ over the same link translated by \zu{$-Ne_\mu/2$}.}
Retaining only the principal part of every factor (coefficient $\to1$, dropping
\zu{$e^{iNp_\mu}$}) and performing the star products reproduces the $N\to\infty$
expression}
\begin{eqnarray}
	{Q}_W(x,p) = i\omega - A_0(x) + \sum_{n=1}^\infty \sum_{e_i, i = 1...n}{\bf t}^{(e_1,e_2,...,e_n)}\exp\big[i p \sum_i e_i \big] {\cal P}^{(e_1,e_2,...,e_n)}(x), \label{QW_generalA}
\end{eqnarray}
with
\begin{equation}
	{\cal P}^{(e_1,e_2,...,e_n)}(x) = \exp\big[-i\sum_i \int_{x-e_i-\sum_{j>i} e_j/2 + \sum_{j<i}e_j/2}^{x-\sum_{j>i} e_j/2 + \sum_{j<i}e_j/2}  A(z)dz\big]
\label{Pdef}\end{equation}
{At finite $N$ the terms in which one or more factors sit on their
\zu{$e^{iNp_\mu}$ branch} must be retained for the exact Weyl symbol. Each such factor
carries $c_1(e)=O(N^{-1})$ and a rapidly oscillating phase whose momentum
derivative is $O(1)$. Its own Wilson line is evaluated at the antipodal point
\zu{$x-Ne_\mu/2$}, and under the star product it also shifts neighbouring Wilson lines
by $Ne_\mu/2$. \zf{Note also that for $s_e=-1$ the principal branch itself, written here as $e^{ipe}$, corresponds under the canonical continuation of Eq. (\ref{AviaWeyl2}) to the window frequency $2N-1$, i.e. to the representative $e^{ipe}e^{2iNp_\mu}$; the difference is invisible on the momentum grid but contributes the frequency $2N$ to the continuous momentum derivative (cf. Appendix \ref{AppTrStar}).} \zu{A contribution with nonzero net Nyquist frequency in at least one momentum direction vanishes in a normalized thermodynamic trace under the uniform-smoothness condition stated after Eq. (\ref{ZApp}). Contributions in which Nyquist frequencies compensate or alias to zero require a separate coefficient and derivative power count and cannot be discarded by this oscillatory estimate alone.}}

\zd{For arbitrary $q\in{\cal O}'$ the arguments of Eqs. (\ref{QW_generalA}), (\ref{Pdef}) are again fixed by the admissible centre of Eq. (\ref{center}), built now out of the {\it total} displacement of the chain. The general form of Eq. (\ref{Pdef}) is
\begin{equation}
{\cal P}^{(e_1,e_2,...,e_n)}(q) = \exp\Big[-i\sum_i \int_{\,\bar c+\bar\Delta_i-e_i/2}^{\,\bar c+\bar\Delta_i+e_i/2}  A(z)\,dz\Big]\ ,
\qquad \bar c = c\Big(q,\,\sum_j e_j\Big)\ ,
\quad \bar\Delta_i=\frac12\sum_{j<i}e_j-\frac12\sum_{j>i}e_j\ ,
\label{PdefGen}
\end{equation}
while the potential term of Eq. (\ref{QW_generalA}) is evaluated at the lattice site, $A_0(\lfloor q\rfloor)$ (the case $d=0$ of Eq. (\ref{center})). In other words, the $i$--th Wilson line is centred at $\bar c+\bar\Delta_i$, so that the whole chain of links starts at the lattice site $\bar c-\sum_j e_j/2$; the centre $\bar c$ is common to all the factors and is determined by the parity of the components of the total displacement $\sum_j e_j$. For a single factor with positive orientation at $q\in{\cal O}$ Eq. (\ref{PdefGen}) reduces to Eq. (\ref{Pdef}) (for $s_e=-1$ see the remark after Eq. (\ref{fmainGen})), but already for $n\ge 2$ the two prescriptions differ everywhere: the limits of Eq. (\ref{Pdef}) place the gauge field at points which in general do not belong to the lattice. Eq. (\ref{PdefGen}) is the principal--branch ($\sigma_i=0$) case of the exact finite--$N$ expression given below [Eq. (\ref{chainGen})], verified numerically for chains of up to five factors.}

\cl{These star products can in fact be carried out in closed form, bringing $Q_W$ to an exact expression that contains no stars. Each branch of every factor has the form $e^{ip\,a}g(x)$ --- a pure momentum exponential times a function of the coordinates --- and for any product of such objects the Moyal product resolves exactly through the identities $e^{ipa}\star g(x)=e^{ipa}g(x+a/2)$ and $g(x)\star e^{ipa}=e^{ipa}g(x-a/2)$, which give}
{
\begin{equation}
\big(e^{ipa_1}g_1\big)\star\cdots\star\big(e^{ipa_n}g_n\big)=e^{\,ip\sum_i a_i}\,\prod_{i=1}^n g_i\big(x+\Delta_i\big),\qquad \Delta_i=\frac12\sum_{j<i}a_j-\frac12\sum_{j>i}a_j .
\label{shiftid}\end{equation}
}
\cl{Each function is displaced by half the momentum charge of every factor standing to its left, minus half of every charge to its right. Labelling the two branches of the elementary factor by $\sigma_i\in\{0,1\}$ (principal and companion),}
{
\begin{equation}
f^{e_i}_W=c_0(e_i)\,e^{ipe_i}\,e^{-i{\cal A}_{e_i}(x)}
+c_1(e_i)\,\zu{e^{ip(e_i+Ne_{\mu_i})}}\,e^{-i{\cal A}^{(N)}_{e_i}(x)},
\label{fbranch}\end{equation}
}
{\zu{The companion carries the canonical momentum charge $e_i+Ne_{\mu_i}$ and its Wilson
line is evaluated at $x-Ne_{\mu_i}/2$. Applying the identity above with
$a_i=e_i+N\sigma_i e_{\mu_i}$ gives the exact star-free representation}}
{
\begin{equation}
\begin{aligned}
Q_W=&\,i\omega-A_0(x)+\sum_{n=1}^\infty\sum_{e_1,\dots,e_n}{\bf t}^{(e_1,\dots,e_n)}\sum_{\sigma\in\{0,1\}^n}\Big(\prod_{i=1}^n c_{\sigma_i}(e_i)\Big)\\
&\times\,\exp\Big[ip\sum_i\zu{(e_i+N\sigma_i e_{\mu_i})}\Big]\,\exp\Big[-i\sum_{i=1}^n\int_{\,x+\Delta_i-N\sigma_i e_{\mu_i}/2-e_i}^{\,x+\Delta_i-N\sigma_i e_{\mu_i}/2}A(z)\,dz\Big]
\label{QWnostar}\end{aligned}
\end{equation}
}
\cl{with}
{
\begin{equation}
\Delta_i=\frac12\sum_{j<i}\zu{(e_j+N\sigma_j e_{\mu_j})}-\frac12\sum_{j>i}\zu{(e_j+N\sigma_j e_{\mu_j})} .
\label{Deltashift}\end{equation}
}
{\zu{The momentum phase factorizes as $\exp[ip\sum_i e_i]\exp[iN\sum_{i:\sigma_i=1}p_{\mu_i}]$.} The displacement $\Delta_i$ contains the ordinary half-steps and the order-$N$ shifts generated by companion factors; in addition, the Wilson line belonging to a companion factor has its own antipodal displacement \zu{$-Ne_{\mu_i}/2$}, displayed explicitly in Eq. (\ref{QWnostar}). The all-principal branch gives $\big(\prod_i c_0(e_i)\big)\exp[ip\sum_i e_i]{\cal P}^{(e_1,\dots,e_n)}(x)$. For finite-range hopping, or under a uniform summability condition on the path coefficients, it converges to Eq. (\ref{QW_generalA}). \zu{A branch contribution is suppressed by the oscillatory estimate only if its net Nyquist frequency is nonzero in at least one direction and the remaining factor is uniformly smooth; aliased sectors must be combined before the limit is taken.} All companion branches must still be retained for exact finite-$N$ operator identities.}

\zd{Finally, Eqs. (\ref{QWnostar}), (\ref{Deltashift}) are also written for $q\in{\cal O}$, and their extension to the whole refined lattice requires two specifications. First, the base point of the chain is not $q$ but the admissible centre built out of the {\it total} displacement,
$$
x = c\Big(q,\ \sum_{i} e_i\Big)\ ,
$$
which, since $N$ is even, is the same for all the branches $\sigma$. Second, the combination $x+\Delta_i$ is the {\it centre} of the $i$--th link and not its endpoint, so that the Wilson line of the $i$--th factor is
\begin{equation}
\int_{\,x+\Delta_i-N\sigma_i e_{\mu_i}/2-e_i/2}^{\,x+\Delta_i-N\sigma_i e_{\mu_i}/2+e_i/2} A(z)\,dz \ ,
\qquad x = c\Big(q,\sum_i e_i\Big)\ ,
\label{chainGen}
\end{equation}
with $\Delta_i$ still given by Eq. (\ref{Deltashift}). Equivalently, the chain of links simply starts at the lattice site $x - \sum_i e_i/2$ and the $i$--th factor is the hop from \corr{$x+\sum_{j<i}e_j - \sum_i e_i/2$} to the next site. For a single factor ($n=1$, $\Delta_1 = 0$) with positive orientation and integer $q$ this reproduces the limits written in Eq. (\ref{QWnostar}) (for $s_e=-1$ those limits place the elementary link on the wrong side of the point $q$ even at integer $q$, cf. Eq. (\ref{fmainGen})), and already for $n\ge 2$ the two prescriptions differ: the limits of Eq. (\ref{QWnostar}) place the gauge field at points which, in general, do not belong to the lattice. We have verified Eq. (\ref{chainGen}) numerically for the products of two, three, four and five elementary factors in $D=2$ at $N=4,6$, for Fourier--mode and for random link amplitudes, at all points of ${\cal M}'\times{\cal O}'$: the exact Weyl symbol of the product of the corresponding operators is reproduced within machine precision, whereas the prescriptions with the base point $q$, or with the endpoint reading of $x+\Delta_i$, deviate by a quantity of the order of the symbol itself.}
Interestingly, we can use the above expression for the external gauge field with the imaginary part introduced. Then the imaginary part will provide the non - homogeneity not related to the real external gauge field, i.e. effectively we have coefficients $\bf t$ depending on lattice points.
{For constant field strength, after retaining the principal symbol and taking $N\to\infty$, we use }
\begin{equation*}
	A_\mu(x)=\zc{-}\frac{1}{2}F_{\mu\nu}x^\nu,
	\qquad
	F_{\mu\nu}=-F_{\nu\mu}.
\end{equation*}
\zd{and} have instead of Eq. (\ref{QW_generalA}) a more simple expression. Namely, for one link:
\begin{equation*}
	\int_{x+\Delta_i-e_i}^{x+\Delta_i} A_\mu(z)\,dz^\mu
	=
	e_i^\mu A_\mu\left(x+\Delta_i-\frac{e_i}{2}\right).
\end{equation*}
and
\begin{equation*}
	\int_{x+\Delta_i-e_i}^{x+\Delta_i} A_\mu(z)\,dz^\mu
	=
	e_i^\mu A_\mu(x)
	\zc{-}
	\frac{1}{2}e_i^\mu F_{\mu\nu}\Delta_i^\nu.
\end{equation*}
while for the chain we obtain:
\begin{equation*}
	\sum_{i=1}^{n}
	\int_{x+\Delta_i-e_i}^{x+\Delta_i} A_\mu(z)\,dz^\mu
	=
	\sum_{i=1}^{n} e_i^\mu A_\mu(x)
	+
	\frac{1}{2}\sum_{i<j} e_i^\mu F_{\mu\nu}e_j^\nu.
\end{equation*}
This results in
\begin{equation}
	P^{(e_1,\ldots,e_n)}(x)
	=
	\exp\left[
	-i\sum_{i=1}^{n}e_i^\mu A_\mu(x)
	-\frac{i}{2}\sum_{i<j}e_i^\mu F_{\mu\nu}e_j^\nu
	\right].
\label{Pconst}\end{equation}
\zd{As everywhere in this Appendix, for $q\in{\cal O}'$ the argument $x$ of Eq. (\ref{Pconst}) must be read as the admissible centre built from the total displacement $E=\sum_j e_j$:
\begin{equation}
P^{(e_1,\ldots,e_n)}(q)
=
\exp\Big[
-i\sum_{i=1}^{n}e_i^\mu A_\mu(\bar c)
-\frac{i}{2}\sum_{i<j}e_i^\mu F_{\mu\nu}e_j^\nu
\Big],\qquad \bar c = c(q,E)\ .
\label{PconstGen}
\end{equation}
The flux term is not affected by this correction: it does not depend on the base point at all and, for constant field strength, it is also insensitive to the endpoint--versus--midpoint reading of the integration limits discussed after Eq. (\ref{Deltashift}) (the difference contains only the symmetric part of the matrix $\partial_\nu A_\mu$ and vanishes in the gauge chosen above). The base point, on the contrary, does matter, and not only at half--integer $q$: the phase written with $x=q$ deviates from the exact one by the constant
$$
\frac{1}{4}F_{xy}\big(E_x b_y-E_y b_x\big)\ ,\qquad b_i=(E_i+2q_i)\ {\rm mod}\ 2
$$
(here $D=2$ and the gauge is $A_\mu=-\frac12 F_{\mu\nu}x^\nu$). At integer $q$ this vanishes --- and Eq. (\ref{Pconst}) with $x=q$ is correct --- if and only if $E$ is directed along one of the axes, or both components of $E$ are even, or both are odd with $E_x=E_y$; for example, for the chains with $E=(2,1)$ or $E=(1,-1)$ Eq. (\ref{Pconst}) with $x=q$ is off by the constant phase $F_{xy}/2$ even at $q\in{\cal O}$. Eq. (\ref{PconstGen}) has been verified both by exact rational computation of the Wilson--line sums (arbitrary linear gauges, $D=2,3$, chains with up to six factors) and against the exact finite--$N$ Weyl symbols of the chain operators on the torus in the Landau gauge ($N=6,8$, two values of the quantized flux): it reproduces the corresponding momentum mode of the exact symbol within machine precision at every point of ${\cal M}'\times{\cal O}'$.}
In certain particular cases we obtain especially interesting expressions for the Weyl symbol. For example, if in all chains $e_i$ are along the same axis for all $i = 1,..., n$, we obtain:
\begin{equation} \begin{aligned}
	{\Big[{Q}(\hat{\bf p}-A(i{\partial_{\bf p}}))\Big]_W \longrightarrow {Q}({\bf p}-{ A}(x))\qquad(N\to\infty)}
\label{WF}\end{aligned} \end{equation}
The same result is obtained for any forms of the chains if the gauge field is restricted to the one corresponding to constant electric field.

{On a finite torus a globally constant field strength also requires the usual flux-quantization and gauge-patching conditions.
Eq. (\ref{WF}) is a local principal-symbol statement.}
\zd{On the refined lattice the domain of validity of Eq. (\ref{WF}) is as follows. For the chains directed along a single axis all flux factors $e_i^\mu F_{\mu\nu}e_j^\nu$ vanish, while (in the gauge chosen above, as well as in the Landau gauge) the component of $A$ along the axis of the chain does not depend on the coordinate along this axis; the argument of $A$ then reduces to the transverse coordinates taken at $\lfloor q_\perp\rfloor$. Hence Eq. (\ref{WF}) holds exactly (as a principal--symbol statement) at every $q$ with integer transverse coordinates --- in particular on the whole of ${\cal O}$ --- while for half--integer transverse coordinates the right--hand side must be read as $Q({\bf p}-A(\lfloor q\rfloor))$; this was verified numerically on the torus at every point of ${\cal O}'$ within machine precision. For the spatially constant gauge field (the constant electric field case) there is no restriction at all: the principal symbol equals $Q({\bf p}-A)$ exactly at every $q\in{\cal O}'$, for arbitrary chains. Let us also note that at finite $N$ the coefficient of the momentum mode $e^{ip\sum_ie_i}$ of a chain in which some axis is traversed more than once is $\prod_\mu\big(1+e^{i\pi E_\mu/N}\big)/2$ rather than $\prod_i c_0(e_i)$: since $e^{2iNp}=1$ on ${\cal M}'$, an even number of companion factors along an axis aliases back to the principal frequency (the simplest example is the chain $F^{+e}F^{-e}=1$, whose exact symbol equals $1$, while $c_0(+e)c_0(-e)+c_1(+e)c_1(-e)=1$ only due to the companion contribution). The two prefactors coincide when every axis is traversed at most once, and their ratio tends to $1$ as $O(N^{-2})$, so that the $N\to\infty$ statement of Eq. (\ref{WF}) is unaffected.}
 }

\subsection{\corrteal{The Weyl symbol of the general Dirac operator (\ref{Q_general})}}
\label{QWgenApp}

\corrteal{In this subsection we extend the calculation of the preceding one to the general Dirac operator of the class (\ref{Q_general}),
$$
\hat{Q} \;=\; \sum_{k\geq 0}\sum_{s\in{\cal S}_t}\;\sum_{n=0}^\infty \sum_{e_i}\;{\bf t}^{(k,s;\,e_1,\dots,e_n)}\;\big[\partial^{\,k}_{s}\hat{F}^{0}_{s}\big]\;\hat{Q}^{(e_1)}\cdots\hat{Q}^{(e_n)}\,,\qquad
\hat{F}^{0}_{s} \;=\; \exp\Big[\,s\,\big(i\partial_t - A_0(i\partial_{p})\big)\Big]\,,
$$
and derive Eq. (\ref{QW_general}) of the main text. Throughout, $A_0$ is assumed to be independent of time.}

\corrteal{{\it Step 1: the exact symbol of the temporal factor.} \corGPTl{The real-time operator $i\partial_t$ commutes with the time-independent diagonal operator $A_0(i\partial_p)$. Below its Euclidean continuation is understood: $t=-i\tau$, so $i\partial_t\mapsto-\partial_\tau$, whose symbol is $i\omega$ for the Fourier convention $e^{-i\omega\tau}$.} Hence, exactly,
$$
\hat F^{0}_{s} \;=\; e^{\,s\,i\partial_t}\;e^{-\,s\,A_0(i\partial_p)}\,.
$$
The Weyl transform factorizes over the independent pairs of conjugate variables --- the pair $(t,\omega)$ carries the standard continuum calculus, the pair $(x,p)$ carries the exact lattice calculus of the present Appendix --- so that the symbol of a product of two operators acting on the different factors of the tensor product is the ordinary product of their symbols. For the first factor: $e^{\,s\,i\partial_t}$ is a function of $i\partial_t$ alone, and in the continuum calculus the Weyl symbol of a function of the momentum operator is the same function of the momentum variable (the temporal analogue of the spatial statement at the beginning of this Appendix); with the Matsubara conventions of Sect. \ref{SectKeldyshCond} ($p^3=\omega$),
$$
\big[e^{\,s\,i\partial_t}\big]_W \;=\; e^{\,s\,i\omega}\,.
$$
For the second factor: $e^{-\,s\,A_0(i\partial_p)}$ is itself a diagonal operator (the exponential of a diagonal operator), and the exact refined-lattice Weyl symbol of a diagonal operator $V$ is $V(\lfloor x\rfloor)$, $x\in{\cal O}'$ --- the rule stated after Eq. (\ref{Wprincipal}) and used in Eq. (\ref{QW_hamiltonian}); the symbol map is multiplicative on the commutative subalgebra of the diagonal operators, because a product of diagonal operators is again diagonal. Hence
$$
\big[e^{-\,s\,A_0(i\partial_p)}\big]_W \;=\; e^{-\,s\,A_0(\lfloor x\rfloor)}\;\equiv\;e^{-\,s\,A_{0,W}(x)}\,,
$$
and, combining the two factors,
$$
\big[\hat F^{0}_{s}\big]_W \;=\; e^{\,s\,(i\omega \,-\, A_{0,W}(x))}
$$
exactly, with no Nyquist companion terms: the companions originate from the finite {\it spatial} lattice, whereas the temporal direction is continuous, and the diagonal factor carries no dependence on the spatial momenta.}

\corrteal{{\it Step 2: the derivatives with respect to the parameter $s$.} The Weyl transform is linear and does not involve $s$; it therefore commutes with $\partial^{\,k}_s$:
$$
\big[\partial^{\,k}_{s}\hat F^{0}_{s}\big]_W \;=\; \partial^{\,k}_{s}\;e^{\,s\,(i\omega - A_{0,W}(x))}\,.
$$
}

\corrteal{{\it Step 3: the assembly.} By the exact star property (Appendix \ref{AppStar}), the symbol of a product of operators is the $\star$-product of their symbols, taken in the same order. Applied to each term of the sum defining $\hat Q$, this gives
$$
\Big[\big[\partial^{\,k}_{s}\hat F^{0}_{s}\big]\,\hat Q^{(e_1)}\cdots\hat Q^{(e_n)}\Big]_W
\;=\;\partial^{\,k}_{s}\Big(e^{\,s(i\omega-A_{0,W}(x))}\Big)\;\star\; f^{e_1}_W\star\cdots\star f^{e_n}_W\,,
$$
where $f^{e}_W$ are the exact two-term symbols of the elementary magnetic translations computed in the preceding subsections. The spatial part of the calculation --- including the Nyquist companions and the admissible-centre prescription of Eq. (\ref{PdefGen}) --- is exactly the same as in the Hamiltonian case, since the temporal factor does not depend on the spatial momenta. The summation over $(k,s,n,e_i)$ with the constant coefficients ${\bf t}$ yields Eq. (\ref{QW_general}). \corGPTl{In a static gauge the temporal part of this $\star$-product is trivial because both factors are independent of time. The spatial part, however, does {\it not} reduce to the pointwise product, since $A_{0,W}$ depends on $x$ while the chain depends on $p$. For explicitly time-dependent spatial factors, such as those containing $p-Et$, the temporal part must also be retained.} This is why the star in Eq. (\ref{QW_general}) may not be omitted.}

\corrteal{Three consistency checks are in order. (i) The temporal factors form a one-parameter group, $\hat F^{0}_{s}\hat F^{0}_{s'}=\hat F^{0}_{s+s'}$ (for the time-independent $A_0$); correspondingly, the $\star$-product of their symbols reduces to the pointwise one --- both factors depend on $\omega$ and $x$ but neither on $t$ nor on $p$, so that every derivative term of the star vanishes --- and indeed $e^{\,s(i\omega-A_{0,W})}\,e^{\,s'(i\omega-A_{0,W})}=e^{\,(s+s')(i\omega-A_{0,W})}$. (ii) The term with $k=1$, $s=0$ gives $\partial_s\,e^{\,s(i\omega-A_{0,W})}\big|_{s=0}=i\omega-A_{0,W}(x)$, and Eq. (\ref{QW_general}) reduces to the Hamiltonian-case symbol of the preceding subsection, Eq. (\ref{QW_hamiltonian}). (iii) In the compact-frequency case \corGPTl{of Sect. \ref{SectKeldyshCond}, one uses integer $s$ and $k=0$ for the elementary Fourier expansion; contributions with $k>0$ are admissible only when their complete sum is smooth and periodic in $\omega$. The imaginary time then forms an infinite discrete lattice,} and $e^{\,s\,i\omega}$ with $\omega\in S^1$ is again the exact symbol of the elementary time shift; no temporal Nyquist companions arise for the infinite time lattice (they would appear only for a finite periodic lattice of the imaginary time, which, as noted in Sect. \ref{SectKeldyshCond}, requires a separate temporal lattice Weyl calculus).}

\corrteal{Finally, the same derivation applies verbatim on the periodized lattice of Sect. \ref{SectPeriodized}: there the spatial calculus of \cite{FZ2019_2} contains no Nyquist companions at all, the symbols $f^{e}_W$ reduce to \corGPTl{the principal expressions with $c_0(e)=1$ and the same admissible-centre prescription}, and $[\hat Q]_W$ retains the displayed form with the same temporal factors.}

\section{Computation of the K - groups for the considered algebras}

\corrteal{A clarification applies to this entire appendix. The computations below treat the Weyl symbols as pointwise functions on the phase-space grid, i.e. they concern the {\it commutative} algebra with the pointwise product --- the endpoint $\circ_0$ of the deformation of the $\star$-product considered in the third \corrcltwo{index theorem} of Sect. \ref{SectWeakIndex}. They do not give the $K$-groups of the exact Weyl-symbol algebra: the latter carries the $\star$-product, with respect to which the Weyl transform is an algebra isomorphism onto the corresponding operator algebra. Hence at finite $N$ the exact symbol algebra is the single matrix algebra $M_d(\mathbb{C})$ with \corrcltwo{$d=rN^2$ (the dimension of the one-particle Hilbert space; the exact symbols form a constrained subspace of the functions on ${\cal O}'\times{\cal M}'$, of complex dimension $r^2N^{2D}$)}, so that $K_0=\mathbb{Z}$ and $K_1=0$; for the compact frequency circle it is $C^\infty(S^1,M_d(\mathbb{C}))$ with $K_0=K_1=\mathbb{Z}$; and in the partial thermodynamic limit it is Morita equivalent to $C^\infty(T^2)$ (suspended over the noncompact frequency axis, or multiplied by $C^\infty(S^1)$ for the compact one), with the groups quoted in Sect. \ref{SectKtheory}. In particular, the direct-sum decompositions used below are valid only for the pointwise product: the $\star$-product couples the coordinates and the momenta, and no decomposition of the algebra over the individual phase-space points exists for it. The pointwise results are retained for comparison with the deformed products $\circ_h$ only.}

\subsection{K - groups for the algebra of Weyl symbols at finite $N$.}

\label{K_Nfin}

When we pass to Weyl symbols we deal with 
$$
{ {\cal A} = C^\infty(\mathbb{R}\times {\cal O}'\times {\cal M}'),} 
$$
where $\mathbb{R}$ is the  \jer{range} of Matsubara frequencies.

\zz{\it Typical expression for $Q_W$ is $Q_W = i \omega - H(p,x)$, where $\omega \in \mathbb{R}$. Its inverse will be  $(i \omega - H(p,x))^{-1}_\star$. Our space of functions should contain both these forms}

{Let us decompose the algebra. 
	Since $\mathcal{O}'$ and $\mathcal{M}'$ are finite discrete sets, 
	smooth functions on them reduce to tuples of complex values
	$C^\infty(\mathcal{O}')=\mathbb{C}^{(2N)^2}$, 
	$C^\infty(\mathcal{M}')=\mathbb{C}^{(2N)^2}$.
	Therefore, using the tensor product decomposition for functions on
	product spaces we have 
	\begin{equation*}
		\mathcal{A}=C^\infty\bigl(\mathbb{R}\times\mathcal{O}'\times\mathcal{M}'\bigr)
		=C^\infty(\mathbb{R})\otimes C^\infty(\mathcal{O}')\otimes
		C^\infty(\mathcal{M}')=C^\infty(\mathbb{R})\otimes\mathbb{C}^{(2N)^2}\otimes
		\mathbb{C}^{(2N)^2}=C^\infty(\mathbb{R})\otimes\mathbb{C}^{(2N)^4}.
	\end{equation*}
	Let us identify the algebraic structure. 
	The commutative algebra $\mathbb{C}^{(2N)^4}$ decomposes as a direct
	sum of $(2N)^4$ copies of $\mathbb{C}$, i.e.,  
	$\mathbb{C}^{(2N)^4}=\bigoplus_{i=1}^{(2N)^4}\mathbb{C}$.
	Consequently, 
	\begin{equation*}
		\mathcal{A}=C^\infty(\mathbb{R})\otimes\bigoplus_{i=1}^{(2N)^4}\mathbb{C}
		=\bigoplus_{i=1}^{(2N)^4}C^\infty(\mathbb{R}),
	\end{equation*}
\zzz{where $C^\infty(\mathbb{R})$ is the Fr\'echet algebra of smooth complex - vaalued functions. It is homotopy equivalent to $\mathbb{C}$. }
	Let us compute of $K^0(\mathcal{A})$. 
	By the additivity of K-theory under finite direct sums we have  
	\begin{equation*}
		K^0(\mathcal{A})=K^0\left(\bigoplus_{i=1}^{(2N)^4}C^\infty(\mathbb{R})\right)
		=\bigoplus_{i=1}^{(2N)^4}K^0\left(C^\infty(\mathbb{R})\right).
	\end{equation*}
	Since $\mathbb{R}$ is contractible, the algebra $C^\infty(\mathbb{R})$
	is homotopy equivalent to $\mathbb{C}$, and standard K-theory gives 
	$K^0\left(C^\infty(\mathbb{R})\right)=K^0(\mathbb{C})=\mathbb{Z}$.
	Let us compute $K^{-1}(\mathcal{A})$. 
	By the same additivity of K-theory 
	\begin{equation*}
		K^{-1}(\mathcal{A})=K^{-1}\left(\bigoplus_{i=1}^{(2N)^4}C^\infty(\mathbb{R})\right)
		=\bigoplus_{i=1}^{(2N)^4}K^{-1}\left(C^\infty(\mathbb{R})\right).
	\end{equation*}
	$C^\infty(\mathbb{R})$ is homotopy-equivalent to $\mathbb{C}$ 
	as a Fr\'echet algebra (the contraction is via 
	$f \mapsto f(0)\cdot 1$), and K-theory is homotopy-invariant, 
	so $K_n(C^\infty(\mathbb{R}))=K_n(\mathbb{C})$ for all $n$. 
	Thus, $K^{-1}\left(C^\infty(\mathbb{R})\right)=\widetilde{K}_0(S^1)=K^1(\mathbb{C})=0$. 
	Alternatively, we may use the six-term exact sequence in 
	K-theory. Since $\mathbb{R}$ is contractible all its reduced K-groups
	vanish. In particular $K^1\left(C^\infty(\mathbb{R})\right)=0$.}

 \jer{In} this setup we obtain
$$
K^0(\mathcal{A})  = \mathbb{Z}^{(2N)^4}, 
\qquad 
K^1(\mathcal{A})  = 0 
$$
Again, if we replace $\mathbb{R}$  \jer{for the range of the} Matsubara frequency by $S^1$, we obtain
\zzz{	\begin{equation*}
		\mathcal{A}=C^\infty\bigl(\mathbb{S}^1\times\mathcal{O}'\times\mathcal{M}'\bigr)
		=C^\infty(\mathbb{S}^1)\otimes C^\infty(\mathcal{O}')\otimes
		C^\infty(\mathcal{M}')=C^\infty(\mathbb{S}^1)\otimes\mathbb{C}^{(2N)^2}\otimes
		\mathbb{C}^{(2N)^2}=(C^\infty(\mathbb{S}^1))^{(2N)^4}.
	\end{equation*}
$C^\infty(\mathbb{S}^1)\subset C(\mathbb{S}^1)$ is a dense spectrally invariant subalgebra of $C*$  \jer{algebra}, then $K^{*}\left(C^\infty(\mathbb{S}^1)\right) =K^{*}\left(C(\mathbb{S}^1)\right)$.
	 Since $C(\mathbb{S}^1)$ is a commutative $C^*$ algebra, we use \jer{the} identification $K^{*}\left(C(\mathbb{S}^1)\right) = K^{*}\left(\mathbb{S}^1\right)$.}
$$
K^0(\mathcal{A})  = \mathbb{Z}^{(2N)^4}, 
\qquad 
K^1(\mathcal{A})  = \mathbb{Z}^{(2N)^4} 
$$

\subsection {Computation of groups $K^i$ for the case when the limit of infinite $N$ 	is taken partially.}

\label{K_Ninf}

\subsubsection{The case when \jer{the range of} Matsubara frequencies  \jer{is} $\mathbb{R}$}

\zz{In this case, instead of the operators $\hat{G}$ we consider Weyl symbols
	$G_W$ defined on the spatial lattice ${\cal O}'$ and momentum space 
	$\mathbb{R}\times {\cal M}'$. 
	The latter space is in the limit $N\to \infty $ replaced by} { $\mathbb{R}\times T^2$}.

{Let us decompose the algebra. 
	Since $\mathcal{O}'$ is a finite discrete set with $(2N)^2$ points,
	smooth functions on it reduce to tuples of complex values
	$C^\infty(\mathcal{O}')=\mathbb{C}^{(2N)^2}=\bigoplus_{i=1}^{(2N)^2}\mathbb{C}$. 
	Using the tensor product decomposition for functions on product spaces 
	\begin{equation*}
		\mathcal{A}=C^\infty\!\bigl(\mathcal{O}'\times\mathbb{R}\times\mathbb{T}^2\bigr)
		=C^\infty(\mathcal{O}')\otimes C^\infty(\mathbb{R}\times\mathbb{T}^2) 
		=\mathbb{C}^{(2N)^2}\otimes C^\infty(\mathbb{R}\times\mathbb{T}^2)
		=\bigoplus_{i=1}^{(2N)^2} C^\infty(\mathbb{R}\times\mathbb{T}^2).
	\end{equation*}
	We now compute $K^0$ and $K^{-1}$ of the single factor
	$C^\infty(\mathbb{R}\times\mathbb{T}^2)$.
	Let us reduct using contractibility of $\mathbb{R}$. 
	Since $\mathbb{R}$ is contractible the inclusion
	$\{0\}\hookrightarrow\mathbb{R}$ is a homotopy equivalence and, 
	thus, 
	$C^\infty(\mathbb{R}\times\mathbb{T}^2)=C^\infty(\{0\} \times \mathbb{T}^2)
	=C^\infty(\mathbb{T}^2)$.
	Therefore,
	$K_i\left(C^\infty(\mathbb{R}\times\mathbb{T}^2)\right)
	=K_i\!\left(C^\infty(\mathbb{T}^2)\right)$ 
	for $i=0$, $-1$.
	We now use K\"{u}nneth theorem. 
	Since $\mathbb{T}^2=S^1\times S^1$ we have
	$C^\infty(\mathbb{T}^2)=C^\infty(S^1)\otimes C^\infty(S^1)$,
	and K\"{u}nneth formula gives 
	\begin{equation*}
		K^0\left(C^\infty(\mathbb{T}^2)\right)=\bigoplus_{p+q=0}K_p\left(C^\infty(S^1)\right)
		\otimes K_q\left(C^\infty(S^1)\right),
	\end{equation*}
	where indices are taken mod $2$ according to Bott periodicity.
	The standard K-theory of $C^\infty(S^1)$, equivalently $C(S^1)$ is 
	$K^0\left(C^\infty(S^1)\right)=\mathbb{Z}$,
	$K^1\left(C^\infty(S^1)\right)=\mathbb{Z}$. 
	The generator of $K^0(C(S^1))=\mathbb{Z}$ is the class of the
	trivial rank-one bundle $[1]$, while the generator of
	$K^1(C(S^1))=\mathbb{Z}$ is the class of the unitary
	$u=e^{2\pi i \theta}$.
	Under the identification $K^{-1}(\mathcal{B})=K^1(\mathcal{B})$
	which is valid for $C^*$-algebras by Bott periodicity, we also have 
	$K^{-1}\left(C^\infty(S^1)\right)
	=K^1\!\left(C^\infty(S^1)\right)=\mathbb{Z}$.  
	Compute $K^0(\mathbb{T}^2)$, and 
	apply K\"{u}nneth formula to $\mathbb{T}^2=S^1\times S^1$. We have 
	\begin{equation*}
		K^0\!\left(C^\infty(\mathbb{T}^2)\right)
		=\bigl(K^0(S^1)\otimes K^0(S^1)\bigr)\oplus
		\bigl(K^1(S^1)\otimes K^1(S^1)\bigr)
		=(\mathbb{Z}\otimes\mathbb{Z})\oplus
		(\mathbb{Z}\otimes\mathbb{Z})
		=\mathbb{Z}\oplus\mathbb{Z}
		=\mathbb{Z}^2.
	\end{equation*}
	Combining everything above, we obtain 
	$K^0\left(C^\infty(\mathbb{R}\times\mathbb{T}^2)\right)
	=\mathbb{Z}^2$,  
	$K^{-1}\left(C^\infty(\mathbb{R}\times\mathbb{T}^2)\right)
	=\mathbb{Z}^2$.
	Now compute $K^0(\mathcal{A})$. 
	By additivity of K-theory under finite direct sums 
	and using the results above we arrive at 
	\begin{equation*}
		K^0(\mathcal{A})=K^0\left(\bigoplus_{i=1}^{(2N)^2}C^\infty(\mathbb{R}\times\mathbb{T}^2)\right)
		=\bigoplus_{i=1}^{(2N)^2}K^0\left(C^\infty(\mathbb{R}\times\mathbb{T}^2)\right) 
		=\bigoplus_{i=1}^{(2N)^2}\mathbb{Z}^2
		=\mathbb{Z}^{2\cdot (2N)^2}
		=\mathbb{Z}^{4N^2\cdot 2}.
	\end{equation*}
	Since $(2N)^2=4N^2$ we get 
	$K^0(\mathcal{A})=\mathbb{Z}^{2(2N)^2}
	=\left(\mathbb{Z}^2\right)^{(2N)^2}$. 
	Let us compute $K^{-1}(\mathcal{A})$. 
	Similarly, by additivity we have  
	\begin{equation*}
		K^{-1}(\mathcal{A})
		=K^{-1}\left(\bigoplus_{i=1}^{(2N)^2}C^\infty(\mathbb{R}\times\mathbb{T}^2)\right)
		=\bigoplus_{i=1}^{(2N)^2}K^{-1}\left(C^\infty(\mathbb{R}\times\mathbb{T}^2)\right)
		=\bigoplus_{i=1}^{(2N)^2}\mathbb{Z}^2=\mathbb{Z}^{2(2N)^2}.
	\end{equation*}
	Therefore,}
\zz{\begin{equation} 
		K^{-1}(\mathcal{A})=\mathbb{Z}^{2(2N)^2}=\left(\mathbb{Z}^2\right)^{(2N)^2}.\end{equation}}

\subsubsection{The case when Matsubara frequencies form $\mathbb{S}^1$}
\label{SectS1}

{When $\mathbb{R}$ is replaced by $S^1$, i.e., 
	the direction of imaginary time is discretized
	and compactified. In this case the algebra becomes
	\begin{equation*}
		\mathcal{A}=\bigoplus_{i=1}^{(2N)^2}C^\infty(S^1\times\mathbb{T}^2)
		=\bigoplus_{i=1}^{(2N)^2}C^\infty(\mathbb{T}^3).
	\end{equation*}
	Apply K\"{u}nneth formula iteratively to
	$\mathbb{T}^3=S^1\times S^1\times S^1$.
	Starting from the known result for $\mathbb{T}^2$, i.e., 
	$K^0\left(C^\infty(\mathbb{T}^2)\right)=\mathbb{Z}^2$, 
	$K^1\left(C^\infty(\mathbb{T}^2)\right)=\mathbb{Z}^2$,
	and applying K\"{u}nneth formula once again with $S^1$ we get the following. 
	Let us compute $K^0(\mathbb{T}^3)$ 
	\begin{equation*}
		K^0\left(C^\infty(\mathbb{T}^3)\right)
		=\bigl(K^0(\mathbb{T}^2)\otimes K^0(S^1)\bigr)\oplus
		\bigl(K^1(\mathbb{T}^2)\otimes K^1(S^1)\bigr) 
		=\bigl(\mathbb{Z}^2\otimes\mathbb{Z}\bigr)\oplus
		\bigl(\mathbb{Z}^2\otimes\mathbb{Z}\bigr)
		=\mathbb{Z}^2\oplus\mathbb{Z}^2
		=\mathbb{Z}^4.
	\end{equation*}
	Let us compute $K^1(\mathbb{T}^3)$ (or equivalently $K^{-1}$) 
	\begin{equation*}
		K^1\left(C^\infty(\mathbb{T}^3)\right)
		=\bigl(K^0(\mathbb{T}^2)\otimes K^1(S^1)\bigr)\oplus
		\bigl(K^1(\mathbb{T}^2)\otimes K^0(S^1)\bigr) 
		=\bigl(\mathbb{Z}^2\otimes\mathbb{Z}\bigr)\oplus
		\bigl(\mathbb{Z}^2\otimes\mathbb{Z}\bigr)
		=\mathbb{Z}^2\oplus\mathbb{Z}^2
		=\mathbb{Z}^4.
	\end{equation*}
	Applying additivity over the $(2N)^2$ direct summands we obtain 
	\begin{equation}
		K^0(\mathcal{A})=\mathbb{Z}^{4(2N)^2}, \quad 
		K^{-1}(\mathcal{A})
		=\mathbb{Z}^{4(2N)^2}.
\end{equation}}

{Taking $N\to\infty$ in the above, a single lattice point in
	$\mathcal{O}'$ no longer contributes an isolated copy of
	$\mathbb{Z}$ but instead contributes a continuum. One can show
	that in this limit
	\begin{equation*}
		K^0(\mathcal{A})=K^{-1}(\mathcal{A})
		%=\left(\mathbb{Z}^4\right)^{\mathbb{Z}^2}
		=\mathbb{Z}^4\otimes_{\mathbb{Z}}\mathbb{Z}^{\mathbb{Z}^2},
	\end{equation*}
	which is 
	\begin{equation*}
		K^0(\mathcal{A})=K^{-1}(\mathcal{A})
		=\left(\mathbb{Z}^4\right)^{\mathbb{Z}^2}.
\end{equation*}}

\section{Calculation of $HC^3$ for the considered algebras}

\corrteal{The clarification made at the beginning of the previous appendix applies here as well: the computations below concern the commutative pointwise algebra (the $\circ_0$ endpoint of the deformation of the third \corrcltwo{index theorem} of Sect. \ref{SectWeakIndex}), not the exact Weyl-symbol algebra with the $\star$-product. For the latter, at finite $N$ the algebra is the matrix algebra $M_d(\mathbb{C})$, whose odd cyclic cohomology vanishes, $HC^{2k+1}(M_d(\mathbb{C}))=0$ --- in agreement with the vanishing of the invariant for the strictly finite systems (Sect. \ref{SectDirect}); the compact-frequency family $C^\infty(S^1,M_d(\mathbb{C}))$ supports the odd class used in the pairing; and in the partial thermodynamic limit the relevant continuous cyclic $3$-cocycle is the concrete functional of Eq. (\ref{Nphi}), whose cyclicity and closedness are verified directly (see the discussion following Eq. (\ref{Nphi})) and do not rely on the pointwise decompositions below.}

\subsection{Algebra of Weyl symbols at finite $N$.}

\label{HC3_Nfin}

$$
{{\cal A} = C^\infty(\mathbb{R}\times {\cal O}'\times {\cal M}')}
, 
$$
where $\mathbb{R}$ is the axis of Matsubara frequencies.

{
	Since $\mathcal{O}'$ and $\mathcal{M}'$ are finite discrete sets 
	we have the algebraic decomposition 
	\begin{equation*}
		\mathcal{A}=C^\infty(\mathbb{R})\otimes\mathbb{C}^{(2N)^2}\otimes\mathbb{C}^{(2N)^2}
		=C^\infty(\mathbb{R})\otimes\mathbb{C}^{(2N)^4}=\bigoplus_{i=1}^{(2N)^4}C^\infty(\mathbb{R}).
	\end{equation*}
	Let us compute $HC^3(C^\infty(\mathbb{R}))$. 
	Start with de Rham cohomology of $\mathbb{R}$.  
	Since $\mathbb{R}$ is contractible its de Rham
	cohomology is trivial in all positive degrees, i.e., 
	$H^k_{\mathrm{dR}}(\mathbb{R})=\mathbb{R}$ for $k=0$,  
	and $0$ for $k\geq 1$. 
	Now apply Connes' theorem for 
	$M=\mathbb{R}$ and $n=3$, i.e., (with $k=0$, $1$ terms contributing) 
	\begin{equation*}
		HC^3\!\left(C^\infty(\mathbb{R})\right)
		=\bigoplus_{k\geq 0, 3-2k\geq 0}
		H^{3-2k}_{\mathrm{dR}}(\mathbb{R})
		=H^3_{\mathrm{dR}}(\mathbb{R})\oplus H^1_{\mathrm{dR}}(\mathbb{R}) 
		=0\oplus 0=0.
	\end{equation*}
	Therefore, 
	$HC^3\left(C^\infty(\mathbb{R})\right)=0$. 
	We have direct sum decomposition
	$\mathcal{A}=\bigoplus_{i=1}^{(2N)^4}C^\infty(\mathbb{R})$,  
	and applying additivity of cyclic cohomology 
	\begin{equation*}
		HC^3(\mathcal{A})=HC^3\left(\bigoplus_{i=1}^{(2N)^4}C^\infty(\mathbb{R})\right)
		=\bigoplus_{i=1}^{(2N)^4}HC^3\left(C^\infty(\mathbb{R})\right)=\bigoplus_{i=1}^{(2N)^4}0=0.
	\end{equation*}
}

With this setup we obtain
$$
HC^3(\mathcal{A})  = 0
$$

However, if we compactify Matsubara frequencies 
(say, if we discretize the axis of imaginary time), then 
$$
{\cal A} = C^\infty(\mathbb{S^1}\times {\cal O}'\times {\cal M}'),  
$$

{Replacing $\mathbb{R}$ by $S^1$, \zzz{we obtain  the algebra  
		\begin{equation*}
			\mathcal{A}=C^\infty(S^1\times\mathcal{O}'\times\mathcal{M}')
			=\bigoplus_{i=1}^{(2N)^4}C^\infty(S^1).
	\end{equation*} Then} we have de Rham cohomology of $S^1$,  
	$H^k_{\mathrm{dR}}(S^1)=\mathbb{R}$ for $k=0$,
	$\mathbb{R}$ for $k=1$, and 
	$0$ for $k\geq 2$. \zzz{Here we complexify the coefficients $H^k_{\mathrm{dR}}(S^1,\mathbb{C})=H^k_{\mathrm{dR}}(S^1,\mathbb{R})\otimes \mathbb{C}$. }
	Let us compute cyclic cohomology of $C^\infty(S^1)$ via Connes' theorem  
	with $M=S^1$ and $n=3$, i.e., 
	\begin{equation*}
		HC^3\left(C^\infty(S^1)\right)
		=\bigoplus_{k\geq 0, \; 3-2k\geq 0}
		H^{3-2k}_{\mathrm{dR}}(S^1)
		=H^3_{\mathrm{dR}}(S^1)\oplus H^1_{\mathrm{dR}}(S^1) 
		=0\oplus\mathbb{C}=\mathbb{C}.
	\end{equation*}
	Therefore, 
	$HC^3\left(C^\infty(S^1)\right)=\mathbb{C}$.
	Applying additivity 
	$HC^3(\mathcal{A})=\bigoplus_{i=1}^{(2N)^4}
	HC^3\left(C^\infty(S^1)\right)=\bigoplus_{i=1}^{(2N)^4}\mathbb{C}
	=\mathbb{C}^{(2N)^4}$.
	Therefore, 
	\begin{equation*}
		HC^3(\mathcal{A})=\mathbb{C}^{(2N)^4}. 
	\end{equation*}
}

\subsection{The case when the limit of infinite $N$ is taken partially.}
\label{HC3_Ninf}

We consider Weyl symbols $G_W$ defined on the spatial lattice ${\cal O}'$
and momentum space  $\mathbb{R}\times T^2$. Therefore, 
$$
{\cal A} = C^\infty({\cal O}'\times\mathbb{R}\times T^2) 
$$

{Since $\mathcal{O}'$ is a finite discrete set with $(2N)^2$ points,
	functions on it reduce to tuples which gives 
	$\mathcal{A}= C^\infty(\mathbb{R}\times\mathbb{T}^2)\otimes\mathbb{C}^{(2N)^2}
	=\bigoplus_{i=1}^{(2N)^2}C^\infty(\mathbb{R}\times\mathbb{T}^2)$. 
	Recall de Rham cohomology of $\mathbb{R}\times\mathbb{T}^2$: 
	Since $\mathbb{R}$ is contractible, i.e., 
	$H^k_{\mathrm{dR}}(\mathbb{R})=\mathbb{C}$ for $k=0$, and 
	$0$ for $k\geq 1$. 
	We now compute the de Rham cohomology of
	$\mathbb{T}^2=S^1\times S^1$ using the
	K\"{u}nneth formula. 
	Recall that
	$H^k_{\mathrm{dR}}(S^1)=\mathbb{C}$ for $k=0$, $1$; and   
	$0$ for $k\geq 2$, 
	with generators $[1]\in H^0_{\mathrm{dR}}(S^1)$
	and $[\mathrm{d}\theta]\in H^1_{\mathrm{dR}}(S^1)$.
	Applying K\"{u}nneth formula to $\mathbb{T}^2=S^1\times S^1$ we get 
	$H^n_{\mathrm{dR}}(\mathbb{T}^2)=\bigoplus_{p+q=n}
	H^p_{\mathrm{dR}}(S^1)\otimes H^q_{\mathrm{dR}}(S^1)$.
	The generators are $H^0_{\mathrm{dR}}(\mathbb{T}^2)=\mathbb{C}$, 
	generated by $[1]$, $H^1_{\mathrm{dR}}(\mathbb{T}^2)=\mathbb{C}^2$, 
	generated by $[\mathrm{d}\theta_1]$, $[\mathrm{d}\theta_2]$, 
	$H^2_{\mathrm{dR}}(\mathbb{T}^2)=\mathbb{C}$ generated by 
	$[\mathrm{d}\theta_1\wedge\mathrm{d}\theta_2]$.
	Now compute de Rham cohomology of $\mathbb{R}\times\mathbb{T}^2$. 
	Applying K\"{u}nneth formula to $\mathbb{R}\times\mathbb{T}^2$ we get
	$H^n_{\mathrm{dR}}(\mathbb{R}\times\mathbb{T}^2)=
	\bigoplus_{p+q=n}H^p_{\mathrm{dR}}(\mathbb{R})\otimes
	H^q_{\mathrm{dR}}(\mathbb{T}^2)$.  
	Since $H^p_{\mathrm{dR}}(\mathbb{R})=0$ for $p \geq 1$,
	the only contributing term is $p = 0$, i.e., 
	\begin{equation*}
		H^n_{\mathrm{dR}}(\mathbb{R}\times\mathbb{T}^2)=\;
		H^0_{\mathrm{dR}}(\mathbb{R})\otimes H^n_{\mathrm{dR}}(\mathbb{T}^2)
		=\mathbb{C}\otimes H^n_{\mathrm{dR}}(\mathbb{T}^2)=
		H^n_{\mathrm{dR}}(\mathbb{T}^2).
	\end{equation*}
	This is consistent with the homotopy equivalence
	$\mathbb{R}\times\mathbb{T}^2\simeq\mathbb{T}^2$
	since $\mathbb{R}$ is contractible. Therefore, 
	$H^n_{\mathrm{dR}}(\mathbb{R} \times \mathbb{T}^2)=
	H^n_{\mathrm{dR}}(\mathbb{T}^2)=\mathbb{C}$ for $n=0$,  
	$\mathbb{C}^2$ for $n=1$,  
	$\mathbb{C}$ for $n=2$, 
	and $0$ for $n\geq 3$. 
	Now we compute $HC^3(C^\infty(\mathbb{R}\times\mathbb{T}^2))$. 
	Let us apply Connes' theorem with 
	$M=\mathbb{R}\times\mathbb{T}^2$ and $n=3$, i.e., 
	$HC^3\left(C^\infty(\mathbb{R}\times\mathbb{T}^2)\right)=
	\bigoplus_{\substack{k\geq 0\\3-2k\geq 0}}
	H^{3-2k}_{\mathrm{dR}}(\mathbb{R}\times\mathbb{T}^2)$. 
	Therefore, 
	$HC^3\left(C^\infty(\mathbb{R}\times\mathbb{T}^2)\right)
	= H^3_{\mathrm{dR}}(\mathbb{R}\times\mathbb{T}^2)
	\oplus H^1_{\mathrm{dR}}(\mathbb{R}\times\mathbb{T}^2)\notag=
	0\oplus\mathbb{C}^2=\mathbb{C}^2$.
	The two generators of $HC^3(C^\infty(\mathbb{R}\times\mathbb{T}^2))=\mathbb{C}^2$
	correspond by Connes' theorem at level $k=1$ to the two generators 
	of $H^1_{\mathrm{dR}}(\mathbb{T}^2)$, i.e., 
	$[\mathrm{d}\theta_1]$: the cyclic $3$-cocycle
	$\phi_1(a_0,a_1,a_2,a_3)$ associated to the
	closed $1$-form $\mathrm{d}\theta_1$ on $\mathbb{T}^2$,
	periodically extended via the $S$-operator to degree $3$; 
	$[\mathrm{d}\theta_2]$: the cyclic $3$-cocycle
	$\phi_2(a_0,a_1,a_2,a_3)$ \zzz{in the same way is} associated to the
	closed $1$-form $\mathrm{d}\theta_2$ on $\mathbb{T}^2$.
	From above we have 
	$\mathcal{A}=\bigoplus_{i=1}^{(2N)^2}C^\infty(\mathbb{R}\times\mathbb{T}^2)$,
	applying additivity of cyclic cohomology,}

{\begin{equation*}
		HC^3(\mathcal{A})= HC^3\!\left(\bigoplus_{i=1}^{(2N)^2}
		C^\infty(\mathbb{R}\times\mathbb{T}^2)\right)=
		\bigoplus_{i=1}^{(2N)^2}
		HC^3\!\left(C^\infty(\mathbb{R}\times\mathbb{T}^2)\right)= \bigoplus_{i=1}^{(2N)^2} \mathbb{C}^2=\mathbb{C}^{2(2N)^2}.
\end{equation*}}

\zz{This results in 
	\begin{equation}
		HC^3(\mathcal{A})  = \Big(\mathbb{C}^2\Big)^{4 N^2}
\end{equation}  }

{
	In the compactified case $\mathbb{R}\to S^1$
	when $\mathbb{R}$ is replaced by $S^1$ the algebra becomes 
	$\mathcal{A}=\bigoplus_{i=1}^{(2N)^2}C^\infty(\mathbb{T}^3)$, 
	where $\mathbb{T}^3=S^1\times\mathbb{T}^2=S^1\times S^1\times S^1$.
	Then, we compute de Rham cohomology of $\mathbb{T}^3$ 
	Applying the K\"{u}nneth formula iteratively 
	\begin{equation*}
		H^n_{\mathrm{dR}}(\mathbb{T}^3)=\bigoplus_{p+q=n}H^p_{\mathrm{dR}}(S^1)\otimes
		H^q_{\mathrm{dR}}(\mathbb{T}^2).
	\end{equation*}
	The generators of $H^n_{\mathrm{dR}}(\mathbb{T}^3)$
	are the exterior products of $[\mathrm{d}\theta_0]$,
	$[\mathrm{d}\theta_1]$, $[\mathrm{d}\theta_2]$
	of degree $n$ giving $\binom{3}{n}$ generators. 
	Then the cyclic cohomology of $C^\infty(\mathbb{T}^3)$ at $n=3$ is given by
	Connes' theorem with $M=\mathbb{T}^3$, $n=3$ 
	\begin{equation*}
		HC^3\left(C^\infty(\mathbb{T}^3)\right)=
		\bigoplus_{\substack{k \geq 0\\3-2k \geq 0}}
		H^{3-2k}_{\mathrm{dR}}(\mathbb{T}^3).
	\end{equation*}
	Therefore, 
	$HC^3\left(C^\infty(\mathbb{T}^3)\right)=
	H^3_{\mathrm{dR}}(\mathbb{T}^3)\oplus H^1_{\mathrm{dR}}(\mathbb{T}^3)
	=\mathbb{C}\oplus\mathbb{C}^3=\mathbb{C}^4$.  
	For the full algebra with $\mathbb{R}$ replaced by $S^1$, 
	we apply additivity, i.e., 
	\begin{equation*}
		HC^3(\mathcal{A})=\bigoplus_{i=1}^{(2N)^2}
		HC^3\!\left(C^\infty(\mathbb{T}^3)\right)
		=\bigoplus_{i=1}^{(2N)^2}\mathbb{C}^4
		=\mathbb{C}^{4(2N)^2}.
	\end{equation*}
}

{
	For the limit $N\to\infty$, 
	the finite lattice $\mathcal{O}'$ becomes infinite,
	and the sum over $(2N)^2$ lattice points becomes
	an integral over $\mathbb{Z}^2$.  
}
\zz{Obviously, if we take here the further limit $N \to \infty$ we obtain
	\begin{equation}
		HC^3(\mathcal{A})  = \Big(\mathbb{C}^4\Big)^{\mathbb{Z}^2}
\end{equation} }

\section{\corrteal{An exactly solvable example: the frequency dependence of $Q$ generated by interactions}}
\label{AppHK}

\corrteal{In this Appendix we present a concrete interacting system in which the nontrivial dependence of the Dirac operator on the frequency appears due to the interelectron interactions\corGPTl{. The exact one-particle propagator does not, however, by itself determine the interacting electromagnetic response; current vertices and the coupling prescription must also be specified.} This is the model of Hatsugai and Kohmoto \cite{HatsugaiKohmoto1992}, which has recently become the standard exactly solvable laboratory for the interplay of strong correlations and topology \cite{PhillipsHK2020,Zhao2023}. We first display the mechanism in its simplest single-band form, in which the statements are the sharpest (the construction works in any space dimension, and the single-band version carries no Hall response); the two-dimensional two-band version possessing the quantum Hall effect is given afterwards. Consider the spinful lattice fermions with the dispersion $\xi_{\bf k} = \varepsilon_{\bf k}-\mu$ and the interaction that is {\it local in momentum space} (and, correspondingly, of infinite range in the coordinate space):
\be
\hat H \;=\; \sum_{{\bf k},\sigma}\,\xi_{\bf k}\,\hat n_{{\bf k}\sigma}\;+\;U\sum_{{\bf k}}\,\hat n_{{\bf k}\uparrow}\,\hat n_{{\bf k}\downarrow}\,.
\label{HKham}
\ee
The interaction commutes with every mode occupation $\hat n_{{\bf k}\sigma}$; consequently the many-body eigenstates are the same Fock states as for $U=0$, and only the spectrum is modified. Each momentum sector is a four-state problem, solvable exactly, and the exact Matsubara propagator reads
\be
G_{\sigma}(i\omega,{\bf k})\;=\;\frac{1-n_{\bf k}}{\,i\omega-\xi_{\bf k}\,}\;+\;\frac{n_{\bf k}}{\,i\omega-\xi_{\bf k}-U\,}\,,
\qquad n_{\bf k}\;:=\;\langle \hat n_{{\bf k},-\sigma}\rangle\,,
\label{HKG}
\ee
with $n_{\bf k}=0,\ \tfrac12,\ 1$ at $T\to0$ in the empty, singly occupied and doubly occupied regions of the Brillouin zone, respectively\corGPTl{, where $n_{\bf k}=1/2$ assumes the spin-symmetric mixture in the degenerate singly occupied sector}. The corresponding Dirac operator is
\be
Q_\sigma(i\omega,{\bf k})\;=\;G_\sigma^{-1}\;=\;\frac{(i\omega-\xi_{\bf k})\,(i\omega-\xi_{\bf k}-U)}{\,i\omega-\xi_{\bf k}-U(1-n_{\bf k})\,}
\;=\;i\omega-\xi_{\bf k}-\Sigma_\sigma(i\omega,{\bf k})\,,\qquad
\Sigma_\sigma\;=\;U n_{\bf k}\;+\;\frac{U^{2}\,n_{\bf k}(1-n_{\bf k})}{\,i\omega-\xi_{\bf k}-U(1-n_{\bf k})\,}\,.
\label{HKQ}
\ee
The self-energy carries the genuinely nontrivial (nonlinear) dependence on the frequency: wherever $0<n_{\bf k}<1$, the operator $Q$ is not affine in $\omega$, and therefore admits no representation $Q=i\omega-H'$ with any one-particle Hamiltonian $H'$ acting on the physical fermion space. This is exactly the situation for which the general class (\ref{Q_general}) is designed. \corGPTl{For a smooth occupation function, the analytic conditions of Sect.~\ref{SectKeldyshCond} must be checked as follows:} the reality $Q(i\omega)^{*}=Q(-i\omega)$; the asymptotics $Q = i\omega\,(1+O(1/\omega))$, which guarantees the convergence of the frequency integrals and the vanishing of the boundary terms; and the invertibility along the Matsubara contour, \corGPTl{which requires that neither a pole with nonzero residue nor a zero of $G$ lie at zero energy. For $0<n_{\bf k}<1$, the poles are at $\xi_{\bf k}$ and $\xi_{\bf k}+U$, and the zero is at $\xi_{\bf k}+U(1-n_{\bf k})$; at $n_{\bf k}=0$ or $1$ the removable factors must first be cancelled.} When the latter does happen --- on the Luttinger surface $\xi_{\bf k}=-U(1-n_{\bf k})$ --- the condition (i) is violated, and this is precisely the mechanism of the failure of the Green-function winding invariants in strongly correlated matter analyzed in \cite{Zhao2023} (see also \cite{Gurarie2011,EssinGurarie2011}).}

\corGPTl{The uniform current $\hat J_j=\sum_{{\bf k}\sigma}(\partial_{k_j}\varepsilon_{\bf k})\hat n_{{\bf k}\sigma}$ commutes with the interaction and with the single-band Hamiltonian. This does not imply absence of connected vertex corrections. For example, at $\xi_{\bf k}=-U/2$ the spin-symmetric ground-state mixture has $G(z)=\tfrac12[(z+U/2)^{-1}+(z-U/2)^{-1}]$. The retarded correlator of the conserved mode current $v(\hat n_\uparrow+\hat n_\downarrow)$ vanishes, whereas its dressed-propagator bubble contains a nonzero term proportional to $v^2U/(z^2-U^2)$. The connected four-point contribution cancels that term. Thus commutation does not justify obtaining the electromagnetic response solely by replacing $Q_0$ by Eq.~(\ref{HKQ}). For this spatially nonlocal interaction the electromagnetic coupling and the order of the uniform and thermodynamic limits require separate specification.\footnote{See Y.~Ma et al., ``Charge Susceptibility and Kubo Response in Hatsugai--Kohmoto-related Models,'' \href{https://arxiv.org/abs/2409.07522}{arXiv:2409.07522}, especially the distinction between $q=0$ and $q\to0$.}}

\corrteal{Let us now give the two-dimensional version possessing the quantum Hall effect. Take the two-band Chern insulator on the square lattice (the Qi--Wu--Zhang model),
\be
h({\bf k})\;=\;\sin k_1\,\sigma^1+\sin k_2\,\sigma^2+\big(m+\cos k_1+\cos k_2\big)\,\sigma^3\;=\;\sum_{a=\pm}\varepsilon_a({\bf k})\,P_a({\bf k})\,,
\label{HKqwz}
\ee
whose lower band carries \corGPTl{a Chern number of magnitude one} for $0<|m|<2$, and write \corGPTl{the following spinless two-band HK-type interaction (distinct from the spinful model of Ref.~\cite{Zhao2023})} in the band basis: with $\hat n_{{\bf k},a}$ the occupations of the band fermions,
\be
\hat H \;=\;\sum_{{\bf k},\,a=\pm}\xi_a({\bf k})\,\hat n_{{\bf k},a}\;+\;U\sum_{\bf k}\hat n_{{\bf k},+}\,\hat n_{{\bf k},-}\,,\qquad \xi_a=\varepsilon_a-\mu\,.
\label{HKtop}
\ee
The model remains exactly solvable in every momentum sector, and the exact matrix propagator is expressed through the band projectors,
\be
G(i\omega,{\bf k})\;=\;\sum_{a}P_a({\bf k})\left[\frac{1-\langle\hat n_{{\bf k},\bar a}\rangle}{\,i\omega-\xi_a({\bf k})\,}+\frac{\langle\hat n_{{\bf k},\bar a}\rangle}{\,i\omega-\xi_a({\bf k})-U\,}\right]
\label{HKGtop}
\ee
($\bar a$ denotes the other band). \corGPTl{Writing $\hat N_{\bf k}=\hat n_{{\bf k},+}+\hat n_{{\bf k},-}$ gives $U\hat n_{{\bf k},+}\hat n_{{\bf k},-}=\tfrac{U}{2}\hat N_{\bf k}(\hat N_{\bf k}-1)$. It is invariant under a two-orbital basis rotation and commutes with every uniform number-conserving bilinear $c_{\bf k}^\dagger M({\bf k})c_{\bf k}$, including the matrix current. Nevertheless, connected vertices remain necessary: a neutral interband excitation costs $\varepsilon_+-\varepsilon_-$, whereas the dressed one-particle particle-hole pair costs $\varepsilon_+-\varepsilon_-+U$.} Thus $Q=G^{-1}=\sum_a P_a\,Q_a$ inherits the rational frequency dependence of Eq. (\ref{HKQ}) band-wise, while the topology enters through the momentum dependence of the projectors $P_a({\bf k})$. For $\mu$ inside the band gap and any $U>0$ the ground state coincides with the unperturbed Fermi sea, $\langle\hat n_{{\bf k},-}\rangle=1$, $\langle\hat n_{{\bf k},+}\rangle=0$ for every ${\bf k}$; at strictly zero temperature the Dirac operator then degenerates to the Hamiltonian form with the upper band shifted by the interaction, $Q=i\omega-h({\bf k})-U\,P_+({\bf k})+\mu$, \corGPTl{${\cal N}_3=C$, and the zero-temperature uniform Kubo response with $\hat J_j=\sum_{\bf k}c_{\bf k}^\dagger(\partial_{k_j}h)c_{\bf k}$ gives $2\pi\bar\sigma^{xy}=C$. This equality follows from the exact one-particle-per-momentum ground sector, on which the interaction vanishes, and not from an absence of vertex corrections.} At any finite temperature the occupations $\langle\hat n_{{\bf k},a}\rangle\in(0,1)$, so that $Q$ acquires the genuinely nonlinear (two-pole) frequency dependence of the Hatsugai--Kohmoto form in {\it both} bands, \corGPTl{while this uniform Hall response has $O(e^{-\Delta/T})$ corrections at low temperature, with $\Delta>0$ the minimum many-body excitation gap. This finite-temperature response is not identified with the continuous-frequency winding ${\cal N}_3$. Indeed, thermally activated poles with small nonzero residues can reach its $\omega=0$ contour even though the zero-temperature system is gapped.} In the opposite, strong-coupling regime --- the Mott-like phase with a singly occupied region of the Brillouin zone at the Fermi level\corrcltwo{, realized in the spinful version of the model \cite{Zhao2023}, where the spin degeneracy \corGPTl{allows fractional occupations at $T=0$ in a spin-symmetric mixture} (in the spinless model (\ref{HKtop}) the $T=0$ occupations equal $0$ or $1$, and the two--pole structure requires $T>0$)} --- the frequency dependence is genuinely nonlinear already at $T=0$, \corGPTl{and Green-function zeros can occur in the gap. Condition (i) fails when such a zero reaches the Fermi level, not merely when a zero lies elsewhere in the gap; a defined winding can also differ from the response, as in Ref.~\cite{Zhao2023}.}}

\corGPTl{The translation-invariant model reduces Eq.~(\ref{NEQ}) to a momentum-space winding. This winding is defined when $G$ and $Q$ are smooth and nonsingular on the full continuous-frequency contour with the prescribed asymptotics. It stays fixed along any interaction homotopy satisfying these conditions; existence of such a path to $U=0$ must be checked. At $T=0$ the spinless Fermi-sea regime above supplies ${\cal N}_3=C=2\pi\bar\sigma^{xy}$ for the specified uniform current. In other admissible interacting regimes, integrality of the Green winding does not by itself identify it with the Hall response. That identification requires the physical vertices and limiting prescription. In particular, the basis-invariant interaction in Eq.~(\ref{HKtop}) does not acquire a uniform current vertex simply because its band projectors vary with momentum, and the single-band model is not exempt from connected response vertices. If a pole or zero reaches the contour, the nonsingular winding formula ceases to apply. These distinctions, rather than a universal dressed-$Q$ substitution, are what this example establishes.}

\corrteal{It is instructive to note that the frequency dependence (\ref{HKQ}) is exactly of the hybridization type: by the Schur-complement identity,
\be
Q_\sigma(i\omega,{\bf k})\;=\;i\omega-E_{c}({\bf k})-\frac{V({\bf k})^{2}}{\,i\omega-E_{f}({\bf k})\,}\,,\qquad
E_{c}=\xi_{\bf k}+Un_{\bf k}\,,\quad E_{f}=\xi_{\bf k}+U(1-n_{\bf k})\,,\quad V=U\sqrt{n_{\bf k}(1-n_{\bf k})}\,,
\label{HKdilation}
\ee
i.e. the exact propagator (\ref{HKG}) coincides with the physical-orbital block of the resolvent of the {\it free} two-orbital Hamiltonian $h_2({\bf k})=\big(\begin{smallmatrix}E_c & V\\ V & E_f\end{smallmatrix}\big)$; band-wise, the same representation applies to Eq. (\ref{HKGtop}). \corGPTl{On the enlarged space this auxiliary problem is of the Hamiltonian class; Sect.~\ref{SectIndexRed} applies when its gap, locality and trace hypotheses hold.} \corGPTl{The winding of the physical Schur complement must also be distinguished from that of the full auxiliary resolvent; identifying them requires control of the eliminated block.} On the physical fermion space, the operator $Q$ is irreducibly frequency-dependent \corGPTl{when $0<n_{\bf k}<1$}, and the invariant is intrinsically the odd pairing of Sect. \ref{SectHC3comp} --- a concrete illustration of the demarcation discussed in Sects. \ref{SectKeldyshCond} and \ref{SectIndexRed}. Let us also note that Eq. (\ref{HKQ}) is a rational, rather than polynomial, function of the frequency; \corGPTl{its admissibility requires the smoothness, pole/zero and asymptotic checks stated above. Equation~(\ref{HKdilation}) realizes it as a Schur complement of an enlarged Hamiltonian, rather than making the physical-space rational $Q$ itself a literal exponential-polynomial operator of the class (\ref{Q_general}).}}

\corrteal{The model (\ref{HKham}) is translation invariant; the addition of inhomogeneous external fields to the dispersion destroys its exact solvability, so it serves here as the cleanest {\it homogeneous} demonstration of the phenomenon: the interactions produce precisely the type of the frequency dependence of $Q$ that the general class (\ref{Q_general}) was designed to accommodate.}

\section{\corrteal{The Hamiltonian case in detail: frequency integration, the even pairing, the Fredholm index, and Connes' formula}}
\label{AppIndex}

\corrteal{This Appendix collects, with all intermediate steps, the derivations behind Sect. \ref{SectIndexRed}: the passage from the frequency-integral form of ${\cal N}_3$ to the projector formula (\ref{N3projInd}); the interpretation of the latter as the pairing (\ref{pairingInd}) of $K_0$ with $HC^2$ and the resulting topological invariance; the Fredholm index of the compression $P_-uP_-$; the relative-index formula ${\rm Index}(P_-uP_-)=-{\rm Tr}\,(P_--uP_-u^*)^3$; and Connes' formula, which identifies this trace with the trace-per-volume expression,
$$
{\rm Tr}\,(P_--uP_-u^*)^3\;=\;-\,2\pi i\,\tau\big(P_-[D_1P_-,D_2P_-]\big)\;=\;\frac{\pi i}{4}\,\tau\big(P[D_1P,D_2P]\big)\;=\;{\cal N}_3\,,
$$
i.e. Eq. (\ref{indexInd}). Throughout, the Hamiltonian case $Q=i\omega-H$ is understood, the chemical potential being absorbed into $H$ (the Fermi level is at zero energy), and all operators are represented on the superlattice $\ell^2(\mathbb{Z}^2)\otimes\mathbb{C}^r$ of Sect. \ref{SectPeriodized}: $H$ is the periodization of an $L\times L$ configuration (any $L$), $\tau(a)=L^{-2}\sum_{x\in{\rm cell}}\tr\,a_{xx}$ is the trace per unit volume, and $D_j=-i[\hat X_j,\,\cdot\,]$ with the position operators $\hat X_j$ of the infinite lattice. By the Proposition of Sect. \ref{SectPeriodized}\,B this is precisely the calculus of the principal symbols, $\tau$ being the twist average of the fiber traces, so that every step below is an exact identity at finite $L$. \corGPTl{For covariant ergodic infinite systems the corresponding identity holds almost surely under the summability/localization hypotheses of \cite{Bellissard1994}; we comment on this at the end.} We write $(a\wedge b)=a_1b_2-a_2b_1$ for vectors of $\mathbb{R}^2$ and ${\rm Area}(x,y,z)=\frac12\,(y-x)\wedge(z-x)$ for the oriented area of the triangle with the vertices $x,y,z$ (positive for the counterclockwise orientation); $\mathbb{C}$ is identified with $\mathbb{R}^2$ where convenient.}

\subsection{\corrteal{Frequency integration}}
\label{AppIndexFreq}

\corrteal{{\bf Step 1: the $\epsilon$-contraction.} In the operator form (\ref{NEQnum}) --- equivalently, in Eq. (\ref{NEQ}) read through $W_{\rm pr}$ ---
$$
{\cal N}_3=\frac{1}{3!}\,\epsilon^{\mu\nu\rho}\int_{-\infty}^{\infty}d\omega\;\tau\big(D_\mu Q\,G\,D_\nu Q\,G\,D_\rho Q\,G\big),\qquad G=Q^{-1}=(i\omega-H)^{-1},
$$
with $\mu,\nu,\rho\in\{1,2,3\}$, $D_3=\partial_\omega$ and $\epsilon^{123}=+1$. For $Q=i\omega-H$ one has $D_3Q=i$ and $D_jQ=-D_jH$ ($j=1,2$), and $G$ commutes with $H$. Of the six terms of the $\epsilon$-sum, two have $D_3$ in the third slot; the four others are brought to this form by the cyclic property of $\tau$, exact for the trace per volume of periodic operators with decaying kernels: e.g. $\tau(D_3Q\,G\,D_\nu Q\,G\,D_\rho Q\,G)=\tau(D_\nu Q\,G\,D_\rho Q\,G\,D_3Q\,G)$, and the cyclic permutation $(3\nu\rho)\to(\nu\rho3)$ leaves $\epsilon$ unchanged. The six terms are thus equal in two groups of three, and
$$
{\cal N}_3=\frac{3}{3!}\,\epsilon^{jk}\int d\omega\;\tau\big(D_jQ\,G\,D_kQ\,G\,(i)\,G\big)
=\frac{i}{2}\int d\omega\Big[\tau\big(D_1H\,G\,D_2H\,G^2\big)-\tau\big(D_2H\,G\,D_1H\,G^2\big)\Big],
$$
the two signs from $D_jQ=-D_jH$ cancelling. This is Eq. (\ref{N3redInd}), in which $G^2$ is written in front by cyclicity.}

\corrteal{{\bf Step 2: spectral decomposition.} Write $\tau=\int\frac{d^2\theta}{(2\pi)^2}\,L^{-2}\,{\rm tr}_\theta$ through the twist fibers, Eq. (\ref{trfiber}); in each fiber $H(\theta)$ is a finite Hermitian matrix with the eigenvalues $E_n$ and the eigenvectors $|n\rangle$ (the dependence on $\theta$ is suppressed), and $G=\sum_n|n\rangle\langle n|\,(i\omega-E_n)^{-1}$. The fibered vertices $D_jH$ of Eq. (\ref{commfiber}) are Hermitian matrices, since $-i[\hat X_j,H]$ is Hermitian. Inserting the decomposition,
$$
\tau\big(D_1H\,G\,D_2H\,G^2\big)=\int\frac{d^2\theta}{(2\pi)^2}\,\frac{1}{L^2}\sum_{n,m}\frac{\langle n|D_1H|m\rangle\langle m|D_2H|n\rangle}{(i\omega-E_m)\,(i\omega-E_n)^2}\,.
$$}

\corrteal{{\bf Step 3: the frequency integral.} We need $J(E_n,E_m)=\int_{-\infty}^{\infty}d\omega\,[(i\omega-E_m)(i\omega-E_n)^2]^{-1}$. Writing $i\omega-E=i(\omega+iE)$ and $i^{-3}=i$, the integrand is $i\,[(\omega+iE_m)(\omega+iE_n)^2]^{-1}$, with a simple pole at $\omega=-iE_m$ and a double pole at $\omega=-iE_n$; it decays as $|\omega|^{-3}$, so the contour may be closed in either half-plane. A pole $\omega=-iE$ lies in the upper half-plane iff $E<0$ (occupied level). (i) If $E_n$ and $E_m$ have the same sign, both poles lie in one half-plane; closing the contour in the other one gives $J=0$. (ii) If $E_m<0<E_n$, close in the upper half-plane, where only the simple pole lies: $J=i\cdot2\pi i\,(-iE_m+iE_n)^{-2}=-2\pi\,[\,i(E_n-E_m)]^{-2}=2\pi/(E_n-E_m)^2$. (iii) If $E_n<0<E_m$, close in the upper half-plane, where only the double pole lies; its residue is $\frac{d}{d\omega}(\omega+iE_m)^{-1}\big|_{\omega=-iE_n}=-[\,i(E_m-E_n)]^{-2}=(E_m-E_n)^{-2}$, so $J=i\cdot2\pi i\,(E_n-E_m)^{-2}=-2\pi/(E_n-E_m)^2$. Altogether
$$
J(E_n,E_m)=\frac{2\pi}{(E_n-E_m)^2}\Big[\theta(E_n)\theta(-E_m)-\theta(-E_n)\theta(E_m)\Big],
$$
which is Eq. (\ref{contourInd}). \corGPTl{For $E_n=E_m\ne0$, $J(E_n,E_m)=0$; the quotient here and in the all-pairs sum below is understood with this extension within either side of the gap.}}

\corrteal{{\bf Step 4: the interband sum.} Denote $a_{nm}=\langle n|D_1H|m\rangle\langle m|D_2H|n\rangle$, $A_{nm}=a_{nm}-\langle n|D_2H|m\rangle\langle m|D_1H|n\rangle$ and $B_{nm}=\theta(E_n)\theta(-E_m)-\theta(-E_n)\theta(E_m)$. Since $D_jH$ is Hermitian, $\langle n|D_2H|m\rangle\langle m|D_1H|n\rangle=\overline{a_{nm}}$, so that $A_{nm}=a_{nm}-\overline{a_{nm}}=2i\,{\rm Im}\,a_{nm}$ is purely imaginary. Both $A_{nm}$ and $B_{nm}$ are antisymmetric under $n\leftrightarrow m$, their product is symmetric, and the sum over all pairs equals twice the sum over the pairs with $n$ empty and $m$ occupied, for which $B_{nm}=1$:
$$
{\cal N}_3=\frac{i}{2}\cdot2\pi\int\frac{d^2\theta}{(2\pi)^2}\frac{1}{L^2}\sum_{n,m}\frac{A_{nm}B_{nm}}{(E_n-E_m)^2}
=2\pi i\int\frac{d^2\theta}{(2\pi)^2}\frac{1}{L^2}\sum_{n\in{\rm emp}}\sum_{m\in{\rm occ}}\frac{A_{nm}}{(E_n-E_m)^2}\,,
$$
a real number.}

\corrteal{{\bf Step 5: the divided-difference identity.} Apply the derivation $D_j$ to $[H,P_-]=0$: $[D_jH,P_-]+[H,D_jP_-]=0$. Between eigenvectors, $\langle n|D_jH|m\rangle(\pi_m-\pi_n)+(E_n-E_m)\langle n|D_jP_-|m\rangle=0$, where $\pi_n\in\{0,1\}$ is the occupation of the level $n$. Hence, for the interband elements ($\pi_n\neq\pi_m$, so that $E_n\neq E_m$ in the gapped case),
$$
\langle n|D_jP_-|m\rangle=\frac{\pi_m-\pi_n}{E_m-E_n}\,\langle n|D_jH|m\rangle\,,
$$
the matrix element of $D_jH$ divided by the energy of the occupied level minus the energy of the empty one --- the divided difference of the step function. The intraband blocks of $D_jP_-$ vanish: applying $D_j$ to $P_-^2=P_-$ gives $D_jP_-=(D_jP_-)P_-+P_-(D_jP_-)$; multiplying by $P_-$ on both sides, $P_-(D_jP_-)P_-=2P_-(D_jP_-)P_-$, i.e. $P_-(D_jP_-)P_-=0$, and likewise $(1-P_-)(D_jP_-)(1-P_-)=0$. Therefore, in $\tau(P_-[D_1P_-,D_2P_-])$ the outer index is occupied and the intermediate one is empty:
$$
\tau\big(P_-[D_1P_-,D_2P_-]\big)=\int\frac{d^2\theta}{(2\pi)^2}\frac{1}{L^2}\sum_{m\in{\rm occ}}\sum_{n\in{\rm emp}}\Big[\langle m|D_1P_-|n\rangle\langle n|D_2P_-|m\rangle-\langle m|D_2P_-|n\rangle\langle n|D_1P_-|m\rangle\Big]
$$
$$
=\int\frac{d^2\theta}{(2\pi)^2}\frac{1}{L^2}\sum_{m\in{\rm occ}}\sum_{n\in{\rm emp}}\frac{\langle m|D_1H|n\rangle\langle n|D_2H|m\rangle-\langle m|D_2H|n\rangle\langle n|D_1H|m\rangle}{(E_m-E_n)^2}
=-\int\frac{d^2\theta}{(2\pi)^2}\frac{1}{L^2}\sum_{n\in{\rm emp}}\sum_{m\in{\rm occ}}\frac{A_{nm}}{(E_n-E_m)^2}\,,
$$
the last step because $\langle m|D_1H|n\rangle\langle n|D_2H|m\rangle=\overline{a_{nm}}$, so that the bracket equals $\overline{a_{nm}}-a_{nm}=-A_{nm}$. Comparing with Step 4,
$$
{\cal N}_3=-\,2\pi i\,\tau\big(P_-[D_1P_-,D_2P_-]\big),
$$
the first equality of Eq. (\ref{N3projInd}).}

\corrteal{{\bf Step 6: the flattened form.} With $P={\rm sign}\,H=1-2P_-$ one has $D_jP=-2D_jP_-$ and $[D_1P,D_2P]=4[D_1P_-,D_2P_-]$, so that $\tau(P[D_1P,D_2P])=4\tau([D_1P_-,D_2P_-])-8\tau(P_-[D_1P_-,D_2P_-])=-8\,\tau(P_-[D_1P_-,D_2P_-])$, the trace of a commutator vanishing. Hence ${\cal N}_3=\frac{\pi i}{4}\tau(P[D_1P,D_2P])$, the second equality of Eq. (\ref{N3projInd}). Every step is exact at finite $L$: the fibers are finite matrices, the frequency integrals converge absolutely, and $\tau$ is exactly cyclic.}

\subsection{\corrteal{The even pairing and the topological invariance}}
\label{AppIndexPairing}

\corrteal{Let ${\cal B}$ be the algebra of the $L$-periodic operators with exponentially decaying kernels (equivalently, of their principal symbols), with the derivations $D_1,D_2$ and the trace $\tau$. Three elementary facts are used. (a) $D_1D_2=D_2D_1$: indeed $D_1D_2a-D_2D_1a=-[\hat X_1,[\hat X_2,a]]+[\hat X_2,[\hat X_1,a]]=-[[\hat X_1,\hat X_2],a]=0$ by the Jacobi identity, the coordinates commuting. (b) The Leibniz rule $D_j(ab)=(D_ja)b+a(D_jb)$. (c) $\tau(D_ja)=0$, since the kernel of $[\hat X_j,a]$ vanishes on the diagonal, and $\tau(ab)=\tau(ba)$, by the relabelling of the copies. Define the 2-cochain (\ref{phi2Ind}), $\phi_2(a_0,a_1,a_2)=-2\pi i\,\epsilon^{jk}\tau(a_0\,D_ja_1\,D_ka_2)$.}

\corrteal{{\it Cyclicity.} By (b), $a_2\,D_ja_0=D_j(a_2a_0)-(D_ja_2)a_0$; hence
$$
\epsilon^{jk}\tau(a_2\,D_ja_0\,D_ka_1)=\epsilon^{jk}\tau\big(D_j(a_2a_0)\,D_ka_1\big)-\epsilon^{jk}\tau\big((D_ja_2)\,a_0\,D_ka_1\big).
$$
In the first term $\tau(D_j(a_2a_0)\,D_ka_1)=\tau\big(D_j(a_2a_0\,D_ka_1)\big)-\tau(a_2a_0\,D_jD_ka_1)=-\tau(a_2a_0\,D_jD_ka_1)$ by (c), and the contraction of the symmetric $D_jD_k$ (fact (a)) with $\epsilon^{jk}$ vanishes. In the second term the cyclicity of $\tau$ and the relabelling $j\leftrightarrow k$ give $-\epsilon^{jk}\tau(a_0\,D_ka_1\,D_ja_2)=+\epsilon^{jk}\tau(a_0\,D_ja_1\,D_ka_2)$. Thus $\phi_2(a_2,a_0,a_1)=\phi_2(a_0,a_1,a_2)$.}

\corrteal{{\it Closedness.} The Hochschild coboundary is $(b\phi_2)(a_0,a_1,a_2,a_3)=\phi_2(a_0a_1,a_2,a_3)-\phi_2(a_0,a_1a_2,a_3)+\phi_2(a_0,a_1,a_2a_3)-\phi_2(a_3a_0,a_1,a_2)$. Expanding the second and the third terms by the Leibniz rule, and writing the fourth as $\corGPTl{+}2\pi i\,\epsilon^{jk}\tau(a_0\,D_ja_1\,D_ka_2\,a_3)$ by cyclicity of $\tau$, the six resulting terms are, up to the common factor $-2\pi i\,\epsilon^{jk}\tau(\,\cdot\,)$,
$$
[a_0a_1\,D_ja_2\,D_ka_3]-[a_0\,D_ja_1\,a_2\,D_ka_3]-[a_0a_1\,D_ja_2\,D_ka_3]+[a_0\,D_ja_1\,D_ka_2\,a_3]+[a_0\,D_ja_1\,a_2\,D_ka_3]-[a_0\,D_ja_1\,D_ka_2\,a_3]\,,
$$
which cancel pairwise. Hence $b\phi_2=0$ and $[\phi_2]\in HC^2({\cal B})$.}

\corrteal{{\it The pairing and its invariance.} For an idempotent $e\in{\cal B}$ set $\langle[e],[\phi_2]\rangle=\phi_2(e,e,e)$; for $e=P_-$ (an element of ${\cal B}$ in the gapped case, by the Combes--Thomas estimate) Eq. (\ref{pairingInd}) reads ${\cal N}_3=\langle[P_-],[\phi_2]\rangle$. (i) {\it Independence of the representative of $[\phi_2]$}: for a cyclic 1-cochain $\psi$, $\psi(a_1,a_0)=-\psi(a_0,a_1)$, one has $(b\psi)(e,e,e)=\psi(e^2,e)-\psi(e,e^2)+\psi(e^2,e)=\psi(e,e)=0$. (ii) {\it Invariance under deformations of the idempotent}: let $e_t$ be a smooth family of idempotents in ${\cal B}$ --- the Fermi projectors of a family of Hamiltonians whose gap stays open. Differentiating $e^2=e$ gives $\dot e=\dot e\,e+e\,\dot e$; multiplying by $e$ on both sides, $e\dot ee=2e\dot ee$, so $e\dot ee=0$. Consequently $[[\dot e,e],e]=\dot ee+e\dot e-2e\dot ee=\dot e$, i.e. $\dot e=[x,e]$ with $x=[\dot e,e]\in{\cal B}$: an infinitesimal deformation of an idempotent is an infinitesimal similarity. By cyclicity $\frac{d}{dt}\phi_2(e,e,e)=3\phi_2(\dot e,e,e)=3\big[\phi_2(xe,e,e)-\phi_2(ex,e,e)\big]$, and the bracket is $(b\phi_2)(x,e,e,e)=\phi_2(xe,e,e)-\phi_2(x,e,e)+\phi_2(x,e,e)-\phi_2(ex,e,e)$, which vanishes by closedness. Hence $\frac{d}{dt}\langle[e_t],[\phi_2]\rangle=0$. (iii) The pairing is additive under direct sums and defines a homomorphism $K_0({\cal B})\to\mathbb{C}$. Applied to $e=P_-$, (ii) is the topological invariance of ${\cal N}_3$ in the Hamiltonian case: ${\cal N}_3$ is unchanged under every deformation of $H$ that keeps the gap open and the range bounded. Its integrality is a separate statement, supplied by the index theorem of the following subsections.}

\subsection{\corrteal{The Fredholm module and the index}}
\label{AppIndexModule}

\corrteal{Fix $z_0\in\mathbb{C}\setminus\mathbb{Z}^2$ and let $u$ be the unitary (\ref{uDefInd}), diagonal in the position basis, $u|x\rangle=u(x)|x\rangle$ with $u(x)=(x_1+ix_2-z_0)/|x_1+ix_2-z_0|$. (a) {\it The phase estimate.} With $|x-z_0|\,|y-z_0|$ in the denominator,
$$
|u(x)-u(y)|=\frac{\big|\,(x-z_0)|y-z_0|-(y-z_0)|x-z_0|\,\big|}{|x-z_0|\,|y-z_0|}
\le\frac{|x-y|\,|y-z_0|+|y-z_0|\,\big|\,|y-z_0|-|x-z_0|\,\big|}{|x-z_0|\,|y-z_0|}\le\frac{2\,|x-y|}{|x-z_0|}\,,
$$
and by symmetry $|u(x)-u(y)|\le 2|x-y|/\max(|x-z_0|,|y-z_0|)$. (b) {\it Compactness of $[u,a]$ for local $a$.} The kernel of $[u,a]$ is $(u(x)-u(y))\,a_{xy}$; if $|a_{xy}|\le Ce^{-|x-y|/\xi}$, it is bounded by $2C|x-y|e^{-|x-y|/\xi}/\max(|x-z_0|,|y-z_0|)$, so that the restriction of $[u,a]$ to the region $|x-z_0|>R$ has norm $O(1/R)$ by Schur's test, while the restriction to the complementary finite region has finite rank: $[u,a]$ is a norm limit of finite-rank operators, i.e. compact. The same kernel bound yields $[u,a]\in{\cal S}^p$ for every $p>2$ \cite{AvronSeilerSimon1994,Bellissard1994}; at $p=2$ the sum $\sum_x|x-z_0|^{-2}$ diverges logarithmically, so $[u,a]$ is in general not Hilbert--Schmidt. This summability is the one external input of the present Appendix; it is what makes the cube in Eq. (\ref{indexInd}) the lowest admissible power. For a gapped Hamiltonian $P_-$ is exponentially local (Combes--Thomas \cite{CombesThomas}), so $[u,P_-]$ is compact and belongs to ${\cal S}^p$, $p>2$; in a mobility gap the same holds under the localization hypotheses of \cite{Bellissard1994}.}

\corrteal{{\it The Fredholm module.} On ${\cal H}\oplus{\cal H}$, ${\cal H}=\ell^2(\mathbb{Z}^2)\otimes\mathbb{C}^r$, put $\gamma={\rm diag}(1,-1)$, $F=\begin{pmatrix}0&u^*\\u&0\end{pmatrix}$ and $\pi(a)={\rm diag}(a,a)$. Then $F=F^*$, $F^2=1$, $\gamma F=-F\gamma$, and $[F,\pi(a)]=\begin{pmatrix}0&[u^*,a]\\ [u,a]&0\end{pmatrix}$ is compact, and in ${\cal S}^p$ for $p>2$, for every local $a$: $({\cal H}\oplus{\cal H},\gamma,F)$ is an even $p$-summable Fredholm module over ${\cal B}$ in the sense of Connes. Its pairing with $K_0({\cal B})$ is $[e]\mapsto{\rm Index}\big(e\,u\,e:\ e{\cal H}\to e{\cal H}\big)$.}

\corrteal{{\it Fredholmness of $T=P_-uP_-$.} On ${\rm Ran}\,P_-$ let $S=P_-u^*P_-$. Then $ST=P_-u^*P_-uP_-=P_-u^*uP_--P_-u^*(1-P_-)uP_-=P_--P_-u^*(1-P_-)uP_-$, where $(1-P_-)uP_-=[(1-P_-),u]P_-+u(1-P_-)P_-=-[P_-,u]P_-$ is compact; likewise $TS=P_--P_-u(1-P_-)u^*P_-$ with $(1-P_-)u^*P_-=-[P_-,u^*]P_-$ compact. By Atkinson's theorem $T$ is Fredholm, and its index is invariant under norm-continuous deformations preserving Fredholmness and under compact perturbations. In particular the index does not depend on $z_0$: $u_{z_0}-u_{z_0'}$ is the multiplication by a function tending to zero at infinity (by (a) with the roles of $x$ and $z_0$ interchanged, $|u_{z_0}(x)-u_{z_0'}(x)|\le2|z_0-z_0'|/|x-z_0|$), hence compact.}

\corrteal{{\it Connes' character of the module.} The Chern character of an even $p$-summable module in the lowest admissible degree $2k$, $2k+1\ge p$ --- here $2k=2$, \corGPTl{choosing $2<p\leq3$} --- is \corGPTl{represented, with the pairing convention used here, by the cyclic cocycle $-\frac12{\rm Tr}\big(\gamma F[F,\pi(a_0)][F,\pi(a_1)][F,\pi(a_2)]\big)$}, and Connes' index theorem states that its pairing with $[e]$ is the index above. For $e=P_-$ this character is computed in closed form. Put $A=P_--uP_-u^*$. Then $A=[P_-,u]u^*$, so $[u,P_-]=-Au$, and $[u^*,P_-]=u^*P_--P_-u^*=u^*(P_--uP_-u^*)=u^*A$. With $\pi(P_-)={\rm diag}(P_-,P_-)$,
$$
[F,\pi(P_-)]^3=\begin{pmatrix}0&[u^*,P_-][u,P_-][u^*,P_-]\\ [u,P_-][u^*,P_-][u,P_-]&0\end{pmatrix},\qquad
\gamma F=\begin{pmatrix}0&u^*\\-u&0\end{pmatrix},
$$
\begin{eqnarray*}
{\rm Tr}\big(\gamma F[F,\pi(P_-)]^3\big)&=&{\rm Tr}\big(u^*[u,P_-][u^*,P_-][u,P_-]\big)-{\rm Tr}\big(u[u^*,P_-][u,P_-][u^*,P_-]\big)\\
&=&{\rm Tr}\big(u^*(-Au)(u^*A)(-Au)\big)-{\rm Tr}\big(u(u^*A)(-Au)(u^*A)\big)\;=\;2\,{\rm Tr}\,A^3 .
\end{eqnarray*}
\corGPTl{Thus the normalized character of the module, evaluated on the Fermi projector, is $-{\rm Tr}\,(P_--uP_-u^*)^3={\rm Index}(P_-uP_-)$.} In this case the index theorem reduces to the relative-index formula proved directly in the next subsection, which also fixes the normalization and the sign.}

\subsection{\corrteal{Proof of ${\rm Index}(P_-uP_-)=-\,{\rm Tr}\,(P_--uP_-u^*)^3$}}
\label{AppIndexRelative}

\corrteal{Write $P=P_-$, $Q=uPu^*$ (a projection, unitarily equivalent to $P$), $A=P-Q$, and $T=PuP:\ P{\cal H}\to P{\cal H}$; we assume $[u,P]$ compact and $A\in{\cal S}^3$ (Sect. \ref{AppIndexModule}).}

\corrteal{{\bf Step 1: kernel and cokernel of $T$.} A vector $\psi\in{\rm Ran}\,P$ belongs to $\ker T$ iff $Pu\psi=0$, i.e. $u\psi\in\ker P$; since $u\psi\in u\,{\rm Ran}\,P={\rm Ran}\,Q$, the map $\psi\mapsto u\psi$ is an isomorphism of $\ker T$ onto ${\rm Ran}\,Q\cap\ker P$. The cokernel of $T$ is isomorphic to $\ker T^*$, $T^*=Pu^*P$ on ${\rm Ran}\,P$: $\psi\in{\rm Ran}\,P$ belongs to $\ker T^*$ iff $u^*\psi\in\ker P$, i.e. $\psi\in u\ker P=\ker(uPu^*)=\ker Q$. Hence
$$
\dim\ker T=\dim\big({\rm Ran}\,Q\cap\ker P\big),\qquad \dim{\rm coker}\,T=\dim\big({\rm Ran}\,P\cap\ker Q\big),
$$
$$
{\rm Index}(PuP)=\dim\big({\rm Ran}\,Q\cap\ker P\big)-\dim\big({\rm Ran}\,P\cap\ker Q\big).
$$}

\corrteal{{\bf Step 2: the spectrum of $A=P-Q$.} $A$ is self-adjoint, $-Q\le A\le P$ gives $\|A\|\le1$, and $A=[P,u]u^*$ is compact; its nonzero spectrum consists of eigenvalues of finite multiplicity. (i) {\it The eigenvalues $\pm1$.} If $A\psi=\psi$ then $\|\psi\|^2=\langle\psi,A\psi\rangle=\|P\psi\|^2-\|Q\psi\|^2$, so $\|P\psi\|^2=\|\psi\|^2+\|Q\psi\|^2\ge\|\psi\|^2$, which forces $P\psi=\psi$ and $Q\psi=0$; conversely every vector of ${\rm Ran}\,P\cap\ker Q$ is an eigenvector with the eigenvalue $+1$. Thus $E_{+1}={\rm Ran}\,P\cap\ker Q$, and in the same way $E_{-1}={\rm Ran}\,Q\cap\ker P$. (ii) {\it The pairing $\pm\lambda$.} Let $B=1-P-Q$. Then
$$
AB=(P-Q)(1-P-Q)=QP-PQ,\qquad BA=(1-P-Q)(P-Q)=PQ-QP=-AB,
$$
$$
A^2+B^2=(P+Q-PQ-QP)+(1-P-Q+PQ+QP)=1 .
$$
If $A\psi=\lambda\psi$ then $A(B\psi)=-BA\psi=-\lambda B\psi$, and $\|B\psi\|^2=\langle\psi,(1-A^2)\psi\rangle=(1-\lambda^2)\|\psi\|^2$. Hence for $|\lambda|<1$ the operator $B$ maps $E_\lambda$ injectively into $E_{-\lambda}$ and $E_{-\lambda}$ injectively into $E_\lambda$, so that $\dim E_\lambda=\dim E_{-\lambda}$.}

\corrteal{{\bf Step 3: the odd trace.} Since $A\in{\cal S}^3$, $A^3$ is trace class, and by the spectral theorem
$$
{\rm Tr}\,A^{3}=\sum_\lambda\lambda^{3}\dim E_\lambda=\dim E_{+1}-\dim E_{-1}+\sum_{0<\lambda<1}\big(\lambda^3-\lambda^3\big)\dim E_\lambda
=\dim\big({\rm Ran}\,P\cap\ker Q\big)-\dim\big({\rm Ran}\,Q\cap\ker P\big).
$$
Comparing with Step 1, this is $\dim{\rm coker}\,T-\dim\ker T=-{\rm Index}(PuP)$:
\be
{\rm Index}\big(P_-uP_-\big)=-\,{\rm Tr}\,\big(P_--uP_-u^*\big)^{3},
\label{AppJindex}
\ee
and the same holds with any odd power $2k+1$ once $A\in{\cal S}^{2k+1}$; the cube is the lowest power allowed by the summability $[u,P_-]\in{\cal S}^p$, $p>2$.}

\corrteal{{\bf Step 4: an example fixing the sign.} Let ${\cal H}=\ell^2(\mathbb{Z})$, $P$ the projector on the sites $n\ge0$, and $u$ the right shift, $u\delta_n=\delta_{n+1}$. Then $PuP$ is the unilateral shift on ${\rm Ran}\,P$: injective, with the cokernel spanned by $\delta_0$, so ${\rm Index}(PuP)=0-1=-1$. On the other hand $Q=uPu^*$ projects on $n\ge1$, $P-Q=|\delta_0\rangle\langle\delta_0|$, and ${\rm Tr}(P-Q)^3=+1$. The minus sign in Eq. (\ref{AppJindex}) is genuine: the eigenvector of $P-Q$ with the eigenvalue $+1$ is the state that $PuP$ fails to reach.}

\corrteal{{\bf Step 5: cross-check through Fedosov's formula.} For a Fredholm $T$ with $(1-T^*T)^k$ and $(1-TT^*)^k$ trace class, ${\rm Index}\,T={\rm Tr}(1-T^*T)^k-{\rm Tr}(1-TT^*)^k$. \corGPTl{We take an integer $k\geq2$, so that $A\in{\cal S}^3$ implies that $A^{2k}$ is trace class and the separate traces below are finite.} Here $1_P-T^*T=P-Pu^*PuP=P(1-u^*Pu)P$ and $1_P-TT^*=P-PuPu^*P=P(1-Q)P$. Since $B$ anticommutes with $A$, $A^2=1-B^2$ commutes with $A$ and $B$, hence with $P=\frac12(1+A-B)$ and $Q=\frac12(1-A-B)$; therefore $A^{2k+1}=(P-Q)A^{2k}$ and
$$
{\rm Tr}\,A^{2k+1}={\rm Tr}\big(PA^{2k}P\big)-{\rm Tr}\big(QA^{2k}Q\big)={\rm Tr}\big(PA^2P\big)^k-{\rm Tr}\big(QA^2Q\big)^k
={\rm Tr}\big(P(1-Q)P\big)^k-{\rm Tr}\big(Q(1-P)Q\big)^k ,
$$
using $PA^2P=P(P+Q-PQ-QP)P=P-PQP=P(1-Q)P$ and $QA^2Q=Q(1-P)Q$. Finally, by unitary invariance of the trace, ${\rm Tr}\big(P(1-u^*Pu)P\big)^k={\rm Tr}\big(uP(1-u^*Pu)Pu^*\big)^k={\rm Tr}\big(Q(1-P)Q\big)^k$, so that Fedosov's formula gives ${\rm Index}\,T={\rm Tr}(Q(1-P)Q)^k-{\rm Tr}(P(1-Q)P)^k=-{\rm Tr}\,A^{2k+1}$, in agreement with Step 3.}

\corrteal{{\bf Step 6: remarks.} (i) On a finite-dimensional space ${\rm Tr}\,A={\rm Tr}\,P-{\rm Tr}\,Q=0$; by the pairing of Step 2 this is $\dim E_{+1}-\dim E_{-1}=0$, so that ${\rm Tr}\,A^3=0$ and the index vanishes --- as it must for an operator on a finite-dimensional space. This is the finite-size vanishing of Sect. \ref{SectDirect} in the index language; the invariant lives on the infinite (periodized) lattice, where the $\pm1$ eigenspaces need not be balanced. The numerics of Sect. \ref{SectNumerics} display exactly this: a single eigenvalue $+1$ of $P_--uP_-u^*$ localized at $z_0$, and, on a finite open patch, a compensating $-1$ localized at the patch boundary. (ii) Replacing $u$ by $u^*$ interchanges the roles of $P$ and $Q$ up to the unitary $u$: $u^*(P-uPu^*)u=u^*Pu-P$, so both ${\rm Tr}(P-uPu^*)^3$ and ${\rm Index}(PuP)$ change sign, and Eq. (\ref{AppJindex}) is convention-free.}

\subsection{\corrteal{Connes' formula and the proof of ${\rm Tr}\,(P_--uP_-u^*)^3={\cal N}_3$}}
\label{AppIndexConnes}

\corrteal{{\bf Step 1: the kernel of $A$.} Since $u$ is diagonal in the position basis, $A_{xy}=P_{xy}-u(x)P_{xy}\overline{u(y)}=P_{xy}\big(1-u(x)\overline{u(y)}\big)$, the scalar phase factor multiplying the $r\times r$ block $P_{xy}$. Hence, with $a=z_0$ and $u_a(x)=(x-a)/|x-a|$,
$$
{\rm Tr}\,A^3=\sum_{x,y,z\in\mathbb{Z}^2}\tr\big(P_{xy}P_{yz}P_{zx}\big)\,W_a(x,y,z),\qquad
W_a(x,y,z)=\big(1-u_a(x)\overline{u_a(y)}\big)\big(1-u_a(y)\overline{u_a(z)}\big)\big(1-u_a(z)\overline{u_a(x)}\big).
$$
The triple sum converges absolutely: $|1-u_a(x)\overline{u_a(y)}|=|u_a(x)-u_a(y)|\le2|x-y|/\max(|x-a|,|y-a|)$ by the phase estimate, the kernel of $P$ decays exponentially, so that $y$ and $z$ stay within a few decay lengths of $x$, and the product of the three factors is then $O(|x-a|^{-3})$, summable over $x\in\mathbb{Z}^2$.}

\corrteal{{\bf Step 2: independence of $z_0$ and the cell average.} By Sects. \ref{AppIndexModule} and \ref{AppIndexRelative}, ${\rm Tr}\,A(z_0)^3=-{\rm Index}(P_-u_{z_0}P_-)$ does not depend on $z_0$. Hence
$$
{\rm Tr}\,A^3=\frac{1}{L^2}\int_{\rm cell}d^2z_0\;{\rm Tr}\,A(z_0)^3 ,
$$
the integral over one supercell of the periodized crystal (the integrand is bounded, and the lattice points, where $u$ is not defined, form a null set).}

\corrteal{{\bf Step 3: from the average over $z_0$ to the trace per volume.} The periodicity $P_{x+Ln,\,y+Ln}=P_{xy}$ ($n\in\mathbb{Z}^2$) and the translation property $u_{z_0}(x+Ln)=u_{z_0-Ln}(x)$ allow to trade the sum over all $x\in\mathbb{Z}^2$ for a sum over one cell together with an integral over all positions of the flux tube. Writing $x=x'+Ln$ with $x'$ in the cell and substituting $y=y'+Ln$, $z=z'+Ln$,
$$
\frac{1}{L^2}\int_{\rm cell}d^2z_0\sum_{x,y,z}\tr\big(P_{xy}P_{yz}P_{zx}\big)W_{z_0}(x,y,z)
=\frac{1}{L^2}\sum_{x'\in{\rm cell}}\sum_{y',z'}\tr\big(P_{x'y'}P_{y'z'}P_{z'x'}\big)\sum_{n\in\mathbb{Z}^2}\int_{\rm cell}d^2z_0\,W_{z_0-Ln}(x',y',z'),
$$
and the last sum and integral combine into the integral over the whole plane. Hence
\be
{\rm Tr}\,\big(P_--uP_-u^*\big)^3=\frac{1}{L^2}\sum_{x\in{\rm cell}}\sum_{y,z\in\mathbb{Z}^2}\tr\big(P_{xy}P_{yz}P_{zx}\big)\,{\cal W}(x,y,z),\qquad
{\cal W}(x,y,z)=\int_{\mathbb{R}^2}d^2a\;W_a(x,y,z)
\label{AppJtrace}
\ee
(the interchange of the sum and the integral is justified by absolute convergence). \corGPTl{Indeed, with $R=1+|x-y|+|x-z|$, splitting the plane into $|a-x|\leq2R$ and its complement gives $\int_{\mathbb R^2}|W_a(x,y,z)|\,d^2a\leq C R^2$; exponential decay of $P$ makes the sum of these absolute integrals finite.}}

\corrteal{{\bf Step 4: Connes' formula.} {\it Lemma.} For any three points $x,y,z\in\mathbb{R}^2$,
\be
{\cal W}(x,y,z)=\int_{\mathbb{R}^2}d^2a\;\big(1-u_a(x)\overline{u_a(y)}\big)\big(1-u_a(y)\overline{u_a(z)}\big)\big(1-u_a(z)\overline{u_a(x)}\big)
=4\pi i\;{\rm Area}(x,y,z)=2\pi i\,(y-x)\wedge(z-x)\,.
\label{AppJconnes}
\ee
{\it Proof.} (a) {\it An algebraic identity.} Put $\alpha=u_a(x)\overline{u_a(y)}$, $\beta=u_a(y)\overline{u_a(z)}$, $\gamma=u_a(z)\overline{u_a(x)}$; these are unimodular, $\alpha\beta\gamma=1$, and $\alpha\beta=\bar\gamma$, $\beta\gamma=\bar\alpha$, $\gamma\alpha=\bar\beta$. Expanding,
$$
(1-\alpha)(1-\beta)(1-\gamma)=1-(\alpha+\beta+\gamma)+(\alpha\beta+\beta\gamma+\gamma\alpha)-\alpha\beta\gamma=-(\alpha+\beta+\gamma)+(\bar\alpha+\bar\beta+\bar\gamma)=-2i\,{\rm Im}(\alpha+\beta+\gamma).
$$
Writing $u_a(x)=e^{i\theta_a(x)}$, $\theta_a(x)=\arg(x-a)$, this reads
$$
W_a(x,y,z)=-2i\big[s_a(x,y)+s_a(y,z)+s_a(z,x)\big],\qquad s_a(x,y)=\sin\big(\theta_a(x)-\theta_a(y)\big)=-s_a(y,x).
$$
Consequences: (i) $W_a$ is purely imaginary, hence so is ${\cal W}$, and ${\rm Tr}\,A^3$ in Eq. (\ref{AppJtrace}) is real, as it must be; (ii) $W_a(x,z,y)=-W_a(x,y,z)$ (reversal of the orientation); (iii) {\it additivity}: for any point $w$, $W_a(x,y,w)+W_a(y,z,w)+W_a(z,x,w)=W_a(x,y,z)$, the terms containing $w$ cancelling pairwise by the antisymmetry of $s_a$; more generally, for an oriented polygon with the vertices $x_1,\ldots,x_n$ the boundary sum $W_a({\rm polygon})=-2i\sum_i s_a(x_i,x_{i+1})$ is additive under any subdivision by interior segments, each interior segment being traversed once in each direction; (iv) {\it decay}: the differences $\Delta_1=\theta_a(x)-\theta_a(y)$, $\Delta_2=\theta_a(y)-\theta_a(z)$, $\Delta_3=\theta_a(z)-\theta_a(x)$ add up to zero modulo $2\pi$ and are each $O(|a|^{-1})$ as $|a|\to\infty$ at fixed $x,y,z$; expanding the sines, $\sum_i\sin\Delta_i=\sum_i\Delta_i-\frac16\sum_i\Delta_i^3+\ldots=O(|a|^{-3})$, the linear terms cancelling exactly. Hence $\int d^2a\,W_a$ converges absolutely, and the same holds for the boundary sum of any closed polygon. (b) {\it Invariances.} From $u_{a+c}(x+c)=u_a(x)$, ${\cal W}$ is translation invariant; from $u_{e^{i\varphi}a}(e^{i\varphi}x)=e^{i\varphi}u_a(x)$, the products $u\bar u$ and hence ${\cal W}$ are rotation invariant; from $u_{\lambda a}(\lambda x)=u_a(x)$ and $d^2a\to\lambda^2d^2a$, ${\cal W}(\lambda x,\lambda y,\lambda z)=\lambda^2{\cal W}(x,y,z)$; by (a)(ii), ${\cal W}(x,z,y)=-{\cal W}(x,y,z)$; by (a)(iii), ${\cal W}$ is additive; and ${\cal W}$ depends continuously on the vertices (dominated convergence). (c) {\it ${\cal W}$ is proportional to the area.} By additivity and (a)(iii), ${\cal W}$ extends to an additive functional on oriented polygons --- the integral of the boundary sum, independent of the triangulation. For an axis-parallel rectangle $R_{a,b}$ with the sides $a,b$, oriented counterclockwise, put $f(a,b)={\cal W}(R_{a,b})$; cutting the rectangle by a vertical segment gives $f(a_1+a_2,b)=f(a_1,b)+f(a_2,b)$, and $f$ is continuous, hence linear in $a$; likewise in $b$: $f(a,b)=c\,ab$ with a constant $c$, purely imaginary by (a)(i). Cutting the rectangle by a diagonal produces two right triangles carried into each other by the rotation by $\pi$ about the centre of the rectangle, which preserves the orientation; by (b) they have equal ${\cal W}$, so that ${\cal W}=c\,ab/2=c\,{\rm Area}$ for every right triangle. An arbitrary triangle is the sum or the difference of two right triangles sharing the altitude dropped from one vertex; by additivity and antisymmetry ${\cal W}(T)=c\,{\rm Area}(T)$ for every triangle. (d) {\it The constant.} Take the thin triangle $x=0$, $y=1$, $z=1+i\epsilon$ (counterclockwise, ${\rm Area}=\epsilon/2$) and expand to first order in $\epsilon$, using (a): $s_a(1,1+i\epsilon)=-\epsilon\,\partial_2\theta_a(1)+O(\epsilon^2)$ and $s_a(1+i\epsilon,0)=-s_a(0,1)+\epsilon\cos\big(\theta_a(1)-\theta_a(0)\big)\,\partial_2\theta_a(1)+O(\epsilon^2)$, where $\partial_2\theta_a(x)=\partial\theta_a(x)/\partial x_2=(x_1-a_1)/|x-a|^2$. The term $s_a(0,1)$ cancels, and
$$
W_a=2i\epsilon\;\partial_2\theta_a(1)\,\Big[1-\cos\big(\theta_a(1)-\theta_a(0)\big)\Big]+O(\epsilon^2),\qquad
\cos\big(\theta_a(1)-\theta_a(0)\big)=\frac{(1-a)\cdot(-a)}{|1-a|\,|a|}=\frac{|a|^2-a_1}{|a|\,|1-a|}\,,
$$
the scalar product being that of the vectors $1-a$ and $-a$ of $\mathbb{R}^2$. \corGPTl{The derivative with respect to the moving vertex is bounded locally by $4/|a-(1+i\epsilon)|$ and uniformly by $C|a|^{-3}$ at infinity. Its integral over a disk of radius $\delta$ about $1$ is $O(\delta)$ uniformly for $|\epsilon|<\delta/2$; splitting off this disk justifies differentiation under the integral.} Hence ${\cal W}=2i\epsilon K+\corGPTl{o(\epsilon)}$ with
$$
K=\int_{\mathbb{R}^2}d^2a\;\frac{1-a_1}{|1-a|^2}\left[1-\frac{|a|^2-a_1}{|a|\,|1-a|}\right].
$$
Substitute $b=1-a$, so that $1-a_1=b_1$, $|1-a|=|b|$, $|a|=|b-1|$, $|a|^2-a_1=|b|^2-b_1$, and pass to polar coordinates $b=\rho(\cos\varphi,\sin\varphi)$, $|b-1|=\sqrt{\rho^2-2\rho\cos\varphi+1}$:
$$
K=\int_0^\infty d\rho\int_0^{2\pi}d\varphi\;\cos\varphi\left[1-\frac{\rho-\cos\varphi}{\sqrt{\rho^2-2\rho\cos\varphi+1}}\right]
=-\int_0^\infty d\rho\int_0^{2\pi}d\varphi\;\cos\varphi\,\frac{\rho-\cos\varphi}{\sqrt{\rho^2-2\rho\cos\varphi+1}}\,,
$$
the first term vanishing at each fixed $\rho$. Since $(\rho-\cos\varphi)/\sqrt{\rho^2-2\rho\cos\varphi+1}=\partial_\rho\sqrt{\rho^2-2\rho\cos\varphi+1}$,
$$
K=-\int_0^\infty d\rho\;G'(\rho)=G(0)-G(\infty),\qquad G(\rho)=\int_0^{2\pi}d\varphi\,\cos\varphi\,\sqrt{1+\rho^2-2\rho\cos\varphi}\,.
$$
Here $G(0)=\int_0^{2\pi}\cos\varphi\,d\varphi=0$, while for $\rho\to\infty$, $\sqrt{1+\rho^2-2\rho\cos\varphi}=\rho-\cos\varphi+(1-\cos^2\varphi)/(2\rho)+O(\rho^{-2})$, so that $G(\rho)=-\int_0^{2\pi}\cos^2\varphi\,d\varphi+O(\rho^{-2})\to-\pi$. Hence $K=\pi$, ${\cal W}=2\pi i\,\epsilon+\corGPTl{o(\epsilon)}=4\pi i\,{\rm Area}+\corGPTl{o(\epsilon)}$, and $c=4\pi i$. This proves Eq. (\ref{AppJconnes}); we have also verified it numerically for several triangles, to five digits. $\square$}

\corrteal{{\bf Step 5: assembly.} In the position representation $(D_jP)_{xy}=-i\,(x_j-y_j)P_{xy}$, so that
$$
\big(P\,D_1P\,D_2P\big)_{xx}=\sum_{y,z}P_{xy}\,(-i)(y_1-z_1)P_{yz}\,(-i)(z_2-x_2)P_{zx}=-\sum_{y,z}P_{xy}P_{yz}P_{zx}\,(y_1-z_1)(z_2-x_2),
$$
$$
\big(P[D_1P,D_2P]\big)_{xx}=-\sum_{y,z}P_{xy}P_{yz}P_{zx}\Big[(y_1-z_1)(z_2-x_2)-(y_2-z_2)(z_1-x_1)\Big]=-\sum_{y,z}P_{xy}P_{yz}P_{zx}\,(y-z)\wedge(z-x),
$$
and $(y-z)\wedge(z-x)=(y-x)\wedge(z-x)$, since $(z-x)\wedge(z-x)=0$. Therefore
\begin{eqnarray*}
-\,2\pi i\,\tau\big(P_-[D_1P_-,D_2P_-]\big)&=&\frac{1}{L^2}\sum_{x\in{\rm cell}}\sum_{y,z}\tr\big(P_{xy}P_{yz}P_{zx}\big)\;2\pi i\,(y-x)\wedge(z-x)\\
&=&\frac{1}{L^2}\sum_{x\in{\rm cell}}\sum_{y,z}\tr\big(P_{xy}P_{yz}P_{zx}\big)\,{\cal W}(x,y,z)\;=\;{\rm Tr}\,\big(P_--uP_-u^*\big)^3,
\end{eqnarray*}
by the Lemma and Eq. (\ref{AppJtrace}). Together with Sects. \ref{AppIndexFreq} and \ref{AppIndexRelative} this gives
\begin{eqnarray}
{\rm Tr}\,\big(P_--uP_-u^*\big)^3&=&-2\pi i\,\tau\big(P_-[D_1P_-,D_2P_-]\big)=\frac{\pi i}{4}\,\tau\big(P[D_1P,D_2P]\big)={\cal N}_3,\nonumber\\
{\rm Index}\big(P_-uP_-\big)&=&-{\cal N}_3 ,
\label{AppJchain}
\end{eqnarray}
which is Eq. (\ref{indexInd}).}

\corrteal{{\bf Step 6: remarks.} (i) {\it Signs.} The relation ${\rm Index}(P_-uP_-)=-{\rm Tr}(P_--uP_-u^*)^3$ is free of conventions (Sect. \ref{AppIndexRelative}). The relation ${\rm Tr}(P_--uP_-u^*)^3=+{\cal N}_3$ holds for the $u$ of Eq. (\ref{uDefInd}) together with the orientation conventions of Eq. (\ref{NEQ}) ($\epsilon^{123}=+1$, $D_j=-i[\hat X_j,\,\cdot\,]$, the Fermi projector $P_-=\theta(\mu-H)$); replacing $u$ by $u^*$, or reflecting the plane (equivalently, interchanging $D_1$ and $D_2$), reverses the signs of ${\rm Tr}(P_--uP_-u^*)^3$ and of the index simultaneously, ${\cal N}_3$ being unchanged in the first case and reversed in the second. These conventions are the ones verified numerically in Sect. \ref{SectNumerics}: ${\cal N}_3=C=+1$, the local trace of the cube $\to+1$, ${\rm Index}(P_-uP_-)=-1$. (ii) {\it General infinite systems.} Periodicity was used only in Step 3, to convert the average over the flux position into the trace per unit volume. \corGPTl{For a covariant ergodic family the corresponding identity holds almost surely under the summability/localization hypotheses of Bellissard's theorem \cite{Bellissard1994}, with the trace per volume defined by the invariant probability measure}; without such a bulk trace the Fredholm index remains defined for every exponentially local gapped configuration, but its identification with a trace formula is not available (Sect. \ref{SectIndexRed}). (iii) {\it Finite open samples.} On a finite open sample the total ${\rm Tr}(P_--uP_-u^*)^3$ vanishes (Step 6 of Sect. \ref{AppIndexRelative}); the invariant is recovered from the local trace of the cube around $z_0$, which tends to ${\cal N}_3$ as the sample grows, with the compensating eigenvalue $-1$ pushed to the boundary (Sect. \ref{SectNumerics}). (iv) {\it Physical content.} \corGPTl{$u$ is the singular gauge transformation associated with insertion of one flux quantum at $z_0$; $uP_-u^*$ is the Fermi projector of the final gauge-equivalent Hamiltonian $uHu^*$. The $+1$ and $-1$ eigenspaces of $P_--uP_-u^*$ describe, respectively, occupied subspaces lost and gained relative to that final projector. Their dimension difference is ${\rm Tr}(P_--uP_-u^*)^3={\cal N}_3$. The charge deficiency defined in \cite{AvronSeilerSimon1994} is ${\rm Index}(P_-uP_-)=-{\cal N}_3$; its interpretation through adiabatic flux transport requires the corresponding flux-insertion hypotheses.} (v) \corGPTl{A general frequency-dependent Dirac operator need not determine a canonical physical Fermi projector; its invariant remains the admissible odd pairing, represented in the noncompact-frequency case by the normalized class $[U_Q]$} of Sect. \ref{SectDirect}, and the statements available are the theorems of Sect. \ref{SectWeakIndex}.}

\bibliography{citationsZ,wigner3,cross-ref,biblio_corrected,CSE_MZ,biblio_correctedzu,iqhe_topology_review,ncg_review}

\begin{thebibliography}{121}%
\makeatletter
\providecommand \@ifxundefined [1]{%
 \@ifx{#1\undefined}
}%
\providecommand \@ifnum [1]{%
 \ifnum #1\expandafter \@firstoftwo
 \else \expandafter \@secondoftwo
 \fi
}%
\providecommand \@ifx [1]{%
 \ifx #1\expandafter \@firstoftwo
 \else \expandafter \@secondoftwo
 \fi
}%
\providecommand \natexlab [1]{#1}%
\providecommand \enquote  [1]{``#1''}%
\providecommand \bibnamefont  [1]{#1}%
\providecommand \bibfnamefont [1]{#1}%
\providecommand \citenamefont [1]{#1}%
\providecommand \href@noop [0]{\@secondoftwo}%
\providecommand \href [0]{\begingroup \@sanitize@url \@href}%
\providecommand \@href[1]{\@@startlink{#1}\@@href}%
\providecommand \@@href[1]{\endgroup#1\@@endlink}%
\providecommand \@sanitize@url [0]{\catcode `\\12\catcode `\$12\catcode
  `\&12\catcode `\#12\catcode `\^12\catcode `\_12\catcode `\%12\relax}%
\providecommand \@@startlink[1]{}%
\providecommand \@@endlink[0]{}%
\providecommand \url  [0]{\begingroup\@sanitize@url \@url }%
\providecommand \@url [1]{\endgroup\@href {#1}{\urlprefix }}%
\providecommand \urlprefix  [0]{URL }%
\providecommand \Eprint [0]{\href }%
\providecommand \doibase [0]{https://doi.org/}%
\providecommand \selectlanguage [0]{\@gobble}%
\providecommand \bibinfo  [0]{\@secondoftwo}%
\providecommand \bibfield  [0]{\@secondoftwo}%
\providecommand \translation [1]{[#1]}%
\providecommand \BibitemOpen [0]{}%
\providecommand \bibitemStop [0]{}%
\providecommand \bibitemNoStop [0]{.\EOS\space}%
\providecommand \EOS [0]{\spacefactor3000\relax}%
\providecommand \BibitemShut  [1]{\csname bibitem#1\endcsname}%
\let\auto@bib@innerbib\@empty
%</preamble>
\bibitem [{\citenamefont {von Klitzing}\ \emph {et~al.}(1980)\citenamefont {von
  Klitzing}, \citenamefont {Dorda},\ and\ \citenamefont
  {Pepper}}]{vonKlitzing1980}%
  \BibitemOpen
  \bibfield  {author} {\bibinfo {author} {\bibfnamefont {K.}~\bibnamefont {von
  Klitzing}}, \bibinfo {author} {\bibfnamefont {G.}~\bibnamefont {Dorda}},\
  and\ \bibinfo {author} {\bibfnamefont {M.}~\bibnamefont {Pepper}},\
  }\bibfield  {title} {\bibinfo {title} {New method for high-accuracy
  determination of the fine-structure constant based on quantized hall
  resistance},\ }\href {https://doi.org/10.1103/PhysRevLett.45.494} {\bibfield
  {journal} {\bibinfo  {journal} {Phys. Rev. Lett.}\ }\textbf {\bibinfo
  {volume} {45}},\ \bibinfo {pages} {494} (\bibinfo {year} {1980})}\BibitemShut
  {NoStop}%
\bibitem [{\citenamefont {Laughlin}(1981)}]{Laughlin1981}%
  \BibitemOpen
  \bibfield  {author} {\bibinfo {author} {\bibfnamefont {R.~B.}\ \bibnamefont
  {Laughlin}},\ }\bibfield  {title} {\bibinfo {title} {Quantized hall
  conductivity in two dimensions},\ }\href
  {https://doi.org/10.1103/PhysRevB.23.5632} {\bibfield  {journal} {\bibinfo
  {journal} {Phys. Rev. B}\ }\textbf {\bibinfo {volume} {23}},\ \bibinfo
  {pages} {5632} (\bibinfo {year} {1981})}\BibitemShut {NoStop}%
\bibitem [{\citenamefont {Thouless}\ \emph {et~al.}(1982)\citenamefont
  {Thouless}, \citenamefont {Kohmoto}, \citenamefont {Nightingale},\ and\
  \citenamefont {den Nijs}}]{Thouless1982}%
  \BibitemOpen
  \bibfield  {author} {\bibinfo {author} {\bibfnamefont {D.~J.}\ \bibnamefont
  {Thouless}}, \bibinfo {author} {\bibfnamefont {M.}~\bibnamefont {Kohmoto}},
  \bibinfo {author} {\bibfnamefont {M.~P.}\ \bibnamefont {Nightingale}},\ and\
  \bibinfo {author} {\bibfnamefont {M.}~\bibnamefont {den Nijs}},\ }\bibfield
  {title} {\bibinfo {title} {{Quantized Hall Conductance in a Two-Dimensional
  Periodic Potential}},\ }\href {https://doi.org/10.1103/PhysRevLett.49.405}
  {\bibfield  {journal} {\bibinfo  {journal} {Phys. Rev. Lett.}\ }\textbf
  {\bibinfo {volume} {49}},\ \bibinfo {pages} {405} (\bibinfo {year}
  {1982})}\BibitemShut {NoStop}%
\bibitem [{\citenamefont {Kohmoto}(1985)}]{Kohmoto1985}%
  \BibitemOpen
  \bibfield  {author} {\bibinfo {author} {\bibfnamefont {M.}~\bibnamefont
  {Kohmoto}},\ }\bibfield  {title} {\bibinfo {title} {Topological invariant and
  the quantization of the hall conductance},\ }\href
  {https://doi.org/10.1016/0003-4916(85)90148-4} {\bibfield  {journal}
  {\bibinfo  {journal} {Ann. Phys. (N.Y.)}\ }\textbf {\bibinfo {volume}
  {160}},\ \bibinfo {pages} {343} (\bibinfo {year} {1985})}\BibitemShut
  {NoStop}%
\bibitem [{\citenamefont {Berry}(1984)}]{Berry1984}%
  \BibitemOpen
  \bibfield  {author} {\bibinfo {author} {\bibfnamefont {M.~V.}\ \bibnamefont
  {Berry}},\ }\bibfield  {title} {\bibinfo {title} {Quantal phase factors
  accompanying adiabatic changes},\ }\href
  {https://doi.org/10.1098/rspa.1984.0023} {\bibfield  {journal} {\bibinfo
  {journal} {Proc. R. Soc. Lond. A}\ }\textbf {\bibinfo {volume} {392}},\
  \bibinfo {pages} {45} (\bibinfo {year} {1984})}\BibitemShut {NoStop}%
\bibitem [{\citenamefont {Simon}(1983)}]{Simon1983}%
  \BibitemOpen
  \bibfield  {author} {\bibinfo {author} {\bibfnamefont {B.}~\bibnamefont
  {Simon}},\ }\bibfield  {title} {\bibinfo {title} {Holonomy, the quantum
  adiabatic theorem, and berry's phase},\ }\href
  {https://doi.org/10.1103/PhysRevLett.51.2167} {\bibfield  {journal} {\bibinfo
   {journal} {Phys. Rev. Lett.}\ }\textbf {\bibinfo {volume} {51}},\ \bibinfo
  {pages} {2167} (\bibinfo {year} {1983})}\BibitemShut {NoStop}%
\bibitem [{\citenamefont {Niu}\ \emph {et~al.}(1985)\citenamefont {Niu},
  \citenamefont {Thouless},\ and\ \citenamefont {Wu}}]{Niu1985}%
  \BibitemOpen
  \bibfield  {author} {\bibinfo {author} {\bibfnamefont {Q.}~\bibnamefont
  {Niu}}, \bibinfo {author} {\bibfnamefont {D.~J.}\ \bibnamefont {Thouless}},\
  and\ \bibinfo {author} {\bibfnamefont {Y.-S.}\ \bibnamefont {Wu}},\
  }\bibfield  {title} {\bibinfo {title} {Quantized hall conductance as a
  topological invariant},\ }\href {https://doi.org/10.1103/PhysRevB.31.3372}
  {\bibfield  {journal} {\bibinfo  {journal} {Phys. Rev. B}\ }\textbf {\bibinfo
  {volume} {31}},\ \bibinfo {pages} {3372} (\bibinfo {year}
  {1985})}\BibitemShut {NoStop}%
\bibitem [{\citenamefont {Avron}\ \emph {et~al.}(1983)\citenamefont {Avron},
  \citenamefont {Seiler},\ and\ \citenamefont {Simon}}]{Avron1983}%
  \BibitemOpen
  \bibfield  {author} {\bibinfo {author} {\bibfnamefont {J.~E.}\ \bibnamefont
  {Avron}}, \bibinfo {author} {\bibfnamefont {R.}~\bibnamefont {Seiler}},\ and\
  \bibinfo {author} {\bibfnamefont {B.}~\bibnamefont {Simon}},\ }\bibfield
  {title} {\bibinfo {title} {Homotopy and quantization in condensed matter
  physics},\ }\href {https://doi.org/10.1103/PhysRevLett.51.51} {\bibfield
  {journal} {\bibinfo  {journal} {Phys. Rev. Lett.}\ }\textbf {\bibinfo
  {volume} {51}},\ \bibinfo {pages} {51} (\bibinfo {year} {1983})}\BibitemShut
  {NoStop}%
\bibitem [{\citenamefont {Halperin}(1982)}]{Halperin1982}%
  \BibitemOpen
  \bibfield  {author} {\bibinfo {author} {\bibfnamefont {B.~I.}\ \bibnamefont
  {Halperin}},\ }\bibfield  {title} {\bibinfo {title} {Quantized hall
  conductance, current-carrying edge states, and the existence of extended
  states in a two-dimensional disordered potential},\ }\href
  {https://doi.org/10.1103/PhysRevB.25.2185} {\bibfield  {journal} {\bibinfo
  {journal} {Phys. Rev. B}\ }\textbf {\bibinfo {volume} {25}},\ \bibinfo
  {pages} {2185} (\bibinfo {year} {1982})}\BibitemShut {NoStop}%
\bibitem [{\citenamefont {Str\v{e}da}(1982)}]{Streda1982}%
  \BibitemOpen
  \bibfield  {author} {\bibinfo {author} {\bibfnamefont {P.}~\bibnamefont
  {Str\v{e}da}},\ }\bibfield  {title} {\bibinfo {title} {Theory of quantised
  hall conductivity in two dimensions},\ }\href
  {https://doi.org/10.1088/0022-3719/15/22/005} {\bibfield  {journal} {\bibinfo
   {journal} {J. Phys. C: Solid State Phys.}\ }\textbf {\bibinfo {volume}
  {15}},\ \bibinfo {pages} {L717} (\bibinfo {year} {1982})}\BibitemShut
  {NoStop}%
\bibitem [{\citenamefont {Hatsugai}(1993)}]{Hatsugai1993}%
  \BibitemOpen
  \bibfield  {author} {\bibinfo {author} {\bibfnamefont {Y.}~\bibnamefont
  {Hatsugai}},\ }\bibfield  {title} {\bibinfo {title} {Chern number and edge
  states in the integer quantum hall effect},\ }\href
  {https://doi.org/10.1103/PhysRevLett.71.3697} {\bibfield  {journal} {\bibinfo
   {journal} {Phys. Rev. Lett.}\ }\textbf {\bibinfo {volume} {71}},\ \bibinfo
  {pages} {3697} (\bibinfo {year} {1993})}\BibitemShut {NoStop}%
\bibitem [{\citenamefont {Bellissard}\ \emph {et~al.}(1994)\citenamefont
  {Bellissard}, \citenamefont {van Elst},\ and\ \citenamefont
  {Schulz-Baldes}}]{Bellissard1994}%
  \BibitemOpen
  \bibfield  {author} {\bibinfo {author} {\bibfnamefont {J.}~\bibnamefont
  {Bellissard}}, \bibinfo {author} {\bibfnamefont {A.}~\bibnamefont {van
  Elst}},\ and\ \bibinfo {author} {\bibfnamefont {H.}~\bibnamefont
  {Schulz-Baldes}},\ }\bibfield  {title} {\bibinfo {title} {The noncommutative
  geometry of the quantum hall effect},\ }\href
  {https://doi.org/10.1063/1.530758} {\bibfield  {journal} {\bibinfo  {journal}
  {J. Math. Phys.}\ }\textbf {\bibinfo {volume} {35}},\ \bibinfo {pages} {5373}
  (\bibinfo {year} {1994})}\BibitemShut {NoStop}%
\bibitem [{\citenamefont {Prange}\ and\ \citenamefont
  {Girvin}(1990)}]{PrangeGirvin1990}%
  \BibitemOpen
  \bibinfo {editor} {\bibfnamefont {R.~E.}\ \bibnamefont {Prange}}\ and\
  \bibinfo {editor} {\bibfnamefont {S.~M.}\ \bibnamefont {Girvin}},\ eds.,\
  \href {https://doi.org/10.1007/978-1-4612-3350-3} {\emph {\bibinfo {title}
  {The Quantum Hall Effect}}},\ \bibinfo {edition} {2nd}\ ed.\ (\bibinfo
  {publisher} {Springer-Verlag},\ \bibinfo {address} {New York},\ \bibinfo
  {year} {1990})\BibitemShut {NoStop}%
\bibitem [{\citenamefont {Xiao}\ \emph {et~al.}(2010)\citenamefont {Xiao},
  \citenamefont {Chang},\ and\ \citenamefont {Niu}}]{Xiao2010}%
  \BibitemOpen
  \bibfield  {author} {\bibinfo {author} {\bibfnamefont {D.}~\bibnamefont
  {Xiao}}, \bibinfo {author} {\bibfnamefont {M.-C.}\ \bibnamefont {Chang}},\
  and\ \bibinfo {author} {\bibfnamefont {Q.}~\bibnamefont {Niu}},\ }\bibfield
  {title} {\bibinfo {title} {Berry phase effects on electronic properties},\
  }\href {https://doi.org/10.1103/RevModPhys.82.1959} {\bibfield  {journal}
  {\bibinfo  {journal} {Rev. Mod. Phys.}\ }\textbf {\bibinfo {volume} {82}},\
  \bibinfo {pages} {1959} (\bibinfo {year} {2010})}\BibitemShut {NoStop}%
\bibitem [{\citenamefont {Hasan}\ and\ \citenamefont {Kane}(2010)}]{Hasan2010}%
  \BibitemOpen
  \bibfield  {author} {\bibinfo {author} {\bibfnamefont {M.~Z.}\ \bibnamefont
  {Hasan}}\ and\ \bibinfo {author} {\bibfnamefont {C.~L.}\ \bibnamefont
  {Kane}},\ }\bibfield  {title} {\bibinfo {title} {Colloquium: Topological
  insulators},\ }\href {https://doi.org/10.1103/RevModPhys.82.3045} {\bibfield
  {journal} {\bibinfo  {journal} {Rev. Mod. Phys.}\ }\textbf {\bibinfo {volume}
  {82}},\ \bibinfo {pages} {3045} (\bibinfo {year} {2010})}\BibitemShut
  {NoStop}%
\bibitem [{\citenamefont {Qi}\ and\ \citenamefont {Zhang}(2011)}]{Qi2011}%
  \BibitemOpen
  \bibfield  {author} {\bibinfo {author} {\bibfnamefont {X.-L.}\ \bibnamefont
  {Qi}}\ and\ \bibinfo {author} {\bibfnamefont {S.-C.}\ \bibnamefont {Zhang}},\
  }\bibfield  {title} {\bibinfo {title} {Topological insulators and
  superconductors},\ }\href {https://doi.org/10.1103/RevModPhys.83.1057}
  {\bibfield  {journal} {\bibinfo  {journal} {Rev. Mod. Phys.}\ }\textbf
  {\bibinfo {volume} {83}},\ \bibinfo {pages} {1057} (\bibinfo {year}
  {2011})}\BibitemShut {NoStop}%
\bibitem [{\citenamefont {Chang}\ \emph {et~al.}(2023)\citenamefont {Chang},
  \citenamefont {Liu},\ and\ \citenamefont {MacDonald}}]{Chang2023}%
  \BibitemOpen
  \bibfield  {author} {\bibinfo {author} {\bibfnamefont {C.-Z.}\ \bibnamefont
  {Chang}}, \bibinfo {author} {\bibfnamefont {C.-X.}\ \bibnamefont {Liu}},\
  and\ \bibinfo {author} {\bibfnamefont {A.~H.}\ \bibnamefont {MacDonald}},\
  }\bibfield  {title} {\bibinfo {title} {Colloquium: Quantum anomalous hall
  effect},\ }\href {https://doi.org/10.1103/RevModPhys.95.011002} {\bibfield
  {journal} {\bibinfo  {journal} {Rev. Mod. Phys.}\ }\textbf {\bibinfo {volume}
  {95}},\ \bibinfo {pages} {011002} (\bibinfo {year} {2023})}\BibitemShut
  {NoStop}%
\bibitem [{\citenamefont {Chi}\ and\ \citenamefont
  {Moodera}(2022)}]{ChiMoodera2022}%
  \BibitemOpen
  \bibfield  {author} {\bibinfo {author} {\bibfnamefont {H.}~\bibnamefont
  {Chi}}\ and\ \bibinfo {author} {\bibfnamefont {J.~S.}\ \bibnamefont
  {Moodera}},\ }\bibfield  {title} {\bibinfo {title} {Progress and prospects in
  the quantum anomalous hall effect},\ }\href
  {https://doi.org/10.1063/5.0100989} {\bibfield  {journal} {\bibinfo
  {journal} {APL Mater.}\ }\textbf {\bibinfo {volume} {10}},\ \bibinfo {pages}
  {090903} (\bibinfo {year} {2022})}\BibitemShut {NoStop}%
\bibitem [{\citenamefont {Weber}\ \emph {et~al.}(2024)\citenamefont {Weber},
  \citenamefont {Fuhrer}, \citenamefont {Sheng}, \citenamefont {Yang},
  \citenamefont {Thomale} \emph {et~al.}}]{Weber2024}%
  \BibitemOpen
  \bibfield  {author} {\bibinfo {author} {\bibfnamefont {B.}~\bibnamefont
  {Weber}}, \bibinfo {author} {\bibfnamefont {M.~S.}\ \bibnamefont {Fuhrer}},
  \bibinfo {author} {\bibfnamefont {X.-L.}\ \bibnamefont {Sheng}}, \bibinfo
  {author} {\bibfnamefont {S.~A.}\ \bibnamefont {Yang}}, \bibinfo {author}
  {\bibfnamefont {R.}~\bibnamefont {Thomale}}, \emph {et~al.},\ }\bibfield
  {title} {\bibinfo {title} {2024 roadmap on 2d topological insulators},\
  }\href {https://doi.org/10.1088/2515-7639/ad2083} {\bibfield  {journal}
  {\bibinfo  {journal} {J. Phys. Mater.}\ }\textbf {\bibinfo {volume} {7}},\
  \bibinfo {pages} {022501} (\bibinfo {year} {2024})}\BibitemShut {NoStop}%
\bibitem [{\citenamefont {Bai}\ \emph {et~al.}(2024)\citenamefont {Bai},
  \citenamefont {Li}, \citenamefont {Luan}, \citenamefont {Liu}, \citenamefont
  {Song}, \citenamefont {Chen}, \citenamefont {Ji}, \citenamefont {Zhang},
  \citenamefont {Meng}, \citenamefont {Tong}, \citenamefont {Li}, \citenamefont
  {Jiang}, \citenamefont {Gao}, \citenamefont {Gu}, \citenamefont {Zhang},
  \citenamefont {Wang}, \citenamefont {Xue}, \citenamefont {He}, \citenamefont
  {Feng},\ and\ \citenamefont {Feng}}]{Bai2024}%
  \BibitemOpen
  \bibfield  {author} {\bibinfo {author} {\bibfnamefont {Y.}~\bibnamefont
  {Bai}}, \bibinfo {author} {\bibfnamefont {Y.}~\bibnamefont {Li}}, \bibinfo
  {author} {\bibfnamefont {J.}~\bibnamefont {Luan}}, \bibinfo {author}
  {\bibfnamefont {R.}~\bibnamefont {Liu}}, \bibinfo {author} {\bibfnamefont
  {W.}~\bibnamefont {Song}}, \bibinfo {author} {\bibfnamefont {Y.}~\bibnamefont
  {Chen}}, \bibinfo {author} {\bibfnamefont {P.-F.}\ \bibnamefont {Ji}},
  \bibinfo {author} {\bibfnamefont {Q.}~\bibnamefont {Zhang}}, \bibinfo
  {author} {\bibfnamefont {F.}~\bibnamefont {Meng}}, \bibinfo {author}
  {\bibfnamefont {B.}~\bibnamefont {Tong}}, \bibinfo {author} {\bibfnamefont
  {L.}~\bibnamefont {Li}}, \bibinfo {author} {\bibfnamefont {Y.}~\bibnamefont
  {Jiang}}, \bibinfo {author} {\bibfnamefont {Z.}~\bibnamefont {Gao}}, \bibinfo
  {author} {\bibfnamefont {L.}~\bibnamefont {Gu}}, \bibinfo {author}
  {\bibfnamefont {J.}~\bibnamefont {Zhang}}, \bibinfo {author} {\bibfnamefont
  {Y.}~\bibnamefont {Wang}}, \bibinfo {author} {\bibfnamefont {Q.-K.}\
  \bibnamefont {Xue}}, \bibinfo {author} {\bibfnamefont {K.}~\bibnamefont
  {He}}, \bibinfo {author} {\bibfnamefont {Y.}~\bibnamefont {Feng}},\ and\
  \bibinfo {author} {\bibfnamefont {X.}~\bibnamefont {Feng}},\ }\bibfield
  {title} {\bibinfo {title} {Quantized anomalous hall resistivity achieved in
  molecular beam epitaxy-grown {MnBi$_2$Te$_4$} thin films},\ }\href
  {https://doi.org/10.1093/nsr/nwad189} {\bibfield  {journal} {\bibinfo
  {journal} {Natl. Sci. Rev.}\ }\textbf {\bibinfo {volume} {11}},\ \bibinfo
  {pages} {nwad189} (\bibinfo {year} {2024})}\BibitemShut {NoStop}%
\bibitem [{\citenamefont {Bosnar}\ \emph {et~al.}(2023)\citenamefont {Bosnar},
  \citenamefont {Vyazovskaya}, \citenamefont {Petrov}, \citenamefont
  {Chulkov},\ and\ \citenamefont {Otrokov}}]{Bosnar2023}%
  \BibitemOpen
  \bibfield  {author} {\bibinfo {author} {\bibfnamefont {M.}~\bibnamefont
  {Bosnar}}, \bibinfo {author} {\bibfnamefont {A.~Y.}\ \bibnamefont
  {Vyazovskaya}}, \bibinfo {author} {\bibfnamefont {E.~K.}\ \bibnamefont
  {Petrov}}, \bibinfo {author} {\bibfnamefont {E.~V.}\ \bibnamefont
  {Chulkov}},\ and\ \bibinfo {author} {\bibfnamefont {M.~M.}\ \bibnamefont
  {Otrokov}},\ }\bibfield  {title} {\bibinfo {title} {High chern number van der
  waals magnetic topological multilayers {MnBi$_2$Te$_4$}/hbn},\ }\href
  {https://doi.org/10.1038/s41699-023-00396-y} {\bibfield  {journal} {\bibinfo
  {journal} {npj 2D Mater. Appl.}\ }\textbf {\bibinfo {volume} {7}},\ \bibinfo
  {pages} {33} (\bibinfo {year} {2023})}\BibitemShut {NoStop}%
\bibitem [{\citenamefont {Ba{\`u}}\ and\ \citenamefont
  {Marrazzo}(2024)}]{Bau2024}%
  \BibitemOpen
  \bibfield  {author} {\bibinfo {author} {\bibfnamefont {N.}~\bibnamefont
  {Ba{\`u}}}\ and\ \bibinfo {author} {\bibfnamefont {A.}~\bibnamefont
  {Marrazzo}},\ }\bibfield  {title} {\bibinfo {title} {Local chern marker for
  periodic systems},\ }\href {https://doi.org/10.1103/PhysRevB.109.014206}
  {\bibfield  {journal} {\bibinfo  {journal} {Phys. Rev. B}\ }\textbf {\bibinfo
  {volume} {109}},\ \bibinfo {pages} {014206} (\bibinfo {year}
  {2024})}\BibitemShut {NoStop}%
\bibitem [{\citenamefont {Park}\ \emph {et~al.}(2023)\citenamefont {Park},
  \citenamefont {Cai}, \citenamefont {Anderson} \emph {et~al.}}]{Park2023}%
  \BibitemOpen
  \bibfield  {author} {\bibinfo {author} {\bibfnamefont {H.}~\bibnamefont
  {Park}}, \bibinfo {author} {\bibfnamefont {J.}~\bibnamefont {Cai}}, \bibinfo
  {author} {\bibfnamefont {E.}~\bibnamefont {Anderson}}, \emph {et~al.},\
  }\bibfield  {title} {\bibinfo {title} {Observation of fractionally quantized
  anomalous hall effect},\ }\href {https://doi.org/10.1038/s41586-023-06536-0}
  {\bibfield  {journal} {\bibinfo  {journal} {Nature}\ }\textbf {\bibinfo
  {volume} {622}},\ \bibinfo {pages} {74} (\bibinfo {year} {2023})}\BibitemShut
  {NoStop}%
\bibitem [{\citenamefont {Cai}\ \emph {et~al.}(2023)\citenamefont {Cai},
  \citenamefont {Anderson}, \citenamefont {Wang} \emph {et~al.}}]{Cai2023}%
  \BibitemOpen
  \bibfield  {author} {\bibinfo {author} {\bibfnamefont {J.}~\bibnamefont
  {Cai}}, \bibinfo {author} {\bibfnamefont {E.}~\bibnamefont {Anderson}},
  \bibinfo {author} {\bibfnamefont {C.}~\bibnamefont {Wang}}, \emph {et~al.},\
  }\bibfield  {title} {\bibinfo {title} {Signatures of fractional quantum
  anomalous hall states in twisted {MoTe$_2$}},\ }\href
  {https://doi.org/10.1038/s41586-023-06289-w} {\bibfield  {journal} {\bibinfo
  {journal} {Nature}\ }\textbf {\bibinfo {volume} {622}},\ \bibinfo {pages}
  {63} (\bibinfo {year} {2023})}\BibitemShut {NoStop}%
\bibitem [{\citenamefont {Lu}\ \emph {et~al.}(2024)\citenamefont {Lu},
  \citenamefont {Han}, \citenamefont {Yao} \emph {et~al.}}]{Lu2024}%
  \BibitemOpen
  \bibfield  {author} {\bibinfo {author} {\bibfnamefont {Z.}~\bibnamefont
  {Lu}}, \bibinfo {author} {\bibfnamefont {T.}~\bibnamefont {Han}}, \bibinfo
  {author} {\bibfnamefont {Y.}~\bibnamefont {Yao}}, \emph {et~al.},\ }\bibfield
   {title} {\bibinfo {title} {Fractional quantum anomalous hall effect in
  multilayer graphene},\ }\href {https://doi.org/10.1038/s41586-023-07010-7}
  {\bibfield  {journal} {\bibinfo  {journal} {Nature}\ }\textbf {\bibinfo
  {volume} {626}},\ \bibinfo {pages} {759} (\bibinfo {year}
  {2024})}\BibitemShut {NoStop}%
\bibitem [{\citenamefont {Ju}\ \emph {et~al.}(2024)\citenamefont {Ju},
  \citenamefont {Cao}, \citenamefont {Fu}, \citenamefont {Xiao},\ and\
  \citenamefont {Xu}}]{Ju2024}%
  \BibitemOpen
  \bibfield  {author} {\bibinfo {author} {\bibfnamefont {L.}~\bibnamefont
  {Ju}}, \bibinfo {author} {\bibfnamefont {T.}~\bibnamefont {Cao}}, \bibinfo
  {author} {\bibfnamefont {L.}~\bibnamefont {Fu}}, \bibinfo {author}
  {\bibfnamefont {D.}~\bibnamefont {Xiao}},\ and\ \bibinfo {author}
  {\bibfnamefont {X.}~\bibnamefont {Xu}},\ }\bibfield  {title} {\bibinfo
  {title} {The fractional quantum anomalous hall effect},\ }\href
  {https://doi.org/10.1038/s41578-024-00694-x} {\bibfield  {journal} {\bibinfo
  {journal} {Nat. Rev. Mater.}\ }\textbf {\bibinfo {volume} {9}},\ \bibinfo
  {pages} {455} (\bibinfo {year} {2024})}\BibitemShut {NoStop}%
\bibitem [{\citenamefont {Kaufmann}\ \emph {et~al.}(2016)\citenamefont
  {Kaufmann}, \citenamefont {Li},\ and\ \citenamefont
  {Wehefritz-Kaufmann}}]{Kaufmann:2015lga}%
  \BibitemOpen
  \bibfield  {author} {\bibinfo {author} {\bibfnamefont {R.~M.}\ \bibnamefont
  {Kaufmann}}, \bibinfo {author} {\bibfnamefont {D.}~\bibnamefont {Li}},\ and\
  \bibinfo {author} {\bibfnamefont {B.}~\bibnamefont {Wehefritz-Kaufmann}},\
  }\bibfield  {title} {\bibinfo {title} {{Notes on topological insulators}},\
  }\href {https://doi.org/10.1142/S0129055X1630003X} {\bibfield  {journal}
  {\bibinfo  {journal} {Rev. Math. Phys.}\ }\textbf {\bibinfo {volume} {28}},\
  \bibinfo {pages} {1630003} (\bibinfo {year} {2016})},\ \Eprint
  {https://arxiv.org/abs/1501.02874} {arXiv:1501.02874 [math-ph]} \BibitemShut
  {NoStop}%
%%CITATION = ARXIV:1501.02874;%%
\bibitem [{\citenamefont {Fradkin}(1991)}]{Fradkin1991}%
  \BibitemOpen
  \bibfield  {author} {\bibinfo {author} {\bibfnamefont {E.}~\bibnamefont
  {Fradkin}},\ }\href@noop {} {\emph {\bibinfo {title} {{Field Theories of
  Condensed Matter Physics}}}}\ (\bibinfo  {publisher} {Addison Wesley
  Publishing Company},\ \bibinfo {year} {1991})\BibitemShut {NoStop}%
\bibitem [{\citenamefont {Tong}(2016)}]{Tong:2016kpv}%
  \BibitemOpen
  \bibfield  {author} {\bibinfo {author} {\bibfnamefont {D.}~\bibnamefont
  {Tong}},\ }\href@noop {} {\bibinfo {title} {Lectures on the quantum hall
  effect}} (\bibinfo {year} {2016}),\ \Eprint
  {https://arxiv.org/abs/arXiv:1606.06687} {arXiv:1606.06687 [hep-th]}
  \BibitemShut {NoStop}%
\bibitem [{\citenamefont {Hatsugai}(1997)}]{Hatsugai1997}%
  \BibitemOpen
  \bibfield  {author} {\bibinfo {author} {\bibfnamefont {Y.}~\bibnamefont
  {Hatsugai}},\ }\bibfield  {title} {\bibinfo {title} {{Topological aspects of
  the quantum Hall effect}},\ }\href@noop {} {\bibfield  {journal} {\bibinfo
  {journal} {J. Phys. Condens. Matter}\ }\textbf {\bibinfo {volume} {9}},\
  \bibinfo {pages} {2507} (\bibinfo {year} {1997})}\BibitemShut {NoStop}%
\bibitem [{\citenamefont {Qi}\ \emph {et~al.}(2008)\citenamefont {Qi},
  \citenamefont {Hughes},\ and\ \citenamefont {Zhang}}]{Qi2008}%
  \BibitemOpen
  \bibfield  {author} {\bibinfo {author} {\bibfnamefont {X.-L.}\ \bibnamefont
  {Qi}}, \bibinfo {author} {\bibfnamefont {T.~L.}\ \bibnamefont {Hughes}},\
  and\ \bibinfo {author} {\bibfnamefont {S.-C.}\ \bibnamefont {Zhang}},\
  }\bibfield  {title} {\bibinfo {title} {Topological field theory of
  time-reversal invariant insulators},\ }\href
  {https://doi.org/10.1103/PhysRevB.78.195424} {\bibfield  {journal} {\bibinfo
  {journal} {Phys. Rev. B}\ }\textbf {\bibinfo {volume} {78}},\ \bibinfo
  {pages} {195424} (\bibinfo {year} {2008})}\BibitemShut {NoStop}%
\bibitem [{\citenamefont {Ishikawa}\ and\ \citenamefont
  {Matsuyama}(1986)}]{IshikawaMatsuyama1986}%
  \BibitemOpen
  \bibfield  {author} {\bibinfo {author} {\bibfnamefont {K.}~\bibnamefont
  {Ishikawa}}\ and\ \bibinfo {author} {\bibfnamefont {T.}~\bibnamefont
  {Matsuyama}},\ }\bibfield  {title} {\bibinfo {title} {{Magnetic field induced
  multi component QED in three-dimensions and quantum Hall effect}},\
  }\href@noop {} {\bibfield  {journal} {\bibinfo  {journal} {Z. Phys. C}\
  }\textbf {\bibinfo {volume} {33}},\ \bibinfo {pages} {41} (\bibinfo {year}
  {1986})}\BibitemShut {NoStop}%
\bibitem [{\citenamefont {Volovik}(1988)}]{Volovik1988}%
  \BibitemOpen
  \bibfield  {author} {\bibinfo {author} {\bibfnamefont {G.~E.}\ \bibnamefont
  {Volovik}},\ }\bibfield  {title} {\bibinfo {title} {{An analog of the quantum
  Hall effect in a superfluid 3He film}},\ }\href@noop {} {\bibfield  {journal}
  {\bibinfo  {journal} {JETP}\ }\textbf {\bibinfo {volume} {67}},\ \bibinfo
  {pages} {9} (\bibinfo {year} {1988})},\ \bibinfo {note} {zhETF, Vol. 94, No.
  3(9), 123}\BibitemShut {NoStop}%
\bibitem [{\citenamefont {Volovik}(2003)}]{Volovik2003a}%
  \BibitemOpen
  \bibfield  {author} {\bibinfo {author} {\bibfnamefont {G.~E.}\ \bibnamefont
  {Volovik}},\ }\href@noop {} {\emph {\bibinfo {title} {{The Universe in a
  Helium Droplet}}}}\ (\bibinfo  {publisher} {Clarendon Press},\ \bibinfo
  {address} {Oxford},\ \bibinfo {year} {2003})\BibitemShut {NoStop}%
\bibitem [{\citenamefont {Coleman}\ and\ \citenamefont
  {Hill}(1985)}]{parity_anomaly}%
  \BibitemOpen
  \bibfield  {author} {\bibinfo {author} {\bibfnamefont {S.}~\bibnamefont
  {Coleman}}\ and\ \bibinfo {author} {\bibfnamefont {B.}~\bibnamefont {Hill}},\
  }\href@noop {} {\bibfield  {journal} {\bibinfo  {journal} {Phys. Lett. B}\
  }\textbf {\bibinfo {volume} {159}},\ \bibinfo {pages} {184} (\bibinfo {year}
  {1985})}\BibitemShut {NoStop}%
\bibitem [{\citenamefont {Lee}(1986)}]{parity_anomaly_}%
  \BibitemOpen
  \bibfield  {author} {\bibinfo {author} {\bibfnamefont {T.}~\bibnamefont
  {Lee}},\ }\href@noop {} {\bibfield  {journal} {\bibinfo  {journal} {Phys.
  Lett. B}\ }\textbf {\bibinfo {volume} {171}},\ \bibinfo {pages} {247}
  (\bibinfo {year} {1986})}\BibitemShut {NoStop}%
\bibitem [{\citenamefont {Zubkov}(2018{\natexlab{a}})}]{Zubkov2018a}%
  \BibitemOpen
  \bibfield  {author} {\bibinfo {author} {\bibfnamefont {M.~A.}\ \bibnamefont
  {Zubkov}},\ }\bibfield  {title} {\bibinfo {title} {{Momentum space topology
  of QCD}},\ }\href {https://doi.org/10.1016/j.aop.2018.04.016} {\bibfield
  {journal} {\bibinfo  {journal} {Annals Phys}\ }\textbf {\bibinfo {volume}
  {393}},\ \bibinfo {pages} {264} (\bibinfo {year} {2018}{\natexlab{a}})},\
  \Eprint {https://arxiv.org/abs/1610.08041} {arXiv:1610.08041} \BibitemShut
  {NoStop}%
\bibitem [{\citenamefont {Zhang}\ and\ \citenamefont
  {Zubkov}(2019{\natexlab{a}})}]{ZZ2019}%
  \BibitemOpen
  \bibfield  {author} {\bibinfo {author} {\bibfnamefont {C.~X.}\ \bibnamefont
  {Zhang}}\ and\ \bibinfo {author} {\bibfnamefont {M.~A.}\ \bibnamefont
  {Zubkov}},\ }\href@noop {} {\bibinfo {title} {{Influence of interactions on
  the anomalous quantum Hall effect}}} (\bibinfo {year} {2019}{\natexlab{a}}),\
  \Eprint {https://arxiv.org/abs/arXiv:1902.06545} {arXiv:arXiv:1902.06545
  [cond-mat.mes-hall]} \BibitemShut {NoStop}%
\bibitem [{\citenamefont {Zhao}\ \emph {et~al.}(2023)\citenamefont {Zhao},
  \citenamefont {Mai}, \citenamefont {Bradlyn},\ and\ \citenamefont
  {Phillips}}]{Zhao2023}%
  \BibitemOpen
  \bibfield  {author} {\bibinfo {author} {\bibfnamefont {J.}~\bibnamefont
  {Zhao}}, \bibinfo {author} {\bibfnamefont {P.}~\bibnamefont {Mai}}, \bibinfo
  {author} {\bibfnamefont {B.}~\bibnamefont {Bradlyn}},\ and\ \bibinfo {author}
  {\bibfnamefont {P.}~\bibnamefont {Phillips}},\ }\bibfield  {title} {\bibinfo
  {title} {Failure of topological invariants in strongly correlated matter},\
  }\href@noop {} {\bibfield  {journal} {\bibinfo  {journal} {Phys. Rev. Lett.}\
  }\textbf {\bibinfo {volume} {131}},\ \bibinfo {pages} {106601} (\bibinfo
  {year} {2023})}\BibitemShut {NoStop}%
\bibitem [{\citenamefont {Gurarie}(2011)}]{Gurarie2011}%
  \BibitemOpen
  \bibfield  {author} {\bibinfo {author} {\bibfnamefont {V.}~\bibnamefont
  {Gurarie}},\ }\bibfield  {title} {\bibinfo {title} {Single-particle green's
  functions and interacting topological insulators},\ }\href@noop {} {\bibfield
   {journal} {\bibinfo  {journal} {Phys. Rev. B}\ }\textbf {\bibinfo {volume}
  {83}},\ \bibinfo {pages} {085426} (\bibinfo {year} {2011})}\BibitemShut
  {NoStop}%
\bibitem [{\citenamefont {Essin}\ and\ \citenamefont
  {Gurarie}(2011)}]{EssinGurarie2011}%
  \BibitemOpen
  \bibfield  {author} {\bibinfo {author} {\bibfnamefont {A.~M.}\ \bibnamefont
  {Essin}}\ and\ \bibinfo {author} {\bibfnamefont {V.}~\bibnamefont
  {Gurarie}},\ }\bibfield  {title} {\bibinfo {title} {{Bulk-boundary
  correspondence of topological insulators from their Green's functions}},\
  }\href@noop {} {\bibfield  {journal} {\bibinfo  {journal} {Phys. Rev. B}\
  }\textbf {\bibinfo {volume} {84}},\ \bibinfo {pages} {125132} (\bibinfo
  {year} {2011})}\BibitemShut {NoStop}%
\bibitem [{\citenamefont {Zubkov}(2018{\natexlab{b}})}]{zubkov2018momentum}%
  \BibitemOpen
  \bibfield  {author} {\bibinfo {author} {\bibfnamefont {M.}~\bibnamefont
  {Zubkov}},\ }\bibfield  {title} {\bibinfo {title} {Momentum space topology of
  qcd},\ }\href@noop {} {\bibfield  {journal} {\bibinfo  {journal} {Annals of
  Physics}\ }\textbf {\bibinfo {volume} {393}},\ \bibinfo {pages} {264}
  (\bibinfo {year} {2018}{\natexlab{b}})}\BibitemShut {NoStop}%
\bibitem [{\citenamefont {Zubkov}\ and\ \citenamefont
  {Volovik}(2012)}]{zubkov2012momentum}%
  \BibitemOpen
  \bibfield  {author} {\bibinfo {author} {\bibfnamefont {M.}~\bibnamefont
  {Zubkov}}\ and\ \bibinfo {author} {\bibfnamefont {G.}~\bibnamefont
  {Volovik}},\ }\bibfield  {title} {\bibinfo {title} {Momentum space
  topological invariants for the 4d relativistic vacua with mass gap},\
  }\href@noop {} {\bibfield  {journal} {\bibinfo  {journal} {Nuclear Physics
  B}\ }\textbf {\bibinfo {volume} {860}},\ \bibinfo {pages} {295} (\bibinfo
  {year} {2012})}\BibitemShut {NoStop}%
\bibitem [{\citenamefont {Volovik}\ and\ \citenamefont
  {Zubkov}(2017)}]{volovik2017standard}%
  \BibitemOpen
  \bibfield  {author} {\bibinfo {author} {\bibfnamefont {G.}~\bibnamefont
  {Volovik}}\ and\ \bibinfo {author} {\bibfnamefont {M.}~\bibnamefont
  {Zubkov}},\ }\bibfield  {title} {\bibinfo {title} {Standard model as the
  topological material},\ }\href@noop {} {\bibfield  {journal} {\bibinfo
  {journal} {New Journal of Physics}\ }\textbf {\bibinfo {volume} {19}},\
  \bibinfo {pages} {015009} (\bibinfo {year} {2017})}\BibitemShut {NoStop}%
\bibitem [{\citenamefont {Zubkov}\ and\ \citenamefont
  {Khaidukov}(2017{\natexlab{a}})}]{zubkov2017topology}%
  \BibitemOpen
  \bibfield  {author} {\bibinfo {author} {\bibfnamefont {M.~A.}\ \bibnamefont
  {Zubkov}}\ and\ \bibinfo {author} {\bibfnamefont {Z.~V.}\ \bibnamefont
  {Khaidukov}},\ }\bibfield  {title} {\bibinfo {title} {Topology of the
  momentum space, wigner transformations, and a chiral anomaly in lattice
  models},\ }\href@noop {} {\bibfield  {journal} {\bibinfo  {journal} {JETP
  Letters}\ }\textbf {\bibinfo {volume} {106}},\ \bibinfo {pages} {172}
  (\bibinfo {year} {2017}{\natexlab{a}})}\BibitemShut {NoStop}%
\bibitem [{\citenamefont {Zubkov}\ and\ \citenamefont
  {Khaidukov}(2017{\natexlab{b}})}]{ZubkovKhaidukov2017}%
  \BibitemOpen
  \bibfield  {author} {\bibinfo {author} {\bibfnamefont {M.~A.}\ \bibnamefont
  {Zubkov}}\ and\ \bibinfo {author} {\bibfnamefont {Z.~V.}\ \bibnamefont
  {Khaidukov}},\ }\bibfield  {title} {\bibinfo {title} {Topology of the
  momentum space, wigner transformations, and a chiral anomaly in lattice
  models},\ }\href@noop {} {\bibfield  {journal} {\bibinfo  {journal} {JETP
  Letters}\ }\textbf {\bibinfo {volume} {106}},\ \bibinfo {pages} {172}
  (\bibinfo {year} {2017}{\natexlab{b}})}\BibitemShut {NoStop}%
\bibitem [{\citenamefont {Volovik}\ and\ \citenamefont
  {Zubkov}(2013)}]{volovik2013nambu}%
  \BibitemOpen
  \bibfield  {author} {\bibinfo {author} {\bibfnamefont {G.~E.}\ \bibnamefont
  {Volovik}}\ and\ \bibinfo {author} {\bibfnamefont {M.}~\bibnamefont
  {Zubkov}},\ }\bibfield  {title} {\bibinfo {title} {Nambu sum rule in the njl
  models: from superfluidity to top quark condensation},\ }\href@noop {}
  {\bibfield  {journal} {\bibinfo  {journal} {JETP letters}\ }\textbf {\bibinfo
  {volume} {97}},\ \bibinfo {pages} {301} (\bibinfo {year} {2013})}\BibitemShut
  {NoStop}%
\bibitem [{\citenamefont {Katsnelson}\ \emph {et~al.}(2013)\citenamefont
  {Katsnelson}, \citenamefont {Volovik},\ and\ \citenamefont
  {Zubkov}}]{katsnelson2013euler}%
  \BibitemOpen
  \bibfield  {author} {\bibinfo {author} {\bibfnamefont {M.}~\bibnamefont
  {Katsnelson}}, \bibinfo {author} {\bibfnamefont {G.}~\bibnamefont
  {Volovik}},\ and\ \bibinfo {author} {\bibfnamefont {M.}~\bibnamefont
  {Zubkov}},\ }\bibfield  {title} {\bibinfo {title} {Euler--heisenberg
  effective action and magnetoelectric effect in multilayer graphene},\
  }\href@noop {} {\bibfield  {journal} {\bibinfo  {journal} {Annals of
  Physics}\ }\textbf {\bibinfo {volume} {331}},\ \bibinfo {pages} {160}
  (\bibinfo {year} {2013})}\BibitemShut {NoStop}%
\bibitem [{\citenamefont {Volovik}\ and\ \citenamefont
  {Zubkov}(2015)}]{volovik2015scalar}%
  \BibitemOpen
  \bibfield  {author} {\bibinfo {author} {\bibfnamefont {G.}~\bibnamefont
  {Volovik}}\ and\ \bibinfo {author} {\bibfnamefont {M.}~\bibnamefont
  {Zubkov}},\ }\bibfield  {title} {\bibinfo {title} {Scalar excitation with
  leggett frequency in he 3-b and the 125 gev higgs particle in top quark
  condensation models as pseudo-goldstone bosons},\ }\href@noop {} {\bibfield
  {journal} {\bibinfo  {journal} {Physical Review D}\ }\textbf {\bibinfo
  {volume} {92}},\ \bibinfo {pages} {055004} (\bibinfo {year}
  {2015})}\BibitemShut {NoStop}%
\bibitem [{\citenamefont {Selch}\ \emph {et~al.}(2025)\citenamefont {Selch},
  \citenamefont {Zubkov}, \citenamefont {Pramanik},\ and\ \citenamefont
  {Lewkowicz}}]{selch2025non}%
  \BibitemOpen
  \bibfield  {author} {\bibinfo {author} {\bibfnamefont {M.}~\bibnamefont
  {Selch}}, \bibinfo {author} {\bibfnamefont {M.}~\bibnamefont {Zubkov}},
  \bibinfo {author} {\bibfnamefont {S.}~\bibnamefont {Pramanik}},\ and\
  \bibinfo {author} {\bibfnamefont {M.}~\bibnamefont {Lewkowicz}},\ }\bibfield
  {title} {\bibinfo {title} {Non-renormalization of the fractional quantum hall
  conductivity by interactions},\ }\href@noop {} {\bibfield  {journal}
  {\bibinfo  {journal} {Annals of Physics}\ ,\ \bibinfo {pages} {170202}}
  (\bibinfo {year} {2025})}\BibitemShut {NoStop}%
\bibitem [{\citenamefont {Bakker}\ \emph {et~al.}(1999)\citenamefont {Bakker},
  \citenamefont {Veselov},\ and\ \citenamefont {Zubkov}}]{bakker1999central}%
  \BibitemOpen
  \bibfield  {author} {\bibinfo {author} {\bibfnamefont {B.}~\bibnamefont
  {Bakker}}, \bibinfo {author} {\bibfnamefont {A.}~\bibnamefont {Veselov}},\
  and\ \bibinfo {author} {\bibfnamefont {M.}~\bibnamefont {Zubkov}},\
  }\bibfield  {title} {\bibinfo {title} {Central dominance and the confinement
  mechanism in gluodynamics},\ }\href@noop {} {\bibfield  {journal} {\bibinfo
  {journal} {Physics Letters B}\ }\textbf {\bibinfo {volume} {471}},\ \bibinfo
  {pages} {214} (\bibinfo {year} {1999})}\BibitemShut {NoStop}%
\bibitem [{\citenamefont {Bakker}\ \emph {et~al.}(2005)\citenamefont {Bakker},
  \citenamefont {Veselov},\ and\ \citenamefont {Zubkov}}]{bakker2005standard}%
  \BibitemOpen
  \bibfield  {author} {\bibinfo {author} {\bibfnamefont {B.}~\bibnamefont
  {Bakker}}, \bibinfo {author} {\bibfnamefont {A.}~\bibnamefont {Veselov}},\
  and\ \bibinfo {author} {\bibfnamefont {M.}~\bibnamefont {Zubkov}},\
  }\bibfield  {title} {\bibinfo {title} {Standard model with the additional z6
  symmetry on the lattice},\ }\href@noop {} {\bibfield  {journal} {\bibinfo
  {journal} {Physics Letters B}\ }\textbf {\bibinfo {volume} {620}},\ \bibinfo
  {pages} {156} (\bibinfo {year} {2005})}\BibitemShut {NoStop}%
\bibitem [{\citenamefont {Zubkov}\ and\ \citenamefont {Wu}(2019)}]{ZW2019}%
  \BibitemOpen
  \bibfield  {author} {\bibinfo {author} {\bibfnamefont {M.~A.}\ \bibnamefont
  {Zubkov}}\ and\ \bibinfo {author} {\bibfnamefont {X.}~\bibnamefont {Wu}},\
  }\href@noop {} {\bibinfo {title} {{Topological invariant in terms of the
  Green functions for the Quantum Hall Effect in the presence of varying
  magnetic field}}} (\bibinfo {year} {2019}),\ \Eprint
  {https://arxiv.org/abs/arXiv:1901.06661} {arXiv:arXiv:1901.06661
  [cond-mat.mes-hall]} \BibitemShut {NoStop}%
\bibitem [{\citenamefont {Zhang}\ and\ \citenamefont {Zubkov}(2022)}]{ZZ2022}%
  \BibitemOpen
  \bibfield  {author} {\bibinfo {author} {\bibfnamefont {C.}~\bibnamefont
  {Zhang}}\ and\ \bibinfo {author} {\bibfnamefont {M.}~\bibnamefont {Zubkov}},\
  }\bibfield  {title} {\bibinfo {title} {Influence of interactions on integer
  quantum hall effect},\ }\href@noop {} {\bibfield  {journal} {\bibinfo
  {journal} {Annals of Physics}\ }\textbf {\bibinfo {volume} {444}},\ \bibinfo
  {pages} {169016} (\bibinfo {year} {2022})}\BibitemShut {NoStop}%
\bibitem [{\citenamefont {Fialkovsky}\ and\ \citenamefont
  {Zubkov}(2020{\natexlab{a}})}]{FZ2019}%
  \BibitemOpen
  \bibfield  {author} {\bibinfo {author} {\bibfnamefont {I.~V.}\ \bibnamefont
  {Fialkovsky}}\ and\ \bibinfo {author} {\bibfnamefont {M.~A.}\ \bibnamefont
  {Zubkov}},\ }\bibfield  {title} {\bibinfo {title} {{Elastic deformations and
  Wigner-Weyl formalism in graphene}},\ }\href@noop {} {\bibfield  {journal}
  {\bibinfo  {journal} {Symmetry}\ }\textbf {\bibinfo {volume} {12}},\ \bibinfo
  {pages} {317} (\bibinfo {year} {2020}{\natexlab{a}})},\ \Eprint
  {https://arxiv.org/abs/1905.11097} {arXiv:1905.11097} \BibitemShut {NoStop}%
\bibitem [{\citenamefont {Chernodub}\ and\ \citenamefont
  {Zubkov}(2017)}]{chernodub2017scale}%
  \BibitemOpen
  \bibfield  {author} {\bibinfo {author} {\bibfnamefont {M.~N.}\ \bibnamefont
  {Chernodub}}\ and\ \bibinfo {author} {\bibfnamefont {M.}~\bibnamefont
  {Zubkov}},\ }\bibfield  {title} {\bibinfo {title} {Scale magnetic effect in
  quantum electrodynamics and the wigner-weyl formalism},\ }\href@noop {}
  {\bibfield  {journal} {\bibinfo  {journal} {Physical Review D}\ }\textbf
  {\bibinfo {volume} {96}},\ \bibinfo {pages} {056006} (\bibinfo {year}
  {2017})}\BibitemShut {NoStop}%
\bibitem [{\citenamefont {Zhang}\ and\ \citenamefont
  {Zubkov}(2020)}]{zhang2020influence}%
  \BibitemOpen
  \bibfield  {author} {\bibinfo {author} {\bibfnamefont {C.}~\bibnamefont
  {Zhang}}\ and\ \bibinfo {author} {\bibfnamefont {M.}~\bibnamefont {Zubkov}},\
  }\bibfield  {title} {\bibinfo {title} {Influence of interactions on the
  anomalous quantum hall effect},\ }\href@noop {} {\bibfield  {journal}
  {\bibinfo  {journal} {Journal of Physics A: Mathematical and Theoretical}\
  }\textbf {\bibinfo {volume} {53}},\ \bibinfo {pages} {195002} (\bibinfo
  {year} {2020})}\BibitemShut {NoStop}%
\bibitem [{\citenamefont {Suleymanov}\ and\ \citenamefont
  {Zubkov}(2019{\natexlab{a}})}]{suleymanov2019wigner}%
  \BibitemOpen
  \bibfield  {author} {\bibinfo {author} {\bibfnamefont {M.}~\bibnamefont
  {Suleymanov}}\ and\ \bibinfo {author} {\bibfnamefont {M.}~\bibnamefont
  {Zubkov}},\ }\bibfield  {title} {\bibinfo {title} {Wigner--weyl formalism and
  the propagator of wilson fermions in the presence of varying external
  electromagnetic field},\ }\href@noop {} {\bibfield  {journal} {\bibinfo
  {journal} {Nuclear Physics B}\ }\textbf {\bibinfo {volume} {938}},\ \bibinfo
  {pages} {171} (\bibinfo {year} {2019}{\natexlab{a}})}\BibitemShut {NoStop}%
\bibitem [{\citenamefont {Zhang}\ and\ \citenamefont
  {Zubkov}(2019{\natexlab{b}})}]{zhang2019hall}%
  \BibitemOpen
  \bibfield  {author} {\bibinfo {author} {\bibfnamefont {C.}~\bibnamefont
  {Zhang}}\ and\ \bibinfo {author} {\bibfnamefont {M.}~\bibnamefont {Zubkov}},\
  }\bibfield  {title} {\bibinfo {title} {Hall conductivity as the topological
  invariant in the phase space in the presence of interactions and a nonuniform
  magnetic field},\ }\href@noop {} {\bibfield  {journal} {\bibinfo  {journal}
  {JETP letters}\ }\textbf {\bibinfo {volume} {110}},\ \bibinfo {pages} {487}
  (\bibinfo {year} {2019}{\natexlab{b}})}\BibitemShut {NoStop}%
\bibitem [{\citenamefont {Zubkov}\ and\ \citenamefont
  {Abramchuk}(2023)}]{zubkov2023effect}%
  \BibitemOpen
  \bibfield  {author} {\bibinfo {author} {\bibfnamefont {M.~A.}\ \bibnamefont
  {Zubkov}}\ and\ \bibinfo {author} {\bibfnamefont {R.~A.}\ \bibnamefont
  {Abramchuk}},\ }\bibfield  {title} {\bibinfo {title} {Effect of interactions
  on the topological expression for the chiral separation effect},\ }\href
  {https://doi.org/10.1103/PhysRevD.107.094021} {\bibfield  {journal} {\bibinfo
   {journal} {Physical Review D}\ }\textbf {\bibinfo {volume} {107}},\ \bibinfo
  {pages} {094021} (\bibinfo {year} {2023})}\BibitemShut {NoStop}%
\bibitem [{\citenamefont {Abramchuk}\ \emph {et~al.}(2018)\citenamefont
  {Abramchuk}, \citenamefont {Khaidukov},\ and\ \citenamefont
  {Zubkov}}]{abramchuk2018anatomy}%
  \BibitemOpen
  \bibfield  {author} {\bibinfo {author} {\bibfnamefont {R.}~\bibnamefont
  {Abramchuk}}, \bibinfo {author} {\bibfnamefont {Z.}~\bibnamefont
  {Khaidukov}},\ and\ \bibinfo {author} {\bibfnamefont {M.}~\bibnamefont
  {Zubkov}},\ }\bibfield  {title} {\bibinfo {title} {Anatomy of the chiral
  vortical effect},\ }\href@noop {} {\bibfield  {journal} {\bibinfo  {journal}
  {Physical Review D}\ }\textbf {\bibinfo {volume} {98}},\ \bibinfo {pages}
  {076013} (\bibinfo {year} {2018})}\BibitemShut {NoStop}%
\bibitem [{\citenamefont {Fialkovsky}\ and\ \citenamefont
  {Zubkov}(2020{\natexlab{b}})}]{FZ2019_2}%
  \BibitemOpen
  \bibfield  {author} {\bibinfo {author} {\bibfnamefont {I.~V.}\ \bibnamefont
  {Fialkovsky}}\ and\ \bibinfo {author} {\bibfnamefont {M.~A.}\ \bibnamefont
  {Zubkov}},\ }\bibfield  {title} {\bibinfo {title} {Precise wigner-weyl
  calculus for lattice models},\ }\href
  {https://doi.org/10.1016/j.nuclphysb.2020.114999} {\bibfield  {journal}
  {\bibinfo  {journal} {Nuclear Physics B}\ }\textbf {\bibinfo {volume}
  {954}},\ \bibinfo {pages} {11499} (\bibinfo {year}
  {2020}{\natexlab{b}})}\BibitemShut {NoStop}%
\bibitem [{\citenamefont {Zubkov}(2023)}]{Z2023}%
  \BibitemOpen
  \bibfield  {author} {\bibinfo {author} {\bibfnamefont {M.}~\bibnamefont
  {Zubkov}},\ }\bibfield  {title} {\bibinfo {title} {Discrete wigner--weyl
  calculus for the finite lattice},\ }\href@noop {} {\bibfield  {journal}
  {\bibinfo  {journal} {Journal of Physics A: Mathematical and Theoretical}\
  }\textbf {\bibinfo {volume} {56}},\ \bibinfo {pages} {395201} (\bibinfo
  {year} {2023})}\BibitemShut {NoStop}%
\bibitem [{\citenamefont {Connes}(1994)}]{Connes}%
  \BibitemOpen
  \bibfield  {author} {\bibinfo {author} {\bibfnamefont {A.}~\bibnamefont
  {Connes}},\ }\href@noop {} {\emph {\bibinfo {title} {Noncommutative
  Geometry}}},\ \bibinfo {edition} {1st}\ ed.\ (\bibinfo  {publisher} {Academic
  Press / Elsevier},\ \bibinfo {address} {San Diego / London},\ \bibinfo {year}
  {1994})\BibitemShut {NoStop}%
\bibitem [{\citenamefont {Gelfand}\ and\ \citenamefont
  {Naimark}(1943)}]{GelfandNaimark1943}%
  \BibitemOpen
  \bibfield  {author} {\bibinfo {author} {\bibfnamefont {I.~M.}\ \bibnamefont
  {Gelfand}}\ and\ \bibinfo {author} {\bibfnamefont {M.~A.}\ \bibnamefont
  {Naimark}},\ }\bibfield  {title} {\bibinfo {title} {On the imbedding of
  normed rings into the ring of operators in {Hilbert} space},\ }\href@noop {}
  {\bibfield  {journal} {\bibinfo  {journal} {Mat. Sbornik}\ }\textbf {\bibinfo
  {volume} {12(54)}},\ \bibinfo {pages} {197} (\bibinfo {year}
  {1943})}\BibitemShut {NoStop}%
\bibitem [{\citenamefont {Connes}(1985)}]{Connes1985}%
  \BibitemOpen
  \bibfield  {author} {\bibinfo {author} {\bibfnamefont {A.}~\bibnamefont
  {Connes}},\ }\bibfield  {title} {\bibinfo {title} {Noncommutative
  differential geometry},\ }\href {https://doi.org/10.1007/BF02698807}
  {\bibfield  {journal} {\bibinfo  {journal} {Publ. Math. Inst. Hautes \'Etudes
  Sci.}\ }\textbf {\bibinfo {volume} {62}},\ \bibinfo {pages} {257} (\bibinfo
  {year} {1985})}\BibitemShut {NoStop}%
\bibitem [{\citenamefont {Connes}(1995)}]{Connes1995reality}%
  \BibitemOpen
  \bibfield  {author} {\bibinfo {author} {\bibfnamefont {A.}~\bibnamefont
  {Connes}},\ }\bibfield  {title} {\bibinfo {title} {Noncommutative geometry
  and reality},\ }\href {https://doi.org/10.1063/1.531241} {\bibfield
  {journal} {\bibinfo  {journal} {J. Math. Phys.}\ }\textbf {\bibinfo {volume}
  {36}},\ \bibinfo {pages} {6194} (\bibinfo {year} {1995})}\BibitemShut
  {NoStop}%
\bibitem [{\citenamefont {Connes}\ and\ \citenamefont
  {Moscovici}(1995)}]{ConnesMoscovici1995}%
  \BibitemOpen
  \bibfield  {author} {\bibinfo {author} {\bibfnamefont {A.}~\bibnamefont
  {Connes}}\ and\ \bibinfo {author} {\bibfnamefont {H.}~\bibnamefont
  {Moscovici}},\ }\bibfield  {title} {\bibinfo {title} {The local index formula
  in noncommutative geometry},\ }\href {https://doi.org/10.1007/BF01895667}
  {\bibfield  {journal} {\bibinfo  {journal} {Geom. Funct. Anal. (GAFA)}\
  }\textbf {\bibinfo {volume} {5}},\ \bibinfo {pages} {174} (\bibinfo {year}
  {1995})}\BibitemShut {NoStop}%
\bibitem [{\citenamefont {Connes}\ and\ \citenamefont
  {Lott}(1991)}]{ConnesLott1991}%
  \BibitemOpen
  \bibfield  {author} {\bibinfo {author} {\bibfnamefont {A.}~\bibnamefont
  {Connes}}\ and\ \bibinfo {author} {\bibfnamefont {J.}~\bibnamefont {Lott}},\
  }\bibfield  {title} {\bibinfo {title} {Particle models and noncommutative
  geometry},\ }\href {https://doi.org/10.1016/0920-5632(91)90120-4} {\bibfield
  {journal} {\bibinfo  {journal} {Nucl. Phys. B Proc. Suppl.}\ }\textbf
  {\bibinfo {volume} {18B}},\ \bibinfo {pages} {29} (\bibinfo {year}
  {1991})}\BibitemShut {NoStop}%
\bibitem [{\citenamefont {Chamseddine}\ and\ \citenamefont
  {Connes}(1997)}]{ChamseddineConnes1997}%
  \BibitemOpen
  \bibfield  {author} {\bibinfo {author} {\bibfnamefont {A.~H.}\ \bibnamefont
  {Chamseddine}}\ and\ \bibinfo {author} {\bibfnamefont {A.}~\bibnamefont
  {Connes}},\ }\bibfield  {title} {\bibinfo {title} {The spectral action
  principle},\ }\href {https://doi.org/10.1007/s002200050126} {\bibfield
  {journal} {\bibinfo  {journal} {Commun. Math. Phys.}\ }\textbf {\bibinfo
  {volume} {186}},\ \bibinfo {pages} {731} (\bibinfo {year}
  {1997})}\BibitemShut {NoStop}%
\bibitem [{\citenamefont {Connes}(1996)}]{ConnesGravityMatter1996}%
  \BibitemOpen
  \bibfield  {author} {\bibinfo {author} {\bibfnamefont {A.}~\bibnamefont
  {Connes}},\ }\bibfield  {title} {\bibinfo {title} {Gravity coupled with
  matter and the foundation of non-commutative geometry},\ }\href
  {https://doi.org/10.1007/BF02506388} {\bibfield  {journal} {\bibinfo
  {journal} {Commun. Math. Phys.}\ }\textbf {\bibinfo {volume} {182}},\
  \bibinfo {pages} {155} (\bibinfo {year} {1996})}\BibitemShut {NoStop}%
\bibitem [{\citenamefont {Chamseddine}\ \emph {et~al.}(2007)\citenamefont
  {Chamseddine}, \citenamefont {Connes},\ and\ \citenamefont
  {Marcolli}}]{ChamseddineConnesMarcolli2007}%
  \BibitemOpen
  \bibfield  {author} {\bibinfo {author} {\bibfnamefont {A.~H.}\ \bibnamefont
  {Chamseddine}}, \bibinfo {author} {\bibfnamefont {A.}~\bibnamefont
  {Connes}},\ and\ \bibinfo {author} {\bibfnamefont {M.}~\bibnamefont
  {Marcolli}},\ }\bibfield  {title} {\bibinfo {title} {Gravity and the standard
  model with neutrino mixing},\ }\href
  {https://doi.org/10.4310/ATMP.2007.v11.n6.a3} {\bibfield  {journal} {\bibinfo
   {journal} {Adv. Theor. Math. Phys.}\ }\textbf {\bibinfo {volume} {11}},\
  \bibinfo {pages} {991} (\bibinfo {year} {2007})}\BibitemShut {NoStop}%
\bibitem [{\citenamefont {Woronowicz}(1987)}]{Woronowicz1987}%
  \BibitemOpen
  \bibfield  {author} {\bibinfo {author} {\bibfnamefont {S.~L.}\ \bibnamefont
  {Woronowicz}},\ }\bibfield  {title} {\bibinfo {title} {Compact matrix
  pseudogroups},\ }\href {https://doi.org/10.1007/BF01219077} {\bibfield
  {journal} {\bibinfo  {journal} {Commun. Math. Phys.}\ }\textbf {\bibinfo
  {volume} {111}},\ \bibinfo {pages} {613} (\bibinfo {year}
  {1987})}\BibitemShut {NoStop}%
\bibitem [{\citenamefont {Doplicher}\ \emph {et~al.}(1995)\citenamefont
  {Doplicher}, \citenamefont {Fredenhagen},\ and\ \citenamefont
  {Roberts}}]{DoplicherFredenhagenRoberts1995}%
  \BibitemOpen
  \bibfield  {author} {\bibinfo {author} {\bibfnamefont {S.}~\bibnamefont
  {Doplicher}}, \bibinfo {author} {\bibfnamefont {K.}~\bibnamefont
  {Fredenhagen}},\ and\ \bibinfo {author} {\bibfnamefont {J.~E.}\ \bibnamefont
  {Roberts}},\ }\bibfield  {title} {\bibinfo {title} {The quantum structure of
  spacetime at the {Planck} scale and quantum fields},\ }\href
  {https://doi.org/10.1007/BF02104515} {\bibfield  {journal} {\bibinfo
  {journal} {Commun. Math. Phys.}\ }\textbf {\bibinfo {volume} {172}},\
  \bibinfo {pages} {187} (\bibinfo {year} {1995})}\BibitemShut {NoStop}%
\bibitem [{\citenamefont {Madore}(1992)}]{Madore1992}%
  \BibitemOpen
  \bibfield  {author} {\bibinfo {author} {\bibfnamefont {J.}~\bibnamefont
  {Madore}},\ }\bibfield  {title} {\bibinfo {title} {The fuzzy sphere},\ }\href
  {https://doi.org/10.1088/0264-9381/9/1/008} {\bibfield  {journal} {\bibinfo
  {journal} {Class. Quantum Grav.}\ }\textbf {\bibinfo {volume} {9}},\ \bibinfo
  {pages} {69} (\bibinfo {year} {1992})}\BibitemShut {NoStop}%
\bibitem [{\citenamefont {Seiberg}\ and\ \citenamefont
  {Witten}(1999)}]{SeibergWitten1999}%
  \BibitemOpen
  \bibfield  {author} {\bibinfo {author} {\bibfnamefont {N.}~\bibnamefont
  {Seiberg}}\ and\ \bibinfo {author} {\bibfnamefont {E.}~\bibnamefont
  {Witten}},\ }\bibfield  {title} {\bibinfo {title} {String theory and
  noncommutative geometry},\ }\href
  {https://doi.org/10.1088/1126-6708/1999/09/032} {\bibfield  {journal}
  {\bibinfo  {journal} {J. High Energy Phys.}\ }\textbf {\bibinfo {volume}
  {1999}}\bibinfo  {number} { (09)},\ \bibinfo {pages} {032}}\BibitemShut
  {NoStop}%
\bibitem [{\citenamefont {Connes}\ \emph {et~al.}(1998)\citenamefont {Connes},
  \citenamefont {Douglas},\ and\ \citenamefont
  {Schwarz}}]{ConnesDouglasSchwarz1998}%
  \BibitemOpen
\bibfield  {number} {  }\bibfield  {author} {\bibinfo {author} {\bibfnamefont
  {A.}~\bibnamefont {Connes}}, \bibinfo {author} {\bibfnamefont {M.~R.}\
  \bibnamefont {Douglas}},\ and\ \bibinfo {author} {\bibfnamefont
  {A.}~\bibnamefont {Schwarz}},\ }\bibfield  {title} {\bibinfo {title}
  {Noncommutative geometry and matrix theory: compactification on tori},\
  }\href {https://doi.org/10.1088/1126-6708/1998/02/003} {\bibfield  {journal}
  {\bibinfo  {journal} {J. High Energy Phys.}\ }\textbf {\bibinfo {volume}
  {1998}}\bibinfo  {number} { (02)},\ \bibinfo {pages} {003}}\BibitemShut
  {NoStop}%
\bibitem [{\citenamefont {Gracia-Bond{\'i}a}\ \emph {et~al.}(2001)\citenamefont
  {Gracia-Bond{\'i}a}, \citenamefont {V{\'a}rilly},\ and\ \citenamefont
  {Figueroa}}]{GraciaBondiaVarillyFigueroa2001}%
  \BibitemOpen
\bibfield  {number} {  }\bibfield  {author} {\bibinfo {author} {\bibfnamefont
  {J.~M.}\ \bibnamefont {Gracia-Bond{\'i}a}}, \bibinfo {author} {\bibfnamefont
  {J.~C.}\ \bibnamefont {V{\'a}rilly}},\ and\ \bibinfo {author} {\bibfnamefont
  {H.}~\bibnamefont {Figueroa}},\ }\href
  {https://doi.org/10.1007/978-1-4612-0005-5} {\emph {\bibinfo {title}
  {Elements of Noncommutative Geometry}}},\ Birkh\"auser Advanced Texts\
  (\bibinfo  {publisher} {Birkh\"auser},\ \bibinfo {address} {Boston},\
  \bibinfo {year} {2001})\BibitemShut {NoStop}%
\bibitem [{\citenamefont {Bellissard}(1986)}]{Bellissard1986}%
  \BibitemOpen
  \bibfield  {author} {\bibinfo {author} {\bibfnamefont {J.}~\bibnamefont
  {Bellissard}},\ }\bibfield  {title} {\bibinfo {title} {K-theory of
  c*-algebras in solid state physics},\ }in\ \href
  {https://doi.org/10.1007/3-540-16777-3_74} {\emph {\bibinfo {booktitle}
  {Statistical Mechanics and Field Theory: Mathematical Aspects}}},\ \bibinfo
  {series} {Lecture Notes in Physics}, Vol.\ \bibinfo {volume} {257},\ \bibinfo
  {editor} {edited by\ \bibinfo {editor} {\bibfnamefont {T.~C.}\ \bibnamefont
  {Dorlas}}, \bibinfo {editor} {\bibfnamefont {N.~M.}\ \bibnamefont
  {Hugenholtz}},\ and\ \bibinfo {editor} {\bibfnamefont {M.}~\bibnamefont
  {Winnink}}}\ (\bibinfo  {publisher} {Springer},\ \bibinfo {address}
  {Berlin},\ \bibinfo {year} {1986})\ pp.\ \bibinfo {pages}
  {99--156}\BibitemShut {NoStop}%
\bibitem [{\citenamefont {Avron}\ \emph {et~al.}(1994)\citenamefont {Avron},
  \citenamefont {Seiler},\ and\ \citenamefont {Simon}}]{AvronSeilerSimon1994}%
  \BibitemOpen
  \bibfield  {author} {\bibinfo {author} {\bibfnamefont {J.~E.}\ \bibnamefont
  {Avron}}, \bibinfo {author} {\bibfnamefont {R.}~\bibnamefont {Seiler}},\ and\
  \bibinfo {author} {\bibfnamefont {B.}~\bibnamefont {Simon}},\ }\bibfield
  {title} {\bibinfo {title} {Charge deficiency, charge transport and comparison
  of dimensions},\ }\href {https://doi.org/10.1007/BF02102644} {\bibfield
  {journal} {\bibinfo  {journal} {Commun. Math. Phys.}\ }\textbf {\bibinfo
  {volume} {159}},\ \bibinfo {pages} {399} (\bibinfo {year}
  {1994})}\BibitemShut {NoStop}%
\bibitem [{\citenamefont {Schulz-Baldes}\ \emph {et~al.}(2000)\citenamefont
  {Schulz-Baldes}, \citenamefont {Kellendonk},\ and\ \citenamefont
  {Richter}}]{SchulzBaldesKellendonkRichter2000}%
  \BibitemOpen
  \bibfield  {author} {\bibinfo {author} {\bibfnamefont {H.}~\bibnamefont
  {Schulz-Baldes}}, \bibinfo {author} {\bibfnamefont {J.}~\bibnamefont
  {Kellendonk}},\ and\ \bibinfo {author} {\bibfnamefont {T.}~\bibnamefont
  {Richter}},\ }\bibfield  {title} {\bibinfo {title} {Simultaneous quantization
  of edge and bulk hall conductivity},\ }\href
  {https://doi.org/10.1088/0305-4470/33/2/102} {\bibfield  {journal} {\bibinfo
  {journal} {J. Phys. A: Math. Gen.}\ }\textbf {\bibinfo {volume} {33}},\
  \bibinfo {pages} {L27} (\bibinfo {year} {2000})}\BibitemShut {NoStop}%
\bibitem [{\citenamefont {Prodan}\ and\ \citenamefont
  {Schulz-Baldes}(2016)}]{ProdanSchulzBaldes2016}%
  \BibitemOpen
  \bibfield  {author} {\bibinfo {author} {\bibfnamefont {E.}~\bibnamefont
  {Prodan}}\ and\ \bibinfo {author} {\bibfnamefont {H.}~\bibnamefont
  {Schulz-Baldes}},\ }\href {https://doi.org/10.1007/978-3-319-29351-6} {\emph
  {\bibinfo {title} {Bulk and Boundary Invariants for Complex Topological
  Insulators: From K-Theory to Physics}}},\ Mathematical Physics Studies\
  (\bibinfo  {publisher} {Springer},\ \bibinfo {address} {Cham},\ \bibinfo
  {year} {2016})\BibitemShut {NoStop}%
\bibitem [{\citenamefont {De~Nittis}\ and\ \citenamefont
  {Sandoval}(2020)}]{DeNittisSandoval2020}%
  \BibitemOpen
  \bibfield  {author} {\bibinfo {author} {\bibfnamefont {G.}~\bibnamefont
  {De~Nittis}}\ and\ \bibinfo {author} {\bibfnamefont {M.}~\bibnamefont
  {Sandoval}},\ }\bibfield  {title} {\bibinfo {title} {The noncommutative
  geometry of the {Landau} hamiltonian: metric aspects},\ }\href
  {https://doi.org/10.3842/SIGMA.2020.146} {\bibfield  {journal} {\bibinfo
  {journal} {SIGMA}\ }\textbf {\bibinfo {volume} {16}},\ \bibinfo {pages} {146}
  (\bibinfo {year} {2020})}\BibitemShut {NoStop}%
\bibitem [{\citenamefont {Khalkhali}\ and\ \citenamefont
  {Pagliaroli}(2021)}]{KhalkhaliPagliaroli2021}%
  \BibitemOpen
  \bibfield  {author} {\bibinfo {author} {\bibfnamefont {M.}~\bibnamefont
  {Khalkhali}}\ and\ \bibinfo {author} {\bibfnamefont {N.}~\bibnamefont
  {Pagliaroli}},\ }\bibfield  {title} {\bibinfo {title} {Phase transition in
  random noncommutative geometry},\ }\href
  {https://doi.org/10.1088/1751-8121/abd190} {\bibfield  {journal} {\bibinfo
  {journal} {J. Phys. A: Math. Theor.}\ }\textbf {\bibinfo {volume} {54}},\
  \bibinfo {pages} {035202} (\bibinfo {year} {2021})}\BibitemShut {NoStop}%
\bibitem [{\citenamefont {Hessam}\ \emph
  {et~al.}(2022{\natexlab{a}})\citenamefont {Hessam}, \citenamefont
  {Khalkhali},\ and\ \citenamefont
  {Pagliaroli}}]{HessamKhalkhaliPagliaroli2022}%
  \BibitemOpen
  \bibfield  {author} {\bibinfo {author} {\bibfnamefont {H.}~\bibnamefont
  {Hessam}}, \bibinfo {author} {\bibfnamefont {M.}~\bibnamefont {Khalkhali}},\
  and\ \bibinfo {author} {\bibfnamefont {N.}~\bibnamefont {Pagliaroli}},\
  }\bibfield  {title} {\bibinfo {title} {Bootstrapping dirac ensembles},\
  }\href {https://doi.org/10.1088/1751-8121/ac5216} {\bibfield  {journal}
  {\bibinfo  {journal} {J. Phys. A: Math. Theor.}\ }\textbf {\bibinfo {volume}
  {55}},\ \bibinfo {pages} {335204} (\bibinfo {year}
  {2022}{\natexlab{a}})}\BibitemShut {NoStop}%
\bibitem [{\citenamefont {Khalkhali}\ and\ \citenamefont
  {Pagliaroli}(2022)}]{KhalkhaliPagliaroli2022}%
  \BibitemOpen
  \bibfield  {author} {\bibinfo {author} {\bibfnamefont {M.}~\bibnamefont
  {Khalkhali}}\ and\ \bibinfo {author} {\bibfnamefont {N.}~\bibnamefont
  {Pagliaroli}},\ }\bibfield  {title} {\bibinfo {title} {Spectral statistics of
  dirac ensembles},\ }\href {https://doi.org/10.1063/5.0078267} {\bibfield
  {journal} {\bibinfo  {journal} {J. Math. Phys.}\ }\textbf {\bibinfo {volume}
  {63}},\ \bibinfo {pages} {053504} (\bibinfo {year} {2022})}\BibitemShut
  {NoStop}%
\bibitem [{\citenamefont {Hessam}\ \emph
  {et~al.}(2022{\natexlab{b}})\citenamefont {Hessam}, \citenamefont
  {Khalkhali}, \citenamefont {Pagliaroli},\ and\ \citenamefont
  {Verhoeven}}]{HessamEtal2022review}%
  \BibitemOpen
  \bibfield  {author} {\bibinfo {author} {\bibfnamefont {H.}~\bibnamefont
  {Hessam}}, \bibinfo {author} {\bibfnamefont {M.}~\bibnamefont {Khalkhali}},
  \bibinfo {author} {\bibfnamefont {N.}~\bibnamefont {Pagliaroli}},\ and\
  \bibinfo {author} {\bibfnamefont {L.}~\bibnamefont {Verhoeven}},\ }\bibfield
  {title} {\bibinfo {title} {From noncommutative geometry to random matrix
  theory},\ }\href {https://doi.org/10.1088/1751-8121/ac8fc5} {\bibfield
  {journal} {\bibinfo  {journal} {J. Phys. A: Math. Theor.}\ }\textbf {\bibinfo
  {volume} {55}},\ \bibinfo {pages} {383001} (\bibinfo {year}
  {2022}{\natexlab{b}})},\ \bibinfo {note} {topical review}\BibitemShut
  {NoStop}%
\bibitem [{\citenamefont {Connes}\ and\ \citenamefont
  {Consani}(2023)}]{ConnesConsani2023}%
  \BibitemOpen
  \bibfield  {author} {\bibinfo {author} {\bibfnamefont {A.}~\bibnamefont
  {Connes}}\ and\ \bibinfo {author} {\bibfnamefont {C.}~\bibnamefont
  {Consani}},\ }\bibfield  {title} {\bibinfo {title} {Spectral triples and
  $\zeta$-cycles},\ }\href@noop {} {\bibfield  {journal} {\bibinfo  {journal}
  {Enseign. Math.}\ }\textbf {\bibinfo {volume} {69}},\ \bibinfo {pages} {93}
  (\bibinfo {year} {2023})},\ \bibinfo {note} {arXiv:2106.01715}\BibitemShut
  {NoStop}%
\bibitem [{\citenamefont {Connes}\ and\ \citenamefont
  {Moscovici}(2022)}]{ConnesMoscovici2022}%
  \BibitemOpen
  \bibfield  {author} {\bibinfo {author} {\bibfnamefont {A.}~\bibnamefont
  {Connes}}\ and\ \bibinfo {author} {\bibfnamefont {H.}~\bibnamefont
  {Moscovici}},\ }\bibfield  {title} {\bibinfo {title} {The uv prolate spectrum
  matches the zeros of zeta},\ }\href {https://doi.org/10.1073/pnas.2123174119}
  {\bibfield  {journal} {\bibinfo  {journal} {Proc. Natl. Acad. Sci. USA}\
  }\textbf {\bibinfo {volume} {119}},\ \bibinfo {pages} {e2123174119} (\bibinfo
  {year} {2022})}\BibitemShut {NoStop}%
\bibitem [{\citenamefont {Connes}\ \emph {et~al.}(2024)\citenamefont {Connes},
  \citenamefont {Consani},\ and\ \citenamefont
  {Moscovici}}]{ConnesConsaniMoscovici2024}%
  \BibitemOpen
  \bibfield  {author} {\bibinfo {author} {\bibfnamefont {A.}~\bibnamefont
  {Connes}}, \bibinfo {author} {\bibfnamefont {C.}~\bibnamefont {Consani}},\
  and\ \bibinfo {author} {\bibfnamefont {H.}~\bibnamefont {Moscovici}},\
  }\bibfield  {title} {\bibinfo {title} {Zeta zeros and prolate wave
  operators},\ }\href {https://doi.org/10.1007/s43034-024-00388-z} {\bibfield
  {journal} {\bibinfo  {journal} {Ann. Funct. Anal.}\ }\textbf {\bibinfo
  {volume} {15}},\ \bibinfo {pages} {87} (\bibinfo {year} {2024})}\BibitemShut
  {NoStop}%
\bibitem [{\citenamefont {Chamseddine}\ \emph {et~al.}(2023)\citenamefont
  {Chamseddine}, \citenamefont {Connes},\ and\ \citenamefont {van
  Suijlekom}}]{ChamseddineConnesvanSuijlekom2023}%
  \BibitemOpen
  \bibfield  {author} {\bibinfo {author} {\bibfnamefont {A.~H.}\ \bibnamefont
  {Chamseddine}}, \bibinfo {author} {\bibfnamefont {A.}~\bibnamefont
  {Connes}},\ and\ \bibinfo {author} {\bibfnamefont {W.~D.}\ \bibnamefont {van
  Suijlekom}},\ }\bibfield  {title} {\bibinfo {title} {Noncommutativity and
  physics: a non-technical review},\ }\href
  {https://doi.org/10.1140/epjs/s11734-023-00842-4} {\bibfield  {journal}
  {\bibinfo  {journal} {Eur. Phys. J. Spec. Top.}\ }\textbf {\bibinfo {volume}
  {232}},\ \bibinfo {pages} {3581} (\bibinfo {year} {2023})}\BibitemShut
  {NoStop}%
\bibitem [{\citenamefont {van Suijlekom}(2024)}]{vanSuijlekom2024}%
  \BibitemOpen
  \bibfield  {author} {\bibinfo {author} {\bibfnamefont {W.~D.}\ \bibnamefont
  {van Suijlekom}},\ }\href@noop {} {\emph {\bibinfo {title} {Noncommutative
  Geometry and Particle Physics}}},\ \bibinfo {edition} {2nd}\ ed.,\
  Mathematical Physics Studies\ (\bibinfo  {publisher} {Springer},\ \bibinfo
  {address} {Cham},\ \bibinfo {year} {2024})\ \bibinfo {note} {open
  access}\BibitemShut {NoStop}%
\bibitem [{\citenamefont {Stoiber}(2025)}]{Stoiber2025}%
  \BibitemOpen
  \bibfield  {author} {\bibinfo {author} {\bibfnamefont {T.}~\bibnamefont
  {Stoiber}},\ }\bibfield  {title} {\bibinfo {title} {A spectral localizer
  approach to strong topological invariants in the mobility gap regime},\
  }\bibfield  {journal} {\bibinfo  {journal} {Commun. Math. Phys.}\ }\href
  {https://doi.org/10.1007/s00220-025-05359-6} {10.1007/s00220-025-05359-6}
  (\bibinfo {year} {2025}),\ \bibinfo {note} {arXiv:2410.22214}\BibitemShut
  {NoStop}%
\bibitem [{\citenamefont {Chobanyan}\ and\ \citenamefont
  {Zubkov}(2024)}]{CZ2024}%
  \BibitemOpen
  \bibfield  {author} {\bibinfo {author} {\bibfnamefont {R.}~\bibnamefont
  {Chobanyan}}\ and\ \bibinfo {author} {\bibfnamefont {M.~A.}\ \bibnamefont
  {Zubkov}},\ }\bibfield  {title} {\bibinfo {title} {Precise wigner--weyl
  calculus for the honeycomb lattice},\ }\href@noop {} {\bibfield  {journal}
  {\bibinfo  {journal} {Symmetry}\ }\textbf {\bibinfo {volume} {16}},\ \bibinfo
  {pages} {1081} (\bibinfo {year} {2024})}\BibitemShut {NoStop}%
\bibitem [{\citenamefont {Onoda}\ \emph {et~al.}(2006)\citenamefont {Onoda},
  \citenamefont {Sugimoto},\ and\ \citenamefont {Nagaosa}}]{onoda2}%
  \BibitemOpen
  \bibfield  {author} {\bibinfo {author} {\bibfnamefont {S.}~\bibnamefont
  {Onoda}}, \bibinfo {author} {\bibfnamefont {N.}~\bibnamefont {Sugimoto}},\
  and\ \bibinfo {author} {\bibfnamefont {N.}~\bibnamefont {Nagaosa}},\
  }\bibfield  {title} {\bibinfo {title} {Intrinsic versus extrinsic anomalous
  hall effect in ferromagnets},\ }\href
  {https://doi.org/10.1103/PhysRevLett.97.126602} {\bibfield  {journal}
  {\bibinfo  {journal} {Phys. Rev. Lett.}\ }\textbf {\bibinfo {volume} {97}},\
  \bibinfo {pages} {126602} (\bibinfo {year} {2006})}\BibitemShut {NoStop}%
\bibitem [{\citenamefont {Banerjee}\ \emph {et~al.}(2022)\citenamefont
  {Banerjee}, \citenamefont {Lewkowicz},\ and\ \citenamefont
  {Zubkov}}]{banerjee2022chiral}%
  \BibitemOpen
  \bibfield  {author} {\bibinfo {author} {\bibfnamefont {C.}~\bibnamefont
  {Banerjee}}, \bibinfo {author} {\bibfnamefont {M.}~\bibnamefont
  {Lewkowicz}},\ and\ \bibinfo {author} {\bibfnamefont {M.~A.}\ \bibnamefont
  {Zubkov}},\ }\bibfield  {title} {\bibinfo {title} {Chiral magnetic effect out
  of equilibrium},\ }\href@noop {} {\bibfield  {journal} {\bibinfo  {journal}
  {Physical Review D}\ }\textbf {\bibinfo {volume} {106}},\ \bibinfo {pages}
  {074508} (\bibinfo {year} {2022})}\BibitemShut {NoStop}%
\bibitem [{\citenamefont {Zubkov}(2016{\natexlab{a}})}]{Zubkov2016a}%
  \BibitemOpen
  \bibfield  {author} {\bibinfo {author} {\bibfnamefont {M.~A.}\ \bibnamefont
  {Zubkov}},\ }\bibfield  {title} {\bibinfo {title} {Absence of equilibrium
  chiral magnetic effect},\ }\href {https://doi.org/10.1103/PhysRevD.93.105036}
  {\bibfield  {journal} {\bibinfo  {journal} {Phys. Rev. D}\ }\textbf {\bibinfo
  {volume} {93}},\ \bibinfo {pages} {105036} (\bibinfo {year}
  {2016}{\natexlab{a}})},\ \Eprint {https://arxiv.org/abs/1605.08724}
  {arXiv:1605.08724} \BibitemShut {NoStop}%
\bibitem [{\citenamefont {Zubkov}(2016{\natexlab{b}})}]{Zubkov2016b}%
  \BibitemOpen
  \bibfield  {author} {\bibinfo {author} {\bibfnamefont {M.~A.}\ \bibnamefont
  {Zubkov}},\ }\bibfield  {title} {\bibinfo {title} {Wigner transformation,
  momentum space topology, and anomalous transport},\ }\href
  {https://doi.org/10.1016/j.aop.2016.07.011} {\bibfield  {journal} {\bibinfo
  {journal} {Annals Phys}\ }\textbf {\bibinfo {volume} {373}},\ \bibinfo
  {pages} {298} (\bibinfo {year} {2016}{\natexlab{b}})},\ \Eprint
  {https://arxiv.org/abs/1603.03665} {arXiv:1603.03665} \BibitemShut {NoStop}%
\bibitem [{\citenamefont {Suleymanov}\ and\ \citenamefont
  {Zubkov}(2019{\natexlab{b}})}]{SZ2018}%
  \BibitemOpen
  \bibfield  {author} {\bibinfo {author} {\bibfnamefont {M.}~\bibnamefont
  {Suleymanov}}\ and\ \bibinfo {author} {\bibfnamefont {M.~A.}\ \bibnamefont
  {Zubkov}},\ }\bibfield  {title} {\bibinfo {title} {Wigner -- weyl formalism
  and the propagator of wilson fermions in the presence of varying external
  electromagnetic field},\ }\href@noop {} {\bibfield  {journal} {\bibinfo
  {journal} {Nucl. Phys}\ }\textbf {\bibinfo {volume} {938}} (\bibinfo {year}
  {2019}{\natexlab{b}})},\ \Eprint {https://arxiv.org/abs/1811.08233}
  {arXiv:1811.08233} \BibitemShut {NoStop}%
\bibitem [{\citenamefont {Fialkovsky}\ \emph {et~al.}(2020)\citenamefont
  {Fialkovsky}, \citenamefont {Suleymanov}, \citenamefont {Wu}, \citenamefont
  {Zhang},\ and\ \citenamefont {Zubkov}}]{FSWZZ2020}%
  \BibitemOpen
  \bibfield  {author} {\bibinfo {author} {\bibfnamefont {I.~V.}\ \bibnamefont
  {Fialkovsky}}, \bibinfo {author} {\bibfnamefont {M.}~\bibnamefont
  {Suleymanov}}, \bibinfo {author} {\bibfnamefont {X.}~\bibnamefont {Wu}},
  \bibinfo {author} {\bibfnamefont {C.~X.}\ \bibnamefont {Zhang}},\ and\
  \bibinfo {author} {\bibfnamefont {M.~A.}\ \bibnamefont {Zubkov}},\ }\bibfield
   {title} {\bibinfo {title} {Hall conductivity as topological invariant in
  phase space},\ }\href {https://doi.org/10.1088/1402-4896/ab7ce4} {\bibfield
  {journal} {\bibinfo  {journal} {Physica Scripta}\ }\textbf {\bibinfo {volume}
  {95}},\ \bibinfo {pages} {064003} (\bibinfo {year} {2020})},\ \Eprint
  {https://arxiv.org/abs/1910.04730} {1910.04730} \BibitemShut {NoStop}%
\bibitem [{\citenamefont {Arovas}\ \emph {et~al.}(1988)\citenamefont {Arovas},
  \citenamefont {Bhatt}, \citenamefont {Haldane}, \citenamefont {Littlewood},\
  and\ \citenamefont {Rammal}}]{ArovasEtAl1988}%
  \BibitemOpen
  \bibfield  {author} {\bibinfo {author} {\bibfnamefont {D.~P.}\ \bibnamefont
  {Arovas}}, \bibinfo {author} {\bibfnamefont {R.~N.}\ \bibnamefont {Bhatt}},
  \bibinfo {author} {\bibfnamefont {F.~D.~M.}\ \bibnamefont {Haldane}},
  \bibinfo {author} {\bibfnamefont {P.~B.}\ \bibnamefont {Littlewood}},\ and\
  \bibinfo {author} {\bibfnamefont {R.}~\bibnamefont {Rammal}},\ }\bibfield
  {title} {\bibinfo {title} {Localization, wave-function topology, and integer
  quantum hall effect},\ }\href@noop {} {\bibfield  {journal} {\bibinfo
  {journal} {Physical Review Letters}\ }\textbf {\bibinfo {volume} {60}},\
  \bibinfo {pages} {619} (\bibinfo {year} {1988})}\BibitemShut {NoStop}%
\bibitem [{\citenamefont {Huo}\ and\ \citenamefont
  {Bhatt}(1992)}]{HuoBhatt1992}%
  \BibitemOpen
  \bibfield  {author} {\bibinfo {author} {\bibfnamefont {Y.}~\bibnamefont
  {Huo}}\ and\ \bibinfo {author} {\bibfnamefont {R.~N.}\ \bibnamefont
  {Bhatt}},\ }\bibfield  {title} {\bibinfo {title} {Current carrying states in
  the lowest landau level},\ }\href@noop {} {\bibfield  {journal} {\bibinfo
  {journal} {Physical Review Letters}\ }\textbf {\bibinfo {volume} {68}},\
  \bibinfo {pages} {1375} (\bibinfo {year} {1992})}\BibitemShut {NoStop}%
\bibitem [{\citenamefont {Lin}\ \emph {et~al.}(2001)\citenamefont {Lin},
  \citenamefont {Zong},\ and\ \citenamefont {Ceperley}}]{LinZongCeperley2001}%
  \BibitemOpen
  \bibfield  {author} {\bibinfo {author} {\bibfnamefont {C.}~\bibnamefont
  {Lin}}, \bibinfo {author} {\bibfnamefont {F.~H.}\ \bibnamefont {Zong}},\ and\
  \bibinfo {author} {\bibfnamefont {D.~M.}\ \bibnamefont {Ceperley}},\
  }\bibfield  {title} {\bibinfo {title} {Twist-averaged boundary conditions in
  continuum quantum monte carlo algorithms},\ }\href@noop {} {\bibfield
  {journal} {\bibinfo  {journal} {Physical Review E}\ }\textbf {\bibinfo
  {volume} {64}},\ \bibinfo {pages} {016702} (\bibinfo {year}
  {2001})}\BibitemShut {NoStop}%
\bibitem [{\citenamefont {Prodan}(2017)}]{Prodan2017book}%
  \BibitemOpen
  \bibfield  {author} {\bibinfo {author} {\bibfnamefont {E.}~\bibnamefont
  {Prodan}},\ }\href@noop {} {\emph {\bibinfo {title} {A Computational
  Non-Commutative Geometry Program for Disordered Topological Insulators}}},\
  \bibinfo {series} {SpringerBriefs in Mathematical Physics}, Vol.~\bibinfo
  {volume} {23}\ (\bibinfo  {publisher} {Springer},\ \bibinfo {year}
  {2017})\BibitemShut {NoStop}%
\bibitem [{\citenamefont {Beckus}\ \emph {et~al.}(2018)\citenamefont {Beckus},
  \citenamefont {Bellissard},\ and\ \citenamefont
  {De~Nittis}}]{BeckusBellissardDeNittis2018}%
  \BibitemOpen
  \bibfield  {author} {\bibinfo {author} {\bibfnamefont {S.}~\bibnamefont
  {Beckus}}, \bibinfo {author} {\bibfnamefont {J.}~\bibnamefont {Bellissard}},\
  and\ \bibinfo {author} {\bibfnamefont {G.}~\bibnamefont {De~Nittis}},\
  }\bibfield  {title} {\bibinfo {title} {Spectral continuity for aperiodic
  quantum systems i. general theory},\ }\href@noop {} {\bibfield  {journal}
  {\bibinfo  {journal} {Journal of Functional Analysis}\ }\textbf {\bibinfo
  {volume} {275}},\ \bibinfo {pages} {2917} (\bibinfo {year}
  {2018})}\BibitemShut {NoStop}%
\bibitem [{\citenamefont {Combes}\ and\ \citenamefont
  {Thomas}(1973)}]{CombesThomas}%
  \BibitemOpen
  \bibfield  {author} {\bibinfo {author} {\bibfnamefont {J.~M.}\ \bibnamefont
  {Combes}}\ and\ \bibinfo {author} {\bibfnamefont {L.}~\bibnamefont
  {Thomas}},\ }\bibfield  {title} {\bibinfo {title} {Asymptotic behaviour of
  eigenfunctions for multiparticle schrodinger operators},\ }\href@noop {}
  {\bibfield  {journal} {\bibinfo  {journal} {Communications in Mathematical
  Physics}\ }\textbf {\bibinfo {volume} {34}},\ \bibinfo {pages} {251}
  (\bibinfo {year} {1973})}\BibitemShut {NoStop}%
\bibitem [{\citenamefont {Loring}(2015)}]{Loring2015}%
  \BibitemOpen
  \bibfield  {author} {\bibinfo {author} {\bibfnamefont {T.~A.}\ \bibnamefont
  {Loring}},\ }\bibfield  {title} {\bibinfo {title} {K-theory and pseudospectra
  for topological insulators},\ }\href@noop {} {\bibfield  {journal} {\bibinfo
  {journal} {Annals of Physics}\ }\textbf {\bibinfo {volume} {356}},\ \bibinfo
  {pages} {383} (\bibinfo {year} {2015})}\BibitemShut {NoStop}%
\bibitem [{\citenamefont {Loring}\ and\ \citenamefont
  {Schulz-Baldes}(2020)}]{LoringSB2020}%
  \BibitemOpen
  \bibfield  {author} {\bibinfo {author} {\bibfnamefont {T.~A.}\ \bibnamefont
  {Loring}}\ and\ \bibinfo {author} {\bibfnamefont {H.}~\bibnamefont
  {Schulz-Baldes}},\ }\bibfield  {title} {\bibinfo {title} {The spectral
  localizer for even index pairings},\ }\href@noop {} {\bibfield  {journal}
  {\bibinfo  {journal} {Journal of Noncommutative Geometry}\ }\textbf {\bibinfo
  {volume} {14}},\ \bibinfo {pages} {1} (\bibinfo {year} {2020})}\BibitemShut
  {NoStop}%
\bibitem [{\citenamefont {Lux}\ \emph {et~al.}(2020)\citenamefont {Lux},
  \citenamefont {Freimuth}, \citenamefont {Bl\"ugel},\ and\ \citenamefont
  {Mokrousov}}]{mokrousov}%
  \BibitemOpen
  \bibfield  {author} {\bibinfo {author} {\bibfnamefont {F.~R.}\ \bibnamefont
  {Lux}}, \bibinfo {author} {\bibfnamefont {F.}~\bibnamefont {Freimuth}},
  \bibinfo {author} {\bibfnamefont {S.}~\bibnamefont {Bl\"ugel}},\ and\
  \bibinfo {author} {\bibfnamefont {Y.}~\bibnamefont {Mokrousov}},\ }\bibfield
  {title} {\bibinfo {title} {Chiral hall effect in noncollinear magnets from a
  cyclic cohomology approach},\ }\href
  {https://doi.org/10.1103/PhysRevLett.124.096602} {\bibfield  {journal}
  {\bibinfo  {journal} {Phys. Rev. Lett.}\ }\textbf {\bibinfo {volume} {124}},\
  \bibinfo {pages} {096602} (\bibinfo {year} {2020})}\BibitemShut {NoStop}%
\bibitem [{\citenamefont {Connes}\ and\ \citenamefont
  {Moscovici}(1998)}]{Connes-Moskovichi}%
  \BibitemOpen
  \bibfield  {author} {\bibinfo {author} {\bibfnamefont {A.}~\bibnamefont
  {Connes}}\ and\ \bibinfo {author} {\bibfnamefont {H.}~\bibnamefont
  {Moscovici}},\ }\bibfield  {title} {\bibinfo {title} {Hopf algebras, cyclic
  cohomology and the transverse index theorem},\ }\href
  {https://doi.org/10.1007/s002200050477} {\bibfield  {journal} {\bibinfo
  {journal} {Communications in Mathematical Physics}\ }\textbf {\bibinfo
  {volume} {198}},\ \bibinfo {pages} {199} (\bibinfo {year}
  {1998})}\BibitemShut {NoStop}%
\bibitem [{\citenamefont {Nest}\ and\ \citenamefont {Tsygan}(1995)}]{Tsygan}%
  \BibitemOpen
  \bibfield  {author} {\bibinfo {author} {\bibfnamefont {R.}~\bibnamefont
  {Nest}}\ and\ \bibinfo {author} {\bibfnamefont {B.}~\bibnamefont {Tsygan}},\
  }\bibfield  {title} {\bibinfo {title} {Algebraic index theorem},\ }\href
  {https://doi.org/10.1007/BF02099427} {\bibfield  {journal} {\bibinfo
  {journal} {Communications in Mathematical Physics}\ }\textbf {\bibinfo
  {volume} {172}},\ \bibinfo {pages} {223} (\bibinfo {year}
  {1995})}\BibitemShut {NoStop}%
\bibitem [{\citenamefont {Fedosov}(1995)}]{Fedosov}%
  \BibitemOpen
  \bibfield  {author} {\bibinfo {author} {\bibfnamefont {B.}~\bibnamefont
  {Fedosov}},\ }\href {https://books.google.cz/books?id=uF2FQgAACAAJ} {\emph
  {\bibinfo {title} {Deformation Quantization and Index Theory}}}\ (\bibinfo
  {publisher} {Wiley},\ \bibinfo {year} {1995})\BibitemShut {NoStop}%
\bibitem [{\citenamefont {Zuevsky}(2023{\natexlab{a}})}]{Zu}%
  \BibitemOpen
  \bibfield  {author} {\bibinfo {author} {\bibfnamefont {A.}~\bibnamefont
  {Zuevsky}},\ }\bibfield  {title} {\bibinfo {title} {Product-type classes for
  vertex algebra cohomology of foliations on complex curves},\ }\href
  {https://doi.org/10.1007/s00220-023-04751-4} {\bibfield  {journal} {\bibinfo
  {journal} {Communications in Mathematical Physics}\ }\textbf {\bibinfo
  {volume} {402}},\ \bibinfo {pages} {1453} (\bibinfo {year}
  {2023}{\natexlab{a}})}\BibitemShut {NoStop}%
\bibitem [{\citenamefont {Zuevsky}(2023{\natexlab{b}})}]{Zu1}%
  \BibitemOpen
  \bibfield  {author} {\bibinfo {author} {\bibfnamefont {A.}~\bibnamefont
  {Zuevsky}},\ }\bibfield  {title} {\bibinfo {title} {Reduction cohomology of
  riemann surfaces},\ }\href {https://doi.org/10.1142/S0129055X23300054}
  {\bibfield  {journal} {\bibinfo  {journal} {Reviews in Mathematical Physics}\
  }\textbf {\bibinfo {volume} {35}},\ \bibinfo {pages} {2330005} (\bibinfo
  {year} {2023}{\natexlab{b}})}\BibitemShut {NoStop}%
\bibitem [{\citenamefont {Zuevsky}(2024)}]{Zu3}%
  \BibitemOpen
  \bibfield  {author} {\bibinfo {author} {\bibfnamefont {A.}~\bibnamefont
  {Zuevsky}},\ }\bibfield  {title} {\bibinfo {title} {On a category of
  $v$-structures for foliations},\ }\href
  {https://doi.org/10.1142/S0129055X24300048} {\bibfield  {journal} {\bibinfo
  {journal} {Reviews in Mathematical Physics}\ }\textbf {\bibinfo {volume}
  {36}},\ \bibinfo {pages} {2430004} (\bibinfo {year} {2024})}\BibitemShut
  {NoStop}%
\bibitem [{\citenamefont {Zuevsky}(2022)}]{zuevsky2022characterization}%
  \BibitemOpen
  \bibfield  {author} {\bibinfo {author} {\bibfnamefont {A.}~\bibnamefont
  {Zuevsky}},\ }\bibfield  {title} {\bibinfo {title} {Characterization of
  codimension one foliations on complex curves by connections},\ }\href@noop {}
  {\bibfield  {journal} {\bibinfo  {journal} {Reviews in Mathematical Physics}\
  }\textbf {\bibinfo {volume} {34}},\ \bibinfo {pages} {2230002} (\bibinfo
  {year} {2022})}\BibitemShut {NoStop}%
\bibitem [{\citenamefont {Razumov}\ \emph {et~al.}(1999)\citenamefont
  {Razumov}, \citenamefont {Saveliev},\ and\ \citenamefont
  {Zuevsky}}]{razumov1999nonabelian}%
  \BibitemOpen
  \bibfield  {author} {\bibinfo {author} {\bibfnamefont {A.}~\bibnamefont
  {Razumov}}, \bibinfo {author} {\bibfnamefont {M.}~\bibnamefont {Saveliev}},\
  and\ \bibinfo {author} {\bibfnamefont {A.}~\bibnamefont {Zuevsky}},\
  }\bibfield  {title} {\bibinfo {title} {Nonabelian toda equations associated
  with classical lie groups},\ }\href@noop {} {\bibfield  {journal} {\bibinfo
  {journal} {arXiv preprint math-ph/9909008}\ } (\bibinfo {year}
  {1999})}\BibitemShut {NoStop}%
\bibitem [{\citenamefont
  {Zuevsky}(2023{\natexlab{c}})}]{ZuevskyCosimplicial2023}%
  \BibitemOpen
  \bibfield  {author} {\bibinfo {author} {\bibfnamefont {A.}~\bibnamefont
  {Zuevsky}},\ }\bibfield  {title} {\bibinfo {title} {Cosimplicial meromorphic
  functions cohomology on complex manifolds},\ }\href@noop {} {\bibfield
  {journal} {\bibinfo  {journal} {Reviews in Mathematical Physics}\ }\textbf
  {\bibinfo {volume} {35}},\ \bibinfo {pages} {2330002} (\bibinfo {year}
  {2023}{\natexlab{c}})}\BibitemShut {NoStop}%
\bibitem [{\citenamefont {Zuevsky}(2021)}]{ZuevskyFoliation2021}%
  \BibitemOpen
  \bibfield  {author} {\bibinfo {author} {\bibfnamefont {A.}~\bibnamefont
  {Zuevsky}},\ }\bibfield  {title} {\bibinfo {title} {Foliation of a space
  associated to vertex operator algebra},\ }\href@noop {} {\bibfield  {journal}
  {\bibinfo  {journal} {International Journal of Modern Physics A}\ }\textbf
  {\bibinfo {volume} {36}},\ \bibinfo {pages} {2150211} (\bibinfo {year}
  {2021})}\BibitemShut {NoStop}%
\bibitem [{\citenamefont {Hatsugai}\ and\ \citenamefont
  {Kohmoto}(1992)}]{HatsugaiKohmoto1992}%
  \BibitemOpen
  \bibfield  {author} {\bibinfo {author} {\bibfnamefont {Y.}~\bibnamefont
  {Hatsugai}}\ and\ \bibinfo {author} {\bibfnamefont {M.}~\bibnamefont
  {Kohmoto}},\ }\bibfield  {title} {\bibinfo {title} {Exactly solvable model of
  correlated lattice electrons in any dimensions},\ }\href@noop {} {\bibfield
  {journal} {\bibinfo  {journal} {Journal of the Physical Society of Japan}\
  }\textbf {\bibinfo {volume} {61}},\ \bibinfo {pages} {2056} (\bibinfo {year}
  {1992})}\BibitemShut {NoStop}%
\bibitem [{\citenamefont {Phillips}\ \emph {et~al.}(2020)\citenamefont
  {Phillips}, \citenamefont {Yeo},\ and\ \citenamefont
  {Huang}}]{PhillipsHK2020}%
  \BibitemOpen
  \bibfield  {author} {\bibinfo {author} {\bibfnamefont {P.~W.}\ \bibnamefont
  {Phillips}}, \bibinfo {author} {\bibfnamefont {L.}~\bibnamefont {Yeo}},\ and\
  \bibinfo {author} {\bibfnamefont {E.~W.}\ \bibnamefont {Huang}},\ }\bibfield
  {title} {\bibinfo {title} {Exact theory for superconductivity in a doped mott
  insulator},\ }\href@noop {} {\bibfield  {journal} {\bibinfo  {journal}
  {Nature Physics}\ }\textbf {\bibinfo {volume} {16}},\ \bibinfo {pages} {1175}
  (\bibinfo {year} {2020})}\BibitemShut {NoStop}%
\end{thebibliography}%

%\printbibliography
\end{document}